# A DENOTATIONAL ENGINEERING OF PROGRAMMING LANGUAGES

to make software systems reliable
and user manuals clear, complete and unambiguous

*A book in statu nascendi*

## Andrzej Jacek Blikle
in cooperation with Piotr Chrząstowski-Wachtel

*It always seems impossible
until it's done.*
Nelson Mandela

Warsaw, March 22nd, 2019

# About the current versions of the book

Both versions — Polish and English — are *in statu nascendi* which means that they are both in the process of correction due to my readers' remarks. Since December 2018 both versions, and currently also two related papers, are available in PDF format and can be downloaded from my website:

http://www.moznainaczej.com.pl/what-has-been-done/the-book

as well as from my accounts on ResearchGate, academia.edu and arXiv.org

I very warmly invite all my readers to send their remarks and questions about all aspects of the book. I am certainly aware of the fact that my English requires a lot of improvements and therefore I shall very much appreciate all linguistic corrections and suggestions as well. You may write to me on andrzej.blikle@moznainaczej.com.pl.

All interested persons are also invited to join the project *Denotational Engineering*. For more details see:

http://www.moznainaczej.com.pl/an-invitation-to-the-project

# Acknowledgements to the Polish version

Since June 2018 a preliminary version of the Polish version has been made available to selected readers which resulted with a flow of remarks. Below is the list of readers whose observations contributed to the improvement of my book. The order is chronological according to the dates of receiving the remarks. I hope that this list will keep growing. To all the contributors I express my very sincere thanks.

Piotr Chrząstowski-Wachtel, Stanisław Budkowski, Antoni Mazurkiewicz, Marek Ryćko, Bogusław Jackowski, Ryszard Kubiak, Paweł Urzyczyn, Stefan Sokołowski, Marek Bednarczyk, Wiesław Pawłowski, Jan Madey, Krzysztof Apt, Andrzej Tarlecki, Jarosław Deminet.

# Acknowledgements to the English version

Vacant, so far…

# A technical remark to the reader of the "Word version" of the book

To protect the layout of formulas, set tabulators to 0,5 cm. Tabulator's setting is a local parameter of a document which you set in the section

Main tools / Paragraph

of the tools panel, where at the lower right corner of that panel you should click a small arrow.

Nelson Mandela's quotation on the front page has been taken from

https://www.brainyquote.com/authors/nelson_mandela

# Contents

# Foreword

When in 1990 I decided to run my family business[1] "for a while" — which took me two decades — I already had a plan for my book on denotational models of programming languages. It was the result of my research for nearly thirty years starting in 1962 after I graduated from The Department of Mathematics and Physics of Warsaw University. I started my work in a group of young researchers who planned to build mathematical tools for software engineering. At that time there were only a few such groups in Poland and maybe 20-30 in the World. Although our approaches were technically different from each other, we were sharing essentially the same opinion about state of the art in software engineering. Let me try to sum it up now in a few lines.

In each engineering — except software engineering — the designing process of a new product starts with a blueprint supported by mathematical calculations. Both provide a mathematical warranty that the future functionality of the product will satisfy the expectations of the designer and the future user.

In the IT industry, the situation was different. In the place of a blueprint and calculations, programmers (i.e. producers) were given an informal description of the future product in a natural language, like English or Polish. As a consequence, a bulk of the budget for product-development was spent on testing, i.e., removing errors introduced at the stage of coding. Since testing may only discover errors but never gives a guarantee of their absence, the remaining bugs were passed on to the user to be removed later under the name of "maintenance". In some cases, these situations were leading to spectacular catastrophes. Here are a few examples:

- the death of six patients in  US hospitals as a result of a wrong computer-computations of radiation dosage (1985),

- the catastrophe of an American lander of the Venus planet (the 1980-ties),

- the catastrophe of an oil platform in a Norwegian fiord (1991),

- Airbus crash in Warsaw (1993)[2],

- an overlooking of Lothar hurricane by German meteorological services (1999),

---

[1] I was borne in a family of Warsaw's confectioners who's firm was established in 1869. The business survived two world wars and 45 years of communist time, hence when our country became independent again in 1989, I decides to develop our family business according the European standards. My father passed away many years ago and by son was too young to take the business over. My preliminary plan was to stay in the business for a few years only and then to come back to my beloved research. The life turn out, however, more difficult than I expected.

[2] In this case, although the cause of the accident had its origin in the software, this error was not due to programmers, but to the aircraft engineers, who did not anticipate certain specific aerodynamic conditions that may occur during the landing of the aircraft. In effect, they passed a wrong specification to programmers. For this information, I am thankful to Jarosław Deminet.

- a rounding error in Intel's microprocessor (1995).

That was the situation in the past. And how is it today? Today software products are a few orders of magnitude larger, and the number of their users grows exponentially. However, the problems mentioned above have not disappeared. The following statistics concerns software products of a total value of 250 billion USD (see [1]):

- 88% of projects exceeded the planned realisation time and/or budget,

- the average overrun of the assumed budget was 189%,

- the average overrun of assumed realisation-time was 222%.

It is also a well-known fact that every user of a software application has to accept a disclaimer. Here is a typical example dating from 2018:

*There is no warranty for the program, to the extent permitted by applicable law. Except when otherwise stated in writing the copyright holders and/or other parties provide the program "as is" without warranty of any kind, either expressed or implied, including, but not limited to, the implied warranties of merchantability and fitness for a particular purpose. The entire risk as to the quality and performance of the program is with you. Should the program prove defective, you assume the cost of all necessary servicing, repair or correction.*

Is it thinkable that a producer of a car, a dishwasher or a building could request such a disclaimer from his client? Why then is the software industry an exception?

In my opinion, the cause of this situation is a lack of such mathematical models and tools for software engineers that would guarantee the functional reliability of products based on the way they have been designed and manufactured. The lack of mathematical models for programming languages also affects user-manuals of these languages which again contributes to a low quality of programs.

In the field of user manuals, I do not see progress either. A published in 1960 report on Algol 60 (see [5]) — a language, which largely influenced the development of several generations of programming languages — far surpassed today's manuals regarding not only the precision and completeness of languages' descriptions but also their compactness[3].

First, their syntax was described by generative Chomsky's grammars rather than — as today — by (usually unclear) examples.

Second, their semantics although defined without any mathematical means (they were not known at that time) was described with the use of well-defined technical concepts such as *variable, block, variable-visibility, procedure, procedure-parameter, recursion*. Ten years later the manual of Pascal [47] was written in a similar style[4].

Unfortunately, one cannot say the same about today's manuals where the authors do not distinguish expressions from instructions, and instructions from declarations.

The described situation is common not only for programming languages but also for Content Management System such as e.g., Joomla! or Drupal, prove the growing popularity of support forums, where desperate users exchange their own experiences. Manuals are rarely used because they are not only imprecise and incomplete but highly unreadable which is due to both the language lacking conceptual apparatus, as well as to their length. For instance, Algol 60 manual contained 237 pages and Pascal manual — 166 pages, whereas in the case of Phyton

---

[3] Similar remarks can be made about a Polish manual [61] of Algol 60.
[4] Similar remarks are true for a Polish manual [53] of Pascal.

[59] we have 696 pages, for Access [68] — 952 pages and the manual of Delphi that was supposed to become the universal language of programming of all time exceeds 2000 pages.

The users' forums are therefore filled up with questions like "Hey, does anyone know how to ...?", to which most frequently nobody answers. From my practice, for three questions asked by me, two remain unanswered. I only find related questions asked by others, which convince me that I am not alone with my problem.

# 1  Introduction

## 1.1    Reverse the traditional order of things

The problem of mathematically-provable program-correctness appeared for the first time in a work of Alan Turing [66] published in conference-proceedings *On High-Speed Calculating Machines*, which took place at Cambridge University in 1949. Later for several decades, that subject was investigated usually as *proving program correctness,* but the developed methods never became everyday tools of software engineers. Finally, all these efforts were abandoned what has been commented in 2016 by the authors of a monography *Deductive Software Verification* [2]:

> *For a long time, the term formal verification was almost synonymous with functional verification. In the last years, it became more and more clear that full functional verification is an elusive goal for almost all application scenarios. Ironically, this happened because of advances in verification technology: with the advent of verifiers, such as KeY, that mostly cover and precisely model industrial languages and that can handle realistic systems, it finally became obvious just how difficult and time-consuming the specification of the functionality of real systems is. Not verification but specification is the real bottleneck in functional verification.*

In my opinion, the failure of constructing a practical system for proving programs correct has two sources.

The first lies in the fact that in building a programming language we start from syntax and only later — if at all — define its semantics. The second source is somehow similar but concerns programs: we first write a program and only then try to prove it correct.

To build a logic of programs for a programming language, one must first define its semantics on a mathematical ground. Since 1970-ties it was rather clear for mathematicians that such semantics to be "practical" must be compositional, i.e., the meaning of a whole must be a composition of the meanings of its parts. Later such semantics were called *denotational* — the meaning of a program is its *denotation* — and for about two decades researchers investigated the possibilities of defining denotational semantics for existing programming languages. Two most complete such semantics were written in 1980 for Ada [12] and for CHILL [30] in using a metalanguage VDM [10]. A little later, but in the same decade, a minor exercise in this field was semantics of a subset of Pascal written in MetaSoft [21], the latter based on VDM.

Unfortunately, none of these attempts resulted in the creation of software-engineering tools that would be widely accepted by the IT industry. In my opinion that was unavoidable since for the existing programming languages a full denotational semantics simply cannot be defined (see Sec. 4). That was, in turn, the consequence of the fact that historically syntaxes were coming first and only later researchers were trying to give them a mathematical meaning. In other words — the decision of how to describe things was before what to describe.

In addition to that, two more issues were complicating denotational models of programming languages. They were related to two mechanisms considered important in 1960-ties but ten years later abandoned and forgotten. One was a common *jump instruction* **goto**, the other — specific procedures that may take themselves as parameters (Algol 60, see [61]). The former had led to the *continuations* (see [44]), the latter to *reflexive domains* (see [63]). Both contributed to the technical complexity of denotational models which was discouraging not only for practitioners but also for mathematicians.

The second group of problems followed from a tacit assumption that in the development of mathematically correct programs the development of programs should precede the proofs of their correctness. Although this order is quite obvious in mathematics — first theorem and then its proof — it is rather unusual for an engineer who first performs all necessary calculations (the proof) and only then builds his bridge or aeroplane.

The idea "first a program and correctness-proof later" seems not only irrational but also practically rather unfeasible for two reasons.

First reason follows from the fact that a proof of a theorem is usually longer than the theorem itself. Consequently, proofs of program correctness should contain thousands if not millions of lines. It makes "hand-made proofs" rather unrealistic. On the other hand, automated proofs were not available by the lack of formal semantics for existing programming languages.

Even more important seem, however, the fact that programs that are supposed to be proved correct are usually incorrect! Consequently, correctness proofs are regarded as a method of detecting errors in programs. It means that we are first doing things wrong to correct them later. Such an approach does not seem very rational either.

As an attempt to cope with the mentioned problems I show in the book some mathematical methods that may be suitable for designing programming languages with denotational semantics. To illustrate the method an exemplary programming language, **Lingua** is developed from denotations to syntax (first publication of that method in [22]). In this way, the decision of what to do (denotations) precedes the decision of how to express that (syntax).

Mathematically both the denotations and the syntaxes constitute many-sorted algebras (Sec. 2.11), and the associated semantics is the homomorphism from syntax to denotations. As turns out there is a simple method — to a large extend algorithmizable — of deriving syntax from (the description of) denotations and the semantics from both of them.

At the level of data structures, **Lingua** covers Booleans, numbers, texts, records, arrays and their arbitrary combinations plus SQL databases. It is also equipped with a relatively rich mechanism of types, e.g. covering SQL-like integrity constraints, and with tools allowing the user to define his/her own types in a structural way. At the imperative level, this language contains structured instructions, type definitions, procedures with recursion and multi-recursion and some preliminaries of object programming.

The issue of concurrency is not tackled in the book since the development of a "fully" denotational semantics for concurrent programs (if at all possible) would require separate research[5].

Of course, **Lingua** is not a real language since otherwise, the book would become unreadable. It is only supposed to illustrate the method which (hopefully) may be used in the future to design and implement a real language of sequential programming.

Ones we have a language with denotational semantics, we can define program-construction rules that guarantee the correctness of programs developed in using these rules. This method was for the first time sketched in my paper [18] and in this book is described in Sec. 8. It consists in developing so-called *metaprograms* which are programs that syntactically include their specifications. The method guarantees that if we compose two or more correct programs into a new program or if we transform a correct program, we get a correct program again. The correctness proof of a program is hence implicit in the way the program is developed.

---

[5] There exist mathematical semantics of concurrency which can be said to be only "partially denotational". An example of such a solution is a "component-based semantics" (cf. [10]), where the denotations of programs' components are assigned to programs in a compositional way (i.e. the denotation of a whole is a composition of the denotations of its parts), but the denotations themselves are so called fucons whose semantics is defined operationally.

Basic mathematical tools used in this book are the following:

1. fixed-point theory in partially ordered sets,

2. the calculus of binary relations,

3. formal-language theory and equational grammars,

4. fixed-point domain-equations based on so-called *naive denotational semantics* (cf. [28]),

5. many-sorted algebras,

6. abstract errors as a tool for the description of error-handling mechanisms,

7. three-valued predicate calculi of McCarthy and Kleene,

8. the theory of total correctness of programs with clean termination.

All these tools are described in Sec. 2 and Sec. 3. Hence the reader does not need to be acquainted with them. The reader is only expected to be familiar with the preliminaries of set theory and mathematical logic and to have a basic experience in programming.

In constructing **Lingua,** I assumed three priorities regarding the choice of programming mechanisms:

- the priority of the simplicity of the model, i.e., the simplicity of denotations, syntax, and semantics; e.g., the resignation from `goto` instruction and self-applicative procedures,

- the priority of the simplicity of metaprogram construction rules; e.g., the assumption that the declarations of variables and procedures, as well as the definitions of types, should always be located at the beginning of a program,

- the priority of protection against "oversight errors" of a programmer; e.g., the resignation of global variables in all types of procedures and of side-effects in functional procedures.

All these commitments forced me to give up some programming constructions which — although denotationally definable — would lead to complicated descriptions and even more complicated program-construction rules. It is worth mentioning in this place that the priority of simplicity is not new in the history of programming languages. For that very reason, programming-language designers abandoned `goto`-s as well as self-applicative procedures.,

## 1.2    What is in the book

I am deeply convinced that one can talk about programming in a precise and clear way. I also believe that taking responsibility for their products by software engineers should be possible in the same way as it is in the case of the engineers of cars, bridges or aeroplanes. On the other hand, I am aware of the fact that the existing tools for software engineers do not allow for the realisation of any of these goals.

As I mentioned already in the Foreword, the book contains many thoughts developed in the years 1960-1990 that later have been abandoned. One of the few teams developing these ideas was working in the Institute of Computer Science of the Polish Academy of Sciences, and I had the pleasure to chair it. At that time we were developing a semi-formal metalanguage called **MetaSoft** dedicated to formal definitions of programming languages (cf. [21]). This language is used in the book as a definitional vehicle for denotational models.

The book starts (Sec. 2) with the introduction of all mathematical tools that are listed in Sec. 1.1 except program-correctness issue.

Sec. 3 is devoted to the general theory of partial and total correctness of programs. These concepts are formulated in the language of binary relations which allows concentrating on the main subject without technical details of a programming language.

The remaining part of the book is devoted to the construction of denotational models for successive programming mechanisms.

Sec. 4 contains a general discussion of algebraic and denotational models of programming languages that are later exploited in the subsequent sections of the book.

Sec. 5 is devoted to the development of a general model of data structures and types which can be used to describe data- and type-mechanisms of a sufficiently large class of algorithmic programming languages. In this model, a type is a pair that consists of a *body* that describes the structure of a data, e.g., a list of records, and a *yoke* that describes other properties, e.g., that in each of these records the sum of numbers assigned to attributes salary and commission should be less than 10.000. Such yokes are typical in SQL though are not named in this way. A language covering these mechanisms is called **Lingua-A** (A stands for "applicative"). It consists of expressions only, i.e., contains neither declarations nor instructions. It is not a prototype of an applicative programming language, but only an applicative fundament of a general-purpose programming language.

Sec. 6 contains a model of **Lingua-1** that covers the whole **Lingua-A** plus structured instructions, variable declarations and some mechanism allowing programmers to build types in a bottom-up way. Types may be given names to store them in the memory.

In Sec. 7 **Lingua-1** is enriched to **Lingua-2** by introducing procedures both imperative and applicative. Recursion and multi-recursion are covered as well.

Sec. 8 is devoted to the idea and techniques of *validating programming* which was my main scientific research area in the years 1970/80. As was already explained (see Sec.1.1) it consists in building metaprograms by using constructors that guarantee metaprogram's correctness. The language for validating programming in **Lingua-2** is called **Lingua-V2** (V for "validating").

Sec. 9 and 10 contain a sketch of an expansion of **Lingua-2** to **Lingua-3** that offers tools for object programming.

Sec. 11 and 12 are devoted to the extension of **Lingua-3** by mechanisms including relational databases such as in SQL. That version of **Lingua** is called **Lingua-SQL**.

I am aware of the fact that the content of the book represents a very restricted part of the world of today's programming languages. Something had to be chosen, however, to begin. **Lingua** contains, therefore, a selection of programming tools that have been known for many years and that are still in use. In the future, I shall try to complete my models with those vehicles that my readers will consider important. I also hope that maybe some of you will undertake this challenge. Please feel invited to cooperate.

## 1.3     What this book is not offering

As I explained in the Foreword and in Sec. 1.1, the reason why I have written this book is the lack of mathematical tools that would allow software producers to take such responsibility for their products as is usual in many other industries such as, e.g. automotive or aircraft industry or in the industry of civil engineering. It does not mean, however, that the book offers a tool ready to be used in the software industry. What I am trying to offer is only a suggestion of where to research for such tools and an associated mathematical framework.

To better explain what I mean let me refer to the concept of *product quality* as understood in the field of *Total Quality Management*. By the quality of a product, we understand the degree of the satisfaction of the product's user. Product quality is usually measured by the number of faults in the product — the fewer faults, the higher the quality — where a fault is any such product property that the user "has the right not to expect". E.g. if we are ordering a beer, we have the right not to expect it to be warm, unless we are ordering a mulled beer.

The quality of a product is therefore not an immanent property of a product, but rather a relation between the product and the expectations of its user. Paradoxically we can increase the quality of a product without changing the product itself but in honestly describing all its faults. This is not a usual practice, however, since such an approach would decrease the changes to sell the product.

In the case of software, user expectations are described by a specification that a program should fulfil. The quality of a program consists therefore in:

1. the compatibility of program specification with the expectations of its user,

2. the compatibility of the program itself with its specification.

In my book, I am tackling only the second aspect. My choice is not caused by the fact that the first problem is less important, or that it has been already solved, but only because the second problem was the main subject on my research for two decades and therefore I dare to talk about it now[6].

In the end, I have to very strongly emphasise that my virtual language **Lingua** is not regarded neither as a practical programming language nor even as a standard of such a language although maybe such a language will grow from **Lingua** in the future. At present, it only offers a platform where to explain the constructions and the models discussed in the book. I have tried to cover in it the most common tools that are present in languages which are known to me.

## 1.4     What is new in my approach

By "my approach" I understand the ideas and techniques described in my early papers from [15] to [25], which have been summarised and extended in this book. All these ideas base on concepts well-known for years:

- denotational semantics of D. Scott's and Ch. Strachey's (cf. [63], [64]),

- generative grammars of N. Chomsky's (cf. [34], [41]),

- Hoare's logic of programs (cf. [46]),

- on many-sorted algebras introduced to the mathematical foundations of computer science by J. A Goguen, J.W, Thatcher, E.G Wagner and J.B Wright (cf. [43]),

- three-valued propositional calculus of J. McCarthy (cf. [55]).

What — I believe is new in my approach — is the following:

1. **Programming language design and development**:

---

[6] I am convinced that the first problem is equally fascinating as the second. I would very much welcomed any initiative of a cooperation in this field.

1.1. A formal, and to a large extend an algorithmic method of a systematic development of syntax from denotations and of a denotational semantics from both of them.

1.2. The idea of a colloquial syntax which allows making syntax user-friendly without damaging a denotational model.

1.3. Systematic use of error-elaboration in programs supported by a three-valued predicate calculus.

1.4. Denotational model based on set-theory rather than on D. Scott's reflexive domains which makes the model much simpler and easy to be formalized.

1.5. A model of data-types that covers not only structured and user-defined types but also SQL integrity constraints.

**2. The development of correct programs**

2.1. A method of systematic development of correct programs with their specifications, rather than an independent development of programs and specifications followed by program-correctness proof.

2.2. The use of three-valued predicates to extend Hoare's logic by a clean termination property.

**3. General mathematical tools**

3.1. Equational grammars applied in defining the syntax of programming languages.

3.2. A three-valued calculus of predicates applied in designing programming languages and in defining sound program constructors for such languages.

## 1.5    Lingua from bird's-eye view

To structure my presentation, the final version of **Lingua** is built layer-by-layer as explained in Sec.1.2. Below I present a condensed and only half-formalised description of the language without entering into technical details which may be found in the main part of the book. I also refrain from describing the process of language development and concentrate on its target version. I address this section to the readers who wish to grasp the idea of **Lingua** without reading the whole book.

### 1.5.1   Notational conventions

Below I shall use the following notation (full description and justification in Sec. 2.1):

- $a : A$ means that $a$ is an element of the set $A$; ; according to the denotational dialect "sets" are most frequently called "domains",

- $f.a$ denotes $f(a)$, and $f.a.b.c$ denotes $((f(a))(b))(c)$; intuitively $f$ takes $a$ as an argument and returns the value $f(a)$ which is a function which takes $b$ as an argument and returns the value $(f(a))(b)$, which is again a function…

- $A \rightarrow B$ denotes the set of all *partial functions* from $A$ to $B$, i.e., functions possibly undefined for some elements of $A$,

- $A \mapsto B$ denotes the set of all *total functions* from $A$ to $B$, i.e., functions undefined for all elements of $A$; of course, each total function is a particular case of a partial function,

- A $\Rightarrow$ B denotes the set of all function from A to B defined for only finite subsets of A; such functions are called *mappings,* and of course, each mapping also is a particular case of a partial function,

- A|B denotes the set-theoretic union of A and B,

- A x B denotes the Cartesian product of A and B,

- tt and ff denote logical values „true" and „false" respectively,

- many-character symbols like dom, bod, com denote metavariables running over domains and if written with parentheses as 'abdsr' denote themselves, i.e., metaconstants.

In order to distinguish between meta level of phrases written in **MetaSoft** and the level of phrases written in **Lingua**, the former level will be written with Arial and the latter with Courier New.

## 1.5.2     Data and (their) types

Data in **Lingua** may be split into three groups:

- *simple data* including Booleans, numbers, and words (finite strings of characters),

- *structural data* including list, many-dimensional arrays, records, trees, and their arbitrary combinations,

- *SQL-data* including rows and tables that carry simple data and databases that carry tables.

Structural data may „carry" simple data as well as other structural data. That means that we may build "deep" data structures, e.g., records that carry lists of trees with arrays in their nodes. Lists and tables carry elements of the same type whereas records and trees are not restricted in this way.

Lists and records are defined in a rather traditional way, although list elements and data assigned to the attributes of records may be arbitrary simple or structural data but not SQL data.

Arrays are formally one-dimensional, but since their elements may be other arrays, we may construct arrays of arbitrary dimensions.

Trees are defined as pairs consisting of a parent and a tuple of children, hence are of the form (parent, (child$_1$,…,child$_n$)). Both a parent and a child may be an arbitrary simple or structural data and even a tree.

Databases are — simplifying a little — records of tables, i.e., finite functions from identifiers into tables, tables are — simplifying again — lists of rows and rows are records that carry simple data.

All these data with the appropriate constructors constitute a many-sorted algebra of data. Many-sorted algebras of data, types, denotations, and syntax make the fundaments of our denotational model. Sections from 2.10 to 2.14 are devoted to a short introduction into the theory of many-sorted algebras.

**Lingua** has been equipped with a mechanism of types that covers typical mechanism of programming languages. By a "mechanism of types" I understand programming tools that allow a programmer to define his/her types for future use either in defining new types or in declaring variables. This mechanism is described in Sec. 5.2.

As we are going to see, types are pairs consisting of a *body* and a *yoke*. With every type, there is associated a set of data called the *clan* of this type.

Intuitively a body describes the "internal structure of a data" — e.g., that a data is a number, a list or a record — and formally is a combination of tuples and mappings. The bodies of simple data are one-element tuples of words: ('Boolean'), ('number') or ('word'). The bodies of lists and arrays are respectively of the form ('L', body) or ('A', body) where body is shared by all the elements of a list/array and where the *initials* 'L' and 'A' indicate that we are dealing with a list or with an array respectively. A record body is of the form ('R', body-record) where body-record is a record of bodies such as, e.g.:

| | | |
|---|---|---|
| Ch-name ; | ('word'), | |
| fa-name ; | ('word'), | |
| birth-year ; | ('number'), | |
| award-years ; | ('A', ('number')), | (*) |
| salary ; | ('number'), | |
| bonus ; | ('number') | |

The words on the left-hand-side of semicolons are identifiers called *attributes*. The first three attributes and the last two have simple bodies, whereas the fourth one — an array body. For the sake of further discussions, this record-body will be referred to as employee.

With every body bod, we associate the set of data denoted by CLAN-Bo.bod. The function CLAN-Bo is defined inductively relative to the structure of bodies. E.g., the set CLAN-Bo.employee contains records with numbers, words, and one-dimensional number arrays assigned to the attributes.

Next important concept from the "world" of data and types is a *composite* that is a pair (dat, bod) consisting of a data and its body such that:

dat : CLAN-Bo.bod

Composites are created in the course of the data-expressions evaluation (see a little later). All data operations in **Lingua** are defined as operations on composites which permits to describe the mechanism of checking if the arguments "delivered" to an operation are of an appropriate type. E.g., if we try to put a word on a list of numbers, the corresponding operation will generate an error message.

Having defined composites, we can define *transfers* and *yokes*. Transfers are one-argument functions that transform composites into composites and *yokes* are transfers with Boolean composites as values. By a *Boolean composite* I mean (tt, ('Boolean')) or (ff, ('Boolean')). Transfers may also assume abstract errors as values (see later).

Mathematically yoks are close to one-argument predicates on composites[7]. An example of a yoke that describes a property of composites whose body is employee may be the following inequality:

salary + bonus < 10000,

---

[7] They "are closed to predicates" rather than simply "are predicates" since they assume as values composites and abstract errors rather than just Boolean values tt and ff. Consequently their logical constructors **and**, **or** and **not** are not the classical constructors but three-valued constructors of a calculus defined by John McCarthy (Sec. 2.9).

This yoke is satisfied whenever its argument is a record composite with (at least) the attributes salary and bonus and the data corresponding to these attributes satisfy the corresponding inequality. In this example

salary + bonus

is a transfer which is not a yoke. It transforms record composites into number composites.

Yokes understood in our way appear in SQL and are called *integrity constraints*. As a matter of facts they have been introduced into our model in order to cope with SQL data.

Transfers have merely a technical role. We need them only to define an algebra where yokes may be created. With every transfer we associate its clan:

CLAN-Tr.tra = (com | tra.com = (tt, ('Boolean'))}

which consists of composites that satisfy that transfer. Of course, clans of transfers that are not yokes, are empty.

A pair that consists of a body and a yoke is called a *type*. For technical reasons, however, types are defined as pairs consisting of a body and an arbitrary transfer. With every type typ = (bod, tra) we associate its *clan* which is the set of such composites whose data belong to the body-clan CLAN-Bo.bod and which satisfy the transfer. Formally:

CLAN-Ty.(bod, tra) = {(dat, bod) | dat : CLAN-Bo.bod **and** (dat, bod) : CLAN-Tr.tra}

The last concept associated with data and types is a *value*, also called *typed data*. A value is a pair (dat, typ), i.e. (dat, (bod, tra)), which we can also write as ((dat, bod), tra). A value may be regarded, therefore either as a pair *data-type* or as a pair *composite-transfer*.

Values are assigned in memory states to the identifiers of variables. An assignment instruction — i.e., an instruction that assigns values to variables — may only change the data assigned to a variable, and in some special cases its body, but never its yoke. Yokes may be only changed by special yoke-oriented instruction.

Let us sum up the list of objects associated with the concepts of data and their types:

- *data* are basic objects processed by programs,

- *bodies* are objects that describe "internal structures" of data,

- *composites* are pairs (dat, bod), where dat : CLAN-Bo.bod; *data-expressions* evaluate to composites,

- *transfers* are one-argument functions on composites and *yokes* are transfers that return Boolean composites or abstract error as values,

- *types* are pairs that consist of a body and a transfer (in fact a yoke); as we are going to see later, *type expressions* evaluate to types, and in memory states they are assigned to *type constants*,

- *values* are pairs consisting of a data and (its) type; in states, data are assigned to *variable identifiers*.

Similarly, as in many programming languages (although not in all) types in **Lingua** have been introduces for four reasons:

1. to define a type of a variable when it is declared, and to assure that this type remains unchanged (with some exceptions)[8] during program executions,

2. to ensure that a data which is assigned to a variable by an assignment is of the type consistent with the type of that variable,

3. to ensure that a similar consistency takes place when sending actual parameters to a procedure or when returning reference parameters by a procedure,

4. to ensure that in evaluating an expression, an error message is generated whenever data "delivered" to that expression are of an inappropriate type, e.g., when we try to add a word to a number.

### 1.5.3     Abstract errors

An important feature of **Lingua** is the inclusion of error message in its model. For this purpose, the domains (sets) of bodies, composites, and types are "equipped" with the elements that are called *errors*. Mathematically errors may be anything, but in **Lingua** they are words, e.g.

'division-by-zero' or

'record-expected'

that intuitively describe the cause of an error. All operations on composites, bodies, and types are also defined on errors, and the majority of them are *error-transparent* which means that if an argument of an operation is an error, then the resulting value is the first error that appears in the course of a computation. Intuitively this corresponds to a situation where program execution aborts and an error message is displayed on a monitor. It may also happen, however, that the appearance of an error causes the execution of an *error handling procedure* (see Sec. 6.1.8 and Sec. 12.7.6.4).

A special case of error-handling operations are Boolean operations (Sec. 2.9) that handle errors along the lines of McCarthy's propositional calculus. For instance:

ff and ee = ff

ee and ff = ee

where ee represents an error or a non-terminating computation. The arguments of conjunction are evaluated from left to write and if the first argument evaluates to ff, then the evaluation of the second argument is skipped. In this way, we maybe avoid an error message or a non-terminating evaluation. E.g. the Boolean expression

```
x ≠ 0 and 1/x > 10
```

assumes the value ff for x=0 even though `1/x > 10` would generate an error or would loop indefinitely. In McCarthy's calculus whenever `x = 0,` then the evaluation of `1/x > 10` is postponed.

A special cases where errors are signalized are overflows. Formally for every domain of data, a predicate is defined that "reacts" to overflows.

---

[8] These exceptions take place e.g. when we add a new attribute to a record or to a database table or if we remove such attribute.

## 1.5.4     Expressions

Expressions are syntactic object and their *denotations*, i.e. their semantic meanings, are functions from states to composites (*data expressions*) or from states to types (*type expressions*). In order to define these concepts we have to start with the definition of a domain of *states*. Here so called *domain equations* come into play:

| | | |
|---|---|---|
| State | = Env x Store | (state) |
| Env | = TypEnv x ProEnv | (environment) |
| Store | = Valuation x (Error \| {'OK'}) | (store) |
| Valuation | = Identifier $\Rightarrow$ Value | (valuation) |
| TypEnv | = Identifier $\Rightarrow$ Type | (type environment) |
| ProEnv | = Identifier $\Rightarrow$ Procedure \| Function | (procedure environment) |

where Error is some fixed set of errors. As we see, states are storing data, types, procedures, and functions (functional procedures) assigned to identifiers as well as errors stored in a "dedicated register". If a state does not carry an error then this register stores the word 'OK'.

The *denotations of data expressions* and the *denotations of type expressions* are the elements of an *algebra of expression denotations* from which a syntactic *algebra of expressions* has been derived. A function from expressions into their denotation is called the *semantics of expressions*[9].

Data-expression denotations are partial functions from states into composites or error messages:

DatExpDen = State $\rightarrow$ Composite \| Error

For every operation on data there is an operation on composites, and for every operation on composites, there is a constructor of data-expression denotations. E.g., the denotation of the expression

```
x + y
```

is a function that for a given state sta first successively checks the following conditions:

- If sta carries an error?

- If there are values assigned to the identifiers x and y in sta?

- If these values are numbers?

- If their sum does not go beyond the set of numbers representable in the current implementation?

If all these checks terminate positively, then the function creates a composite (dat, ('number')), where dat is the sum of the numbers assigned to x and y. If some of these checks do not terminate successively then an appropriate error message is generated, e.g.,

'number-expected'

and the computation terminates. In particular, if the input state carries an error, then this error becomes the result of the computation.

---

[9] This "function" is in fact a homomorphism from the algebra of expressions into the algebra of expression denotations.

Notice that data expressions represent partial functions since they may call functional procedures whose executions may loop indefinitely.

Contrary to that type-expression denotations are total functions from states into types or error messages:

TypExpDen = State ⟼ Type | Error

The constructors of such denotations are defined similarly as for data expressions although now they base on type operations rather than on data operations. E.g., the denotation of the following type expression:

```
record-type

   Ch-name         as word,

   fa-name         as word,

   birth-year      as number,

   award-years     as number-array ee

   salary          as number

   bonus           as number

ee
```

is a function on states that creates a record type or generates an error. This expression refers to two built-in types `word` and `number` and one user-defined type `number-array` (arrays of numbers). A typical case of a type-expression evaluation generating an error may be an operation of the removal of an attribute of a record-type if this attribute does not appear in the record.

Data-expression denotations and type-expression denotation together with their constructors constitute an *expression-denotation algebra* (Sec. 5.3). From that algebra, we derive it syntactic counterpart — an *expression algebra* (Sec. 5.4).

## 1.5.5    Instructions

Expressions belong to the *applicative part* of our language. Their denotations take states as arguments but neither create them nor change. The latter tasks are performed by *instructions, variable declaration, procedure- and function declarations* and by *type definitions*. All of them belong to the *imperative layer of the language*.

*Instruction denotations* are partial functions from states to states:

InsDen = State → State

Contrary to expression denotations that may generate an error, instruction denotations write an error into the error register which is a component of a state. The denotations of the majority of instructions are *transparent* relative to error-carrying states, i.e., they do not change such states but only pass them to the subsequent part of the program. However, an error may also cause an error-handling action.

The basic instruction is, of course, the *assignment* of a value to a variable identifier. Syntactically assignment instructions are of the form:

```
identifier := data-expression
```

The denotation of an assignment changes an input state into an output state in the following steps:

1. checking if the input state does not carry an error, and if this is the case, then the input state becomes the output state, and the execution terminates; in the opposite case

2. checking if the identifier has been declared, i.e., if in the input state it is bound to a value or to a pseudo value (see later); if this is not the case then an error message 'identifier-undeclared' is loaded to the error register; in the opposite case

3. trying to evaluate the data expression; if this attempt generates an error then this error is loaded to the error register; in the opposite case

4. checking if the composite computed from the expression has a body conformant with the body of the identifier's type, and if that is not the case then an error message is loaded to the error register; in the opposite case

5. checking if the composite computed from the expression satisfies the yoke assigned to the identifier in its type, and if that is not the case then an error message is loaded to the error register; in the opposite case

6. the computed composite is assigned to the identifier with the yoke remaining unchanged.

The remaining instructions belong to one of the following six sorts where the first three are *atomic instructions,* and the remaining three are *structural instructions*, i.e., instructions composed of atomic instructions and expression:

1. the replacement of a yoke assigned to a variable by another one,

2. the activation of an error-handling procedure,

3. the call of an imperative procedure,

4. the sequence of instructions,

5. the conditional composition of instructions of the form `if-then-else-fi`,

6. the loop `while-do-od`.

Of all these instruction only procedure calls have to be explained.

When a procedure is called it gets two lists of *actual parameters*: *value parameters* and *referential parameters,* the values of which are assigned to the corresponding *formal parameters* also value- and referential. After the execution of a procedure body, the values of formal referential parameters are passed to the corresponding actual referential parameters.

The mechanism of parameter passing is the only communication channel between procedure body and its hosting program. Inside a procedure body, only local variables are available and these variables are not available outside the procedure. It is to be emphasised that this decision has not been "forced" by the fact that we are building a denotational model. It has been taken — like many others — for pragmatic reasons that I shall try to justify in the following parts of the book.

Contrary to variables, all types and procedures declared in the preamble of a program (see later) have a global character, i.e., they are visible inside all procedure bodies. In a procedure body, one may also declare local variables, procedures, and types that are not available outside procedure body. It is again a pragmatic (engineering) decision rather than a denotational necessity.

For imperative procedures, there is a mechanism of both a *direct recursion* (a procedure calls itself) and an *indirect recursion* (procedure A calls procedure B which calls procedure C which calls… procedure A).

The denotations of data- and type expressions, of instructions, of variable- and procedure declarations, of preambles and programs (see sections that follow) constitute a common for all of them many-sorted algebra of denotations which is described in Sec. 6 (without procedures) and in Sec. 7 (with procedures). In these sections also the corresponding syntactic algebras are described.

### 1.5.6    Variable declaration and type definitions

*Variable-declaration denotations* are total functions that map states into states:

VarDecDen = State ⟼ State

assigning types to identifiers and leaving their data undefined. More formally, they assign *pseudo-values* which are pairs of the form $(\Omega, \mathsf{typ})$, where $\Omega$ is an abstract element called a *pseudo-data*. Syntactically a single declaration is of the form:

**let** identifier **be** type-expression **tel**

Variable declarations are similar to assignments with the difference that in the former case an error is signalized whenever the identifier is bound in the input state to a pseudo-value, a value, a type, or a procedure. In each such case the error message 'identifier-not-free' is generated. It means that a variable may be declared in a program only once. Subsequently, its value may be changed only be changing its composite and possibly the yoke. Bodies may be changed only in the case or records and database tables and only if we add new attributes or if we remove existing attributes (an engineering decision).

The denotations of type definitions are similar to these of variable declarations with the difference they assign types rather than pseudo-values to identifiers.

TypDefDen = State ⟼ State

An identifier that is bound to a type in a state is called a *type constant*. Notice that "a constant" rather than "a variable" since the type assigned to it, cannot be changed in the future (an engineering decision).

Similarly as in the case of assignments, also type definitions, and variable declarations may be combined sequentially using semicolon.

### 1.5.7    Procedures' declarations

Procedures may be imperative or functional. The former are functions that receive two lists of actual parameters — value parameters and reference parameters — and return partial functions on stores[10]. Functional procedures take only value parameters and return partial functions on states:

ipr  : ImpPro = ActPar x ActPar ⟼ Store → Store

fpr  : FunPro = ActPar          ⟼ State → (Composite | Error)

In these equations, ActPar is a domain of *actual-parameter lists*. Notice that we do not talk here about procedure denotations but about procedures as such since they are "purely denotational" concepts, i.e. that they do not have syntactic counterparts. At the level of syntax, we have only *procedure declarations* and *procedure calls* which, of course, have their denotations. Syntactically an imperative-procedure declaration is of the form:

```
proc Identifier (val for-val-par ref for-ref-par)

   program

end proc
```

where `program` is a program (see later). Both parameter lists are lists of variable identifiers each followed by a type expression, e.g.

---

[10] The fact that procedures transform stores rather than states is a technical trick that allows to avoid a selfapplication of a function, i.e. a situations where a function takes itself as an argument. More about that problem in Sec. 4.1. Of course, procedure calls are instructions and therefore transform states into states.

```
(val age, weight as number, name as word ref patient as pa-
tient-record)
```

Expressions different from single-identifier expressions are not allowed as value parameters since this would complicate the model as well as program-construction rules (an engineering decision).

If we want to declare a group of mutually recursive procedures then we use a *multiprocedure declaration* of the form:

```
begin multiproc

   DekPro-1;

   DekPro-2;

   …

   DekPro-n

end multiproc
```

The denotations of functional procedures are partial functions that transform states into composites (as expressions). They also do not have syntactic counterparts, and their declarations are of the form:

```
fun identifier (for-parameters)

   program

return expression as type-expression
```

The call of such a procedure first executes the program and then evaluates the expression in the output state of the program. The composite generated by that expression becomes the result of the procedure call. Such a call has no *side-effects*, i.e., it never modifies a state (an engineering decision). In particular case, the program may be a trivial one that "does nothing", and the expression may be a single identifier.

Procedures discussed above accept as parameters only variable identifiers, i.e., identifiers that bind values. All types and procedures defined in the program are visible in procedure bodies as global entities, and therefore they do not need to be passed as parameters (an engineering decision).

In the version of **Lingua** described above procedures cannot take other procedures as parameters. It is shown in (Sec. 7.6) how to overcome this restriction by constructing a hierarchy of procedures that can take as parameters only procedures of a lower rank than themselves. This protects procedures from taking themselves as parameters since this leads to models which are not denotational in our sense[11].

## 1.5.8    Object programming

Object-programming tools are only sketched in this book by showing some general ways of building their denotational model. Very briefly speaking an *object* — which similarly to procedures has no representation in syntax — is a state-to-state transformation that "loads" some procedures, functions, and types into a state. Syntactically an *object definition* is of the form:

> **set-object** `identifier` **as** `object-expression` **tes-object**

where `object-expression` is a sequence of type definitions and procedure declarations both imperative and functional.

Objects are stored in a dedicated part of a memory which I call an *object library*. Objects are made accessible in programs by *object calls*. Object calls may appear not only in programs but also in the definitions of objects which leads to a heritage mechanism. More on objects in Sec.9.

## 1.5.9    SQL programming

In typical programming languages that give access to SQL tools — known in the literature as *Application Programming Interface* or *Call Level Interface* — the interpreter of the hosting language activates procedures of an existing SQL engine. In our case, however, such a philosophy would not be acceptable. If we intend to equip **Lingua** with the constructors of correct programs, we have to build our SQL engine based on a denotational model.

Of course, we have to make sure that our database constructors are close enough to SQL standard. We cannot think, however, about full compatibility since first there is no one SQL standard and second — none of the existing ones is defined in a sufficiently precise way. We have to make sure only that **Lingua-SQL** programs can process SQL databases created in other implementations.

Section 12 contains a denotational model of basic SQL constructors and in particular of queries, cursors, and views.

## 1.5.10    Programs

Programs in **Lingua** are composed of two consecutive parts:

1. a *preamble* that consists of an arbitrary number of sequentially composed variable, and procedure declarations, and type definitions,

2. an instruction which, of course, may be arbitrarily complex.

This structure — first all declarations and definitions, and only then all instructions — is not a "denotational necessity" but contributes to the simplicity of program-construction rules.

## 1.5.11    Validating programming

Very briefly, validating-programming consists of building *metaprograms* that are composed of two mutually nested layers:

---

[11] Formally speaking this decision is forced more by set-theoretical argument than by the denotationality of our model (see Sec. 4.1).

1. *a programming layer* that is a program in the usual sense,
2. *a descriptive layer* which consists of pre- and post-conditions plus assertions (conditions) that are "nested" in-between instructions.

A metaprogram is said to be *correct,* if for every initial state that satisfies the preconditions the following is true:

1. the program executes without looping or generating an error message,
2. all assertions are satisfied during this execution,
3. the terminal state of the program satisfies the postcondition.

Our correctness is the *total correctness* of C.A.R Hoare in (see [46]) strengthened by the assumption that the program does not hang-up with an error message.

Notice that in the classical theory of program correctness the correctness in always related to a context of a precondition and a postcondition, whereas now we talk about the correctness as such since the pre- and post-conditions are parts of the metaprogram. The inclusion of the descriptive layer allows for the construction of complex correct programs from simple ones.

Below we see a simple example of a metaprogram where `isr(n)` denotes the integer part of the square root of `n`:

```
Q4: let z, x be number
      pre true                                    (precondition)
        z := 1;
        x := 0;
        begin-asr z > 0 and x ≥ 0
          while z² ≤ n do z:=2*z od;
          x := 0;
          while z > 1
            do
              z := z/2;
              if (x+z)² < n then x:=x+z else x:=x fi
            od
        end-asr
      post x = isr(n) and z = 1                   (postcondition)
```

The part of the program between **begin-asr**  con and **end-asr** is called the *range of condition* `con`. If our metaprogram is correct, then the condition is satisfied by all intermediate states in its range.

Correct-program construction starts from simple programs whose correctness is proved in a traditional way. The subsequent programs are constructed from already existing programs in using construction rules that preserve the correctness. Here is an example of such a rule that is used to construct a program consisting of a conditional instruction:

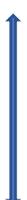

```
(1) prc ⇨ dae or(not dae)
(2) def pam pre (prc and dae) sin-1 post poc
(3) def pam pre (prc and not dae) sin-2 post poc
```

**def** pam **pre** prc **if** dae **then** sin-1 **else** sin-2 **fi post** poc

This rule we read as follows:

> ***If***
>
>> *(1) every state that satisfies the precondition* prc *satisfies either the data expression* dae *or its negation; this assumption means that if the precondition is satisfied, then the branching data-expression* dae *evaluates to a Boolean value, hence neither loops nor generates an error message,*
>>
>> *(2) this metaprogram is correct*
>>
>> *(3) this metaprogram is correct*
>
> ***then***
>
>> *the metaprogram below the line is correct.*

This rule allows for the construction of a correct metaprogram starting from two correct metaprograms. Observe that in the classical predicate calculus the metacondition (1) would be a tautology but in our case — due to the third logical value — it is not so.

# 2  MetaSoft and its mathematics

When in the years 1970 to 1990 I was lecturing mathematical foundations of computer science to practitioners I frequently heard an objection that there is definitely much too much of this mathematics that software engineers have to swallow. Bosses of IT departments expected that their teams could be "trained" in that new mathematics within one weekend and maximally two. Then I was trying to bring to their attention the fact that every future engineer attends two to five semesters of mathematics during his or her university studies. The majority of this mathematics was however created at the borderline between XIX and XX century and is oriented towards physics, astronomy, and classical engineering rather than informatics.

When at the beginning of the second half of XX century mathematicians started to think about mathematical theories for computer science some of the branches of mathematics earlier considered as "unpractical" — e.g., such as set theory, mathematical logic, and abstract algebras — became their very practical tools. A little later new branches started to emerge: theory of abstract automata and formal languages, logics of programs, models of concurrent systems and many others. Today mathematical foundations of computer science is a large and fast-growing new branch of applied mathematics.

Of course in this section, I do not pretend to present even a sketch of that mathematics. I limit my course to only these tools which I shall use in the books. I am conscious of the fact that for some readers going through Sec. 2 may be quite a challenge. However, becoming familiar with **MetaSoft** will allow them to describe complex programming constructions in a way which is relatively simple and — what is especially important — complete and unambiguous.

## 2.1    Basic notational conventions of MetaSoft

**MetaSoft** is a semi-formal (i.e., not fully formalised) mathematical notation oriented towards formal descriptions of programming languages. Since typically formulas that appear in such descriptions oversize everything we know from traditional mathematics, some new notational conventions will be introduced. In particular, when it comes to defining models of programming languages (starting from Sec. 4) instead of using single-letter symbols like a, b, c many-letter symbols are used such as sta (for "state"), den (for "denotation"), etc. To distinguish **MetaSoft** formulas in the text, they are printed in Arial font. At the end of the book, there is a list of most frequently used symbols.

### 2.1.1    General mathematical notation

Logical operators are given traditional names: **and**, **or**, **not**, tt, ff. The two last are logical constants "true" and "false". For quantifiers I shall use:

∀ — *general quantifier* (for all)

∃ — *existential quantifier* (there exists)

Instead of i = 1,2,…,n I frequently write i = 1;n. By "iff" I mean "if and only if".

## 2.1.2   Sets

Symbol Ø denotes the empty set and

{$a_1,...a_n$} or {$a_i$ | i = 1,...,n}

denotes a finite n-element set. I shall also use an abbreviation

i = 1;n to denote i = 1,...,n.

The fact that a is (or is not) an element of A shall be written as

a : A     (a /: A)

and the inclusion of sets shall be written as

A ⊂ B

By

A | B   and   A∩B

we I denote the union and the intersection of sets  A and B. If Fam is a family of sets then

U.Fam

denotes the union of that family. By

A x B

I denote the Cartesian product of sets. The expression:

A x B x C x D

denotes the set of tuples of the form (a, b, c, d)  The expression:

A x (B x C) x D

denotes the set of tuples of the form (a, (b, c), d) and analogously for other combinations of parentheses. For every n ≥ 0 *the n-th Cartesian power* $A^{cn}$ *of a set* A is the set of tuples of the elements of A, i.e.:

$A^{c0}$ = {()}                         — the only element of that set is an empty tuple

$A^{cn}$ = {($a_1,...,a_n$) | $a_i$ : A}      — for n > 0

Using Cartesian power we define two other operations:

$A^{c+}$ = U.{$A^{cn}$ | n > 0}

$A^{c*}$ = $A^{c0}$ | $A^{c+}$

The set of all subsets of A and respectively of all finite subsets of A is denoted by

Sub.A

FinSub.A

The following notations shall be used for sets of relations and functions:

| | |
|---|---|
| Rel.(A,B) | — the set of all binary relations between A and B; i.e., the set of all subsets of A x B; more about binary relations in Sec. 2.6, |
| A → B | — the set of all *partial functions* from A to B, i.e., functions that do not need to be defined for all elements of A, |

A ↦ B        — the set of all *total functions* from A to B, i.e., functions that are defined for all elements of A; notice that each total function is a partial function but not vice-versa,

A ⇒ B        — the set of all *mappings* from A to B, i.e., functions defined for only a finite subset of A.

In accordance with the notation for sets by

  f : A ↦ B

we mean that f is an element of the set A ↦ B, hence a total function from A to B and of course analogously for other operators creating sets of functions. This rule also explains why the traditional a ∈ A is written as a : A.

## 2.1.3    Functions

For practical reasons, the value of a function shall be written as f.a rather than f(a). Why this is practical will be seen a little later. The expression

  f.a = ?                                                                                    (2.1-1)

means that f is not defined for a. It does not mean however that "?" is anything like an "undefined element". The expression f.a = ? stands for

  **not** (∃b)(f.a=b)

Analogously

  f.a = !

stands for (∃b)(f.a=b). For an arbitrary function

  f : A → B

and an arbitrary set C by the *truncation of* f *to* C I mean:

  f trun C = {(a, f.a) | a : A ∩ C}.

The *domain* of f is the set where f is defined, i.e.

  dom.f = {a | a : A **and** f.a = !}

In the sequel I shall also use the notation

  f [a/?] = f trun (dom.f − {a})

Another notation that will be used frequently comes from Haskell Curry and concerns many-argument function whose arguments are taken successively one after another, e.g..

  f : A → (B → (C → (D → E)))                                              (2.1-2)

The value of such a function should be formally written as

  ((((f.a).b).c).d)

 but Curry writes

  f.a.b.c.d

which intuitively means that

- function f takes a as an argument and returns as value a function f.a that belongs to the set $B \to (C \to (D \to E))$ and next

- function f.a takes as an argument an element b and returns as a value function f.a.b that belongs to the set $C \to (D \to E)$, etc.

This notation allows to avoid many parentheses and moreover to define function of "mixed" types like e.g.

$$f : A \to (B \mapsto (C \to (D \Rightarrow E))) \quad \text{or} \tag{2.1-1}$$

$$f : (A \to B) \mapsto (C \to (D \Rightarrow E))$$

Another simplifying convention allows to write

$$f : A \to B \mapsto C \to D \Rightarrow E \tag{2.1-2}$$

instead of

$$f : A \to (B \mapsto (C \to (D \Rightarrow E))) \tag{2.1-3}$$

The expression

$$f : \mapsto A \tag{2.1-4}$$

means that f is a zero-argument function with only one value that belongs to A. That value is denoted by

$$f.()$$

About formulas from (2.1-2) to (2.1-6) we say that they describe *types* of corresponding functions. For instance we say that the function in (2.1-5) *is of the type*

$$A \to B \mapsto C \to D \Rightarrow E$$

For every function

$$f : A \mapsto A,$$

by its *n-th iteration* where $n = 0,1,2,\ldots$ I shall mean the function

$$f^n : A \mapsto A$$

defined in the following way:

$f^0$ is an *identity function* on A, i.e. $f.a = a$ for every $a : A$,

$f^n.a = f.(f^{n-1}.a)$ for $n > 0$.

In mathematical definitions of programming languages I shall frequently use many-level *conditional definitions of functions* with the following scheme:

$$f.x =$$

$$p_1.x \ \blacktriangleright \ g_1.x$$

$$p_2.x \ \blacktriangleright \ g_2.x$$

$$\ldots \tag{2.1-7}$$

$$p_n.x \ \blacktriangleright \ g_n.x$$

where each $p_i$ is a classical predicate, i.e., a total function with logical values tt or ff and each $g_i$ is just a function. The formula (2.1-7) is read as follows:

if $p_1.x$ is true, then $f.x = g_1.x$ and otherwise,

if $p_2.x$ is true, then $f.x = g_2.x$ and otherwise,

…

Intuitively speaking the evaluation of this function goes line by line and stops at the first line where $p_i.x$ is satisfied. Of course, to make such a definition of function $f$ unambiguous the alternative of all predicates $p_i.x$ must evaluate to "true", which means that all these predicates must exhaust all cases. To ensure that condition at the last line, we frequently write **true**, which stands for predicate which is always true. It can also be read as "in all other cases".

In the scheme (2.1-7) I also allow the situation where in the place of a $g_i.x$ we have the undefinedness sign "?" which means that for $x$ that satisfies $p_i.x$ the function $f$ is undefined. This convention allows conditional definitions of partial functions.

In conditional definitions I also use a technique similar to defining local constants in programs. For instance if $f : A \times B \longmapsto C$ we can write

$f.x =$

   $p_1.x$ ➡ $g_1.x$

   **let**

      $(a, b) = x$

   $p_2.a$ ➡ $g_2.x$

   $p_3.b$ ➡ $g_3.x.$

which is read as: *let $x$ be a pair of the form* $(a, b)$. We can also use **let** in the following way:

$f.x =$

   $p_1.x$ ➡ $g_1.x$

   **let**

      $y = h.x$

   $p_2.x$ ➡ $g_2.y$

   $p_3.x$ ➡ $g_3.y.$

All these explanations are certainly not very formal, but the notation should be clear when it comes to applications in the sequel of the book.

A finite total function $f : \{a_1,…,a_n\} \longmapsto \{b_1,…,b_n\}$ defined by the formula:

$f.x =$

   $x=a_1$ ➡ $b_1$

   $x=a_2$ ➡ $b_2$

   …

   $x=a_n$ ➡ $b_n$

I shall write as

   $[a_1/b_1,…,a_n/b_n]$  or alternatively as $[a_i/b_i \mid i = 1;n]$.

The empty function will be denoted by [ ]. Let $f : A \rightarrow B$ and $g : C \rightarrow D$. The *overwriting of* $f$ *by* $g$ is a function denoted by

   $f \blacklozenge g : A|C \rightarrow B|D$

and defined in the following way:

$(f \blacklozenge g).x =$

$g.x = !$ ➡ $g.x$

**true** ➡ $f.x$

In particular this means that if $f.x = ?$ and $g.x = ?$, then $f \blacklozenge g.x = ?$. A particular case of overwriting is an *update of a function* written as $f[a_1/b_1, \ldots, a_n/b_n]$ and defined by the formula

$(f[a_1/b_1, \ldots, a_n/b_n]).x =$

$x = a_1$ ➡ $b_1$

…

$x = a_1$ ➡ $b_n$

**true** ➡ $f.x$

## 2.1.4   Tuples

An expression

$(a_1, \ldots, a_n)$ or alternatively $(a_i \mid i=1;n)$

denotes n-*tuple*. Consequently () denotes an empty tuple. The difference between tuples and finite sets is such that the order of elements in a tuple is relevant and repetitions are allowed which is not the case for sets. E.g.

$\{a, b, c, d\} = \{a, b, a, d, c\}$

$(a, c, b, c, d) \neq (a, c, c, d, b)$

where $a$ to $d$ are different with each other.

Tuples are used as mathematical models for several concepts in theoretical computer science and among others for pushdowns. In this area the following functions shall be used later on in the book:

$push.((a_1, \ldots, a_n), b)$  $= (a_1, \ldots, a_n, b)$   for $n \geq 0$

$pop.(a_1, \ldots, a_n)$  $= (a_1, \ldots, a_{n-1})$ for $n \geq 2$

$pop.(a)$  $= ()$

$pop.()$  $= ()$

$top.(a_1, \ldots, a_n)$  $= a_n$                for $n \geq 1$

$top.()$  $= ?$

Another application of tuples are words in the theory of formal languages (see Sec. 2.4). In that case, we have the function of *concatenation*:

$(a_1, \ldots, a_n) © (b_1, \ldots, b_m) = (a_1, \ldots, a_n, b_1, \ldots, b_m).$

We shall also use two predicates:

$are\text{-}repetitions.(a_1, \ldots, a_n) = tt$  iff there exist $i \neq j$ such that $a_i = a_j$

$no\text{-}repetitions.(a_1, \ldots, a_n) = tt$   iff there are no $i \neq j$ such that $a_i = a_j$

Tuples may also be regarded as functions from natural numbers into their elements i.e.

$(a_1,…,a_n).i = a_i$

Let now for a certain set $A$

Tuple = $A^{c^*}$

be the set of all tuples over $A$. For sets of tuples the following functions shall be used:

remove-repetitions : Tuple ⟼ Tuple

remove-repetitions.$(a\text{-}1,…,a\text{-}n)$ =

| | |
|---|---|
| $n = 0$ | ➔ () |
| $n = 1$ | ➔ $(a_1)$ |
| $a_1 : \{a_2,…,a_n\}$ | ➔ remove-repetitions.$(a_2,…,a_n)$ |
| **true** | ➔ $(a_1)$ © remove-repetitions.$(a_2,…,a_n)$ |

join-without-repetition : Tuple x Tuple ⟼ Tuple

join-without-repetition.$(tup_1, tup_2)$ = remove-repetitions.$(tup_1$ © $tup_2)$

common-part : Tuple x Tuple ⟼ Tuple

common-part.$( (a_1,…,a_n), (b_1,…,b_m) )$ =

| | |
|---|---|
| $n = 0$   ➔ () | |
| $m = 0$   ➔ () | |
| $a_1 : \{b_1,…,b_m\}$ | ➔ $(a\text{-}1)$ © common-part.$( (a_2,…,a_n), (b_1,…,b_m) )$ |
| **true** | ➔ common-part.$( (a_2,…,a_n), (b_1,…,b_m) )$ |

difference : Tuple x Tuple ⟼ Tuple

difference. $( (a_1,…,a_n), (b_1,…,b_m) )$ =

| | |
|---|---|
| $n = 0$ | ➔ () |
| $m = 0$ | ➔ $(a_1,…,a_n)$ |
| $a_1 : \{b_1,…,b_m\}$ | ➔ difference.$( (a_2,…,a_n), (b_1,…,b_m) )$ |
| **true** | ➔ $(a_1)$ © difference.$( (a_2,…,a_n), (b_1,…,b_m) )$ |

The last operation selects these elements of a tuple which satisfy a given predicate. Let then

p : A ⟼ {tt, ff, ee}

be a three-valued predicate. With every such predicate, we associate a *filtering function* which removes from a tuple all elements $a$ that do not satisfy p, i.e., such that p.$a$ : {ff, ee}.

filter.p : Tuple ⟼ Tuple

filter.p.$(a_1,…,a_n)$ =

n = 0        ➜  ()

p.$a_1$ = tt   ➜  ($a_1$) © filter.p.($a_2$,…,$a_n$)

**true**      ➜  filter.p.($a_2$,…,$a_n$)

## 2.2    Partially ordered sets

Let A be an arbitrary set and let

⊑ : Rel(A,A)

be a binary relation in that set. Relation ⊑ is said to be a *partial order* in A if for any a, b, c : A the following conditions are satisfied:

1.  a ⊑ a                                  *reflexivity*

2.  if a ⊑ b and  b ⊑ c then a ⊑ c   *transitivity*

3.  if a ⊑ b and b ⊑ a then a = b    *weak antisymmetricity*

If only 1. and 2. are satisfied then ⊑ is said to be *quasiorder.* In the sequel we shall deal most frequently with partial orders.

If a ⊑ b then we say that a is *smaller* than b or that b *is greater* than a. If additionally a ≠ b then we say that a is *significantly smaller* than b or that b is *significantly greater* than a.

A pair (A, ⊑) is called a *partially ordered set* (abbr. POS) and the set A is called its *carrier.* The word "partial" means that not any two elements of A are comparable with each other. If however for any a and b either a ⊑ b or b ⊑ a,

then we say that this is a *total order.*

Of course, every linear order is partial, and every partial order is quasiorder but not vice versa. An example of a partial order which is not total is the inclusion of sets. Such POS is called *set-theoretic POS.*

Let B be a subset of a partially ordered A and let b : B. In that case

- b is called a *minimal element* in B, if there is no a : B such that a ⊑ b and a ≠ b

- b is called the *least element* in B, if for any a : B holds b ⊑ a,

- b is called a *maximal element* in B, if there is no a : B such that b ⊑ a and a ≠ b,

- b is called the *greatest element* in B, if for any a : B holds a ⊑ b.

There exist partially ordered sets without a minimal element and sets where there is more than one such element. However, if there is a smaller element in the set, then it is the unique minimal element and analogously for maximal and greatest elements.

An *upper bound* of B is such an element of A which is greater than any element of B. Notice that an upper bound of a set does not need to belong to that set, but if it does belong, then it is the greatest element of the set.

If the set of all upper bounds of B has the least element, then this element is called the *least upper bound* of B[12]. If a two-element set {a, b} has the least upper bound, then we denote it by

a ∨ b

In a set-theoretic POS the least upper bound of a family of sets is the set-theoretic union of that family.

A partial order ⊑ in A is said to be *well-founded* if every not empty subset of A has the least element.

## 2.3     Chain-complete sets

Let $(A, ⊑)$ be a partially ordered set. By a *chain* in that set we mean any sequence of elements of A:

$a_1, a_2, a_3, …$

such that $a_i ⊑ a_{i+1}$. If the set of all elements of a chain has the least upper bound, then it is called the *limit* of that chain and is denoted by:

$\lim(a_i \mid i = 1,2,…)$

A POS is said to be *chain-complete partially ordered set* (abbr. CPO) if:

1. every chain in A has a limit,

2. there exists the least element in A.

This least element we shall denote by Φ.

A total function $f : A \longmapsto A$ is said to be *monotone* if $a ⊑ b$ implies $f.a ⊑ f.b$ and we say that it is *continuous* if the following two conditions are satisfied:

1. for any chain $(a_i \mid i = 1,2,…)$ the sequence $(f.a_i \mid i = 1,2,…)$ is also a chain,

2. if the former has a limit, then the latter has a limit as well and

$\lim(f.a_i \mid i = 1,2,…) = f.[\lim(a_i \mid i = 1,2,…)]$.

As is easy to see, every continuous function is monotone, which follows from the fact that if $a ⊑ b$ then $\lim(a, b, b, …) = b$.

Continuous functions satisfy a theorem — due to S.C. Kleene [48] — which we shall frequently use in our applications.

**Theorem** 2.3-1 *If f is continuous in a chain complete set, then the set of all solutions of the equation*

$x = f.x$                                                                (2.3-1)

*is not empty and contains the least element defined by the equation*

$Y.f = \lim(f^n.Φ \mid n = 0,1,2,…)$ ∎

**Proof** of that theorem is very simple:

$f.(Y.f) = f.(\lim(f^n.Φ \mid n = 0,1,2,…)) = \lim(f^n.Φ \mid n = 1,2,…) = \lim(f^n.Φ \mid n = 0,1,2,…)$.

---

[12] The greatest lower bound is defined in an analogous way but that concept will not be used in the book.

The last equality follows from the fact that $f^0.\Phi = \Phi$, hence adding $f^0.\Phi$ to the chain does not change its limit. ∎

The equation (2.3-1) is called a *fixed point equation* and its solution $Y.t$ — the *least fixed point of function* f. It is the least solution of the equation (2.3-1) but is the sequel I will call it simply *the solution* since other solutions will not be concerned.

The concept of a one-argument continuous function may be simply generalised to functions of many arguments. We say that

$$f : A^{cn} \longmapsto A \qquad\qquad (2.3-2)$$

is *continuous with regard to its first element*, if for any tuple $(a_1,\ldots,a_{n-1})$ the function

$$g.a = f.(a, a_1,\ldots,a_{n-1})$$

is continuous. In an analogous way we define the continuity of f with regard to any other of its arguments.

A many-argument function (2.3-2) is called *continuous* if it is continuous in all of its arguments.

As we are going to see soon, continuous functions are fundamental for our applications since due to Kleene's theorem we can recursively define sets and functions. Such definitions will most frequently have the form

$$x_1 = f_1.(x_1,\ldots,x_n)$$

$$\ldots$$

$$x_n = f_n.(x_1,\ldots, x_n)$$

Of course, every such set of equations may be regarded as one equation

$$X = f.X$$

in a POS over a Cartesian product $A_1 \times \ldots \times A_n$ where

$$f.(x_1,\ldots,x_n) = (f_1.(x_1,\ldots,x_n),\ldots, f_n.(x_1,\ldots,x_n))$$

and where the order is define component-wise, i.e.

$$(a_1,\ldots,a_n) \sqsubseteq^n (b_1,\ldots,b_n) \text{ iff } a_i \sqsubseteq b_i \text{ for } i = 1,\ldots,n.$$

As is easy to show, if all $A_i$ are chain-complete, then their Cartesian product is chain-complete with regard to the above order. Besides, if each $f_i$ is continuous, then f is continuous as well.

As turns out, fixed-point sets of equations with continuous functions may be transformed (and reduced) in a way analogous to the case of algebraic equations. It is expressed by two theorems due to Hans Bekić [9] and Jacek Leszczyłowski [51].

**Theorem** 2.3-2 *If* f, g $: A \times A \longmapsto A$ *are continuous, then the set of equations*

$$a = f.(a,b)$$

$$b = g.(a,b)$$

*is equivalent to*

$$a = f.(a,b)$$

$$b = g.( f.(a,b),b) \blacksquare$$

**Theorem** 2.3-3 *If* f, g : A x A $\longmapsto$ A *are continuous, then the set of equations*

    a = f.(a,b)

    b = g.(a,b)

*is equivalent to*

    a = h.b

    b = g.(a,b)

*where* h *is a function that to every* b *assigns the least fixed point of* f.(x,b) *regarded as a one-argument function of* x *running over the set* A. ∎

As we are going to see the theory of fixed-point equations in CPO is an important tool for writing recursive definitions of sets and of functions in denotational models.

## 2.4    The CPOs of formal languages

Grammars of *natural languages* such as English, Polish, French, etc. may be regarded as algorithm allowing to check which sentences are grammatically correct and which are not. In this spirit, Noam Chomsky has developed in early 1960. his model of *generative context-free grammars* or simply *context-free grammars* (see [31], [32], [33] and [34]). Formal languages generable by such grammars have been called *context-free languages*.

Although this model turned out to be too simple for natural languages, it was successfully applied for programming languages. In the early years for Algol 60 and Pascal, later for ADA and CHILL and many others. This contributed to the rapid development of their theory. First internationally recognised monography on that subject was written in 1966 by Seymour Ginsburg [41], and the first Polish monography in 1971 by myself [13]. A year later I have published a paper on *equational grammars* [15] which are equivalent to context-free grammars.

This section contains a short introduction to context-free languages in the context of equational grammars.

Let A be an arbitrary finite set of symbols called an *alphabet*. By a *word* over A, we mean every finite tuple over A including the empty tuple. Traditionally words are written as sequences of characters, e.g., accbda and the *empty word* is denoted by ε.

If x and y are words, then by their *concatenations* — which we denote by x © y or simply by xy — we mean a sequential combinations of these words . E.g.

    abdaa © eaag = abdaaeaag

The function © is called *concatenation* as well. Every set L of words over A is called a *formal language* (or simply a *language*) over A. By Lan(A) we denote the family of all languages over A and Ø — the empty language (empty set). If P and Q are languages, then their *concatenation* is the language defined by the equation:

    P © Q = {p © q | p:P **and** q:Q}.

As we see by © we denote not only a function on words but also on languages. If it does not lead to ambiguities, P © Q is written as PQ. Since concatenation is an associative operation, we can write PQL instead of (PQ)L or P(QL). I shall also assume that concatenation binds stronger than set-theoretic union, hence instead of

    (P © Q) | (R © S)

I shall write

> PQ | RS

It is also easy to see that concatenation is distributive over the union, i.e.

> (P | Q) R = PR | QR.

The n-th *power of a language* P is defined recursively:

> $P^0 = \{\,\varepsilon\,\}$
>
> $P^n = P \, ⓒ \, P^{n-1}$ for $n > 0$

We shall also use two operators called respectively *plus* and *star*:

> $P^+ = U.\{P^i \mid i > 0\}$
>
> $P^* = P^+ \mid P^0$

Hence for an alphabet A, the set $A^+$ is the set of all non-empty words over A, and $A^*$ is the set of all words over A. Languages over A are subsets of $A^*$.

The inclusion of sets is, of course, a partial order in Lan(A) and (Lan(A), ⊂) is a CPO with empty language as the least element. As is easy to show all above operations on languages plus the union of languages are continuous. For any two languages, P and Q their least upper bound is their union P | Q, and the limit of a chain of languages is the union of all elements of the chain.

It should be emphasized that the Cartesian power of sets introduced in Sec. 2.1.2 is different from the power of languages. Notice that if P and Q are languages, then:

> P ⓒ Q = { p ⓒ q | p : P **and** q : Q }
>
> P x Q = { (p, q) | p : P **and** q : Q }

The concatenation of languages is hence still a language, whereas the Cartesian product is not.

## 2.5    Equational grammars

Since all the operations on languages defined in Sec. 2.4 are continuous, they can be used in fixed-point equations (Sec. 2.3) regarded as grammars. This idea is elaborated below.

Consider a simple example of a set of equations that defines the set of identifiers of a programming language. We assume that identifiers always start from a letter:

> Letter     = {a, b, …, z}
>
> Digit      = {0, 1, …, 9}
>
> Character = Letter | Digit
>
> Suffix     = {ε} | Character ⓒ Suffix
>
> Identifier = Letter ⓒ Suffix

Such sets of equations are called *equational grammars,* and their solutions (tuples of languages) are called *many-sorted languages.* In the above case the defined many-sorted language is a tuple of five categories (sorts):

> Letter, Digit, Character, Suffix, Identifier.

The category Suffix has an auxiliary character since its only role is to express the fact that an identifier must start with a letter. Its equation can be eliminated in using the Theorem 2.3-2 and the Theorem 2.3-3. As is easy to prove

Suffix = Character$^*$

hence our grammar may be reduced to a more compact form

Letter      = {a, b, …, z}

Digit        = {0, 1, …, 9}

Identifier   = Letter © (Letter | Digit)$^*$

This grammar defines a many-sorted language which consists of three categories — hence is different from the former — but defines the same set Identifier.

Let us now investigate equational grammars more formally as was described in [15]. Let A be an arbitrary non-empty finite alphabet and let

Fam ⊂ Lan(A)

be an arbitrary family of languages over A. Let Pol(Fam) denotes the least class of functions of the type:

p : Lan(A)$^{cn}$ ⟼ Lan(A)

for all n ≥ 0 which contains:

(1) all projections, i.e. functions of the form f.$(X_1,…,X_n)$ = $X_i$ for i ≤ n,

(2) all functions with constant values in Fam,

(3) the union and concatenation of languages

and is closed over the composition (superposition) of functions.

Functions in Pol(Fam) are called *polynomials over* Fam. Since all functions described in (1), (2) and (3) are continuous and composition preserves continuity, all polynomials are continuous.

By an *atomic language* over A we shall mean any one-element language {w}, where w : A$^*$. Polynomials over an arbitrary set of atomic languages are called *Chomsky's polynomials*[13]. Below a few examples of such polynomials:

$p_1$.(X,Y,Z) = {b}

$p_2$.(X,Y)    = {b}

$p_3$.(X,Y,Z) = X

$p_4$.(X,Y,Z) = ({d}X{b}YY{c} | X) Z

Observe that for a complete identification of a polynomial we have to define its arity. This can be seen on the example of $w_1$ and $w_2$.

---

[13] Noam Chomsky — an American linguist, philosopher and political activist. Professor of linguistics at Massachusetts Institute of Technology, co-creator of the concept of transformational-generative grammars. Chomsky did not introduces the idea of Chomsky's polynomials but his grammars are very close to them.

Polynomials which do not "contain" union — e.g., such as $p_1$, $p_2$, and $p_3$ — are called *monomials*. Since concatenation is distributive over the union, every polynomial may be reduced to a union of monomials.

An *equational grammar* over an alphabet $A$ is every fixed-point set of equations of the form:

$X_1 = p_1.(X_1,…,X_n)$

…

$X_n = p_n.(X_1,…,X_n)$

where all $p_i$ are Chomsky's polynomials over $A$. Since polynomials are continuous, this set of equation has a unique least solution $(L_1,…L_n)$. The languages $L_1,…L_n$ are said to be *defined by our grammar*. We also say that they are *equationally definable*.

As has been proved in [15], the class of equationally-definable languages is identical with the class of *context-free* languages in the sense of Chomsky[14]. Such a class remains the same if we allow the operations "*" and "+" in polynomials and if polynomials are built over arbitrary equationally-definable languages. For proofs of all these facts see [15].

Due to these facts in the sequel of the book equationally-definable languages will be called *context-free*.

## 2.6    The CPOs of binary relations

Let $A$ and $B$ be arbitrary sets. Any subset of their Cartesian product $A \times B$ will be called a *binary relation* or just a *relation* between these sets. Hence

$Rel(A,B) = \{R \mid R \subset A \times B\}$

is the set of all binary relations between $A$ and $B$. Instead of writing $(a,b) : R$ I shall usually write $a\ R\ b$.

If $A = B$, then instead of $Rel(A,A)$ I write $Rel(A)$. For every $A$ I define an *identity relation*:

$[A] = \{(a, a) \mid a:A\}$

By $\emptyset$ I denote the *empty relation*[15]. Let now

Boolean = {tt, ff}                — logical values

$p : A \rightarrow$ Boolean              — a predicate

With every predicate, we assign an identity relation defined by

$Id(p) = [\{a \mid p.a = tt\}]$

If $R : Rel(A,B)$, then

$dom.R = \{a \mid (\exists b:B)\ a\ R\ b\}$ — the *domain* of $R$

$cod.R = \{b \mid (\exists a:A)\ a\ R\ b\}$  — the *codomain* of $R$

Let $P : Rel(A,B)$ and $R : Rel(B,C)$. *Sequential composition* of $P$ and $R$ we call a relation

$P \bullet R : Rel(A,C)$

---

[14] Which means that for each equational grammar there exists an equivalent context-free grammar and vice versa.
[15] The same symbol was used for an empty set which is not an inconsistency since each relation is a set.

defined as follows:

P ● R = {(a, c) | (∃b:B) (a P b & b R c)}

For every two relations, their composition always exists although may be an empty relation. As is easy to check ● is associative i.e.

(P ● R) ● Q = P ● (R ● Q)

It is, therefore, legal to write P ● R ● Q. I shall also write $PR$ instead of P ● R whenever this does not lead to misunderstanding, and I shall assume that composition binds stronger than union, hence instead of

(P ● R) | (Q ● S)

I write

$PR$ | $QS$

In the sequel of the book sequential composition of relations will be most frequently applied in the particular case where the composed relations are function. In that case:

(P ● R).a = R.(P.a)

and therefore

(P ● R ● Q).a = (P ● (R ● Q)).a = Q.(R.(P.a)))

which means that in a sequential composition of functions, the composed functions are "executed" from left to right.

Similarly as for languages also for relations, the operations of power and star are defined. In this case:

$R^0$ = [A] — identity relation in over $A$

$R^n$ = $RR^{n-1}$ for n > 0

$R^+$ = U {$R^n$ | n > 0}

$R^*$ = $R^+$ | $R^0$

The *converse relation* for $R$ is defined as follows

a $R^{-1}$ b    iff    b $R$ a

A relation $R$ is called a *function*, if for any a, b and c, if a $R$ b and a $R$ c, ten  b = c.

If $R$ and $R^{-1}$ are functions, then $R$ is said to be a *convertible function* or a *one-one function*. If P and $R$ are functions, then $PR$ is also a function and

$(PR).a$ = P.(R.a)

hence the composition of functions is their superposition.

The set of relations $Rel(A,B)$ constitutes a CPO with ordering by set-theoretic inclusion and the empty relation as the least element. All of the defined operations on relations are continuous. The future we shall frequently refer to the following theorem:

**Theorem 2.6-1** *For any* P,Q : $Rel(A)$ *the least solutions of equations*

X = P | QX    *and*

X = P | XQ

*are respectively*

X = Q*P     *and*

X = P*Q

*Moreover, if both* P *and* R *are functions with disjoint domains, then both these solutions are also functions.*     ∎

In this place, it is worth noticing that the set of partial functions

A → B

constitutes a chain-complete subset of (Rel(A,B), ⊏) that is closed under the composition of arbitrary functions and union of functions with disjoint domains. Of course, both these operations are continuous.

Due to these facts functions can be defined by fixed-point (recursive) equations. Since A and B are arbitrary this is also true for functions of the type

f : $A_1$ → $A_2$ → … → $A_n$

provided that appropriate constructors are defined. As an example let us consider a recursive definition of the arithmetic function of power that refers to the functions of multiplication and subtraction:

power : Number x Number → Number[16]

power.n.m =

m = 0 ➔ 1

m > 0 ➔ n ∗ power.n.(m-1))

where

Number = {0, 1,….}

I shall show now that this definition can be expressed as a fixed-point equation in the CPO of binary relations:

Rel.(Number x Number, Number)

I shall construct a fixed-point equations whose solution is the function:

power.(n, m) = $n^m$

regarded as a relation from our CPO. let me start from the definitions of a certain operation of composition of functions

F, Q : Rel.(A x A, A)                                                                (**)

By the *composition of* F and Q *on the second argument,* I shall mean the relation

F ❷ Q = {{a,b,c} | (∃d) (a,b,d) : F **and** (a,d,c) : Q}

If F and Q are functions then

---

[F ❷ Q].(a,b) = Q.(a, F.(a,b))

The set of rations (\*\*) is, of course, a CPO with set-theoretic inclusion. One can show that ❷ is continuous on the first argument. Let us consider now the equation:

power = zero | minus ❷ power ❷ times                                    (\*\*)

where:

zero(n, 0)     = 1

minus.(n, m) = m-1    for m > 0, and for m = 0 this function is undefined

times.(n, m)  = n∗m

Since the set-theoretic union and our composition are both continuous in the CPO of relations (\*), Kleene's theorem implies that the solution of (\*\*) is the limit of the chain of relation

$P^0 \subset P^1 \subset P^2 \subset \ldots$                                    (\*\*\*)

which are functions defined in the following way:

$P^0$     = zero

$P^{i+1}$    = (minus ❷ $P^i$) ❷ times     for i ≥ 0

This means that for every i ≥ 0 function $P^i$ is a partial function of power restricted to m ≤ i:

$P^i$.(n, m) =

m ≤ i ➜ $m^i$

**true**  ➜ ?

Since all these functions coincide on the common parts of their domains, the set-theoretic union of the chain (\*\*\*) is a function, and it is the power function defined for arbitrary n,m ≥ 0.

## 2.7    The CPO of denotational domains

One of the main tools of denotational models of software systems are sets traditionally referred to as *domains*. These domains are most frequently defined using equations — and in particular fixed-point equations — based on functions that are listed below. The majority of these functions have been already defined, but I repeat their descriptions just to have a full list of them in one place:

1)  A | B       — set-theoretic union

2)  A ∩ B       — set-theoretic intersection

3)  A x B       — Cartesian product

4)  $A^{cn}$        — Cartesian n-th power

5)  $A^{c+}$        — Cartesian +-iteration

6)  $A^{c*}$        — Cartesian *-iteration

7)  FinSub.A   — the set of all finite subsets

8)  A ⟹ B      — the set of all mappings including the empty mapping

9)  A – B       — set-theoretic difference

10) Sub.A      — the set of all subsets

11) $A \rightarrow B$       — the set of all functions from $A$ to $B$

12) $A \mapsto B$       — the set of all total functions from $A$ to $B$

13) $Rel.(A,B)$    — the set of all relations from $A$ to $B$

These operators will be used in "direct" equations, e.g.

State          $=$ Identifier $\Rightarrow$ Object

Instruction    $=$ State $\rightarrow$ State                                    (2.7-1)

but also in fixed point equations, e.g.:

Record      $=$ Attribute $\Rightarrow$ Object

Object       $=$ Number | Record                                    (2.7-2)

Whereas definition $(2.7\text{-}1)$ does not raises any doubts, in the case of $(2.7\text{-}2)$ the situation is different. Since this is obviously a fixed-point equation we have to prove the continuity of $\Rightarrow$ and |, but the continuity where? What is the CPO of domains? Set-theoretic inclusion is clearly it's partial order, but what is the carrier?

Potentially that carrier should include all domains that we shall define in the future, hence something like the set of all sets. Unfortunately — as it has been known since 1930-ties — such a set does not exist[17]. Despite this fact, our problem can be solved on the base of M.P. Cohn [35] construction. As he has shown, for any collection of sets $B$ (a collection does not need to be a set!) there exists a set of sets $Set.B$ with the following properties:

1. all sets in $B$ belong (as elements) to $Set.B$,

2. $Set.B$ is closed under all our operations from 1) to 14),

3. $Set.B$ is closed under unions of all denumerable families of its elements,

4. the empty set $\emptyset$ belongs to $Set.B$.

Following this construction, we choose as the family $B$, the set of basic domains used later on in domain equations, such as Booleans, numbers, identifiers, characters, etc. Since $(Set.B, \subset)$ is a set-theoretic CPO, we can talk about the continuity of functions defined on sets in $Set.B$. As is easy to show functions from 1) to 8) are continuous, the difference of set is continuous only on the left argument, and the remaining functions are not continuous, and therefore they cannot appear in fixed-point equations[18].

---

[17] Formally speaking the attempt of constructing such a set leads to a contradiction. Indeed, let Z be the set of all sets. Let then Ze be the set of all sets that are their own elements and Zn — the set of all sets that are not their own elements. Since obviously Z = Ze | Zn, set Zn must belong to either Ze or Zn. The first case must be excluded since in that case Zn should belong to Zn. The second case is impossible either, since then Zn must not belong to itself. Intuitively speaking one can say that the collection of all sets is "to large to be a set".

[18] As an example let me show that the operator $\rightarrow$ is not continuous. Let then $A_1 \subset A_2 \subset \ldots$ be an arbitrary chain of mutually different sets, and let B be an arbitrary set. The sequence of domains $A_i \rightarrow B$ constitutes a chain but none of its elements contain any total function on the union $UA_i$, hence any such function belongs to $U(A_i \rightarrow B)$, which means that $U(A_i \rightarrow B) \neq UA_i \rightarrow B$. In an analogous way we may show the non-continuity of the operators $A \mapsto B$ and $Rel.(A,B)$. Notice, however, that $U(A_i \Rightarrow B) = UA_i \Rightarrow B$, and similarly for the right-hand-side argument which means that $\Rightarrow$ is continuous on both arguments.

As we see, therefore, the equation (2.7-2) has a solution (the least solution) defined by the theorem of Kleene (Sec. 2.3). Records defined in that way may "carry" other records, but "lower" than themselves, which can again carry lower records. In the end, we have records carrying numbers. If however, we replace $\Rightarrow$ by $\rightarrow$, then (**) has no solution. A problem of exactly that type encountered mathematicians who in early 1970-ties had been trying to define denotational semantics for Algol 60. More on that subject in Sec. 4.1.

The fact that non-continuous operators cannot be used in fixed-point domain equations does not mean however that they cannot be used in fixed-point equations "at all". For instance, our two sets of equations (*) and (**) can be legally combined into one:

State       = Identifier   $\Rightarrow$ Object

Instruction = State       $\rightarrow$ State

Record      = Attribute   $\Rightarrow$ Object

Object      = Number    |   Record                                        (2.7-3)

Although "as a whole" this is a fixed-point set of equations with one non-continuous operation, the recursion is present in only two last equations where the operators are continuous. This set of equations is therefore legal.

## 2.8    Abstract errors

For practically all expressions appearing in programs their values in some circumstances cannot be computed "successfully". Here are a few examples:

- the value of x/y cannot be computed if y = 0,

- the value of the expression x+1 cannot be computed if x has not been declared in the program.

- the value of x+y cannot be computed if the sum exceeds the maximal number allowed in the language,

- the value of the array expression a[k] cannot be computed if k is out of the domain of array a,

- the query "Has John Smith retired?" cannot be answered if John Smith is not listed in the database.

In all these cases a well-designed implementation should stop the execution of a program and generate an error message.

To describe that mechanism formally, we introduce the concept of an *abstract error*. In a general case abstract errors may be anything, but in our models, I assume that they are texts, such as e.g. 'division-by-zero'. They are closed in apostrophes to distinguish them from metavariables at the level of MetaSoft.

The fact that an attempt of evaluating x/0 raises an error message can be now expressed by the equation:

x/0 = 'division-by-zero'

In the general case with every domain Data, we associate a corresponding domain with abstract errors

DataE = Data | Error

where Error is the set of all abstract errors that are generated by our programs. Consequently every partial operation

op : Data$_1$ x … Data$_n$ → Data

is extended to a total operation

ope : DataE$_1$ x … DataE$_n$ ↦ DataE

Of course ope should coincide with op wherever op is defined, i.e. if d$_1$,…,d$_n$ are not errors and op.(d$_1$,…,d$_n$) is defined, then ope.(d$_1$,…,d$_n$) = op.(d$_1$,…,d$_n$). Now ope will be said to be *transparent for errors* or simply *transparent* if following condition is satisfied:

if d$_k$ is the first error in the sequence d$_1$,…,d$_n$, then ope.(d$_1$,…,d$_n$) = d$_k$

This condition means that arguments of ope are evaluated one-by-one from left to right, and the first error (if it appears) becomes the final value of the computation.

The majority of operations on data that will appear in our models will be transparent. An exception are boolean operations discussed in Sec. 2.9.

Error-handling mechanisms are frequently implemented in such a way, that errors serve only to inform the user that (and why) program evaluation has been aborted. Such a mechanism will be called *reactive*. In some applications, however, the generation of an error results in an action, e.g. of recovering the last state of a database (Sec. 12.7.6.4). Such mechanisms will be called *proactive*.

As we shall see in the sequel of the book, a reactive mechanism may be quite simply enriched to a proactive one. Since, however, the latter is technically more complicated, in the development of our example-language **Lingua**, except **Lingua-SQL**, we shall most frequently apply a reactive model, although with a few exceptions (sections 6.1.8, 7.3.1, 7.3.3 and 12.7.6.4).

A well-defined error-handling mechanism allows avoiding situations where programs stop without any explanation, or even worse — when they do not stop but generate an incorrect result without any warning to the user (see Sec. 12.7.6.4).

## 2.9     A three-valued propositional calculus

Tertium non datur — used to say ancients masters. Computers denied this principle.

In the Aristotelean logic, every sentence is either true or false. The third possibility does not exist. In the world of computers, however, the third possibility is not only possible but just inevitable. In evaluating a boolean expression such as, e.g., x/y>2 an error (see Sec. 2.8) can appear.

To describe the error-handling mechanism of boolean expression, besides the basic domain of Boolean values

Boolean = {tt, ff}

we introduce a domain with a third element

BooleanE = {tt, ff, ee}

where ee stands for "error" , but in this case represents either an error or an infinite computation (a looping). In this section, I assume for simplicity that there is only one error. This assumption does not destroy the generality of the model as long as the error handling mechanism is reactive (see Sec. 2.8). At the same time, it turns out that the transparency of boolean operators would not be an adequate choice. To see that consider a conditional instruction[19]:

```
if x ≠ 0 and 1/x < 10 then x := x+1 else x := x-1 fi
```

We would probably expect that for x=0 one should execute the assignment x:=x-1. If however, our conjunction would be transparent, then the expression

```
x ≠ 0 and 1/x < 10
```

would evaluate to 'division-by-zero' which means that the program aborts. Notice also that the transparency of **and** implies

ff **and** ee = ee

which means that when an interpreter evaluates p **and** q, then it first evaluates both p and q — as in "usual mathematics" — and only later applies **and** to them. Such a mode is called an *eager evaluation*.

An alternative to it is a *lazy evaluation* where if p = ff, then the evaluation of q is abandoned, and the final value of the expression is ff. In such a case:

ff **and** ee = ff

tt **or** ee = tt

A three-valued propositional calculus with the above lazy evaluation was described in 1961 by John McCarthy [55] who defined boolean operators as shown in Tab. 2.9-1

| **or-m** | tt | ff | ee | | **and-m** | tt | ff | ee | | **not-m** | |
|---|---|---|---|---|---|---|---|---|---|---|---|
| tt | tt | tt | tt | | tt | tt | ff | ee | | tt | ff |
| ff | tt | ff | ee | | ff | ff | ff | ff | | ff | tt |
| ee | ee | ee | ee | | ee | ee | ee | ee | | ee | ee |

**Tab. 2.9-1 Propositional operators of John McCarthy**

To see the intuition behind the evaluation of McCarthy's operators consider the expression p **or-m** q noticing that its arguments are computed from left to right[20]:

- If p = tt, then we give up the evaluation of q (lazy evaluation) and assume that the value of the expression is tt. Notice that in this case we maybe avoid an error message that could be generated by q. Therefore **or-m** is not transparent for errors.

- If p = ff, then we evaluate q, and its value becomes the value of the expression.

- If p = ee, then this means that the evaluation of our expression aborts at the evaluation of its first argument, hence the second argument is not evaluated at all. Consequently, the final value of the expression must be ee.

The rule for **and** is analogous. Notice that McCarthy's operators coincide with classical operators on classical values (grey fields in the table). McCarthy's implication is defined classically:

---

[19] Here I anticipate the future syntax of **Lingua** where `Courier New` is used in order to distinguish program texts form statements expressed in MetaSoft.
[20] The suffix "-m" stands for "McCarthy" and is used to distinguish McCarthy's operators not only from classical ones but also from the operators of Kleene, which are discussed later.

p **implies-m** q = (**not-m** p) **or-m** q

As we are going to see, not all classical tautologies remain satisfied in McCarthy's calculus. Among those that remain satisfied we have:

- associativity of alternative and conjunction,
- De Morgan's laws

and among the non-satisfied are:

- **or-m** and **and-m** are not commutative, e.g., ff **and-m** ee = ff but ee **and-m** ff = ee,
- **and-m** is distributive over **or-m** only on the right-hand side, i.e.

  p **and-m** (q **or-m** s)  =  (p **and-m** q) **or-m** (p **and-m** s) however

  (q **or-m** s) **and-m** p  ≠  (q **and-m** p) **or-m** (s **and-m** p) since

  (tt **or-m** ee) **and-m** ff = ff  and  (tt **and-m** ff) **or-m** (ee **and-m** ff) = ee

- analogously **or-m** is distributive over **and-m** only on the right-hand side,
- p **or-m** (**not** p) does not need to be true but is never false,
- p **and-m** (**not** p) does not need to be false but is never true.

On the ground of that calculus, we build in Sec. 8 a much richer calculus of partial predicates[21] to be used in the rules of correct-programs construction. At the level of propositional calculus, the partiality of predicates corresponds to the case where ee represents an error or an infinite execution.

  Notice that McCarthy's calculus understood in this way is — due to its laziness — implementable, which is the consequence of the equations:

ee **and** p = ee      for every p : {tt, ff, ee}

ee **or** p = ee      for every p : {tt, ff, ee}

If a program loops in computing the first argument then it will never proceed to the computation of the second. This property of McCarthy's calculus is not satisfied by an alternative to it calculus defined by S.C. Kleene [48], where

ff **and** ee = ff

ee **and** ff = ff

This means that the falsity of conjunction requires the falsity of at least one of its argument. Operationally this means that to compute p **and** q, we have to compute both p and q. That rule is implementable only if ee does not correspond to infinite computations or if we can compute both arguments in parallel. As we will see in Sec.12 Kleene's calculus is used in SQL queries and, to a certain degree, in validating programming (Sec. 8.2).

## 2.10   Data algebras

Data types that are used in programs are usually described by sets of objects — such as numbers, Booleans, strings, arrays, lists, etc. — and some operations on that objects. For instance, a data type of numbers may be described as a tuple:

---

[21] The partiality of predicates is due to the use of functional-procedure calls in expressions.

AlgNum = (Number, make.no.1, plus, minus, times, divide)          (2.10-1)

This tuple will be called the *algebra of numbers* where Number — called the *carrier of the algebra* — is the set of all numbers and make.no.1, plus, minus, times, divide are functions on numbers called *constructors*. The following formulas define the domains and the codomains of constructors:

| make-nu.1: | | $\mapsto$ Number | |
|---|---|---|---|
| plus | : Number x Number | $\mapsto$ Number | |
| minus | : Number x Number | $\mapsto$ Number | (2.10-2) |
| times | : Number x Number | $\mapsto$ Number | |
| divide | : Number x Number | $\rightarrow$ Number | |

The zero-argument function make-nu.1 (make number one) represents a *constant* of our algebra. This function has no arguments, and its unique value is 1, hence:

make-nu.1.() = 1

If our algebra were a model of a programming language, the presence of this constant would mean, that number 1 may be expressed directly at the level of syntax by writing `make-nu.1`. Notice that the number 2 cannot be expressed similarly. Instead, we have to write e.g.

    plus.(make-nu.1.(), make-nu.1.())

Number 2 is thus created from two ones whereas number 1 — from "nothing". Both

    make-nu.1 .()

and

    plus.(make-nu.1.(), make-nu.1.())

are examples of expressions written in so-called *abstract syntax* (see Sec. 2.12). Since such a syntax is not very user-friendly it is in general simplified to *concrete syntax* (see Sec. 4.5) where we would write respectively 1 and 1+1.

Notice that divide is a partial function since dividing by zero is not allowed.

Our algebra of numbers is an example of *abstract algebra,* and the list of formulas (2.10-2) is called their *signature* (formal definitions in Sec.2.11). The word "abstract" expresses the fact that algebra of numbers is not a branch of mathematics dedicated to solving algebraic equations, but an abstract mathematical object.

Of course in programming languages that operate on numbers we restrict the set of available numbers — i.e. the carrier of the algebra — to a finite set of decimal numbers representable in the arithmetic of our computer[22]. If by NumberR we denote the set of such numbers, then the signature of our algebra would be the following:

| make-nu.num | : | $\mapsto$ NumberR | for num : NumberR |
|---|---|---|---|
| plus | : NumberR x NumberR | $\rightarrow$ NumberR | |
| minus | : NumberR x NumberR | $\rightarrow$ NumberR | |

---

[22] Notice that in user manuals the range of numbers is usually defined as an interval, e.g. from $-2^{63}$ to $2^{63} - 1$ (see [38]) without mentioning that numbers with infinite or with too long binary representations will be truncated.

| | |
|---|---|
| minus | : NumberR x NumberR $\rightarrow$ NumberR |
| divide | : NumberR x NumberR $\rightarrow$ NumberR |

In this algebra we have a finite family of zero-argument constructors indexed by representable numbers:

{make-nu.num | num : NumberR}

Here make-nu is a meta-constructor which is not a constructor of our algebra but is used to generate zero-argument constructors of the algebra.

Notice that in this algebra all constructors except zero-argument constructors are partial functions since each of them may lead out of the domain of representable numbers. This solution has two faults:

- mathematical fault — in the theory of abstract algebra which constitutes a fundament of denotational models (see Sec. 2.11) all constructors are assumed to be total; the introduction of partial constructors is probably possible but would certainly complicate the model.

- informational fault — in our programming language **Lingua** we want to have an error-message mechanism that warns the user about each non-performability of an operation.

To cope with both these problems we introduce abstract errors as described in Sec.2.8 and replace the carrier NumberR by the carrier

NumberE = NumberR | Error

where the set Error contains all error messages that we shall need in our algebra. Now the signature of our algebra is as follows:

| | | | |
|---|---|---|---|
| make-nu.num | : | $\mapsto$ NumberE | for num : NumberR |
| plus | : NumberE x NumberE | $\mapsto$ NumberE | |
| minus | : NumberE x NumberE | $\mapsto$ NumberE | |
| times | : NumberE x NumberE | $\mapsto$ NumberE | |
| divide | : NumberE x NumberE | $\mapsto$ NumberE | |

Passing to another aspect of data-type definitions let us notice that in the majority of programming languages with data-type *number* we associate not only arithmetic operations but also *predicates* such as e.g. less or equal, hence functions with numerical arguments but Boolean values. In order to describe such a structure we need an algebra with two carriers — NumberE and BooleanE — hence

AlgNumBoo = (NumberE, BooleanE,

{make-nu.num | num : NumberR}, plus, minus, times, divide,

less, equal, make-bo.tt, make-bo.ff, or, and, not)

Operations in this algebra constitute the following signature:

| | | | |
|---|---|---|---|
| make-nu.num | : | $\mapsto$ NumberE | for num : NumberR |
| plus | : NumberE x NumberE | $\mapsto$ NumberE | |
| minus | : NumberE x NumberE | $\mapsto$ NumberE | |
| times | : NumberE x NumberE | $\mapsto$ NumberE | |

| divide | : NumberE x NumberE $\mapsto$ NumberE | |
|---|---|---|
| less | : NumberE x NumberE $\mapsto$ BooleanE | (2.10-3) |
| equal | : NumberE x NumberE $\mapsto$ BooleanE | |
| make-bo.tt | : $\mapsto$ BooleanE | |
| make-bo.ff | : $\mapsto$ BooleanE | |
| or | : BooleanE x BooleanE $\mapsto$ BooleanE | |
| and | : BooleanE x BooleanE $\mapsto$ BooleanE | |
| not | : BooleanE $\mapsto$ BooleanE | |

An algebra with two carriers is said to be *two-sorted algebra*. Sometimes signature of many-sorted algebras are visualizes graphically as on Fig. 2.10-1. For simplicity I included only some operation on that figure.

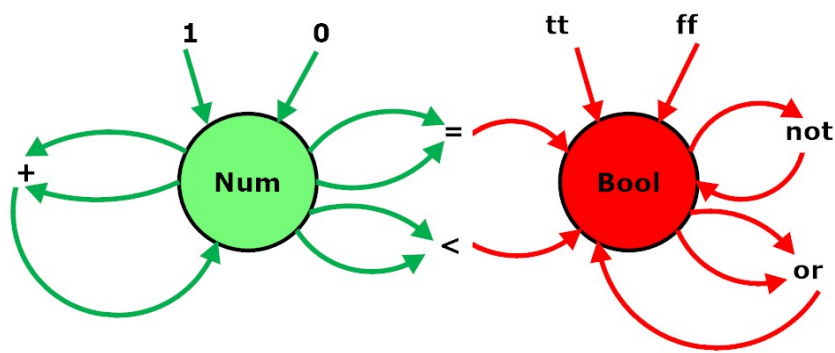

**Fig. 2.10-1 Graphical representation of a two-sorted algebra**

Notice that zero-argument operations of our algebra do not lead out of the set of representable numbers NumberR. Of course, all other operations must satisfy that principle as well. For example the operation of addition should be defined in the following form:

plus.(num-1, num-2) =

| num-1 : Error | ➔ num-1 |
|---|---|
| num-2 : Error | ➔ num-2 |
| **not** +.(num-1, num-2) : NumberR | ➔ 'overloading'[23] |
| **true** | ➔ +.(num-1, num-2) |

where „+" is the arithmetical addition. Notice that plus is not commutative since

plus.(err-1, err-2) ≠ plus.(err-2, err-1)

if only err-1 ≠ err-2.

## 2.11    Many-sorted algebras

Our algebra AlgNumBoo is said to be *two-sorted* since it has two sorts of carriers: NumberE and BooleanE. In the sequel, we shall construct algebras with many sorts therefore called

---

[23] The negation operator **not** in this clause has been written with boldface font since this is not a constructor of our algebra, but a metaconstructor from the level of MetaSoft.

*many-sorter algebras* or simply *algebras*. Formally a many-sorted algebra is the following tuple:

Alg = (Sig, Car, Fun, car, fun)

where

| | | |
|---|---|---|
| Sig = (Cn, Fn, ar, so) | — | is called the *signature of the algebra*, |
| Cn | — | is a finite set of words called the *names of carriers*; the carriers themselves are called *sorts*, |
| Fn | — | is a finite set of words called the *names of functions*; the functions themselves are called *constructors* |
| ar : Fn ⟼ Cn$^*$ | — | to every name of a function fn the function ar assigns a finite (possibly empty) sequence of sorts' names |
| | | ar.fn = (cn$_1$,…,cn$_k$) |
| | | called the *arity* of fn[24] |
| so : Fn ⟼ Cn | — | to every name of a function fn the function so assigns a carrier name so.fn which is called *the sort of* fn, |
| Car | — | a finite set of carriers, |
| Fun | — | a finite set of total functions with arguments and values in carriers; these functions are called *constructors,* |
| car : Cn ⟼ Car | — | to every name cn of a carrier function car assigns a carrier car.cn, |
| fun : Fn ⟼ Fun | — | to every function name fn such that |
| | | ar.fn = (cn$_1$,…,cn$_k$) |
| | | so.fn = cn |
| | | the function fun assigns a total function |
| | | fun : car.cn$_1$ x … x car.cn$_k$ ⟼ car.cn |

The concepts of *arity* and *sort* are applied not only to function names but to the corresponding functions themselves. Functions in the set Fun are traditionally called *constructors*. Zero-argument constructors, i.e., constructors whose arity is an empty sequence are called *constants* of the algebra. If f is such a constant, then we write

f : ⟼ Carrier

and the unique value of f is written as

f.()

It should be emphasised that all constructors must be total functions. It is a technical rule which can always be satisfied by introducing abstract errors as discussed in Sec. 2.8.

As we shall see in the sequel, our lengthy definition of a many-sorted algebra has been introduced to distinguish syntax from denotations (semantics) both in models of data-types as well as in denotational models of languages. For concrete algebras, however, e.g., such as

---

[24] The word „arity" comes from unary, binary, ternary etc.

discussed in Sec.2.10 the signature is implicit in the set of formulas such as (2.10-3). Now consider two algebras:

Alg$_i$ = (Sig$_i$, Car$_i$, Fun$_i$, car$_i$, fun$_i$) for i = 1,2

with signatures

Sig$_i$ = (Cn$_i$, Fn$_i$, ar$_i$, so$_i$) for i = 1,2

We say that Sig$_2$ is an *extension* of Sig$_1$ or that Sig$_1$ is a *restriction* of Sig$_2$, if

1. Cn$_1$ $\subset$ Cn$_2$ and Fn$_1$ $\subset$ Fn$_2$,

2. functions ar$_2$, so$_2$ coincide with ar$_1$, so$_1$ on Fn$_1$.

We say that algebra Alg$_2$ is an *extension of algebra* Alg$_1$, if

1. Sig$_2$ is an extension of Sig$_1$,

2. car$_1$.cn $\subset$ car$_2$.cn for every sort cn : Cn$_1$,

3. fun$_2$.fn coincides with fun$_1$.fn on the appropriate carriers for every fn : Fn$_1$.

In other words, each (nontrivial) extension of an algebra results from that algebra by adding new carriers and/or new constructors and/or new elements to the existing carriers.

Two many-sorted algebras are said *similar* if they have the same signature. In the future, we shall frequently define concrete algebras by defining their carriers and constructors but without showing their signatures explicitly. In that case, we shall say that two algebras are similar if it is possible to construct a common signature for them.

If Alg$_1$ and Alg$_2$ are similar, then we say that Alg$_1$ is a *subalgebra* of Alg$_2$ if:

1. the carriers of Alg$_1$ are subsets of the corresponding carriers of Alg$_2$,

2. the constructors of Alg$_1$ coincide with constructors of Alg$_2$ on the carriers of Alg$_1$.

Therefore every subalgebra of an algebra is a restriction of that algebra but not vice versa. By a *many-sorted homomorphism* from algebra Alg$_1$ into a similar algebra Alg$_2$ where

Alg$_i$ = (Sig$_i$, Car$_i$, Fun$_i$, car$_i$, fun$_i$) for i = 1,2

we call a family of functions

H = {h.cn | cn : Cn}

whose elements called *the components of that homomorphism* map the elements of Alg$_1$ into the elements of Alg$_2$, hence

h.cn : car$_1$.cn $\mapsto$ car$_2$.cn    for all cn : Cn

and where for every constructor name fn : Cn such that

ar.fn = (cn$_1$,…,cn$_n$)    where    n $\geq$ 0

and every tuple of arguments

(a$_1$,…,a$_n$) : car$_1$.cn$_1$ x … x car$_1$.cn$_n$

the following equality holds

h.cn.(fun$_1$.fn.(a$_1$,…,a$_n$)) = fun$_2$.fn.(h.cn$_1$.a$_1$,…,h.cn$_n$.a$_n$)    (2.11-1)

In other words a homomorphic image

of the value of a function $fun_1.fn$ from the first algebra on a (possibly empty) tuple of arguments $(a_1,…,a_n)$

is the value of the corresponding function $fun_2.fn$ from the second algebra on the tuple of homomorphic images of the first tuple i.e. on $(h.cn_1.a_1,…,h.cn_n.a_n)$.

Notice that for $n = 0$ the equality $(2.11\text{-}1)$ has the form

$h.cn.(fun_1.fn.()) = fun_2.fn.()$

The fact that $H$ is a homomorphism from $Alg_1$ into $Alg_2$ shall be written as:

$H : Alg_1 \longmapsto Alg_2$

Our definition of homomorphism implies that if some carriers of the algebra $Alg_1$ are empty, then the corresponding components of the homomorphism have to be empty as well. An algebra where all carriers are empty is called *an empty algebra*.

In the general case homomorphisms do not map algebras onto algebras but into algebras, which means that not every element in $Alg_2$ must be an image of an element form $Alg_1$. For instance an identity homomorphism from integers to numbers

$I2N : (Integer, 1, plus, minus) \longmapsto (Number, 1, plus, minus)$

is of course not "onto", whereas a homomorphism from integers into even integers

$I2E : (Integer, 1, plus, minus) \longmapsto (Even, 1, plus, minus)$

defined by the equality $I2E.int = 2*int$ is of course "onto". In the general case a homomorphism $H : Alg_1 \longmapsto Alg_2$ is called:

- a *monomorphism*    — if all its components are one-to-one functions; e.g., $I2N$ and $I2E$,
- an *epimorphism*    — if all its components are "onto"; e.g., $I2E$
- an *isomorphism*    — if it is both a monomorphism and an epimorphism; e.g., $I2E$.

**Theorem** 2.11-1 *For every homomorphism* $H : Alg_1 \longmapsto Alg_2$ *the image of* $Alg_1$ *in* $Alg_2$, i.e., the restriction of $Alg_2$ *to the images through* $H$ *of* $Alg_1$ *with the appropriate truncation of constructors of* $Alg_2$ *constitutes a subalgebra of* $Alg_2$. ∎

**Proof** To prove the theorem we have to show that the images in $Alg_2$ of the carriers of $Alg_1$ are closed under the operations of $Alg_2$. Let then $(b_1,…,b_n)$ from $Alg_2$, be the image of $(a_1,…,a_n)$ in $Alg_1$, i.e. let:

$(b_1,…,b_n) = (h.cn_1.a_1,…,h.cn_n.a_n)$

Let furthermore for some function name $fn$ be

$fun_2.fn.(b_1,…,b_n) = b$

We have to show that $b$ has a coimage in $Alg_1$. It is indeed the case since on the ground of $(2.11\text{-}1)$:

$fun_2.fn.(b_1,…,b_n) = fun_2.fn.(h.cn_1.a_1,…,h.cn_n.a_n) = h.cn.(fun_1.fn.(a_1,…,a_n))$

hence $h.cn.(fun_1.fn.(a_1,…,a_n))$ is the needed coimage of $b$ in $Alg_1$. ∎

An algebra which is the image of a homomorphism $H : Alg_1 \longmapsto Alg_2$ is called *the kernel of the homomorphism* $H$ in $Alg_2$.

All our considerations about homomorphisms can be generalized to the case where the signatures of two algebras

$Sig_i = (Cn_i, Fn_i, ar_i, so_i)$ for $i = 1,2$

are not identical but are *similar* in the sense that there exist two reversible functions of similarity

$Sn : Cn_1 \longmapsto Cn_2$

$Sf : Fn_1 \longmapsto Fn_2$

such that if

$Sf.fn_1 = fn_2$

$ar_1.fn_1 = cn_{11},\ldots,cn_{1p}$

$ar_2.fn_2 = cn_{21},\ldots,cn_{2m}$

then

$p = m$

$Sn.cn_{1i} = cn_{2i}$     for  $i = 1;p$

In other words two signatures are similar if they the same number of carrier names and function names and the corresponding function names have identical arities and sorts up to the names of carriers.

Now we can generalise the notion of the similarity of algebras: two algebras shall be called *similar* if their signatures are similar. For every fixed functions $Sn$ and $Sf$ the concept of homomorphism and the corresponding theorems remain valid for the generalised similarity.

## 2.12   Abstract syntax and reachable algebras

Every signature

$Sig = (Cn, Fn, ar, so)$

unambiguously determines a certain algebra with that signature and with formal languages as carriers. This algebra is called *abstract syntax over signature* $Sig$ and will be denoted by AbsSy($Sig$)[25]. The elements of its carriers are words of a many-sorted formal language

$\{Lan.cn \mid cn : Cn\}$

defined by an equational grammar (see Sec.2.5) in a way described below.

To every carrier name $cn$ we associate a language denoted by $Lan.cn$. The tuple of all these languages is defined by an equational grammar where for every $cn : Cn$ we have the following equation[26]:

$Lan.cn = \{nf_1\} © \{(\} © Lan.cn_{11} © \{,\} © \ldots © \{,\} © Lan.cn_{1n(1)} © \{)\} \mid$

$\qquad\qquad \ldots$ (2.12-1)

$\qquad \{nf_k\} © \{(\} © Lan.cn_1 © \{,\} © \ldots © \{,\} © Lan.cn_{n(k)} © \{)\}$

Here $nf_i$ for $i = 1;k$ are functions' names with

---

[25] The concept of an abstract syntax regarded as a mathematical idealization of the syntax of programming languages appeared for the first time in papers of J. McCarthy [55] and P. Landin [50] but with abstract algebras was for the first time associated by J.A. Goguen, J.W. Thacher, E.G. Wagner and J.B. Wright [43]. A little later I used that concept in an attempt to give a formal semantics to a subset of Pascal [21].

[26] We assume, of course, that the commas "," and the parentheses "(" and ")" do not appear in the signature as constructors' names.

so.nf$_i$ = cn

and

ar.nf$_i$ = (cn$_{i1}$,…,cn$_{in(i)}$)        for  i = 1;k

We assume that if for a carrier name cn there is no function name fn such that so.nf = cn, then the corresponding language is empty, i.e. its defining equation is:

Lan.cn = ∅

This assumption in necessary in order to make abstract syntax over an algebra over a given signature. For every non-empty Lan.cn its elements are words of the form

fn$_i$(w$_{i1}$,…,w$_{in(i)}$)

i.e. of the form fn$_i$ © ( © w$_{i1}$ © … © w$_{in(i)}$ ©) where © is the concatenation of words and

w$_{ik}$ : Lan.cn$_k$.

In this place, it is worth noticing that if there are no zero-argument functions' names (constants) in the signature, then all languages (carriers) of the corresponding abstract syntax are empty.

Since abstract syntaxes are generated from signatures, they may be associated with arbitrary algebras (through their signatures). If Alg is an algebra with signature Sig, then AbsSy(Sig) will be called *the abstract syntax of algebra* Alg. For instance, if AlgNumBoo is the two-sorted algebra described in Sec.2.10 than the carrier of its abstract syntax are defined by the following equational grammar where NumExp and BooExp are languages of numeric expressions and Boolean expressions respectively

NumExp = 0 |1 |

              plus(NumExp, NumExp) | minus(NumExp, NumExp) |

              times(NumExp, NumExp) | divide(NumExp, NumExp)

                                                                                    (2.12-1)

BooExp = tt | ff |

              less(NumExp, NumExp) | equal(NumExp, NumExp) |

              or(BooExp, BooExp) | and(BooExp, BooExp) | not(BooExp)

In this grammar I use three notational conventions that are assumed as standards for future use:

1. if it does not lead to a confusion a one-element set {a} is written as a,

2. for each zero-argument constructor named nk, instead of nk() I write nk, e.g. 1 instead of 1(),

3. the concatenation sign © is omitted, e.g. I write ab instead of a © b,

Examples of a numeric and Boolean expression are the following:

plus(plus(minus(1,0),1),plus(1,1))

or(less(plus(plus(minus(1,0),1),plus(1,1)),plus(1,1)),ff)

As we see the expressions of our languages do not contain variables and are written in a *prefix notation* where function symbols always precede their arguments. E.g., we write plus(1,1) instead of (1 plus 1). The latter style is called *infix-notation*.

In the syntactic algebra defined by that grammar, the elements of carriers are numeric and Boolean expressions without variables and constructors correspond to constructors' names from our signature. For instance, with a constructor name plus we associate a constructor [plus] of the algebra AbsSy(Sig) defined by the equation

[plus].[num-exp1, num-exp2] = plus(num-exp1,num-exp2)[27]

It is a constructor which given two expressions num-exp1 and num-exp2 returns the expression plus(num-exp1,num-exp2).

Now we can formulate a theorem with fundamental importance for denotations models of programming languages.

**Theorem 2.12-1** *For every many-sorted algebra* Alg *with a signature* Sig *there is exactly one homomorphism* H : AbsSy(Sig) ⟼ Alg. ∎

**Proof** Every homomorphism H : AbsSy(Sig) ⟼ Alg must (from the definition) satisfy the equation:

H.cn.[fn($w_1$ , … , $w_n$ )] = fun.fn.[H.$cn_1$.$w_1$,…,H.$cn_n$.$w_n$]

where

ar.fn = ($cn_1$,…,$cn_n$)

so.fn = cn

$w_i$ : Lan.$cn_i$    for    i = 1;n

Since every word in abstract syntax is of a unique (for it) form fn($w_1$ , … , $w_n$), the above equations (for all fn) define the family {H.cn | cn : Cn} in an unambiguous way. In the case of empty carriers of AbsSy(Sig) the corresponding components of H are empty. ∎

The unique homomorphism from AbsSy(Sig) to Alg will be called *the designating homomorphism* since in a certain way it designates the algebra Alg. For instance, if by {N, B} we denote the designating homomorphism for AlgNumBoo, then this homomorphism maps Boolean expression less(plus(1,1), times(1,1)) into the Boolean value ff:

B.[less(plus(1,1),times(1,1))] =

fun.less.(N.[plus(1,1)], N.[times(1,1)]) =

fun.less.(fun.plus.(N.[1],N.[1]), fun.times.([N.[1], N.[1])) =

fun.less (fun.plus(1,1), fun.times(1,1)) = ff

On the ground of theorems 2.11-1 and 2.12-1, in every algebra Alg, the is a unique subalgebra which is the kernel of the designating homomorphism of Alg. That algebra is called *the reachable subalgebra* of Alg. This name expresses the fact that every element of that algebra can be constructed (reached) from the constants of the algebra in using the constructors of the algebra. For instance, the reachable subalgebra of the algebra

(Number, 1, plus, divide)

is the algebra of positive rational numbers

(PosRat, 1, plus, divide)

since only such numbers can be constructed from 1 in using plus and divide. Notice that if we remove 1 from the algebra of numbers, then its reachable algebra becomes empty and consequently so becomes also its algebra of abstract syntax.

---

[27] The meta-parentheses "[" and "]" are introduced in order to distinguish them from parentheses that belong to the defined language.

An algebra is called *reachable* if it coincides with its reachable subalgebra. In particular, every algebra of abstract syntax is reachable. Reachable is also every empty algebra. Now we can formulate two important theorems.

**Theorem 2.12-2** For any two similar algebras Alg₁ and Alg₂, *if* Alg₁ *is reachable, then there is at most one homomorphism*

H : Alg₁ ⟼ Alg₂,

*and if this is the case, then the image of* Alg₁ *in* Alg₂ *is reachable.* ∎

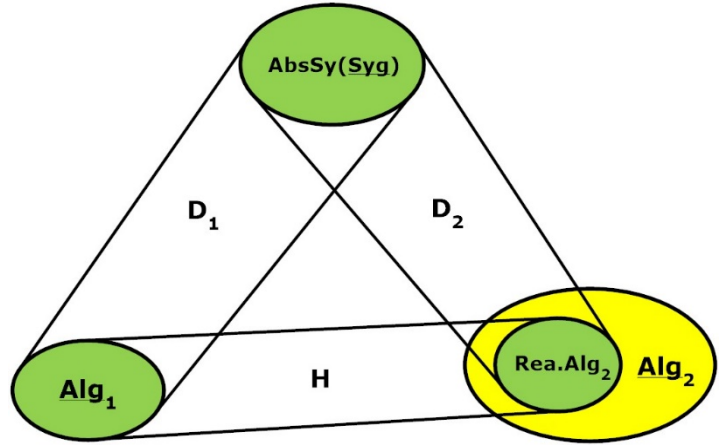

**Fig. 2.12-1 Reachable algebras**

**Proof**. This theorem and its proof are illustrated in Fig. 2.12-1. Since Alg₁ and Alg₂ are similar, they must have a common signature Sig and a common abstract syntax AbsSy(Sig). Therefore — on the ground of Theorem 2.12-1 — there exist two unambiguously defined designating homomorphisms

D₁ : AbsSy(Sig) ⟼ Alg₁ and

D₂ : AbsSy(Sig) ⟼ Alg₂

Now if there exists a homomorphism H : Alg₁ ⟼ Alg₂, then the composition

D₁ ● H : AbsSy(Sig) ⟼ Alg₂.

defined as the composition of their components, is a homomorphism. Since D₂ is the unique homomorphism between these algebras, we have

D₁ ● H = D₂,

and since Alg₁ is reachable, the above equation defines H unambiguously, because otherwise, we could define another homomorphism from AbsSy(Sig) into Alg₂ which would contradict Theorem 2.12-1. It proves that the image of Alg₁ in Alg₂ is reachable. ∎

As an immediate consequence of this theorem we have another theorem:

**Theorem 2.12-3** *For every nonempty algebra* Alg *over signature* Sig *the following claims are equivalent:*

(1) Alg *is reachable,*

(2) *every homomorphism of the type* H : Alg₁ ⟼ Alg *(for an arbitrary* Alg₁*) is onto,*

(3) *the designating homomorphism* D : AbsSy(Sig) ⟼ Alg *is onto.* ∎

**Proof** Let Alg be reachable and let for some Alg₁ similar to Alg there exist a homomorphism

H : Alg₁ ⟼ Alg,

and let

D : AbsSy(Sig) ⟼ Alg₁

be the designating homomorphism for Alg₁. In that case

D ● H : AbsSy(Sig) ⟼ Alg

is the designating homomorphism for Alg, hence since Alg is reachable then D ● H must be *onto*, and therefore also H must be *onto*. Hence (1) implies (2). Now (3) follows from (2) as its particular case and (2) implies (1) by the definition of reachability. ∎

At the end of this section one more useful theorem:

**Theorem 2.12-4** *An algebra has a nonempty reachable subalgebra if and only if it contains at least one constant, i.e., a zero-argument constructor.* ∎

**Proof** If there is a constant in the algebra, then it belongs to its reachable part hence this part is not empty. If however such o constant does not exist, then in the grammar corresponding to that algebra there are no constant monomials, and therefore all the carriers of abstract syntax are empty. Therefore the reachable part of Alg is an empty algebra. ∎

Abstract syntaxes are in general not very convenient in practical programming, and therefore they are usually replaced by more user-friendly syntaxes historically called *concrete*. In such a case elements of abstract syntax may be regarded as *parsing trees* of concrete expressions, a concept that since 1960-ties plays a fundamental role in the theory compilation of programming languages (see, e.g. [3]).

## 2.13   Ambiguous and unambiguous algebras

An algebra Alg with a signature Sig is said to be *unambiguous* if its designating homomorphism

D : AbsSy(Syg) ⟼ Alg

is a one-to-one function, i.e. if for every carrier Car.cn of Alg and every element of that carrier e there is at most one word w : Lan.cn in the abstract syntax AbsSy(Syg) such that

D.cn.w = e

Algebras which are not unambiguous are called *ambiguous*.

Algebras of denotations of programming languages are most frequently ambiguous. For instance, the algebra AlgNum described in 2.10 (if supplied with abstract errors to make their constructor total) is ambiguous since, e.g., two different words plus(plus(1,1),1) and plus(1,plus(1,1)) correspond to the same number 3.

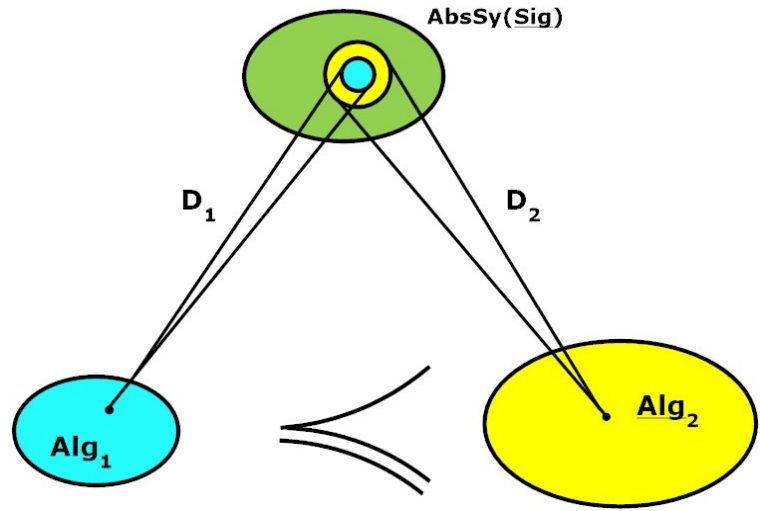

**Fig. 2.13-1 Two ambiguous algebras**

Now consider two algebras $Alg_1$ and $Alg_2$ with a common signature $Sig$ hence also with a common abstract syntax $SkAbs(Sig)$. Let

$D_1 : SkAbs(Sig) \longmapsto Alg_1$

$D_2 : SkAbs(Sig) \longmapsto Alg_2$

be two corresponding designating homomorphisms.

Algebra $Alg_1$ is said to be *less (or equally) ambiguous than* algebra $Alg_2$, what shall be written as

$Alg_1 \leqslant Alg_2$

if the homomorphism $D_2$ is glueing not more than $D_1$ (Fig. 2.13-1), i.e., if for any two words $w_1$ and $w_2$ in abstract syntax that belong to the same carrier $Car.cn$ the following implication holds:

if      $D_1.cn.w_1 = D_1.cn.w_2$      then      $D_2.cn.w_1 = D_2.nn.w_2$

Intuitively speaking if an element of $Alg_1$ may be constructed in two different ways — the two ways are $w_1$ and $w_2$ — than the two ways lead to the same element in $Alg_2$.

Ambiguous algebras play an important role in the theory of programming languages since for the majority of existing programming languages, their corresponding algebras of concrete syntax — if ever formally described — would turn out to be ambiguous. To explain this fact assume that $AbsSy(Sig)$ is defined by the grammar

NumExp = 0 | 1 | +(NumExp, NumExp),

$Alg_1$ is an algebra of infix expressions without parentheses defined by the grammar

NumExp = 0 | 1 | NumExp + NumExp

and $Alg_2$ is the algebra of integers. Let now $D_1$ replaces prefixes by infixes and removes parentheses.

Anticipating the considerations of Sec. 4 the algebra of numbers will be called the *algebra of denotations* (of meanings) for both our algebras of numeric expressions and the homomorphism $D_2$ will be called the *denotational homomorphism* (the *semantics*) of the algebra of abstract syntax. In this place, a question may be raised if there exists a denotational homomorphism form parentheses-free expressions into numbers.

To answer this question notice that for such algebras and their corresponding homomorphisms the following equalities hold:

$D_1.[+(+(1,1),1)] = 1+1+1$          $D_2.[+(+(1,1),1)] = 3$

$D_1.[+(1,+(1,1)] = 1+1+1$          $D_2.[+(1,+(1,1)] = 3$

As we see $D_1$ is glueing not more than $D_2$. In "practical mathematics", hence also in programming languages, we frequently omit "unnecessary parentheses" whenever we deal with associative operations. The corresponding algebras are in general ambiguous and therefore the denotational homomorphism $D_2$ need not exist. If however, they are not more ambiguous than the algebras of denotations, then such a homomorphism exist which follows from the following theorem:

**Theorem 2.13-1** *If $Alg_1$ and $Alg_2$ are similar and $Alg_1$ is reachable, then the (unique) homomorphism* $D : Alg_1 \longmapsto Alg_2$ *exists if and only if* $Alg_1 \preccurlyeq Alg_2$. ■

This unique homomorphism may be constructed as (intuitively speaking) the composition of the inverse of $D_1$ with $D_2$, hence

$D = D_1^{-1} \bullet D_2$.

Although the inverse of $D_1$ maps the elements of $Alg_1$ into sets of abstract expressions, yet all these expressions are mapped by $D_2$ into the same element of $Alg_2$. For formal proof of this theorem see [24].

The usability of ambiguous grammars also from the perspective of parsing was investigated in 1972 by A.V. Aho and J.D. Ullman in [3].

## 2.14   Algebras and grammars

The first step in the process of programming-language construction consists in defining an algebra of denotations from which we unambiguously derive a corresponding algebra of abstract syntax. Since the latter is usually not very user-friendly, we transform it to a concrete syntax (cf. Sec. 2.12) in using a homomorphism which does not "glue too much". In the user manual, the latter should be described by an equational grammar which leads to the question whether for each algebra of syntax a corresponding grammar exists. To treat this problem formally, we need the concepts of *a skeleton function*.

A function $f$ on words over an alphabet $A$ is said to be a *skeleton function* if there exists a tuple of words $(w_1,…,w_k, w_{k+1})$ over $A$, called *the skeleton of this function* such that

$f.(x_1,…,x_k) = w_1x_1…w_kx_nw_{k+1}$

An example of a skeleton function may be

$f.(exp\text{-}b,ins_1,ins_2) =$ **if** $exp\text{-}b$ **then** $ins_1$ **else** $ins_2$ **fi**

The skeleton of this function is (**if**, **then**, **else**, **fi**). Notice that the function

$f.(ins_1,ins_2,exp\text{-}b) =$ **if** $exp\text{-}b$ **then** $ins_2$ **else** $ins_1$ **fi**

is not a skeleton function since the order of arguments on the left-hand side of the equation do not coincide with the order on its right-hand side.

In particular cases, a skeleton function may have more than one skeleton. E.g. the one-argument function $f : \{a\}^* \longmapsto \{a\}^*$ defined by equation

$f.(x) = x\, a$

has two skeletons $(\varepsilon,a)$ and $(a,\varepsilon)$, since it may be equivalently defined by the equation

$\quad$ f.(x) = a x

However, if we change the type of the function into f : {a,b}* $\longmapsto$ {a,b}* without changing the defining equation, then this function has only one skeleton $(\varepsilon,a)$.

A many-sorted algebra will be called a *syntactic algebra* if it is a reachable algebra of words.

A many-sorted algebra will be called a *context-free algebra* if all its constructors are skeleton operations. Typical examples of context-free algebras are algebras of abstract syntax. As was shown in Sec. 2.12 for such algebras we can to build equational grammars that define their carriers and operations. Since that construction may be easily applied to any context-free algebra we can formulate the following theorem:

**Theorem 2.14-1** *For every context-free algebra there is an equational grammar that generates is carriers.* ∎

The following theorem is also true:

**Theorem 2.14-2** *For every equational grammar there is a context-free algebra with carriers defined by that grammar.* ∎

**Proof** Let

$\quad$ $X_1 = pol_1.(X_1,\ldots,X_n)$

$\quad$ …

$\quad$ $X_n = pol_1.(X_1,\ldots,X_n)$

be an equational grammar with $(L_1,\ldots,L_n)$ as the (unique) solution. Assume that polynomials of that grammar are expressed as unions of monomials. The corresponding algebra

$\quad$ Alg = (Sig, Car, Fun, car, fun),

is defined in the following way:

- Sig = (Nc, Nf, ar, so)

- Nc = $\{cn_1,\ldots,cn_n\}$ — carriers' names are arbitrary, but the number of these names must be equal to the number of equations in the grammar,

- Nf = $\{fn_1,\ldots,fn_m\}$ — functions' names are arbitrary, but the number of these names must be equal to the number of monomial occurrences in the grammar,

- ar and so are defined in that way that they correspond to the arities and sorts of monomials in the grammar,

- Car = $\{L_1,\ldots,L_n\}$,

- Fun — the set of all monomials in our grammar,

- $car.cn_i = L_i$    for    i = 1,…,n

Notice now that every monomial in our grammar is (from the definition) a Chomsky's mononomial (see Sec. 2.5), hence satisfies the equation:

$\quad$ $car.cn_i(x_1,\ldots,x_n) = \{s_1\}\ x_1\ \ldots\ \{s_k\}\ x_k\ \{s_{k+1}\}$

This completes the definition of our algebra. Observe that the defined algebra is unique up to the names of carriers and constructors.

Now we have to show that the carriers of Alg are closed over all its constructors and that the algebra is reachable. For this proof see [24]. ∎

Here is a simple example showing how to construct an algebra from a grammar. Consider the following grammar of a two-sorted language

Number = 1 | x | Number + Number

Boolean = Number < Number | Boolean x Boolean

For simplicity curly brackets for function, names have been dropped out. The operations of our grammar are defined by the following equations (the symbols of concatenation © has been omitted as well) where n-exp and b-exp with indexes denote numerical and Boolean expressions respectively:

one.() = 1

variable.() = x

plus.($n$-$exp_1$, $n$-$exp_2$) = $n$-$exp_1$ + $n$-$exp_2$

less.( $n$-$exp_1$, $n$-$exp_2$) = $n$-$exp_1$ < $n$-$exp_2$

and.( $b$-$exp_1$, $b$-$exp_2$) = $b$-$exp_1$ & $b$-$exp_2$

An equational grammar is said to be *unambiguous* (resp. *ambiguous*) if the corresponding algebra is unambiguous (resp. ambiguous). Intuitively an algebra is ambiguous if there exists a word w that can be generated by that grammars in two different ways. These "different ways" are different elements of abstract syntax that are coimages of w with regard to the designating homomorphism (see Sec. 2.12). For instance, the word 1+1+1 may be generated in two different ways:

plus(1,plus(1,1)

plus(plus(1,1),1)

As has been already mentioned, the syntax of a programming language will be constructed as a homomorphic image of abstract syntax. Since these syntaxes will be described by equational grammars, it is important to know which homomorphisms of syntactic algebras do not lead out of the class of context-free algebras.

Let us start with an example of a homomorphism that destroys the context-freeness of an algebra. Let Alg be a one-sorted algebra with the carrier $\{a\}^+$ and with two operations:

h.() = a

f.(x) = x a

This algebra is of course context-free. Now consider a similar algebra with a carrier

$\{a^n b^n c^n \mid n = 1,2,…\}$

and constructors

h.() = abc

f.($a^n b^n c^n$) = $a^{n+1} b^{n+1} c^{n+1}$

This algebra is not context-free since its carrier is not a context-free language (see [41]) but it is isomorphic with our former algebra with the homomorphism, and in fact isomorphism:

I.$a^n = a^n b^n c^n$  for every $n \geq 1$

As is easy to see this homomorphism is not a skeleton function.

A homomorphism H between two syntactic algebras is called a *skeleton homomorphism* — I recall that such a homomorphism, if it exists, is unique (see Theorem 2.12-3) — if for every constructor fun.fn of the source algebra, for which

so.fn = cn

ar.fn = $(cn_1,…,cn_n)$

there exists a skeleton $(s_1,…,s_{n+1})$, such that

H.fn.$(fun_1.fn.(x_1,…,x_n))$ = $s_1$ $x_1…s_n x_n s_{n+1}$

In other words a homomorphic image of every constructor of source algebra is a skeleton constructor in the target algebra.

**Theorem 2.14-3** For every syntactic algebra Alg the following facts are equivalent:

(1) Alg *is context-free*,

(2) *every homomorphism into* Alg *is a skeleton homomorphism*,

(3) *there exists a skeleton homomorphism into* Alg.

For proof see [24].

Let us consider now a simple example of a process of constructing a syntax for a given algebra[28]. Let it be a one-sorted algebra of numbers with three operations:

create-nu.1   :                    $\longmapsto$ Number

plus          : Number x Number  $\longmapsto$ Number

times         : Number x Number  $\longmapsto$ Number

The corresponding abstract syntax, denote it by Syn-0, is defined by the following grammar with only one equation, where Expression denotes a language of numerical expressions with constant values:

Exp = create-nu.1.() | plus(Exp, Exp) | times(Exp, Exp)

For simplicity, I assume the same notation as in the algebra of numbers. The first step on our way to the final syntax consists in:

- replacing create-nu.1 by 1,
- replacing plus and times by + and *,
- replacing prefix notation with infix notation.

This step corresponds to the following homomorphism:

H.[create-nu.1.()] = 1

H.[plus$(exp_1,exp_2)$] = (H.[$exp_1$] + H.[$exp_2$])

H.[times$(exp_1,exp_2)$] = (H.[$exp_1$] ∗ H.[$exp_2$])

This is of course a skeleton homomorphism and the corresponding context-free grammars is the following:

Exp = 1 | (Exp + Exp) | (Exp ∗ Exp)

---

[28] In more general terms such processes will be discussed in Sec. 4.5.

In the second and the last step of syntax construction we would like to allow dropping out "unnecessary parentheses", e.g. writing 1+1+1 instead of (1+(1+1)) and analogously for multiplication. This, however, turns out to be impossible since each homomorhism which removes parentheses has to satisfy the equations:

H.[(exp$_1$ + exp$_2$)] = H.[exp$_1$] + H.[exp$_2$]

H.[(exp$_1$ ∗ exp$_2$)] = H.[exp$_1$] ∗ H.[exp$_2$]

but this would mean that it glues expressions with different denotations, e.g.

H.[((1+1)∗(1+1))] = H.[((1+(1∗1))+1)] = 1+1∗1+1

Although H is a skeleton homomorphism, which implies that its target grammar

Exp = 1 | Exp + Exp | Exp * Exp

is context-free, the corresponding algebra is more ambiguous than the algebra of numbers, hence a denotational semantics of this syntax into the algebra of numbers does not exist.

A known traditional way of solving this problem as e.g. in Algol [61] or in Pascal [47] consists in reconstructing the whole model of the language by introducing to the algebras of denotations and of syntax three carriers Com (component), Fac (factor) and Exp (numeric expression) and the following signature:

| | | | |
|---|---|---|---|
| c-to-e | : Com | ⟼ Exp | *component to expression identically* |
| + | : Com x Exp | ⟼ Exp | *addition* |
| f-to-c | : Fac | ⟼ Com | *factor to component identically* |
| * | : Fac x Com | ⟼ Com | *multiplication* |
| 1 | : Fac | ⟼ Fac | *the generation of 1 as a factor* |
| e-to-c | : Exp | ⟼ Fac | *expression to factor identically* |

The corresponding grammar of abstract syntax is the following:

Exp  = c-to-e(Com) | +(Com, Exp)

Com = f-to-c(Fac) | *(Fac, Com)

Fac  = 1 | (Exp)

and for the first (isomorphic to it) transformed syntax:

Exp  = (Com) | (Com + Exp)

Com = (Fac) | (Fac * Com)

Fac  = 1 | (Exp)

In this grammar names of identity functions have been omitted, which however does not destroy the unambiguity of the grammar since these names appear in elements of different carriers.

Now we can define a skeleton homomorphism that removes parentheses in each of three sorts of expressions:

E.[(com)]          = com

E.[(com + exp)] = S.[com] + E.[exp]

C.[(fac)]             = C.[fac]

C.[(fac ∗ com)]  = F.[fac] ∗ C.[com]

F.[1]         = 1

F.[(exp)]       = (exp)

This leads to the following context-free grammar

Exp  = Com | Com + Exp

Com = Fac | Fac ∗ Com

Fac  = 1 | (Exp)

This grammar may be also written in a direct way in using the constructor of iteration:

Exp  = Com [+ Exp]*       *an expression is a sum of components*

Com = Fac [∗ Com]*        *a component is a multiplication of factors*

Fac  = 1 | (Exp)          *a factor is a constant or an expression in parentheses*

Observe that the parentheses-removal homomorphism is not an isomorphism, since it glues (1+(1+)) and ((1+1)+1) into 1+1+1 and similarly for multiplication. However it does not glue "to much" since addition and multiplication are associative. On the other hand from expression ((1+1)*(1+1)) it removes only external parentheses.

The denotational homomorphism for our grammar is now the following:

Se.[com]        = Ss.[com]

Se.[com + exp]  = Sc.[com] + Se.[exp]

Ss.[fac]        = Sc.[fac]

Ss.[fac ∗ com]  = Sc.[fac] ∗ Ss.[com]

Sc.[1]         = 1

Sc.[(exp))]      = Se.[exp]

Notice that the above equations express the school rules of priority of multiplication over addiction.

---

**Commentary 2.14-1**

The reader to whom I have promised that denotational models of programming languages will offer readable definitions may have some doubts at this moment. So far the simple language of arithmetic expressions that is very well known to every ground-school student has been described in a rather complex way and in addition with the use of advanced mathematics. This, of course, requires a commentary.

First, what we can say to a student in a simple way, when "talking" to a computer we have to express in a way appropriate for the interpreter. That appropriate way is denotational homomorphism which may be mapped one-to-one into a code of an interpreter.

Second, the discussed language serves only to illustrate the denotational method on a very simple example. The real advantage of the method will be better understood when we introduce more advanced programming mechanisms such as declarations, types, instructions, recursive procedures, objects, etc. whose definitions require advanced mathematical tools.

Third, in writing a user's manual for our language, we may directly refer to our acquaintance with school mathematics by saying that numerical expressions can be written and are calculated in a "normal way", which means that their grammar is not shown to the user at all. However, as we shall in Sec. 4.5 there are better solutions to that problem called *colloquial syntax*.

Two following lessons may be learned from our exercise:

First, the description of the simple operation of dropping out unnecessary parentheses requires rather complicated and not very intuitive grammar. Such a grammar is necessary for the implementor but not for the user, who can be simply informed that numerical expressions are written and understood in a "usual" way.

Second, the idea of dropping parentheses came out only at the level of second syntactic algebra, when the two former have been already defined. Therefore, to implement that idea one has to start the construction of the model from scratch. In our simple example this does not lead to too much work, but in real situations, things may look different. To avoid such problems, one should think about syntax as early as on the level of the algebra of denotations. This, however, contradicts the philosophy "from denotations to syntax" and also ruins the principle that denotations should be constructed in a maximally simple way.

The above problems had been investigated in [22], [24] and [29]. A solution suggested there consists in assuming that the programmer's syntax, that will be called *colloquial syntax,* does not need to be a homomorphic image of concrete syntax. In our example concrete syntax would be defined by the grammar:

Exp = 1 | (Exp + Exp) | (Exp ∗ Exp)

and colloquial syntax — which allows for (although does not force) the omission of parentheses — would be defined by the grammar:

Exp = 1 | (Exp + Exp) | (Exp ∗ Exp) | Exp + Exp | Exp ∗ Exp

Observe that the algebra of colloquial syntax is not only not homomorphic to the former but is even not similar since it has a different signature. On the other hand, it is easy to define a transformation that would map our colloquial syntax into concrete syntax by adding the "missing" parentheses. Such a transformation will be called a *restoring transformation.* In practice, this leads to a user manual which contains a formal definition of concrete syntax (a grammar) plus an informal rule which says, e.g., that parentheses may be omitted in the "usual way"[29].

In the general case, a restoring transformation may be described formally or informally according to the complexity of colloquialisation. Its formal definition is, however, always necessary for implementors who have to write a procedure that converts each colloquial program into its concrete version.

More on colloquial syntax as such in Sec. 4.5, and on colloquialisms in **Lingua** in Sec. Sec. 5.4.3, 6.2.3, 7.8.3, and 12.9.

In the end, one methodological remark seems necessary. Languages discussed in this section covered only expressions without variables. Such a case has, of course, no practical value, and it was chosen only to make examples of algebras and corresponding grammars possibly simple. Starting from Sec. 4 I shall discuss methods of constructing denotational models for more realistic languages.

---

[29] As we are going to see in Sec.5.4.3 the situation may a little more complicated.

# 3  The semantic correctness of programs

## 3.1     Historical remarks

The semantic correctness of programs, historically called *program correctness,* was a subject of investigations from the very beginning of computer's era. The earliest paper in this field—today practically forgotten — has been published by the British mathematician Alan Turing in 1949 [66]. Nearly twenty years later in the year 1967, the same ideas were investigated again by American scientist Richard Floyd [39]. In 1978 Association of Computing Machinery established an annual Turing Price for outstanding achievements in informatics. One of the first winners of that price in 1978 was… Richard Floyd.

As far as I know, it has never been found out if Floyd new Turing's work. In the 1980-ties I had written on that subject to Cambridge University, but the only answer was a very categorical advise that I should not try to build "yet another myth about Turing"[30].

The work of Floyd introduced a very important concept of *an invariant of a program* and concerned programs represented by graphical forms called *flow diagrams.* Two years later a British scientist C.A.R Hoare (also a Turing Price winner) published a paper concerning Floyd's ideas applied to *structured programs*, i.e., programs constructed with the help of sequential composition, branching if-then-else and while loops. This approach called later *Hoare's Logic* had given rise to a large field of research in the future. See also Edsger W. Dijkstra [37].

Research devoted to program correctness was also developed in Poland. The first paper on that subject (although in an approach different to Hoare's) was published in 1971 by Antoni Mazurkiewicz [52]. A year later during the first conference in a series of conferences on *Mathematical Foundations of Computer Science*[31] Antoni and I have presented a common paper on a similar subject based on an algebra of binary relations and covering recursive programs and nondeterminism. Nearly ten years later I have published a paper [20] with a complete model of program-correctness rules for programs corresponding to arbitrary flow-diagrams without procedures and recursion. Contrary to many papers in this field and in particular to papers developing so-called Hoare's logic, I assumed that program failure might correspond not only to infinite computations but also to program abortion.

---

[30] Alan Turing (1912-1954) was one of the creators of the theory of computability. His model known today as *Turing machine* is regarded as one of fundamental concepts of this theory. Due to his work "On Computable Numbers, With an Application to the Entscheidungsproblem" Turing was considered as one of the greatest mathematicians of the world. Unfortunately he was also subject to a homophobic discrimination. When in 1952 police has learned about his homosexuality he was forced to choose between prison or hormonal therapy. He has chosen the latter but committed a suicide.

[31] This conference was organized in 1972 by a group of young researchers form the Institute of Computer Science of the Polish Academy of Sciences and the Department of Mathematics and Mechanics of Warsaw's University. Next year a similar conference was organized in Czechoslovakia witch gave rise to a long series of MFCS conferences. Since 1974 proceedings of these conferences have been published by Springer Verlag in the series Lecture Notes in Computer Science.

In this place, it would also be appropriate to mention two fields of research developed at Warsaw University. The first one is a formalised approach to program correctness based on a specific algorithmic logic [8] where programs appear in logical formulas. The second [52] is much more engineering-oriented and splits into three areas: *grammatical deduction, performance-analysis of computing systems* and *formal specification of software requirements*. An interesting application of the third approach is described in a paper by D.L. Parnas, G.J.K. Asmis, J. Madey [60] devoted to a safety assessment of the software for the shutdown systems of the Darlington Nuclear Power Generating Station (Canada).

The idea of proving programs correct — despite its undoubted scientific importance — was never widely applied in software engineering. In my personal opinion that was due to the tacit assumption that programs come first and their proofs come later. This order is quite natural in mathematics where theorem precedes its proof is rather unusual in engineering. Imagine an engineer who first constructs a bridge and only later performs all the necessary calculations. Such a bridge would probably collapse even before has been finished, and this is what happens with programs. The first version of code usually does not work as expected, hence a large part of program development budget is spent on testing and "debugging", i.e., on removing bugs introduced at the stage of writing the code. It is a well-known fact that all bugs can never be identified and removed by testing, hence the remaining bugs are removed on user's expense under the name of "maintenance".

In this book (Sec. 3 and Sec. 8) I am trying to develop ideas sketched earlier in my papers [18] and [19] where instead of proving programs correct, a programmer develops correct programs using rules that guarantee the correctness. In such a framework a software engineer can work as an engineer who builds bridges, cars or aeroplanes where products are developed by using rules that guarantee the correctness of these products.

Since rules for the development of correct programs are derived from the rules of proving programs correct I start below from the latter. The discussion is carried out on the ground of an algebra of binary relations since this leads to a relatively simple model where all technicalities of programming languages can be omitted. Of course, to apply these rules in a practical environment, they have to be expressed on the ground of a mathematical model of a programming language. A language **Lingua** with such a model is constructed in Sec. 6 and Sec 7. In Sec. 8 correct-program-development rules for **Lingua** are shown.

## 3.2    Iterative programs

Each program and also each of its instructions defines a certain binary *input-output relation* which transforms input states into output states. In a deterministic case, this relation is a function. In this simple model, one can express quite a few ideas associated with program correctness. I will start then with them.

Let $S$ be an arbitrary possibly infinite set of objects called *states*. In our applications, states are mappings (finite functions) which map identifiers into their values such as Booleans, numbers, strings, records, databases, etc. In the abstract case, however, we do not need to assume anything about $S$. In this section we shall consider binary relations over $S$, i.e., elements of the set:

$$Rel(S) = \{R \mid R \subset S \times S\}$$

Relations represent possibly nondeterministic programs and their components. The fact that

$$a \; R \; b \quad \text{for } a, b : S$$

means that there exists an execution of program R which starts in a and terminates in b. The lack of such a b means that the trial of running R with the initial state a results in either abortion or an infinite execution. In the future language **Lingua**, it is assumed that each abortion results in a final state that "carries" an error signal. In such a case the lack of a final state always means an infinite computation. In the general model, however, I do not make such an assumption.

In "prehistoric" informatics, i.e., in the years 1940/1950, programs were sequences of labelled instructions executed sequentially one after another unless a *jump instruction* goto interrupted that flow. With jump instruction and conditional instruction if-then, one could build an arbitrary graph of elementary instructions called a *flow-diagram*. Early papers on program correctness were devoted to such programs that later have been called *iterative programs*.

The most general relational model of an iterative program is the following fixed-point set of equations:

$$X_1 = R_{11} X_1 \mid \ldots \mid R_{1n} X_n \mid R_1$$

$$\ldots \qquad\qquad\qquad\qquad\qquad\qquad\qquad\qquad (3.2\text{-}1)$$

$$X_n = R_{n1} X_1 \mid \ldots \mid R_{nn} X_n \mid R_n$$

that corresponds to a graph whose nodes are numbers $1,\ldots,n$ and relations $R_{ij}$ are assigned to edges. Since here coefficients of variables stand on their left-hand side such equations are called *left-hand-side linear*. The codes of corresponding programs may be written as a sequences of labelled instructions of the form:

i : **do** $R_{ij}$ **goto** j.

Since $R_{ij}$ are not necessarily functions, such programs may have a non-deterministic character. For (3.2-1) to be deterministic two conditions must be satisfied:

- all $R_{ij}$ are functions,
- for every i, all $R_{i1},\ldots,R_{in}$ have disjoint domains.

If $(P_1,\ldots,P_n)$ is the least solution of (3.2-1), then $P_i$ is the input-output relation on the path from node 1 to node i. Therefore, if we assume that 1 represents the initial node and n is the final node, then $P_n$ is the resulting relation of our program. The class of iterative programs understood in that way together with their correctness-proof rules had been investigated in [20][32].

Programmers of the decade 1950/1960 were competing with each other in building more and more complicated flowchart programs that usually nobody except them was able to understand. Unfortunately quite frequently the authors themselves were not able to predict the behaviour of such programs.

As a reaction to these problems, first algorithmic programming languages such as Fortran and Algol were created. They were offering tools for *structured programming* such as sequential composition, if-then-else, and while[33]. Such programs were much easier to understand and also allowed for significant simplification of program-correctness proof rules.

In the sequel, we shall restrict our discussion to only three basic *structural constructors* since the other (e.g., for) may be defined with their help:

---

[32] This paper includes also a discussion of *right-hand-linear* equations i.e. of the general form X = XR | Q.

[33] The author who introduced the term "structured programming" was a Dutch computer scientist Edsger Dijkstra (see [36] and [37]).

1. sequential constructor denoted by a semicolon ";",

2. conditional constructor if-then-else-fi,

3. loop constructor while-do-od.

The sequential composition is the composition of relations (functions) as defined in Sec. 2.6. To define the remaining constructors, we have to introduce additional concepts because in our case boolean expressions are three-valued predicates (Sec. 2.9) rather than classical ones.

First, observe that every three-valued predicate on states may be represented by two disjoint set of states:

C = {s | p.s = tt}

¬C = {s | p.s = ff}

Of course, if p is a two-valued predicate then C | ¬C = S but for three-valued predicates, this is not the case. Here I recall that the third logical value corresponds in our model to a non-termination which means that program either aborts or loops indefinitely.

Let now P and Q represent arbitrary programs and the pair of disjoint sets of states (C,¬C) — an arbitrary three-valued predicate. Our three programs' constructors are now defined in the following way:

1. P ; Q                          = P Q
2. **if** (C,¬C) **then** P **else** Q **fi**   = [C] P  |  [¬C] Q
3. **while** (C,¬C) **do** P **od**       = ([C] P)* [¬C]

where [C] denotes the identity relation defined on C. The third constructor may also be defined by the fixed-point equation

X = [C] PX  |  [¬C]

Notice that the two remaining constructors can also be regarded as defined by (trivial) fixed-point equations without variables on the right-hand side:

X = P Q

X = [C] P  |  [¬C] Q

## 3.3    Procedures and recursion

Next step towards the development of programming techniques was the introduction of procedures and in particular — recursive procedures. In the most general case a procedure is a relation:

P : Rel(S)

which was given a name allowing to call it when running a program. Therefore in procedural languages, two new constructions appeared:

- procedure declaration,

- procedure call (instruction).

The former constituted a new sort of objects; the latter was just another type of instruction.

Procedures were declared by defining a certain macroinstruction, called *procedure body*, equipped with mechanisms of passing and returning parameters. At the level of our general model of procedures, I neglect these mechanisms since they can be represented simply by transformations of states into states, hence can be regarded as a prefix respectively suffix of a procedure body.

Since procedure calls could appear within instructions, they were also allowed in procedure bodies. In the beginning, procedures could not call themselves. E.g., that was the case in early algorithmic languages such as early Fortran and SAKO[34].

The option of calling procedures by themselves appeared for the first time in Algol 60 [61] and was referred to as *recursive procedures*. A decade later it has been built into Pascal [47].

On the ground of the algebra of relations recursive procedures may be regarded as solutions of fixed-point *polynomial equations* of the form:

$X_1 = \Psi_1.(X_1,…,X_n)$

…

$X_1 = \Psi_n.(X_1,…,X_n)$

where each $\Psi_i(X_1,…,X_n)$ is a combination of variables and constants by composition and union of relations. Such sets of equations may be regarded as single fixed-point equations in a CPO (Sec. 2.3) of relational vectors ordered component-wise, i.e., in the CPO over the carrier:

$Rel(S)^{cn} = \{(R_1,…,R_n) \mid R_i : Rel(S)\}$

Every such a set of polynomial equations defines, therefore, a vectorial function:

$\Psi : Rel(S)^{cn} \mapsto Rel(S)^{cn}$

$\Psi.(R_1,…,R_n) = (\Psi_1.(R_1,…,R_n),…, \Psi_n.(R_1,…,R_n))$

If each $\Psi_i$ is continuous in all their variables, then $\Psi$ is continuous as well, and therefore Kleene's theorem holds (Sec. 2.3).

Whereas the program correctness problem was widely investigated for iterative programs in the years 1960-1980, for programs with recursive procedures that was not the case. In my opinion, the situation was caused by the lack of structured constructors covering recursion. To partially cope with this problem I shall investigate in Sec. 3.5.2 and Sec. 3.6.2 a simple scheme of a recursive procedure with only one procedural call that corresponds to an equation of the form:

$X = HXT \mid E$ (3.3-2)

where H, T, E : Rel(S) are relations called the *head* the *tail* and the *exit* of procedure respectively. Although this is certainly not a general scheme for a recursive procedure, it seems quite common in practice. This scheme will be referred to as *simple recursion*.

Notice that (3.3-2) covers the case of the iterative instruction while with H = [C]P, T = [S] and E = [¬C].

At the end of this section, one methodological remark is necessary. Although in **Lingua** all programs are deterministic, hence correspond to functions rather than relations in the general theory of program correctness I restrict my investigations to functions only in the case of while, since in other cases determinism does not simplify the proof rules.

---

[34] SAKO was a programming language built in the "Department of Mathematical Apparatuses" of Polish Academy of Sciences and then implemented on a computer called XYZ and constructed also in that department.

## 3.4    Two concepts of program correctness

To express program correctness on the ground of binary relations, we shall use two operations of a composition of a relation with a set. Both are similar to sequential compositions of relations as defined in Sec. 2.6. In the sequel A, B, C,… will denote subsets of the set of states S and P, Q, R,… will denote relations in Rel(S). Both operations are denoted by the same symbol "•":

A•R = {b | (∃a:A) a R b}    — *left composition*; the image of A wrt R

R•B = {a | (∃b:B) a R b}    — *right composition*; the coimage of B wrt R.

In the sequel the symbol of composition "•" will be most frequently omitted; hence we shall write AR and RA.

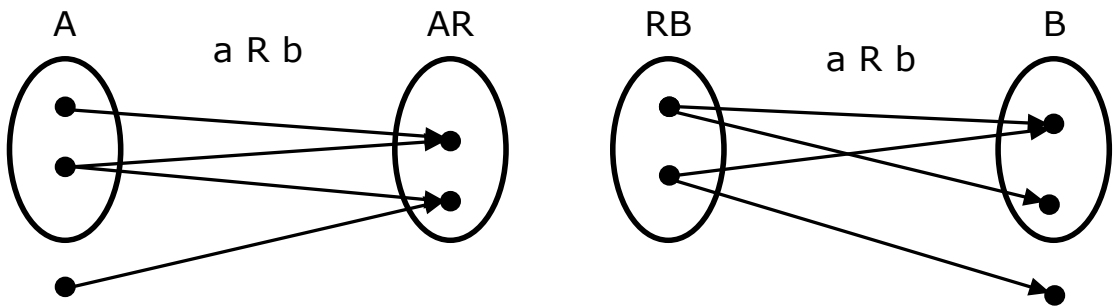

**Fig. 3.4-1 Left- and right composition of a set with a relation**

Intuitively speaking:

AR is the set of all final states of executions of R that start in A; notice however that due to the nondeterminism of R some of them may be at the same time final states of executions that start outside A,

RB is the set of all initial states of executions of R that terminate in B, but due to the nondeterminism of R, some of them may at the same time generate execution that terminates outside B.

Both compositions of a relation with a set have properties similar to that of the composition of relations. For instance, they are commutative:

A(RQ) = (AR)Q

(RQ)B = R(QB)

and distributive over unions of sets and relations:

(A | B) R  = (AR) | (BR)

A (R | Q) = (AR) | (AQ)

…

They are also monotone in each argument:

if  A ⊂ B   then   AR ⊂ BR

if  R ⊂ Q   then   AR ⊂ AQ

and analogously for right-hand-side composition. In fact, both operations are continuous in each argument. In the sequel, we shall assume that composition binds stronger than union hence we shall write

AR | BR instead of (AR) | (BR)

Now let us recall (Sec. 2.6) that

[A] = {(a, a) | a:A}

denotes a subset of identity relation (i.e., function) on sets restricted to A

**Lemma 3.4-1** *For any* A,B,C ⊂ S, *and* R : Rel(S) *the following equalities hold:*

(1) [A]B = A∩B

(2) A[B] = A∩B

(3) (A∩B)R = A [B] R

(4) R(A∩B) = R [A] B

(5) (A∩B)R ⊂ C *is equivalent to* A[B]R ⊂ C

(6) *if* A ⊂ [B]RC *then* (A∩B) ⊂ RC ∎

Proofs are left to the reader.

Now we are ready to define two fundamental concepts concerning the correctness of programs: *partial correctness* and *total correctness*. Both these concepts express the fact that if the input data of a program satisfy certain conditions, then the output data have expected properties. For instance, we may expect that a list-sorting program when given an appropriate list (precondition) will return a sorted list (postcondition).

Since with every property of states, we can unambiguously associate a set of states with that property, the correctness of a program P wrt a precondition A and postcondition B may be easily expressed in the algebra of relations and sets:

AP ⊂ B — *partial correctness* wrt precondition A and postcondition B

A ⊂ PB — *total correctness*   wrt precondition A and postcondition B

Partial correctness means that every execution that starts in A, if it terminates, then it terminates in B. Partial correctness is written as

[ParPre A] P [ParPost B]

A is called *partial precondition* and B is called *partial postcondition*.

Total correctness means that for every initial state in A there is an execution that terminates in B. Total correctness is written as

[TotPre A] P [TotPost B]

A is called *total precondition* and B is called *total postcondition*.

Notice that AP is, of course, the least set B such that AP ⊂ B. This set is called then the *strongest partial postcondition* for precondition A and program P. It represents the strongest postcondition that can be expected to be satisfied by executions that start in A.

Analogously PB is called the *weakest total precondition* for postcondition A and program P. It is the weakest precondition that guarantees the existence of an execution that terminates in B.

The definitions of partial and total correctness written using quantifiers are the following:

| | | |
|---|---|---|
| AP ⊂ B | — | (∀ a:A) if (∃ b) aPb then b:B |
| | | every execution of P that starts with a : A, if it terminates at all, terminates in B, but there may be no such execution, |
| A ⊂ PB | — | (∀ a:A) (∃ b:S) if aPb then b:B |
| | | every execution of P that starts with a, if it terminates, then it terminates in B, but there may be such executions that start with a but do not terminate in B. |

As we see, none of these properties is stronger than the other one. In the deterministic case, however (i.e., if P is a function) total correctness means that for every a : A the (unique) execution that starts with a terminates in B, hence every execution that starts in A terminates in B. Therefore if F is a function its total correctness implies its partial correctness:

$$\text{if } A ⊂ FB \text{ then } AF ⊂ B \tag{3.4-1}$$

The following implication is also true:

$$\text{if } AF ⊂ B \text{ and for every } a : A, F.a \text{ is defined, then } A ⊂ FB \tag{3.4-2}$$

In other words, if F is a total function, then its partial correctness implies its total correctness. Indeed, let a : A. Then there is b such that b = F.a. However since AF ⊂ B we have b : B and therefore a : FB. In this way we have proved the following theorem:

**Theorem 3.4-2** *Deterministic program* F *is totally correct wrt* A *and* B *iff it is partially correct wrt* A *and* B *and for every* a : A *its final state is defined.* ∎

As a consequence a proof of total correctness of a deterministic program F may be split into two steps:

    I.     proof of partial correctness, i.e., AF ⊂ B,

    II.    proof that F is defined on A, i.e., A ⊂ FS

Let us observe now that if in a programming language we introduce a mechanism of abstract errors, then using partial correctness we may express the fact that with a chosen precondition no finale state (if it exists) carry an error message, i.e., that the program will not hang up. In such a case the only way F may be undefined is where program loops indefinitely.

If F.a is defined, then we say that program represented by F satisfies in the state a *the termination property.*

In many practical programs, the termination property is so obvious that its proof may be safely skipped. For instance, if all loops in a program are **while** instructions that run over finite sets of data, then every loop must eventually terminate.

It is very important to know however that there exist programs where the proof of termination may be extremely difficult. One such example has been displayed on the front of Warsaw's University Library. The hypothesis of the total correctness of this program, i.e., of its termination, is as follows:

**TotPre** `n > 0`

    `x := n;`

```
while x > 1 do;
    if x mod 2 = 0 then x := x/2 else x := 3x + 1
TotPost x = 1
```

This hypothesis is just another formulation of *Collatz hypothesis* published in 1937 and not solved until today even though it was investigated by many distinguished mathematicians including Statnisław Ulam. So far it has been proved only that the hypothesis is true for all numbers $n < 5*2^{60}$.

A similar situation concerns *Fermat's theorem*[35] that was announced in the year 1637 and proved only in 1994 by a British mathematician Andrew Wiles. His proof is 100 pages long and uses an advanced topological theory of elliptic curves.

On the ground of the theory of computability, it has been proved (Alan Turing) that there is no algorithm which for every program and every input state could effectively — i.e., in a finite number of steps — decide whether this program stops for this input state.

**Theorem 3.4-2** *In the general case the termination property of programs is not decidable.* ∎

In the sequel, proof rules for program correctness will be expressed by showing in which way the correctness of composed programs may be proved by proving the correctness of their components. In the most general case such rules will be written in the following form:

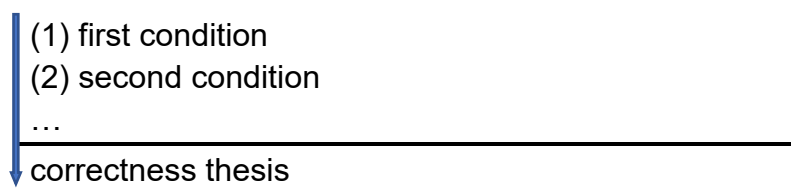

(1) first condition
(2) second condition
…
correctness thesis

where the arrow shows the direction of implication. In some rules, we have both-sided arrows which mean that the implication is of the iff type. As we shall see a little later, top-down-arrow rules show how to build correct programs from correct components.

## 3.5    Partial correctness

When defining program correctness proof rules, it is worth distinguishing between two classes of program constructors: *simple constructors* which do not introduce repetition mechanisms and *recursive constructors* which introduce such mechanisms. The formers are defined by simple combinations of the composition and union of relations, the latter require fixed-point equations. From this perspective, iteration is a particular case of recursion.

### 3.5.1    Sequential composition and branching

The most frequently used simple constructors of programs are sequential composition and branching.

---

[35] This theorem claims that for no integer n > 2 there exist three positive integers x, y, z that satisfy the equality $x^n + y^n = z^n$. That theorem had been written by Fermat on the margin of a book who also wrote that he found a "marvellously simple proof" of the theorem which was however too long to fit to the margin.

**Rule 3.5.1-1 Partial correctness of a composition**

*For arbitrary* A,D ⊂ S *and* P,Q : Rel(S) *the following rule is satisfied:*

> *there exist conditions* B *and* C *such that:*
> (1) [ParPre A]  P  [ParPost B]
> (2) [ParPre C]  Q  [ParPost D]
> (3) B ⊂ C
> ─────────────────────────────────
> (3) [ParPre A]  P ; Q  [ParPost D]

**Proof** The assumptions above the line expressed in the language of relations are:

   (1) AP ⊂ B

   (2) CQ ⊂ D

Therefore and from the monotonicity of composition

   (AP) Q ⊂ CQ ⊂ D

hence from the associativity of composition

   A (PQ) ⊂ D.

To prove the bottom-to-top implication, it is sufficient to set

   B = C = AP

Hence AP ⊂ B and BQ = APQ ⊂ D  ∎

**Rule 3.5.1-2 Partial correctness of if-then-else**

*For arbitrary* A,D,C,¬C ⊂ S *and* P,Q : Rel(S), *if* C ∩ ¬C = Ø, *then the following rule is satisfied:*

> (1) [ParPre A ∩ C]   P  [ParPost B]
> (2) [ParPre A ∩ ¬C]  Q  [ParPost B]
> ─────────────────────────────────
> [ParPre A]   **if** (C,¬C) **then** P **else** Q **fi**   [ParPost B]

**Proof.** Observe that (1) and (2) can be written as:

   A [C] P ⊂ B

   A [¬C] Q ⊂ B

and if we add these inclusions side by side we get

   A ([C] P | [¬C] Q) ⊂ B  ∎

At the end three more rules which follow directly from the monotonicity of composition of a set with a relation.

**Rule 3.5.1-3 Strengthening a partial precondition**

*For every* P : Rel(S) *and any* A,B,C ⊂ S *the following rule holds:*

$$\begin{array}{l} \text{[ParPre A]   P   [ParPost B]} \\ \text{C} \subset \text{A} \\ \hline \text{[ParPre C]  P  [ParPost B]} \end{array}$$

**Rule 3.5.1-4 Weakening a partial postcondition**

*For every* $\text{P} : \text{Rel(S)}$ *and any* $\text{A,B,C} \subset \text{S}$ *the following rule holds:*

$$\begin{array}{l} \text{[ParPre A]   P   [ParPost B]} \\ \text{B} \subset \text{C} \\ \hline \text{[ParPre A]  P  [ParPost C]} \end{array}$$

**Rule 3.5.1-5 The conjunction of pre- and postconditions**

*For every* $\text{P} : \text{Rel(S)}$ *and any* $\text{A,B,C,D} \subset \text{S}$ *the following rule holds:*

$$\begin{array}{l} \text{[ParPre A]   P   [ParPost B]} \\ \text{[ParPre C]   P   [ParPost D]} \\ \hline \text{[ParPre A}\cap\text{C]  P  [ParPost B}\cap\text{D]} \end{array}$$

As we see a proof of partial correctness of a structured program without recursion consists of three steps:

A.  finding the appropriate intermediate pre- and postconditions,

B.  proving partial correctness of the components of the program,

C.  proving implications between appropriate pre- and postconditions.

The pre- and postconditions that appear in correctness proofs are called *assertions*.

Basic problem in proving programs correct consists in finding the appropriate assertions. It may be a mathematical problem — what sort of properties should they express[36] — as well as a practical problem — how to express in readable form conditions with usually very many variables. In this place, it is worth mentioning also that assertions included in programs may play the role of programs' specifications.

In the current section, I have omitted the problem of proving properties of atomic components of programs such as, e.g. assignments or variable declarations. It was the consequence of the fact that in the language of binary relations between abstract states such rules cannot be expressed. This issue is postponed to Sec. 8 where a programming language will come to the play.

## 3.5.2   Recursion and iteration

In order to formulate proof rules for mutually recursive procedures I generalise the operation of composition of relations with relations and with sets to the case of vectors of respectively relations and sets:

---

[36] The fact that Collatz hypothesis has not been proved yet means that an appropriate assertion has not been found which could be used as an invariant of the loop. About invariants in Sec. 3.5.2 and Sec. 3.6.2

$(P_1,...,P_n) (R_1,...,R_n) = (P_1R_1,...,P_nR_n)$

and analogously for the composition of a relations with sets. In an obvious way we can also generalise the inclusion of sets to the inclusion of vectors:

$(A_1,...,A_n) \subset (B_1,...,B_n)$ means $A_1 \subset B_1$ and ... and $A_n \subset B_n$

For simplicity, the inclusion between vectors of sets is denoted by the same symbol as the inclusion of sets. In the sequel vectors of sets and relations as well as operations on them will be written with boldface characters.

A vector of relations **R** is said to be *partially correct* relative to the vectors of sets **A** and **B** (with appropriate numbers of elements) iff **A R** $\subset$ **B** what shall be written as

[ParPre **A**]  **R**  [ParPost **B**]

and analogously for *total correctness*. The notion of a continuous function is generalised to the case of vectorial functions in an obvious way.

Now we can formulate partial-correctness proof rule in the general case of such vectors of relations which are fixed-points of arbitrary continuous function. Although this case covers polynomial equations the assumption that an equation is polynomial would not contribute to the simplicity of the rule. For concrete, simple polynomials, such rules will be shown a little later in this section.

### Rule 3.5.2-1 Partial correctness of a vector of relations defined by a fixed-point equation

*For every continuous function* **Ψ** : Rel(S)$^{cn}$ $\mapsto$ Rel(S)$^{cn}$, *if* **R** *is the least solution of the equation* **X = Ψ.X**, *then for any* **A,B** : S$^{cn}$ *the following rule holds, where* **Ø** = (Ø,..., Ø) *is the* n*-element vector of empty relations:*

> *there exists a family of conditions* {**B**$_i$ | i ≥ 0} *such that*
> (1) (∀i ≥ 0) [ParPre **A**] **Ψ**$^i$**.Ø** [ParPost **B**$_i$]
> (2) **U**{**B**$_i$ | i ≥ 0} $\subset$ **B**
> ─────────────────────────────────────
> (3) [ParPre **A**]  **R**  [ParPost **B**]

**Proof** Form Kleene's theorem (Sec. 2.3)

**R** = **U** {**Ψ**$^i$**.Ø** | i ≥ 0}

Adding the components of (1) sidewise we obtain

**A U**{**Ψ**$^i$**.Ø** | i ≥ 0} $\subset$ **U**{**B**$_i$ | i ≥ 0}

hence together with (2), we have (3). To prove the bottom-to-top implication, we assume

**B**$_i$ = **A** (**Ψ**$^i$**.Ø**) for i ≥ 0 ∎

From this rule, we obtain immediately a rule for single recursion, i.e., where n = 1:

### Rule 3.5.2-2 Partial correctness of a relation defined by a fixed-point equation

*For every continuous function* Ψ : Rel(S) $\mapsto$ Rel(S), *if* R *is the least solution of the equation* X = Ψ.X, *then for any* A,B $\subset$ S *the following rule holds:*

> *there exists a family of conditions* $\{B_i \mid i \geq 0\}$ *such that*
> (1) $(\forall i \geq 0)$ [ParPre A] $\Psi^i.\emptyset$ [ParPost $B_i$]
> (2) $U\{B_i \mid i \geq 0\} \subset B$
> ──────────────────────────────
> (3) [ParPre A]   R   [ParPost B]

We can also formulate more specific rules for each particular polynomial function, e.g. for the simple-recursion constructor as defined in Sec. 3.3. Below two versions of such a rule:

**Rule 3.5.2-3 Partial correctness of a relation defined by simple recursion (version 1)**

*For any* $H,T,E : \text{Rel}(S)$, *if the relation* R *is the least solution of the equation*

   $X = HXT \mid E$

*then for any* $A,B \subset S$ *the following rule holds:*

> *there exists a family of conditions* $\{B_i \mid i \geq 0\}$ *such that*
> (1) $(\forall i \geq 0)$ A $H^i$ E $T^i \subset B_i$
> (2) $U\{B_i \mid i \geq 0\} \subset B$
> ──────────────────────────────
> (3) [ParPre A]   R   [ParPost B]

**The proof** follows immediately from Rule 3.5.2-2 and from the fact that, as is easy to prove,

   $R = U\{H^i \text{ E } T^i \mid i \geq 0\}$ ∎

The following one-directional rule with a stronger assumption may be useful as well:

**Rule 3.5.2-4 Partial correctness of a relation defined by simple recursion (version 2)**

*For any* $H,T,E : \text{Rel}(S)$, *if the relation* R *is the least solution of the equation*

   $X = HXT \mid E$

*then for any* $A,B \subset S$ *the following rule holds:*

> (1) $(\forall Q)$ AQ $\subset$ B *implies* A HQT $\subset$ B
> (2) AE $\subset$ B
> ──────────────────────────────
> (3) [ParPre A]   R   [ParPost B]

**Proof** From (1) and (2) we can prove by induction that for every $i \geq 0$:

   A $(H^i \text{ E } T^i) \subset B$

and therefore by side-wise summation we get (3). ∎

   This rule may be written in an alternative way which was pointed to me by Andrzej Tarlecki.

**Rule 3.5.2-4A Partial correctness of a relation defined by simple recursion (version 3)**

*For any* H,T,E : Rel(S), *if the relation* R *is the least solution of the equation*

$$X = HXT \mid E$$

*then for any* A,B ⊂ S *the following rule holds:*

> (1) AH ⊂ A *and* BT ⊂ B
> (2) AE ⊂ B
> ─────────────────────────
> (3) [ParPre A]  R  [ParPost B]

**Proof** (1) and (2) imply the inclusions A (H$^i$ E T$^i$) ⊂ AE T$^i$ ⊂ B T$^i$ ⊂ B. It may be also proved that (1A) implies (1). ∎

Setting H = [C]P, T = [S] and E = [¬C] from both these rules we can draw rules for while-do-od iteration:

## Rule 3.5.2-5 Partial correctness of while-do-od loop (version 1)

*For every relation* P : Rel(S) *and any disjoint* C, ¬C ⊂ S, *if relation* R *is the least solution of equation*

$$X = [C]PX \mid [¬C],$$

*then for any* A,B ⊂ S *the following rule holds:*

> *there exists a family of conditions* {B$_i$ | i ≥ 0} *such that*
> (1) (∀ i ≥ 0) A ([C]P)$^i$ [¬C] ⊂ B$_i$
> (2) U{B$_i$ | i ≥ 0} ⊂ B
> ─────────────────────────
> (3) [ParPre A]  R  [ParPost B]

## Rule 3.5.2-5 Partial correctness of while-do-od loop (version 2)

*For every relation* P : Rel(S) *and any disjoint* C, ¬C ⊂ S, *if relation* R *is the least solution of the equation*

$$X = [C]PX \mid [¬C],$$

*then for any* A,B ⊂ S *the following rule holds:*

> (1) (∀ Q) AQ ⊂ B *implies* A [C]QR ⊂ B
> (2) A[¬C] ⊂ B
> ─────────────────────────
> (3) [ParPre A]  R  [ParPost B]

∎

In the literature, the following rule is also well known. This time it is written with the pre- and postcondition notation:

## Rule 3.5.2-5 Partial correctness of while-do-od loop (version 3)

*For every relation* $P : Rel(S)$ *and any* $A, B, C, \neg C \subset S$, *if* $C \cap \neg C = \emptyset$, *then the following rule is satisfied:*

> *there exists* $N \subset S$ *(called loop invariant) such that:*
> (1) [ParPre N∩C]  P  [ParPost N]
> (2) A ⊂ N
> (3) N [¬C] ⊂ B
> ─────────────────────────────────────────
> (4) [ParPre A]  **while** (C, ¬C) **do** P **od**   [ParPost B]

         ∎

**Proof** Let (1) – (3) be satisfied. From (1) by induction we can prove:

$$N([C]P)^i \subset N \text{ for all } i \geq 0$$

Therefore and from (2)

$$A([C]P)^i \subset N \text{ for all } i \geq 0$$

hence from (3)

$$A([C]P)^i[\neg C] \subset N[\neg C] \subset B \text{ for all } i \geq 0$$

In summing these inclusions sidewise, we get (4). Now assume that (4) is satisfied and let us denote:

$$(5) \; N = A([C]P)^*$$

Therefore and from (4) we get $N[\neg C] \subset B$, hence (3). In turn (5) is equivalent to

$$N = A \mid A([C]P)^+,$$

hence (2). To prove (1) notice that:

$$(N \cap C)P = N[C]P = A[C]P \mid A([C]P)^*[C]P = A([C]P)^+ \subset N \quad \blacksquare$$

## 3.6    Total correctness

Rules for total correctness are used to prove that if some preconditions are satisfied, then at least one program's execution terminates with postconditions being satisfied. Remember that in the general case the undefinedness of a final state can mean both an error signal and an infinite computation. Remember also that in the case of nondeterminism total correctness of $R$ expressed by

$$[TotPre A] R [TotPost B] \quad \text{i.e.} \quad A \subset RB$$

only means that for any $a : A$ there exists an execution that starts in $a$ and ends in $B$, but there may also be executions which either end outside of $B$ or do end at all. If however, $R$ is a function, then total correctness means that for any $a : A$ the (unique) execution which starts with $a$ terminates in $B$.

## 3.6.1    Sequential composition and branching

**Rule 3.6.1-1 Total correctness of a composition**

*For any* A,D ⊂ S *and* P,Q : Rel(S) *the following rule holds:*

> *there exist conditions* B *and* C *such that*
> (1) [TotPre A]   P  [TotPost B]
> (2) [TorPre C]   Q  [TotPost D]
> (3) B ⊂ C
> ―――――――――――――――――――――――――――
> (3) [TotPre A]   PQ   [TotPost D]

∎

**Proof.** Two first assumptions above the line written in algebraic form are

   (1) A ⊂ P B

   (2) C ⊂ Q D.

Therefore immediately:

   A ⊂ P B ⊂ P C ⊂ P (Q D) = (P;Q) D.

Now assume that A ⊂ (P;Q) D, which means that A ⊂ P (Q D). Assuming B = C = QD we get (1) and (2). ∎

**Rule 3.6.1-2 Total correctness of** if-then-else[37]

*For any* A,D,C,¬C ⊂ S *and* P,Q : Rel(S), *if* C ∩ ¬C = Ø, *then the following rule is satisfied:*

> (1) [TotPre A ∩ C]    P  [TotPost B]
> (2) [TotPre A ∩ ¬C]  Q  [TotPost B]
> (3) A ⊂ C | ¬C
> ―――――――――――――――――――――――――――
> (4) [TotPre A]   **if** (C, ¬C) **then** P **else** Q **fi**   [TotPost B]

**Proof.** Let:

   (1) A ∩ C ⊂ PB

   (2) A ∩ ¬C ⊂ QB

   (3) A ⊂ C | ¬C

Therefrom:

   [C] (A ∩ C)    ⊂ [C] PB

   [¬C] (A ∩ ¬C) ⊂ [¬C] QB

Adding the inclusions sidewise:

   [C] (A ∩ C)  | [¬C] (A ∩ ¬C) ⊂ [C] PB | [¬C] QB = ([C]P | | [¬C] Q) B

The following equalities are also true

―――――――――――――――――――――――――――

[37] Notice that in case of two-valued predicates, i.e. defined for every state, condition (3) is not necessary, since C | ¬C = S.

[C] (A ∩ C) = A ∩ C

and analogously for ¬C. Hence and from (3)

[C] (A ∩ C)  | [¬C] (A ∩ ¬C) = (A ∩ C)  | (A ∩ ¬C) = A

and finally

(4) A  ⊂ [C] PB | [¬C] QB

In turn (4) implies A ⊂ C | ¬C, and from (4) and the fact that C and ¬C are disjoint, follow (1) and (2). ∎

At the end three more rules for pre- and postconditions analogous to the respective rules for partial correctness.

### Rule 3.6.1-3 The strengthening of a total precondition

*For every* P : Rel(S) *and any* A,B,C ⊂ S *the following rule holds:*

> [TotPre A]  P  [TotPost B]
> C ⊂ A
> ───────────────────────────
> [TotPre C]  P  [TotPost B]

### Rule 3.6.1-4 The weakening of a total postcondition

*For every* P : Rel(S) *and any* A,B,C ⊂ S *the following rule holds:*

> [TotPre A]  P  [TotPost B]
> B ⊂ C
> ───────────────────────────
> [TotPre A]  P  [TotPost C]

### Rule 3.6.1-5 The conjunction of conditions

*For every* P : Rel(S) *and any* A,B,C,D ⊂ S *the following rule holds:*

> [TotPre A]  P  [TotPost B]
> [TotPre C]  P  [TotPost D]
> ───────────────────────────────
> [TotPre A∩C]  P  [TotPost B∩D]

## 3.6.2   Recursion and iteration

Similarly, as in the case of partial correctness, we start from the case of a general recursive operator.

### Rule 3.6.2-1 Total correctness of a vector defined by a fixed-point equation

*For every continuous function* **Ψ** : Rel(S)$^{cn}$ ↦ Rel(S)$^{cn}$, *if* **R** *is the least solution of* **X = Ψ.X**, *then the following rule holds, where* **∅** = (∅,…,∅):

> *there exists a family of conditions* $\{A_i \mid i \geq 0\}$ *such that*
> (1) $(\forall i \geq 0)$ [TotPre $A_i$] $\Psi^i.\varnothing$ [TotPost **B**]
> (2) $A \subset U\{A_i \mid i \geq 0\}$
> ──────────────────────────────
> (3) [TotPre **A**]   **R**  [TotPost **B**]

**Proof**  If **R** is the least fixed point of $\Psi$, then from the continuity of $\Psi$

(4) $R = U\{\Psi^i.\varnothing \mid i \geq 0\}$

Adding sidewise inclusions (1) we have

$U\{A_i \mid i \geq 0\} \subset U\{\Psi^i.\varnothing \mid i \geq 0\}$ **B**

hence from (2) we have (3). Now assume that $A \subset RB$ which means that

$A \subset U\{\Psi^i.\varnothing \mid i \geq 0\}$ **B**

Let for $i \geq 0$

$A_i = (\Psi^i.\varnothing)$ **B**

Then obviously $A_i \subset (\Psi^i.\varnothing)$ **B**. ∎

From this rule for $n = 1$ we immediately have

### Rule 3.6.2-2 Total correctness of a relation defined by a fixed-point equation

*For every continuous function* $\Psi : \mathsf{Rel(S)} \mapsto \mathsf{Rel(S)}$, *if* $R$ *is the least solution of an equation* $X = \Psi.X$, *then the following rule holds:*

> *there exists a family of conditions* $\{A_i \mid i \geq 0\}$ *such that*
> (1) $(\forall i \geq 0)$ [TotPre $A_i$] $\Psi^i.\varnothing$ [TotPost B]
> (2) $A \subset U \{A_i \mid i \geq 0\}$
> ──────────────────────────────
> (3) [TotPre A]   R  [TotPost B]

### Rule 3.6.2-3 Total correctness of a procedure defined by simple recursion (version 1)

*If relation* $R$ *is the least solution of the equation* $X = H \: X \: T \mid E$ *then the following rule holds:*

> *there exists a family of conditions* $\{A_i \mid i \geq 0\}$ *such that*
> (1) $(\forall i > 0)$ $A_i \subset H^i \: E \: T^i$ B
> (2) $A \subset U \{A_i \mid i \geq 0\}$
> ──────────────────────────────
> (3) [TotPre A]  R  [TotPost B]

**Proof** immediately from rule 3.6.1-2 and the fact that

$R = U\{H^i \: E \: T^i \mid i \geq 0\}$     ∎

### Rule 3.6.2-4 Total correctness of a procedure defined by simple recursion (version 2)

*If relation* $R$ *is the least solution of the equation* $X = H \: X \: T \mid E$ *then the following rule holds:*

(1) $(\forall Q)$ $(A \subset Q \; B \; \text{implies} \; A \subset HQT \; B)$
(2) $A \subset EB$
___________________________________________
(3) [TotPre A]  R  [TotPost B]

**Proof.** From (1) and (2) we can prove by induction, that for every $i \geq 0$

$A \subset (H^i \; E \; T^i) \; B$

and by sidewise summation, we get  (3).  ∎

To discuss total correctness rules for while-do-od we introduce a new notion. We say that the components a loop $(C, \neg C, P)$ satisfy *termination condition* in $D$, if

$D \subset ([C]P)^*[\neg C]S$

Notice that if $P$ is a function, then the termination condition means that every execution of the instruction **while** $(C, \neg C)$ **do** $P$ **od** that starts in $D$ will terminate. Indeed, if $s : D$, then

$s : ([C]P)^*[\neg C]S$

and in that case, there exists $n \geq 0$, such that $s : ([C]P)^n[\neg C]S$. Since $P$ is a function and $C$ is disjoint with $\neg C$, the power index $n$ is determined unambiguously, and hence the unique execution of our loop that starts with $s$ corresponds to $n$ executions of $[C]P$ followed by one execution of $[\neg C]$.

Observe that if $P$ represents a program with internal loops, then the termination condition guarantees the termination of all these loops as well. Termination condition may also be written as

[TotPre D]  **while** $(C, \neg C)$ **do** $P$ **od**  [TotPost S]

The rule for the total condition of while will be restricted to deterministic programs since this leads to its significant simplification.

### Rule 3.6.2-5 Total correctness of a while-do-od loop

*If* $F : Rel(S)$ *is a function, then for any* $A, B, C, \neg C \subset S$, *where* $C \cap \neg C = \varnothing$ *the following rule holds:*

there exists a condition $N$ *(an invariant) such that*
(1) [TotPre N ∩ C]  F  [TotPost N]
(2) $N \subset C \mid \neg C$
(3) $A \subset N$
(4) $N \cap \neg C \subset B$
(5) $(C, \neg C, F)$ *satisfies termination condition in* $N$
___________________________________________
(6) [TotPre A]  **while** $(C, \neg C)$ **do** F **od**  [TotPost B]

**Proof** Assume that (1) – (5) are satisfied. We have to prove                    ∎

(6) $A \subset ([C]F)^*[\neg C]B$

Basing on (3) and (4) this problem may be reduced to

(7) $N \subset ([C]F)^*[\neg C](N \cap \neg C)$

However, since $[\neg C](N \cap \neg C) = [\neg C]N$, (7) is equivalent to

(8) $N \subset [\neg C]N \mid ([C]F)^*[\neg C]N$

Let then $s : N$. From that and from (2) $s : N \cap C$ or $s : N \cap \neg C$. If $s : N \cap \neg C$, then $s : [\neg C]N$, what completes the proof. Let $s : N \cap C$. Basing on (5)

$s : ([C]F)^*[\neg C]S$.

hence there exists a state $s_1$, such that $s ([C]F)^n s_1$ holds for a certain $n > 0$. Now observe that since $F$ is a function, then from total correctness expressed by (1) we may draw the conclusion about corresponding partial correctness which can be written as

$(N \cap C)F \subset N$

or as

$N[C]F \subset N$

Therefore $s_1 : N$, hence $s : ([C]F)^*[\neg C]N$, what terminates the top-down proof. Now assume the satisfaction of (6) i.e.

$A \subset ([C]F)^*[\neg C]B$

Let's denote:

$N = ([C]F)^*[\neg C]B$.

In that case:

(3) follows directly from (6),

(2) follows directly from the definition of $N$,

(5) follows from the definition of $N$ and the inclusion $B \subset S$,

(4) follows from the fact that $C$ and $\neg C$ are disjoint and from the equations

$N \cap \neg C = ([C]F)^*[\neg C]B \cap \neg C = [\neg C]B \subset B$

(1) follows from the equations

$N \cap C = ([C]F)^*[\neg C]B \cap C = ([C]F)^+[\neg C]B = [C]F([C]F)^*[\neg C]B = [C]PN \subset PN$     ∎

Notice that since $F$ is a function, then the condition $N$ is an invariant of the loop in the sense of partial correctness.

In many practical situations, it is not very convenient to prove termination condition directly from its definition. In that case a useful vehicle may a lemma using the concept of a chain-restricted set.

Let $(U, >)$ be a set with a binary relation defined in it. We shall say that this set is *chain-restricted* if there is no infinite sequence $u_1, u_2, \ldots$ in it such that

$u_i > u_{i+1}$ for $i = 1, 2, \ldots$.

As is easy to show, if $(U, >)$ is chain-restricted, then the relation $>$ is:

*antireflexive,* i.e. no $u$ satisfies $u > u$,

*antisymmetric,* i.e. for any $u, w$ if $u > w$, then $w > u$ does not hold.

**Lemma 3.6.2-1** *If* $F$ *is a function,* $N \cap C \subset FN$, $C \cap \neg C = \emptyset$ *and* $N \subset C \mid \neg C$, *then the components of the loop* $(C, \neg C, F)$ *satisfy termination condition in* $N$, *iff there exists a chain-restricted set* $(U, >)$ *and a total function*

$$K : S \longmapsto U$$

*such that* $N \subset \text{dom}.K$ *and for any* $a, b : S$

*if* $a \; F \; b$ *then* $K.a > K.b$  ∎

   **Proof** Assume that the assumptions of our lemma is satisfied but the inclusion

$$N \subset ([C]F)^*[\neg C]S.$$

does not hold. In that case, there exists $s_0 : N$, i.e. $s_0 : C \mid \neg C$, that does not belong to $([C]F)^*[\neg C]S$, hence it does not belong to $\neg C$, and therefore $s_0 : N \cap C$. Consequently, $s_1 = F.s_0$ is defined and belongs to $N$. Therefore $s_1 : C \mid \neg C$. However, $s_1$ cannot belong to $\neg C$, since then $s_0$ would belong to

$$[C]F[\neg C]S$$

which is a subset of $([C]F)^*[\neg C]S$. Reasoning in this way we could prove the existence of such a sequence $(s_i : i = 0, 1, \ldots)$ such that

$$s_i \; F \; s_{i+1} \text{ for } i = 0, 1, \ldots$$

This however would imply the existence of a sequence

$$K.s_i > K.s_{i+1} \text{ for } i = 0, 1, \ldots$$

which is not possible.

   Assume now that $N \cap C \subset PN$ and $N \subset C \mid \neg C$ and that the components of the loop $(C, \neg C, F)$ satisfy terminating condition in $N$. In that case $(N, F)$ must be chain-restricted since otherwise the existence of an infinite chain

$$s_i \; F \; s_{i+1} \text{ for } i = 0, 1, \ldots$$

that starts in $N$ would mean that $s_0$ does not belong to $([C]F)^*[\neg C]S$. Therefore $U = N$ and $K$ is an identity.  ∎

# 4  General remarks about denotational models

This section introduces the reader into the general theory of denotational models based on the theory of abstract algebras. In the sequel of the book we shall see how using these models we may construct programming languages with two basic categories of programming tools:

1. *applicative tools* covering datalogical and typological expressions whose denotations are functions from states into data and into types respectively,

2. *imperative tools* covering instructions and declarations whose denotations are functions from states into states.

## 4.1    How did it happen?

Mathematicians working on mathematical models for programming languages were usually assuming — as in mathematical logic — that a programming language should be described by three mathematical objects:

1. Syn — *syntax* which in this book is a context-free syntactic algebra,

2. Den — *denotations* which in this book constitute an algebra with the same signature as the corresponding algebra of syntax,

3. Sem : Syn ↦ Den — *semantics* that associates denotations to syntactic objects and in this book is a many-sorted homomorphism between two mentioned algebras.

Intuitively speaking a denotational semantics describes the meaning of every complex syntactic object as a composition of the meanings of it parts. This property of semantics — called *compositionality* — permits for the descriptions of complex objects by means of so called *structured induction.*

At this point, it should be mentioned that denotational (compositional) models of semantics, which for mathematicians have been always an obvious choice, have not been chosen in the first formal model of a programming languages. Similarly to the prototypes of sewing machines that were mechanical arms repeating movements of a tailor and to the first steamboat engine droving oars, the first formal definition of a programming language was a description of a virtual computer executing programs[38].

---

[38] First metalanguage used to write such semantics was developed by IBM laboratory Vienna and was called Vienna Definition Language (VDL). Later some members of that team have created a lab on the Danish Technical University in Lyngby with the aim of writing "more denotational" semantics in a meta-language called Vienna Development Method (VDM) [10]. This language was used among others applications to describe the semantics of programming languages ADA and CHILL. In the case of the former, which was expected to become a universal programming language of all times, the process of writing its semantics resulted in repairing many inaccuracies of the language, and in developing first Ada compiler. Unfortunately, both Chill and Ada were excessively complex, and hence were fairly quickly forgotten.

This model of semantics, called later *operational semantics*, was abandoned after a few years of experiments, because the description of the virtual machine was not less complex then the code of a real compiler and still it was not a description of the actual machine[39].

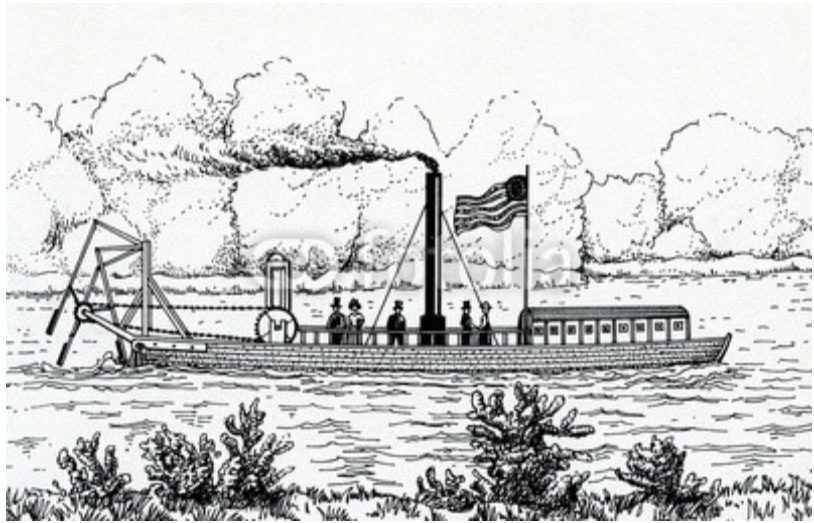

**Fig. 4.1-1 Steamboat moving oars**

The road to denotational semantics was however not simple either. As I mentioned earlier, the first denotational models of programming languages were characterized by great mathematical complexity. Technically this was the consequence of the assumption that two following mechanisms are undisputable features of high-level programming languages:

1. the jump instruction `goto` that could transfer program execution from any line of code to any other; this mechanism was present in virtually all programming languages in the years 1960/70, and was inherited from low-level languages, where it was the only tool for building logical structures of programs,
2. procedures that may take themselves as parameters; this construction was present in Algol 60 frequently considered at that time as an untouchable standard.

The requirement to describe `goto` led to technically quite a complex model called *continuations*[40]. In this model, each instruction was seen not as a transformation of states, but as an operation adding its own effect to the effect of all instructions that follow it in the program (called its continuation). The meaning of a program was then built starting from the end. Consequently, continuation semantics was not only technically complex but above all far from programmers' intuition. Independently of this at the turn of 1960-ties to 1970-ties, IT professionals began to be aware of the danger imposed by the use of goto instruction (see [36]). Programs with goto had been difficult to understand what caused that often behave not as expected by programmers. As a result, structured programming (see Sec. 3.2) based on if-the-else and while was becoming more and more popular.

Continuation model, although technically complex, was based on traditional mathematics. This cannot be said about the model of procedures which may take themselves as parameters.

---

[39] To be precise this remark is true for sequential programming only (without concurrent processes), i.e. such that we shall deal with in this book. An operational semantics for concurrent programs was developed by Plotkin ???.

[40] First author who introduced that concept — although under a different name of *tail functions* — was Antoni Mazurkiewicz [52]

Here it should be made clear that we are not talking about recursive procedures that call themselves in their bodies — such constructions can be described by fixed-point equations — but constructions of type f(f), where a function takes itself as an argument. Such functions were not known to mathematicians, because they cannot be described on the ground of classical set theory, not to mention the fact that mathematicians never needed such functions. It is also worth noting that if by F we would denote a set of self-applicable functions, then such a set had to satisfy the fixed-point domain equation:

$$F = (F \mid A) \rightarrow B$$

where $\rightarrow$ is not continuous. Therefore the existence of its solutions is not guaranteed by Kleene's theorem (Sec. 2.3).

In Algol 60 the construction f(f) was implemented in such a way, that procedure f was receiving as a parameter not exactly itself, but a copy of its own code inserted into its body during compilation. Such an operation was called *copy rule*. Mathematicians of the decade of 1960-ties were initially fascinated by this construction because it was challenging the existing concept of a function. As a result of this fascination, the theory of so-called *reflexive domains* was created by Dana Scott and Christopher Strachey [56] and was later described in details by J.E. Stoy in monograph [63][41]. A group of mathematicians have been developing research in this direction, however, for software engineers, reflexive domains were even more difficult and less intuitive then continuations. Pretty soon it turned out also that the ability to upload procedure to itself as a parameter leads to even greater dangers in practice of programming than the use of goto. Consequently, in later programming languages, self-applicable procedures were abandoned. Unfortunately, this has led to abandoning denotational semantics as well.

The denotational model that is described in this book emerged as a modification of two earlier models. Instead of describing instructions by continuations I assume that they represent state-to-state functions where states are mappings assigning values to variables also called *valuations*.

The concept of a valuation of variables was well known to mathematicians from mathematical logic since the pioneering work of Alfred Tarski [65]. In those times the meanings of expressions were described as functions mapping the valuations of variables:

$$v : \text{Valuation} = \{x, y, z\} \rightarrow \text{Value}$$

into values. E.g. the meaning of the expression

$$2x+4y$$

is a function

$$F[2x+4y] : \text{Valuation} \rightarrow \text{Number}$$

such that

$$F[2x+4y].v = 2 * v(x) + 4 * v(y)$$

From there only one step to an observation that the meaning of an instruction

---

[41] To my colleagues mathematicians I may explain that the idea of reflexive domains was in fact a direct realization of copy rule. The authors of this model used the fact that functions definable by programs are computable, hence can be "numbered" with natural numbers, i.e. each function f may be given a unique number n(f). In this model f(f) meant f(n(f)) which can be modelled on the ground of classical set theory. That was in fact a mathematical application of copy rule since n(f) may be regarded as the code of f.

```
x := 2x + 4y
```

is such a transformation of valuation that the value of `x` in the new valuation is the value of the expression `2x+4y` in the previous one. This idea was applied in my paper [14] published in 1971 where I described a prototype-denotational semantics of a very simple programming language.

In turn, the inspiration to abandon the model of continuations came to me from the book of Michael Gordon [40], in which the author treats Scott's recursive domains as "usual sets" with the following commentary on page 29:

*"We shall not discuss the mathematics involved in Scott's theory at all; our approach to recursive domains is similar to an engineering approach to differential equations, namely we assume they have solutions but don't bother with the mathematical justification."*

I have read this book in the year 1981 during a train ride from Copenhagen to Århus, where I was going to meet Peter Mosses a strong proponent of the theory of Scott. The book was for me an important break-through since for the first time I was reading a semantics of a programming language with the understanding not only of its mathematics but also of the informatic content. It is true that the greater part of the book was dedicated to continuation semantics. However, the very treatment of reflexive domains as "usual sets" was a serious simplification. I also get the impression that this informal treatment did not lead to any mathematical problems. Only later I realised that Gordon did not actually deal with self-applicable functions.

The approach of Michael Gordon, although intuitively simple, was mathematically not quite acceptable since the assumption that reflexive domains are usual sets is simply not true. It wasn't therefore quite clear if his model did not lead to inconsistencies which is certainly critical when building a model with the aim of developing a logic of programs.

To cope with this problem Andrzej Tarlecki and myself published in 1983 a paper [25], in which we showed a denotational model of programming languages, where domains of denotations are sets in the sense of classic set theory, and the denotations of instructions are state-to-state transformations. This approach stimulated in 1980-ties the creation of a metalanguage **MetaSoft** [18] in the Institute of Computer Science of the Polish Academy of Sciences. And this is the approach I have chosen to base my book on.

## 4.2    From denotations to syntax

All early works on the semantics of programming languages concerned building semantics for existing languages. That has led to a tacit assumption that syntax should come first and denotations are defined later. Of course, there is a certain logic in this way of thinking since how can we build a model for something that does not yet exist? After all, astronomers were describing the mechanics of celestial bodies when the Sun and the planet were already there.

This way of thinking has, however, a certain vulnerability because computer science — what I have already mentioned previously — should not be compared to astronomy, physics, or biology, where we describe the world around us. Building a programming language is an engineering task such as constructing a bridge or an aeroplane. Would any engineer ever think of first building a bridge basing on common sense and only then making all necessary calculations? Such a bridge would certainly collapse, as I wrote already in Sec. 1.1

In my approach, I decided to reverse the traditional order where we first build the syntax and only later its mathematical model, i.e. denotations. I will show how to build a language starting from its algebra of detonation and only in the second step generating a syntax adequate for these denotations.

A sample programming language that will be built in this book has been named **Lingua**. I have chosen this name to commemorate the circumstances under which from October to December 1969 I wrote my first denotational semantics of a very simple programming language. This work was later published in Dissertationes Mathematicae [11] as my habilitation (postdoctoral) thesis. By three months as a scholar of the Italian Government, I was working in the Istituto di Elaborazione dell'Informazione in Pisa. I didn't yet know the works of Dana Scott or the concept of denotational semantics, and I constructed my language and its semantics on a model theory known in mathematical logic. Only eighteen years later, in the year 1987, I described (in [19]) the idea of proceeding from detonation to syntax.

## 4.3    Languages of the family Lingua

As has been announced in Sec.4.2 the method of building a denotational model of a programming language will be shown on the example of **Lingua**. This language will be constructed layer-by-layer starting with applicative mechanisms and enriching them by successive imperative constructions. Each successive layer will constitute an enrichment of the former by new mechanisms:

| | |
|---|---|
| **Lingua-A** | an applicative part of the future language including datalogical and typological expressions hence the models of data and types; |
| **Lingua-1** | structural instructions, declarations of variables and definitions of types; |
| **Lingua-2** | imperative procedures with mutual recursion and functional procedures with simple recursion; |
| **LinguaV-2** | tools for building correct (validated) programs in **Lingua-2;** |
| **Lingua-3** | object-oriented programming; |
| **Lingua-SQL** | application programming interface (API) for SQL databases. |

From the algebraic perspective, the algebra of detonation of each of these languages will be an extension (in the sense as defined in Sec. 2.11) of the preceding algebra in the series. In other words, each of our languages will be constructed from the former by adding new elements to the existing carrier, and/or new carriers, and/or new constructors. This scalability of algebras should lead to the scalability of possible implementations.

In this place, I should emphasise that **Lingua** is not regarded as a future standard of a denotations-based language but only as a field of experiments in which to show how such a standard could possibly be built.

## 4.4    Why do we need denotational models of programming languages?

Denotational model of a programming language serves as a starting points for the realisation of three tasks:

1.  building the implementation of the language, i.e. its parser and interpreter or compiler,

2. creating rules of building correct specified programs in this language,

3. writing a user manual.

In building a language in this way we should observe one very important (although not quite formal) principle:

---

**The principle of simplicity**

*A programming language should be as simple and easy to use as possible, however without damaging its functionality, mathematical clarity and completeness of its description. The same applies to the manual of the language and to the rules of building correct programs.*

---

This principle shall be realised by caring to make:

1. the syntax of the language as close as possible to the language of intuitive mathematics, for example, whenever this is common, we allow for infix notation and the omission of "unnecessary" parentheses,

2. the structure of the language (i.e. programs' constructors) leading to possibly simple rules of constructing correct programs (Sec. 8),

3. the semantics of the language easy to understand by the user rather than convenient for the builder of implementation; for the latter an implementation-oriented equivalent model should be written (what is meant by that will be explained later in the book).

Special attention should be given to point 2 because the simplicity of the rules of building correct programs leads to a better understanding of programs by programmers. This fact was realised already in the years 1970 and has led to the elimination of goto instructions. This decision resulted in a major simplification of programs' structures, which increased their reliability. As turned out this change did not limit the functionality of programming languages.

Following point 3, I will sometimes — as common in mathematics — "forget" about the difference between syntax and denotations. E.g. I will talk about the value of an expression x + y, rather than about the value of its detonation. I would say that the instruction x:=y+1 modifies variable x, instead of saying that the denotation of this instruction modifies the memory state at variable x, etc. Of course, at the model's level syntax will be precisely distinguished from denotations.

## 4.5    Five steps to a denotational model

Building up **Lingua** I refer to an algebraic model as described in Sec. 2.10 to Sec. 2.14. This model corresponds to the diagram of three algebras shown in Fig. 4.5-1. We build it in such a way that the equation:

As = Co ● Cs

is satisfied, which guarantees the existence of a denotational semantics of our language.

The construction of a denotational model begins with an algebra of detonation Den. Its constructors unambiguously determine the reachable subalgebra ReDen, from which we unambiguously derive the *abstract syntax algebra* AbsSy. The first of these steps is creative since it comprises all the major decisions about the future language. Contrary to it the second step can be performed algorithmically[42].

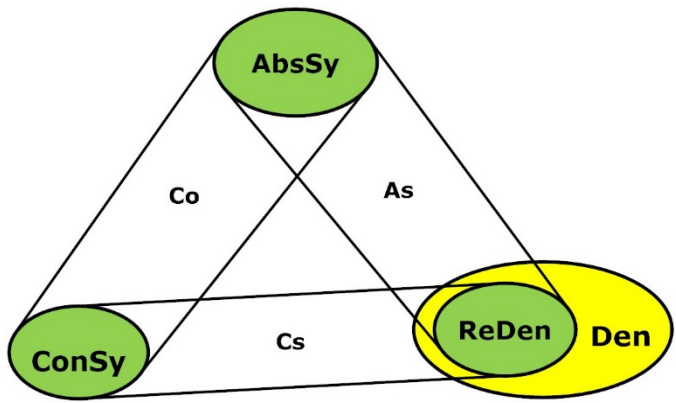

**Fig. 4.5-1 An algebraic model of a programming language**

As we saw in Sec. 2.12 abstract syntax is not very convenient for programmers. To make it more user-friendly, we build a *concrete syntax* ConSy. In typical situations, this is done by replacing prefix notation by infix notation and skipping some "unnecessary" parentheses. A very simple example of such a transformation was showed in Sec. 2.13, where concrete syntax was a homomorphic image of abstract syntax. The corresponding homomorphism Co (*con-cretisation*) was glueing not more than As (*abstract semantics*), and therefore there existed a unique homomorphism:

Cs : ConSy ⟼ ReDen

(*concrete semantics*), that was the semantics of concrete syntax. In this way, we have created the main components of our denotational model. Notice that the step from abstract syntax to concrete syntax is creative — although rather simple.  An example of a transition from abstract syntax to concrete syntax has been shown in Sec. 2.13 and consisted in skipping the parentheses in arithmetic expressions such as, e.g. ((a + b) + c). This construction, however, was possible only because expressions in that language contained only addition which is commutative[43], i.e.

(a + b) + c = a + (b + c).

In Sec 2.14 we have seen an expression language with two operations — addition and multiplication — which forced us to build a new algebra of denotations in order to allow the "usual" omission of parentheses. In addition, this model was not very intuitive.

Such a solution was used in context-free grammars of Algol 60 [61] and Pascal [43] with semantics described in an informal way. At that times language designers were assuming that grammars should serve both programmers and implementors.  As we have seen in Sec. 2.13, however, this requirement forces language designers to think about concrete syntax when building denotations. This interferes with our philosophy "from detonation to syntax", where we first decide about the content of the language and only then about how to express it by means of syntax.

---

[42] Of course a corresponding algorithm does not take an abstract algebra as an input, but its signature described in a metalanguage — in our case in **MetaSoft**. This technique will be explained in details in Sec. 5.4.1

[43] As we are going to see in Sec. 5.4.3.3, the addition of numbers in a computer is not commutative, which is due to the effect of overload. Here we use an abstract addition only to explain the idea of a colloquial syntax.

An alternative to the above procedure, which has been already mentioned in Sec. 2.14, is a syntax with specific notational conventions which we call *colloquialisms* (Fig. 4.5-2). The introduction of colloquialisms into concrete syntax ConSy leads to *colloquial syntax* ColSy which most frequently is not homomorphic to concrete syntax, and even has a different signature. There must be, however, an implementable transformation

Rt : ColSy ⟼ ConSy

which removes colloquialisms, e.g. by adding the missing parentheses. Such a transformation is called *restoring transformation* and of course, is not a homomorphism.

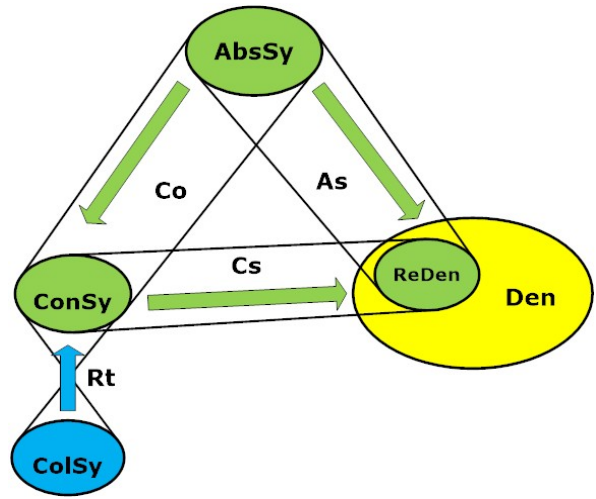

**Fig. 4.5-2 An algebraic model of a language with colloquial syntax**

In such a case in programmer's manual of a language, concrete-syntax — which constitutes the core of the syntax — is defined by a context-free grammar and colloquialisms are described informally. For instance, we explain that in writing arithmetic expressions we can skip parentheses while maintaining the priority of multiplication and division over addition and subtraction.

In such a model the builder of implementation receives a standard denotational model of a language plus a formal definition (algorithm) of restoring transformation. In such a case the execution of programs consists of three steps:

1. a pre-treatment of the source code by a restoring transformation,

2. a parsing the resulting concrete-syntax code into an abstract-syntax code,

3. an interpretation or compilation of the abstract-syntax code.

The construction of the a full denotational model of a language proceeds in five steps called the *five-step method*.

1. In the first step, we build an algebra of detonations Den that includes objects of the future language as well as their constructors. In that step, major decisions are taken about the functionality of the language. Language designer must specify the repertoire of constructors in Den (of functions between carriers) in such a way that the corresponding (unique) reachable subalgebra ReDen contains all the objects that we want to access through syntax.

2. The signature of algebra Den determines uniquely the algebra of abstract syntax AbsSy and the corresponding homomorphism (abstract semantics) As. The step from ReDen to AbsSy can be performed in a fully algorithmic way. From the perspective of language designer, this step does not require any creativity and may be assisted by a software tool.

3. The abstract syntax is usually not very user-friendly since is restricted to prefix notation and requires the use of many "unnecessary" parentheses. To cope with these inconveniences, a concrete syntax ConSy is created which much closer to programmers' syntax. Algebraically this syntax is a homomorphic image of abstract syntax and must be built in such a way that the corresponding homomorphism Co (*concretisation*), glue no more than As. Constructing a concrete syntax is thus the second creative task of language designer and usually is performed in a series of steps which consists in creating successive concretisations of abstract syntax.

4. If Co glues no more than Sa, then by Theorem 2.13-1 there is a unique homomorphism Cs from concrete syntax into detonation and more specifically onto its reachable subalgebra. This is *concrete semantics* or simply the *semantics* of our language. Its description can be algorithmically generated from the metalanguage descriptions of As and Co.

5. In the last step, we introduce colloquialisms and describe the restoring transformation. This step is creative.

As we can see, creative tasks of language designers take place in the first, third and fifth step. The steps second and fourth can be performed algorithmically.

After having built a denotation model of a language, one can proceed to the definitions of correct-program constructors (see Sec. 8). This step corresponds to a historic task of building programs' logic in Hoare's style.

## 4.6    Notational conventions of our metalanguage

In the description of our sample language **Lingua** we use three levels of formalisation each associated with different fonts:

1. at the level of the concrete and colloquial syntax of **Lingua,** we use `Courier New`,

2. at the level of formal definitions of remaining components of our model, i.e. the algebras of detonation and abstract syntax with the corresponding semantics, we use Arial while notational conventions come from half-formal language **MetaSoft**, which was already mentioned in Sec. 4.1,

3. at the level of informal descriptions and comments, we use Times New Roman.

Indices, which in traditional mathematics are written with a reduced font and at a lowered level for example $a_i$, will be treated as arguments of functions by writing $a.i$, where $a$ is regarded as a function and $i$ — as its argument.

Due to a great variety of symbols occurring in software's definitions, in place of typical one-character symbols as in usual mathematics, e.g. $a$, $b$, $c$, ... we use many-character symbols like ide, sta, sto,... which is a technique well known to programmers.

The names of sets always start with a capital letter, for example Number or InsDen (instructions' denotations) and the names of their elements with small letters.

Following the convention used in VDM (the Vienna Development Method; see [10]) the metavariables that run over domains are "announced" in the definitions of domains by writing, e.g.:

ide : Identifier   = Letter © Character*

val : Valuation   = Identifier ⟹ Data

which means that ide runs over Identifier and val over Valuation. At the end of the book, there is a list of all most frequently used alphanumeric symbols.

As has been mentioned already in Sec. 2.8, values which are strings of characters are closed in apostrophes to distinguish them from metavariables. E.g. ide is a metavariable that runs over the domain of identifiers, and 'abcd' is a concrete string of four letters.

In order to shorten conditional definitions of functions that in full version are written as a list of if-then-otherwise clauses:

condition-1   ➜ value-1

…

condition-n   ➜ value-n

we also allow a compact notation:

condition-i   ➜ value-i      for i = 1;n

This will be clear when it comes to examples.

# 5  Algebraic-denotational model of data structures

## 5.1    The general idea of the model

In early programming languages such as Fortran, Algol 60, Pascal or Cobol the concept of a type was introduced in the first place to allocate appropriate memory space to variables. With boolean variables, single-bit registers were assigned, with numeric variables — many-bit registers and finally with array variables — a larger memory spaces depending on the size of this array. Over time it turned out, however, that assigning types to variables allows not only for a better management of memory space but also contributes to a deeper understanding of programs' functionality by programmers.

Today, when memory management is no longer so critical (except for, e.g. databases), this second aspect is still important. It also happens that type varies when the corresponding data is changed. For example, adding a new attribute to a record changes not only the content of the record but also its type. In such situations we need a type-tracking mechanism synchronised with data processing.

In the denotational models of **Lingua** this mechanism is implemented by building two algebras:

- an algebra of composites which are pairs consisting of a data and of its structure called *body*[44],

- an algebra of types which are pairs consisting of a body and a predicate called *yoke*.

Intuitively types are associated with sets of data. A numeric type is the the set of numbers and list type — a collection of lists. However, handling types understood in this way would not be very practical. Therefore in our model types are independent mathematical beings uniquely determining sets of data called the *clans of types*. This model allows programmers not only to define their own types but also to store types in computer memory for later use. This also allows to build complex types in a bottom-up way, i.e. by composing simple types into complex ones.

In the course of investigating types understood in this way — and to tell the truth, after many failed attempts which took me nearly a year of work — I've come to the conclusion that to use types as the descriptions of data sets, it is convenient to regard them as pairs consisting of:

1. a description of data structure such as *number, word, array, list, record* or *tree*; formally such descriptions are tuples, mapping and their combination and are called the *bodies* of types,

---

[44] Some inspiration for the introduction of this model was for me the idea used in the definition of programming language Ada written in a metalanguage VDM (see [10] and [12]). In that case however there were two semantics: a dynamic semantics to compute data and a static semantics to compute types. The former was describing program execution, the latter a procedures carried out at compile time. This can be convenient for the implementator of a language, but seems rather far from programmer's perspective which I am trying to stick to in this book.

2.  a description of other properties of data, e.g. that a number belongs to a certain interval, that array consists of integers only or that the sum of values assigned to two chosen attributes of a record does not exceed a certain value; formally each such descriptions are predicate on data and are called the *yokes* of types.

The first way of describing types is quite common for most programming languages, the second is specific to database languages, e.g. based on SQL. As a matter of fact, in the latter case, both ways are used.

To introduce the described concepts into our denotational model, we define five algebras which constitute an *algebraic model of data structures*:

1.  *data algebra* — an algebra of numbers, Booleans, words, lists, records, etc.,

2.  *bodies algebra* — an algebra of structures corresponding to all sorts of data; e.g. structures of numbers are different from the structures of number arrays,

3.  *composites algebra* — an algebra of pairs (data, body) where body describes the structure of data,

4.  *transfers algebra* — an algebra of one-argument functions mapping composites into composites; the earlier mentions yokes are transfers that return Boolean composites as their values,

5.  *types algebra* — an algebra of pairs (body, yoke); in that case, no relation between body and yoke are assumed.

Next, to the indicated by our algebras five sorts of objects, we create a sixth one called a *value* which is a pair (data, type) where data is of the type type. Values may also be regarded as triples (data, body, yok) where:

*   the structure of data is described by body,

*   the composite (data, body) satisfies yok.

For values, no algebra is defined since its operations shall be implicitly described in the algebra of expression detonations. This is only a technical decision aimed at the simplification of our model.

Next two algebras that we shall need in our model are the algebra of *data expression denotations* and of *type-expression denotations* with elements that are functions mapping states into data and types respectively:

ded : DatExpDen = State → Composite | Error

ted : TypExpDen = State ↦ Type | Error

The assumption that data-expression denotations are partial functions is the consequence of the fact that in Sec. 7.5 we shall introduce expressions which are functional-procedures calls and therefore may generate infinite executions.

We shall assume that states assign values and types to identifiers. In the case of values, we talk about *data variables* and in the case of types — about *type constants.* Variables may change their values during program executions whereas types assigned to constants remain unchanged.

Anticipating future definitions, it should be pointed out in this place that although data variables store values, the expression will evaluate to composites rather than to values. Hence if a value (data, body, yok) is assigned to an identifier ide then the execution of the assignment

```
ide := expression
```

leads to the creation of a new composite (data-1, body-1) and a new value (data-1, body-1, yoke) that is assigned to ide only under the condition that the new composite (data-1, body-1) satisfies the inherited yoke yoke. If yoke is not satisfied, then an error signal is generated. As we see, yokes are not engaged in expression evaluation and serve only to protect type discipline on the level of instructions.

In this place, the reader may ask a question why bodies have not been treated analogously as yokes, but have been included in composites. Well, the reason for that decision was to have an easily implementable mechanism that allows checking if data constructors in an expression "receive appropriate arguments" and to generate an error signal otherwise. E.g. we can check if two data that are to be added are numbers rather than words, records or lists (details in Sec. 5.2.3)

In the subsequent sections, we shall see a simplified version of our model of data structures which should, however, allow seeing how to expand it to most constructions occurring in existing programming languages or (possibly) how to create new solutions.

At the end, it should be emphasised that **Lingua-A** which we are going to construct here, is not a prototype of a stand-alone applicative programming language, but only an example of an applicative part of an imperative language. In **Lingua-A** we have a mechanism for expression evaluation but not for state transformation. The latter mechanisms will be introduced in **Lingua-1** (Sec. 6).

## 5.2    The algebras of data structures

### 5.2.1    The algebra of data

Let us start from defining a certain standard family of data domains that will constitute the fundament for a future algebra of expression denotations. It should be emphasised that domains which are defined below are supersets of future domains of data generable by programs in **Lingua**. Due to that assumption, we can define our domains by simple domain equations (Sec. 2.7). Here is the list of that domains:

| | | |
|---|---|---|
| boo | : Boolean | = {tt, ff} |
| num | : Number | — the set of all numbers with finite decimal representations |
| ide | : Identifier | — a fixed finite subset of the domain Alphabet$^+$ |
| wor | : Word | = {'}Alphabet*{'} |
| lis | : List | = Data$^{c*}$ |
| arr | : Array | = Number $\Longrightarrow$ Data |
| rec | : Record | = Identifier $\Longrightarrow$ Data |
| dat | : Data | = Boolean \| Number \| Word \| List \| Array \| Record |

Alphabet is a fixed finite set of characters (except quotation marks), while Identifier is a finite fixed set of non-empty strings over Alphabet. A word is a string (possibly empty) of the elements of Alphabet closed by apostrophes. We assume that the sets Alphabet and Identifier are the parameters of our model and have been fixed once and for all.

Notice that identifiers are not included in data. Identifiers that appear in records will be called *record attributes*.

The first three domains on our list will be called *simple domains* and their elements — *simple data*. We introduce therefore yet another domain:

dat : SimpleData = Boolean | Number | Word

Lists, arrays and records constitute *structural domains,* and their elements are called *structural data*. The latter split again into two classes: tuples (lists) and mappings (arrays and records).

All domains which are defined above, except Data and Identifier, will be referred to as *data sorts,* e.g. *numeric sort*, *word sort, array sort* etc.

Notice that as of now a list may contain data of different sorts, the domain of indices of an array may be any finite set of numbers, not necessarily integers, a record may associate any data with its the attributes. All data can also be "arbitrarily large". In the future, many of these features will be restricted by choosing appropriate constructors in algebras. Of course, constructors as such do not restrict the carriers of algebras, but they restrict reachable subsets of carriers (Sec. 2.12), i.e. sets of these elements of algebras that are generated in the course of programs' executions.

List, arrays and records may be empty. This assumption has only a technical character.

Notice also that arrays are one-dimensional. On the other hand, since their elements can be arrays as well, we can create arrays of any dimension. For example, a two-dimensional array is a one-dimensional array of one-dimensional arrays.

Once we have specified data domains, we have to define data constructors. Their choice is a key engineering decision taken in the first phase of language design. Below we specify an example of a list of constructors for **Lingua-A**. It is rather modest to keep our model simple enough and should be regarded as a parameter of the model.

The operations that are defined below will be referred to as *theoretic operations* since they cannot be "fully" implemented. For instance, the theoretical division:

divide : Number x Number → Number

may generate an arbitrarily large number or a number with an arbitrarily large decimal representation, hence in both cases not representable in a computer. Meanwhile, in most programming languages — and this is going to be the case in **Lingua-A** as well — the value of the expression:

x / y

must not exceed a certain fixed number (or a certain fixed "length" of a decimal representation) and must be "computable" in any situation, i.e. even if x and y are not numbers or if y is zero. Of course, in such situations, the expression should generate an error message. All these restrictions will be built into our model on subsequent stages of its creation.

The list of theoretical operations starts from zero-argument constructors. Their idea was explained in Sec. 2.10.

| create-id.ide | : ↦ Identifier | for ide | : Identifier |
|---|---|---|---|
| create-bo.boo | : ↦ Boolean | for boo | : Boolean |
| create-nu.num | : ↦ Number | for num | : NumberS |
| create-wo.wor | : ↦ Word | for wor | : WordS |

In these formulas NumberS and WordS are subsets of Number and Word respectively with elements syntactically representable, i.e. such that they can be "typed to programs from the keyboard" (S stands for "syntax"). What are these sets, will depend on the implementation. At the general level, we assume only that they are finite. We don't assume, however, but also do not rule out, that the remaining constructors of numbers and words may generate data only from those sets. We accept the situation that not all numbers and words that may be generated during programs' executions have to be elements of NumberS or WordS respectively.

The domain of identifiers contains only those elements that can be entered from the keyboard and therefore in this case the suffix "S" is not needed.

The remaining constructors have arities and types as defined below. At this stage, we accept partial constructors. Although that decision is not quite consistent with the definition of many-sorted algebras (Sec. 2.11), it does not lead to any problems, since data algebra has an auxiliary character and its constructors are only used to define composite constructors which are going to be total.

| | | | |
|---|---|---|---|
| and | : Boolean x Boolean | $\mapsto$ Boolean | |
| or | : Boolean x Boolean | $\mapsto$ Boolean | |
| not | : Boolean | $\mapsto$ Boolean | |
| | | | |
| equal | : SimpleData x SimpleData | $\mapsto$ Boolean | |
| less | : Number x Number | $\mapsto$ Boolean | |
| | | | |
| add | : Number x Number | $\mapsto$ Number | |
| divide | : Number x Number | $\rightarrow$ Number | (partial function) |
| | | | |
| glue | : Word x Word | $\mapsto$ Word | |
| | | | |
| create-li | : Data | $\mapsto$ List | |
| push-li | : Data x List | $\mapsto$ List | |
| top-li | : List | $\rightarrow$ Data | (partial function) |
| pop-li | : List | $\rightarrow$ List | (partial function) |
| | | | |
| create-ar | : Data | $\mapsto$ Array | |
| put-to-ar | : Array x Data | $\mapsto$ Array | |
| change-in-ar | : Array x Number x Data | $\rightarrow$ Array | (partial function) |
| get-from-ar | : Array x Number | $\rightarrow$ Data | (partial function) |
| | | | |
| create-re | : Identifier x Data | $\mapsto$ Record | |
| put-to-re | : Identifier x Data x Record | $\mapsto$ Record | |
| get-from-re | : Record x Identifier | $\rightarrow$ Data | (partial function) |

| cut-from-re | : Record x Identifier | → Record | (partial function) |
|---|---|---|---|
| change-in-re | : Record x Identifier x Data | → Record | (partial function) |

Notice that at this stage we do not need abstract errors since we are dealing with operations which can be partial. We assume that Boolean and numerical constructors are defined "in the usual way". Three-valued Boolean constructors will be introduced only in the algebra of composites, where we are going to have abstract errors. The Boolean constructor equal is restricted to simple data, which is an engineering decision rather than a mathematical necessity.

The glue constructor is such a concatenation of strings that removes internal apostrophes. The remaining constructors can be easily defined using operations on tuples and mappings described in Sec. 2.1. Their definitions are the following:

| create-li.dat | = (dat) |
|---|---|
| push-li.(dat, lis) | = lis © (dat) |
| top-li.lis | = top.lis |
| pop-li.lis | = pop.lis |

| create-ar.dat | = [1/dat] | | |
|---|---|---|---|
| put-to-ar.(arr, dat) | = arr[ind/dat] | where | ind = max.(dom.arr) + 1 |
| get-from-ar.(arr, num) | = arr.num | | |

Notice that arr.num may be undefined in which case also get-from-ar.(arr, num) is undefined.

As is apparent from the above definitions, all reachable arrays, i.e. arrays created during the execution of programs, will be mappings the domains of which are intervals of the form [1,…, n].

Since records, similarly as arrays, are mappings, the definitions of their constructors will be similar:

| create-re.(ide, dat) | = [ide/dat] |
|---|---|
| put-to-re.(ide, dat, rec) | = rec[ide/dat] |
| get-from-re.(rec, ide) | = rec.ide |
| cut-from-re.(ide, rec) | = rec[ide/?] |
| change-in-re.(rec, ide, dat) | = |

rec.ide= ? ➔ ?

**true**      ➔ rec[ide/dat]

The operation change-in-re has been defined in such a way that it is undefined whenever the indicated attribute is not present in the record. This is, of course, an engineering decision, rather than a mathematical necessity, because rec[ide/dat] is also defined when rec.ide is undefined[45]. In **Lingua** any attempt of exchanging exchange the value of a non-existent attribute will lead to an error signal.

For technical reasons, that will be explained a bit later, we assume that data algebra is built over two carriers:

  ide   : Identifier

  dat   : Data

and that all its constructors are regarded as partial operations on Cartesian products of such carriers. For instance, the division function is now

  divide : Data x Data → Data

that "typologically" is an operation on arbitrary data but its domain of definedness (i.e. domain as defined in 2.1.3) is restricted to numbers with an additional assumption that the second argument must be different from zero. In turn, the operation of getting a data from a record will be a partial function:

  get-from-re : Data x Identifier → Data

that returns a data only whenever its first argument is a record and the second argument appears as an attribute of the first argument. In this way, all the constructors of our data algebra are partial functions.

The assumption that data algebra is two-sorted is a consequence of the fact that the future algebra of detonation is going to be two-sorted. This is explained later at the end of Sec. 5.3.2.

As has been mentions already, we are dealing here with a somewhat generalised concept of an algebra since all its constructors are partial functions. This generalisation does not prevent, however, to talk about the signatures of such algebras, and hence also about the similarities of algebras, of which one or both are partial. As a matter of fact, in further considerations, the data algebra — which we shall denote by DatAlg — will be the only algebra with partial constructors.

The choice of carriers and constructors of data algebra is one of the most important engineering decision when creating a programming language. At that step, we decide about the applicative part of the language, i.e. about its algebras of composites, bodies, yokes, types, expression denotations and at the end — about its abstract syntax.

At the end one methodological comment. As a matter of fact, the two-sortedness of data algebra has been assumed for the convenience of language designer rather than of a programmer. However, at the level of programmer's manual, we can still show a many-sorted signature as it was the case at the beginning of the present section.

## 5.2.2   The algebra of bodies

*Bodies* describe "internal structures of data" and are used in the definitions of types. For each sort of data we define the corresponding sort of bodies. Bodies are going to be tuples,  records and their combinations. The domain of bodies is defined by following equation:

  bod : Body   =

---

[45] I assume such a solution based on a guess that once a programmer has used the exchange operations to change the value of a non-existent attribute, then he/she probably thought that this attribute occurs in the record. If therefore this is not the case, an warning error should be generated.

| | |
|---|---|
| {('Boolean')} \| {('number')} \| {('word')} \| | (*simple bodies*) |
| {'L'} x Body \| | (*list bodies*) |
| {'A'} x Body \| | (*array bodies*)     (5.2.2-1) |
| {'R'} x (Identifier $\Rightarrow$ Body) | (*record bodies*) |

Bodies of simple data are one-element tuples of words. Symbols 'L', 'A' and 'R' are called *body initials* and serve to distinguish between bodies of structural data and bodies of their elements. E.g. ('A ', ('number')) is the body of numeric arrays, and ('L', ('A ', ('number'))) is the body of lists, whose elements are numeric arrays.

In the case of a list-body ('L ', bod) we say that bod is the *inner body* of the list-body. A similar convention is assumed for array-bodies. The elements of the domain

bor : BodRec = Identifier $\Rightarrow$ Body

are called *body-records*. Hence every record-body is of the form ('R', bor).

The definitions of body domains anticipate the future principle that all elements of a list will have a common body.  The same rule will be true for arrays, whereas in the case of records to each attribute a different body can be assigned.

At this point, it should be underlined that restrictions imposed on bodies are due to an engineering decision rather than a mathematical necessity. This decision seems consistent with fairly common standards of universal programming languages and also permits to describe mechanisms of SQL, which will be discussed in Sec. 12.[46]

Notice that array body does not specify the number of array elements. The introduction of such restrictions will be possible, however, with the help of yokes (see Sec. 5.2.4).

A little later with every operation on data, we shall assign an operation on bodies defined in such a way that in calculating a "new" data its body may be calculated "in parallel". Since operations on bodies will generate error messages, we introduce a universal set of errors:

err : Error

with the only assumption that its elements are words over a certain alphabet closed in apostrophes. An example of such an error may be 'no-such-attribute'. At this moment we do not need to define the set of errors more specifically since the constructors defined in the sequel will only transfer errors independently on their "content" (c.f. reactive error-handling mechanism in Sec. 2.8). Now we introduce the domain of bodies with errors:

bod : BodyE = Body \| Error

and assume that in the algebra of bodies denoted by BodAlg we have only two carriers:

ide  : Identifier

bod : BodyE

The constructors of this algebra are defined in such a way that to every data constructor (theoretical operation) ope we associate a body constructor Bc[ope] that "computes" the body of the result of ope using the bodies of its arguments.

---

[46] In the denotational model for SQL database tables will be (simplifying a little) lists of records with equal bodies, and databases will be records of tables with different bodies.

To associate data with bodies, we assign to each body a set of data called the *clan of this body*. Since domains of bodies are disjoint, we can define a function of body-claning CLAN-Bo that to each body assign its clan:

CLAN-Bo : BodyE $\mapsto$ Sub.Data

This function is defined by structural induction

CLAN-Bo.err           = $\varnothing$                    for every error err : Error

CLAN-Bo.('Boolean')    = Boolean

CLAN-Bo.('number')     = Number

CLAN-Bo.('word')       = Word

CLAN-Bo.('L', bod)     = (CLAN-Bo.bod)$^{c^*}$

CLAN-Bo.('A', bod)     = Number $\Rightarrow$ CLAN-Bo.bod

CLAN-Bo.('R', [ide-1/bod-1,…,ide-n/bod-n]) =

           { [ide-1/dat-1,…,ide-n/dat-n] | dat-i : CLAN-Bo.bod-i    for i = 1;n}

We assume that the clan of an empty record-body (i.e. the body where n = 0) is the one-element set consisting of an empty record and similarly for arrays.

As we can see, the clans of different bodies are disjoint. However, their union does not exhaust data domain Data, which means that not all data have bodies. For example, a list of numbers mixed with words does not have a body. As we will see later, expressions in **Lingua-A** will generate only such data that have bodies. For future use of bodies in definitions of expression denotations we introduce a partial function:

BOD : Data $\rightarrow$ Body

that is defined only for data that have bodies and which to every data assigns its body, i.e.

for every bod : Body, if dat : CLAN-Bo.bod  then  BOD.dat = bod

For instance:

BOD.2 = ('number')

BOD.[fa-name/'Smith', ch-name/'Adam'] =

    ('R', [fa-name/('word'), ch-name/('word')]).

For technical reasons that will be clear later, we assume that BOD is also defined for identifiers and that it is an identity in that case, i.e.

BOD.ide = ide

Since clans of bodies are disjoint, the function BOD is well defined.

Consider now an arbitrary theoretical operation on data and identifiers, i.e. a constructor of the type

ope : DatIde-1 x … x DatIde-n $\rightarrow$ Data

where each DatIde is either DataE or Identifier. To every such data constructor we assign a transparent (see Sec. 2.8) body constructor:

Bc[ope] : BodIde-1 x … x BodIde-n $\mapsto$  BodyE

where

if  DatIde-i = Identifier        then      BodIde-i = Identifier   and vice versa

with such a property that

**if**

| | | |
|---|---|---|
| ope.(arg-1,…,arg-n) | is defined | **and** |
| BOD.(ope.(arg-1,…,arg-n)) | is defined | **and** |
| BOD.arg-i | are defined for i = 1;n | **and** |
| Bc[ope].(BOD.arg-1,…,BOD.arg-n) | is not an error | |

**then**

BOD.(ope.(arg-1,…,arg-n)) = Bc[ope].(BOD.arg-1,…,BOD.arg-n)

If the above implication is satisfied then we say that the constructor Bc[ope] is *adequate for* ope. We can say therefore that for mutually adequate constructors function BOD "behaves" as a partially defined homomorphism.

For each constructor of data algebra, we define now an adequate to it constructor of body algebra. To do this, we introduce an auxiliary function:

sort : BodyE ↦ {('Boolean'), ('number'), ('word'), 'L', 'A', 'R'}

sort.bod =

| | | |
|---|---|---|
| bod : Error | ➔ | bod |
| bod = ('Boolean') | ➔ | ('Boolean') |
| bod = ('number') | ➔ | ('number') |
| bod = ('word') | ➔ | ('word') |
| bod : {'L'} x Body | ➔ | 'L' |
| bod : {'A'} x Body | ➔ | 'A' |
| bod : {'R'} x (Identifier ⟹ Body) | ➔ | 'R' |

Now we can define constructors of bodies. Their first group consists of zero-argument constructors of identifiers the same as in data algebra:

create-id.ide  : ↦ Identifier   for ide  : Identifier

In this case we obviously omit the context Bc[...] in their names. The second group of body constructors begins with three zero-argument constructors:

| | | |
|---|---|---|
| Bc[create-bo.boo] | : ↦ ('Boolean') | for boo : Boolean |
| Bc[create-nu.num] | : ↦ ('number') | for num : NumberS |
| Bc[create-wo.wor] | : ↦ ('word') | for wor : WordS |

In fact, we are dealing here with three indexed families of constructors, that within each family are identical with each other. This "algebraic prodigality" is assumed only to gain the similarity between algebras of data and of bodies, which we shall need in the future.

A body constructor that corresponds to the addition of numbers is now defined as follows:

Bc[add].(bod-1, bod-2) =

    bod-i : Error             ➔ bod-i      for i = 1;2

    bod-1 = bod-2 = ('number")  ➔ ('number')

    **true**                    ➔ 'number-expected'

and similarly for other operations on the data including the predicate equal. It is clear that all operations defined in that way are adequate. Operations on list bodies are defined as follows:

  Bc[create-li].bod =

    bod : Error     ➔ bod

    **true**         ➔ ('L', bod)

  Bc[push-li].(bod-e, bod-l) =                          (e – element, l – list)

    bod-i : Error     ➔ bod-i     for i = e, l

    sort.bod-l ≠ 'L'  ➔ 'list-expected'

    **let**

      ('L', bod-in) = bod-l                 (in – internal body of the list)

    bod-in ≠ bod-e ➔ 'inconsistent-bodies'

    **true**          ➔ bod-l

  Bc[top-li].bod =

    bod : Error   ➔ bod

    sort.bod ≠ 'L' ➔ 'list-expected'

    **let**

      ('L', bod-e) = bod

    **true**         ➔ bod-e

  Bc[pop-li].bod =

    bod : Error   ➔ bod

    sort.bod ≠ 'L' ➔ 'list-expected'

    **true**         ➔ bod

Array-body constructors are defined in a similar way:

  Bc[create-ar].bod =

    bod : Error     ➔ bod

    **true**           ➔ ('A', bod)

Bc[put-to-ar].(bod-a, bod-d) =
    bod-i : Error      ➔ bod-i      for i = a, d
    sort.bod-a ≠ 'A'   ➔ 'array-expected'
    **let**
       ('A', bod-e) = bod-t
    bod-e ≠ bod-d    ➔ 'inconsistent-bodies'
    **true**           ➔ bod

Bc[get-from-ar].bod =
    bod : Error   ➔ bod
    sort.bod ≠ 'A'➔ 'array-expected'
    **let**
       ('A', bod-e) = bod
    **true**         ➔ bod-e

The last group concerns record-bodies:

Bc[create-re].(ide, bod) =
    bod : Error   ➔ bod
    **true**          ➔ ('R', [ide/bod])

Bc[put-to-re].(bod-r, ide, bod-d) =
    bod-i : Error      ➔ bod-i      for i = r, d
    sort.bod-r ≠ 'R'   ➔ 'record-expected'
    **let**
       ('R', bor) = bod-r
    bor.ide = !      ➔ 'attribute-not-free'
    **true**         ➔ ('R', bor[ide/bod-d])

Bc[get-from-re].(ide, bod-r) =
    bod-r : Error     ➔ bod-r
    sort.bod-r ≠ 'R'   ➔ 'record-expected'
    **let**

   ('R', bor) = bod-r

  bor.ide = ?   ➔ 'unknown-attribute'

  **true**     ➔ bor.ide

Bc[cut-from-re].(ide, bod-r) =

  bod-r : Error   ➔ bod-r

  sort.bod-r ≠ 'R'  ➔ 'record-expected'

  **let**

   ('R', bor) = bod-r

  bor.ide = ?   ➔ 'unknown-attribute'

  **true**     ➔ ('R', bor[ide/?])

Bc[change in-re].(bod-r, ide, bod-d) =

  bod-i : Error   ➔ bod-i   for i = r, d

  sort.bod-r ≠ 'R'  ➔ 'record-expected'

  **let**

   ('R', bor) = bod-r

  bor.ide = ?    ➔ 'unknown-attribute'

  bod-d ≠ bor.ide  ➔ 'inconsistent-bodies'

  **true**      ➔ bod-r[ide/bod-d]

At this point, a significant engineering decision has been taken in assuming that when assigning new data to an attribute of a record, the new body must be identical with the previous one. Notice that this decision does not follow from the principle of adequacy. Notice also that our operation returns either an error or input record-body.

The task of verifying that the defined operations are adequate is left to the reader.

## 5.2.3 The algebra of composites

Using bodies, we can describe properties of data reflecting their "internal structure". This will allow us to introduce (in **Lingua-1**)variables declared in such a way that all their future values have a common body.

By a *structured data,*[47] we shall mean a pair consisting of a data and a body. The domain of such data is defined therefore by the equation:

 sda : StrDat = Data x Body

A structured data (dat, bod) is said to be *well-structured* if

 dat : CLAN-Bo.bod i.e. if  BOD.dat = bod.

---

[47] This should not be confused with *structural data* as defined in Sec. 5.2.1

Well-structured data will be called *composites*. Hence another domain:

com : Composite = {(dat, bod) | dat : CLAN-Bo.bod}                    (5.2.3-1)

A composite (dat, bod) is said to *carry* the data dat and the body bod. Notice that composites do not carry errors[48].

A composite that carries a simple data is called *simple composite* an analogously are understood *structural composites*. We shall also talk about *Boolean composites, numerical composites, list composites* etc. Since a special role will play Boolean composites, we introduce their domain:

com : BooComposite = {(boo, ('Boolean')) | boo : {tt, ff}}

In the sequel, composites are going to be the values of data expressions. For that sake we expand the earlier introduces function sort (Sec. 5.2.2) onto composites and identifiers:

sort.(dat, bod)   = sort.bod

sort.ide           = ide

We also introduce two new selection functions:

data.(dat, bod)  = dat

body.(dat, bod)  = bod

data.ide          = ide

body.ide          = ide

Notice that for simple composites functions sort coincides with body, but for structural composites, this is not the case. Now the domains of composites are supplemented with errors:

com : CompositeE        = Composite | Error

com : BooCompositeE   = BooComposite | Error

Over composites, we build the *algebra of composites* ComAlg which is similar (the same signature) to BodAlg and DatAlg. This is a two-sorted algebra with carriers:

ide   : Identifier

com : CompositeE.

With each constructor ope in data algebra we assign a constructor of composites Cc[ope] which on data performs ope (whenever data belong to its domain) and on bodies performs Bc[ope]. Composite constructors are going to be total function generating error whenever data do not belong to the domain of ope. These constructors should also "care" about data representability. For this sake we introduce a universal predicate:

oversized : Composite ⟼ Boolean

---

[48] In this place Andrzej Tarlecki asked a question, why I introduce bodies, if every data has a unique body unambiguously defined by the function BOD? Due to that we could operate on explicitly given data with implicitly assigned bodies. From a pure mathematical point of view that would be, of course, quite correct. I decided, however, otherwise in order to show explicitly how the modification of data contributes to the modification of their bodies. This approach suggests a certain way of the construction of **Lingua** implementation, and is also — in my opinion — useful when we define types and type constructors (Sec. 5.2.5).

which assumes value tt whenever its argument is too large to be acceptable by the "current implementation"[49]. I do not define it explicitly since I regard it as a parameter of our model. Of course for each sort of data its maximal acceptable size may be different.

In this place, it is worth mentioning that maximal size, in general, cannot be simply described by a predicate "not larger than". E.g. for numbers the corresponding size limit should define not only the interval, where representable numbers must belong but also their maximal number of decimal digits. From that perspective, the number 1/3 is oversized in any implementation. In turn, the corresponding norm for lists cannot be defined entirely by the length of the list only, but also has to take into account the size of its elements. The same is true for arrays and records.

The only universal assumption about the predicate oversize is such that zero-argument constructors create-bo, create-nu, create-wo do not generate oversized data. Practically this means that syntax analyser shall react with an error message if a programmer tries to write "a too large data" into his program. We also assume that

<div align="center">

**all composite constructors shall be defined in such a way
that all reachable composites are not oversized.**

</div>

Besides the oversize predicate we assume to have in our model a universal rounding operation:

round : Data ⟼ Data

that "truncates" numbers with too long or infinite decimal representations and for all other data is the identity function. Notice that without such function 1/3 could be rejected as oversized.

All constructors of <u>non-Boolean</u> composites will be defined as "Cartesian products" of corresponding data constructors and body constructors supplemented by all necessary checks (constructors for Boolean composites are defined a little later). Let then:

ope : DatIde-1 x … x DatIde-n ⟼ Data

be such a constructor. The corresponding composites constructor is defined in the following way (ComIde-i is understood analogously to BodIde-i):

Cc[ope] : ComIde-1 x … x ComIde-n ⟼ CompositeE

Cc[ope].(arg-1,…,arg-n) =

    arg-i : Error                ➔ arg-i             for i = 1;n

    **let**

        dat-i     = data.arg-i     for i = 1;n

        bod-i     = body.arg-i     for i = 1;n

---

[49] Stefan Sokołowski mentioned to me that in some applications dealing with the predicate **oversized** in the proofs of total correctness of programs may lead to technically complicated calculations. He suggested, therefore, that it may we worth considering a two-stage program development: at the first stage we do not care about overloads, and the second we analyse the developed programs from the perspective of possible overloads. On the ground of our model such a solution is, of course, quite feasible. Except a "full" semantics we may define a "simplified" semantics where the **oversized** predicate is always false. This may have sense not only for a two-stage programming but also in such applications where it is clear that overload "practically" does not happen, e.g. in many business-applications or database-applications. On the other hand in other applications the issue of an overload may be quite critical. A good example are arithmetic microprograms. In 1995 Intel Corporation had to replace hundreds of thousands of microprocessors on the market due to an error connected with overload.

    bod     = Bc[ope].(bod-1,…,bod-n)

bod : Error                  ➔ bod     (*)

ope.(dat-1,…,dat-n) = ?     ➔ error-message                          (5.2.3-2)

**let**

    dat   = round.(ope.(dat-1,..,dat-n))

    com  = (dat, bod)

oversized.com                ➔ 'overload'

**true**                         ➔ com

The execution of Cc[ope] starts from making sure that none of its arguments is an error. If this is not the case, then the first of these errors becomes the final result (transparency).

Otherwise, an attempt is made to compute the resulting body, and if it is not an error, we examine whether the result of applying data constructor is defined. If this is not the case, then the appropriate error message is generated, e.g. 'division-by-zero'. If these tests do not lead to an error, the resulting data is calculated by applying the operations ope and round. Notice that for data that are not numbers, round is an identity function.

The verification whether the resulting body is not an error — clause (*) — implements the principle (Sec. 5.1) that before performing data operations we check, whether its arguments are "appropriate".

The next step is to check whether the resulting data does not exceed the acceptable size. This step, however, should not be taken literally by assuming that we first create a "too large" composite, and only then generate an error message. The presence of predicate oversized means that the implementation of the language is equipped with a mechanism to predict that oversize will happen.

If also this test is successful, the resulting composite is accepted. It is well structured due to the fact that the operation Bc[ope] is adequate for ope.

The transformation Bc defined in this way applies also to zero-argument constructors, i.e. for $n = 0$.

At this point a methodological comment is necessary. The general form of the definition of Cc[ope] may raise a certain doubt since it is known from computability theory (see Sec. 3.4) that in the general case of partial functions, the predicate

    ope.(dat-1,…,dat-n) = ?

is not computable. In our case, however, the predicates that correspond to data operations are easily implementable. For instance, in the case of division we only check if its second argument is different from zero:

Cc[divide].(com-1, com-2) =

    com-i : Error    ➔ com-i                  for i = 1,2

    **let**

        (dat-i, bod-i)  = com-i                  for i = 1,2

        bod            = Bc[divide].(bod-1, bod-2)

    bod : Error    ➔ bod

dat-2 = 0       ➔ 'division-by-zero'

**true**         ➔ (round.(divide.(dat-1, dat-2)), bod)

Our general scheme applies also to zero-argument constructors. E.g.

Cc[create-nu.128] = (128, ('number'))

In defining Boolean constructors of composites, we shall follow McCarthy's philosophy (Sec. 2.9) which makes these constructors not transparent. Consequently, the scheme (5.2.3-2) is not applicable in that case, hence Boolean constructors must be defined independently. Composite constructor that corresponds to McCarthy's conjunction is defined as follows (here -C stands for "composite"):

and-C.(com-1, com-2) =

com-1 : Error                ➔ com-1

sort.com-1 ≠ ('Boolean')     ➔ 'Boolean-expected'

data.com-1 = ff              ➔ (ff, ('Boolean'))

com-2 : Error                ➔ com-2

sort.com-2 ≠ ('Boolean')     ➔ 'Boolean-expected'                   (*)

**true**                     ➔ (data.com-2, ('Boolean'))

Notice that whenever the "execution" of this definition reaches the clause (*), we can conclude that **data.com-1 = tt**, hence the resulting data is equal to **data.com-2**. Our constructor is of course adequate.

The negation constructor is quite obvious:

not-C.com =

com : Error                  ➔ com

sort.com ≠ ('Boolean')       ➔ 'Boolean-expected'

com = (tt, ('Boolean'))      ➔ (ff, ('Boolean'))

com = (ff, ('Boolean'))      ➔ (tt, ('Boolean',))

The constructor for alternative is defined in a way which makes it satisfy De Morgan's laws, hence:

or-C.(com-1, com-2) =

not-C.(and-C.(not-C.com-1, not-C.com-2))

## 5.2.4   The algebra of transfers

The concept of a body allows expressing these features of data, which in many programming languages exhaust the concept of a type, e.g., the type of Booleans, numbers, lists, arrays, etc. Some languages, however, offer a higher expressiveness of types. For instance, in SQL one may declare types of such tables, where types associated to columns refer not only to bodies but also to other data properties such as **small-number** or even to properties of whole columns such as unique (no repetitions). A table type may also include a predicate that must be satisfied by each row of the table. In turn, database types may include information about the subordination relations between tables.

To build such types in the languages of **Lingua** series, we have to introduce predicates on bodies. This, in turn, requires the introduction of a more general concept of a *transfer*.

By a *transfer,* we mean every one-argument function that maps composites and errors into composites and errors. We introduce therefore the domain:

tra : Transfer = CompositeE ⟼ CompositeE

A particular case of transfers are *yokes* that map arbitrary composites into boolean composites. Their domain is, therefore:

yok : Yoke = CompositeE ⟼ BooCompositeE

Constructors of transfer will be defined in such a way that all reachable transfers will be transparent wrt errors, i.e. will satisfy the equation:

tra.err = err    for every err : Error

We say that a *composite* com *satisfies a transfer (yoke)* tra, if

tra.com = (tt, ('Boolean')

Yokes are therefore one-argument predicates on composites. Anticipating future syntax of our language the yoke expression

```
value < 10
```

represents a yoke that is satisfied whenever the input composite carries a number and that number is less than 10. In this expression `value` is not a variable identifier, but a key word representing the data carried by the input composite. Another example may be the yoke expression

```
value + 2 < 10
```

which expresses the fact that if the value of data carried by the current composite is incremented by 2, then the result is less than 10. Denotationally this yoke is a composition of the former yoke with a transfer that increments by 2 the data carried by the input composite. In turn the expression:

```
record.salary + record.commission < 7000
```

correspond to a yoke that is satisfied if its argument-composite carries a record with numeric attributes `salary` and `commission` whose sum is less than 7000[50].

Analogously as in the case of data, bodies and composites we construct now a two-sorted algebra of transfers TraAlg:

ide : Identifier

tra : Transfer

Notice that the carriers of that algebra do not contain errors, but contains transfers (and yokes) that may return errors as their values. This algebra belongs to a different level than the former algebra since transfers belong to the level of constructors of the algebra of composites.

The majority of transfer constructors — similarly to the constructors of bodies and composites — are derived from data operations, although not necessarily from only such operations and not necessarily from all these operations. By Tc[ope] we denote the transfer constructor associated with data operation ope.

---

[50] From a mathematical viewpoint we could omit the key words in the syntax of composites, e.g. in writing „< 10" or „+2<10", but such syntax would be very unintuitive.

Beside data operations described in Sec. 5.2.1, we shall also use some additional operations which are useful in building transfers. To keep further investigations not too long we restrict them to the four:

| | | | |
|---|---|---|---|
| sum | : Number$^{c+}$ | $\mapsto$ Number | — the sum of numbers on a list |
| max | : Number$^{c+}$ | $\mapsto$ Number | — the maximal number on a list |
| small-nu | : Number | $\mapsto$ Boolean | — e.g. a number in [-9999, 9999] |
| increasing-nu | : Number$^{c+}$ | $\mapsto$ Boolean | — increasingly ordered list of numb. |

The first two are called in SQL *aggregation operations*. The third is a typical predicate describing a table field. The fourth has been introduced just to show that in our model we may go with types and yokes a step further than in typical programming languages.

The fact that these operations have been called "additional" means that they are not going to be included in the repertoire of "ordinary" expression, but will be available only in transfers. This is of course (again ) not a mathematical must but an engineering choice assumed in accordance with SQL standard. Below the list of transfers in TraAlg split into six groups:

### Constructors of identifiers

create-id.ide : $\mapsto$ Identifier    for ide : Identifier

### Constructors of transfers processing simple data

| | | | |
|---|---|---|---|
| Tc[create-nu.num] : | | $\mapsto$ Transfer | for num : NumberS |
| Tc[create-wo.wor] : | | $\mapsto$ Transfer | for wor : WordS |
| Tc[add] | : Transfer x Transfer | $\mapsto$ Transfer | |
| Tc[divide] | : Transfer x Transfer | $\mapsto$ Transfer | |
| Tc[sum] | : Transfer | $\mapsto$ Transfer | |
| Tc[max] | : Transfer | $\mapsto$ Transfer | |

### Constructors of yokes

| | | |
|---|---|---|
| Tc[equal] | : Transfer x Transfer | $\mapsto$ Transfer |
| Tc[less] | : Transfer x Transfer | $\mapsto$ Transfer |
| Tc[small-nu] | : Transfer | $\mapsto$ Transfer |
| Tc[increasing-nu] | : Transfer | $\mapsto$ Transfer |

| | | | |
|---|---|---|---|
| Tc[create-bo.boo] : | | $\mapsto$ Transfer | for boo  : Boolean |
| and-T | : Transfer x Transfer | $\mapsto$ Transfer | |
| or-T | : Transfer x Transfer | $\mapsto$ Transfer | |
| not-T | : Transfer | $\mapsto$ Transfer | |

**Constructors of quantified yokes**

> all-on-li : Transfer                        $\mapsto$ Transfer
>
> all-in-ar : Transfer                        $\mapsto$ Transfer

**Constructors of selection-transfers from list, arrays and records**

> Tc[get-from-li]   :                        $\mapsto$ Transfer
>
> Tc[get-from-ar]  : Transfer                $\mapsto$ Transfer
>
> Tc[get-from-re]  : Identifier              $\mapsto$ Transfer

**Identity transfer**

> pass          :                        $\mapsto$ Transfer

The constructors of identifiers are well-known from former algebras.

Next two constructors create transfers with fixed values (except errors). E.g. in the case of numbers to each number num corresponds one constructor, e.g.

> Tc[create-nu.2] : $\mapsto$ Transfer
>
> Tc[create-nu.2].().com =
>
> > com : Error   ➔ com
> >
> > **true**   ➔ Cc[create-nu.2].()

This definition may be written also in a direct form, i.e. without referring to the constructor of composites:

> Tc[create-nu.2].().com =
>
> > com : Error   ➔ com
> >
> > **true**   ➔ (2, ('Number'))

Zero-argument constructors for words and Booleans are defined analogously. Each of these constructors creates a transfer the value of which is a fixed composite independent of transfer's input unless it is an error.

Four further transfer constructors that correspond to arithmetic operations are defined according to a standard scheme: for every data operation ope we define the corresponding transfer constructor Tc[ope] by the equations:

> Tc[ope].(tra-1,…, tra-n).com = Cc[ope].(tra-1.com,…,tra-n.com)          (5.2.4-1)

This scheme shall be applied to all (future) constructors of simple data. Notice that if all tra-i are transparent wrt errors and so is Cc[ope], then also Tc[ope].(tra-1,…, tra-n) must be transparent.

First four yokes' constructors are defined as in (5.2.4-1) which covers also two zero-argument constructors:

> Tc[create-bo.boo].() = Cc[create-bo.boo].() = (boo, ('Boolean'))   for boo : Boolean

Here we introduce two new symbols:

TT = Tc[create-bo.tt]

FF = Tc[create-bo.ff]

For the remaining Boolean operations we assume constructors of Kleene's logic (see Sec. 2.9) rather than that of McCarthy as in the case of composites. In the case of conjunction such a definition is the following (-T stands for "transfer"):

and-T.(tra-1, tra-2).com =

| com : Error | ➔ com | |
|---|---|---|
| **let** | | |
| com-i = tra-i.com | | for i = 1;2 |
| com-i = (ff, ('Boolean')) | ➔ (ff, ('Boolean')) | for i = 1;2 |
| com-i : Error | ➔ com-i | for i = 1;2 |
| sort.com-i ≠ ('Boolean') | ➔ 'Boolean expected' | for i = 1;2 |
| **true** | ➔ (tt, ('Boolean')) | |

As we see, to falsify this conjunction it is enough that arbitrary of its arguments carry ff. If this is not the case, then the result is either an error or a composite carrying tt. Constructor not-T coincides with McCarthy's one, and or-T is defined in such a way as to satisfy De Morgan's laws.

Notice that in the case of composites McCarthy's calculus was assumed, since then — as we shall see in Sec. 7.5 — expressions may generate infinite executions, which is due to the fact, that they may contain functional-procedure calls. Since we do not allow functional procedures in transfers, we can assume a "more lazy" Kleene's calculus. This calculus has also been assumed in SQL standard (Sec. 12)

Notice however that the yoke created by and-T is — according to the general assumption about transfers — transparent wrt errors. The "laziness" of and-T concerns only errors generated by argument-transfers tra-1 and tra-2. The same comment applies to the alternative.

The names of Boolean constructors do not have the form Tc[ope] since they do not refer to any data-algebra constructors. The general-quantifier constructors for lists and arrays have the same property. Similarly to Boolean constructors, they create yokes from yokes, but formally are applicable to arbitrary transfers.

| all-on-li : Transfer ↦ Transfer | | (all on list) |
|---|---|---|
| all-on-li.yok.com = | | |
| com : Error | ➔ com | |
| sort.com ≠ 'L' | ➔ 'list expected' | |
| **let** | | |
| (dat-1,…,dat-n) = data.com | | |
| ('L', bod)    = body.com | | (list elements have all the same body) |
| com-i    = tra.(dat-i, bod) | | for i = 1;n |
| com-i : Error | ➔ com-i | for i = 1;n |

      $(\forall\ i = 1\ ;n)$ com-i = (tt, ('Boolean'))  ➔ (tt, ('Boolean'))

      **true**                              ➔ (ff, ('Boolean'))

After an initial processing of the input composite, we check if all elements of the carried list satisfy transfer tra. In an analogous way we define a constructor for arrays:

   all-in-ar : Transfer ⟼ Transfer                                       (all in array)

   all-in-ar.tra.com =

      com : Error                        ➔ com

      sort.com ≠ 'A'                   ➔ 'array-expected'

      **let**

         [num-1/dat-1,…,num-n/dat-n]   = data.com

         ('A', bod)             = body.com      (elements of array have all the same body)

         com-i                  = tra.(dat-i, bod)     for i = 1;n

      com-i : Error                    ➔ com-i

      $(\forall\ i = 1\ ;n)$ com-i = (tt, ('Boolean'))  ➔ (tt, ('Boolean'))

      **true**                              ➔ (ff, ('Boolean'))

Notice that if the array is many-dimensional, then this yoke refers to the element of the first level, but since this construction may be iterated, we may describe properties of arbitrary arrays.

Three consecutive constructors correspond to selections related to structural data. They are derived from data operations which reflects in their names. The first one creates a transfer of getting the top element of a list:

   Tc[get-from-li] : ⟼ Transfer

   Tc[get-from-li].() = Cc[get-from-li]

hence:

   Tc[get-from-li].().com = Cc[get-from-li].com

where Cc[get-from-li] is defined according to the scheme (5.2.3-2). This constructor creates, therefore, a transfer that is a constructor of the algebra of composites. Of course Cc "will check", if com is a list-composite and if it is not, will generate an error.

Notice that the definition of Tc[get-from-li] does not correspond to schema (5.2.4-1) since the list composite is the argument of Tc[get-from-li] is not processed by any transfer. This is again an engineering decision to the effect that:

                    **yokes shall not modify "internally" structural composites**

The only case in our model where a transfer is processing a structural composite is the selection of an element. The same restriction concerns two successive constructors. For that reason all such constructors have one argument less than the corresponding data constructors. E.g. Cc[get-from-li] gets only one argument which is a list composite, and therefore Tc[get-from-li] is zero-argument.

Let us recall that in the case of simple data such restriction has not been assumed that is illustrated by the yoke expression `value`+2<10. The corresponding transfer first modifiers the input composites employing the addition operation and only then checks the inequality. In our future syntax of transfer expressions, we cannot describe a transfer which first modifies a record by adding an attribute to it and later evaluates the condition, e.g.

> `record`.salary + `record`.bonus < 7000

The second constructor from this group creates a transfer that selects an array element pointed out by a given index, the latter computed by its only argument:

> Tc[get-from-ar] : Transfer ⟼ Transfer

> Tc[get-from-ar].tra.com = Cc[get-from-ar].(com, tra.com)

Here again, all checks are performed by composite constructor Cc[get-from-ar], that also checks if com carries an array and tra.com — a number. It also generates an error, if com is an error.

Observe that since com must carry an array and tra must generate a number, with the current repertoire of transfers' constructors, tra must be a constant-value transfer with a numeric value. Hence tra must be the result of the constructor

> Tc[create-nu.num]

If, however, we would have a transfer such as, e.g. count-ele-ar, then the transfer

> Tc[get-from-ar].count-ele-ar

would return the last element of the array.

The third constructor creates, in an analogous way, a constructor selecting a composite from a record

> Tc[get-from-re] : Identifier ⟼ Transfer

> Tc[get-from-re].ide.com = Cc[get-from-re].(ide, com)

Since the last constructor called pass do not correspond to any data operation, its name again does not contain the context Tc[…]. That zero-argument constructor creates the identity transfer:

> pass.().com = com

We need this constructor to make some arguments tra-i.com in the schema (5.2.4-1) equal to com. If at the level of concrete syntax we write pass as `value` (get a value), then the yoke expression

> `value` < 10,

corresponds accordingly to (5.2.4-1) to the yoke

> Tc[less-nu].(pass.(), Tc[create-nu.10].()).

Unfolding this expression with the assumption that com = (num, ('Number')), we get:

> Tc[less-nu].(pass.(), Tc[create-nu.10].()).com =
>
>   Cc[less-nu].(pass.().com, Cc[create-nu-10].().com) =
>
>   Cc[less-nu].(com, (10, ('number')) =
>
>   (less-nu.(num, 10), ('Boolean'))

Now, similarly as for bodies, also for transfers we introduce a claning function. By *the clan of a transfer* we shall mean the set of composites that satisfy this transfer, i.e.

CLAN-Tr : Transfer ⟼ Sub.Composite

CLAN-Tr.tra = {com | tra.com = (tt, ('Boolean'))}

Of course, the clans of non-boolean transfers are empty[51].

Consider now the already known example of a yoke expression that describes properties of a record:

**record**.salary + **record**.commision < 7000

The corresponding yoke is:

Tc[less].(

    Tc[add].(Tc[get-from-re].(create-id.salary.()),

       Tc[get-from-re].(create-id.commision.() ),

    Tc[create-nu].7000.() )

We can also describe a property of record list where all element must have that property:

**all-list**(**record**.salary + **record**.commision < 7000)

The corresponding yoke is

all-on-li.(

    Tc[less].(

       Tc[add].(Tc[get-from-re].(create-id.salary.()),

          Tc[get-from-re].(create-id.commision.()) ),

       Tc[create-nu].7000.()

           )

         )

We can also combine this local property of a list with its global property:

**all-list**(**record**.salary + **record**.commision < 7000) **and**

**sum**(**record**.commision) < 100.000

Of course to do that we have to introduce a function that computes the sum of all elements of a numeric list.

---

[51] In this place one can rise a question, why we define clans for transfers rather than only for yokes. This question has a larger context, however, namely — why we introduce transfers at all if we are interested in yokes only. The answer follows from the fact that our model bases on algebras and an algebra of yoks without transfer would be very poor. We would have a similar situation if we try to build an algebra of Boolean expression without arbitrary expressions. In turn, as we are going to see in Sec. 5.2.5, the algebras of transfers and of bodies will be used in the construction of the algebra of types. Therefore, since we cannot give up non-Boolean transfers, it is convenient to define CLAN-Tr for arbitrary transfers.

## 5.2.5   The algebra of types

By a *type,* we mean a pair consisting of a body and a transfer. The domain of types is therefore defined by the equation[52]:

   typ : Type = Body x Transfer

About a type (bod, tra) we say that it *carries* body bod and transfer tra. We say that it is *Boolean, numeric, wordy, etc.* if bod is of the corresponding sort. Similarly we understand the notions of *simple types* and *structural types*. If the yoke of a type is TT, then we say that the type is *yokeless*. Notice that types — similarly as composites — do not carry errors.

With every type, we associate a set of composites called the *clan of the type*. We, therefore, define the function:

   CLAN-Ty : Type ⟼ Sub.Composite

   CLAN-Ty.(bod, tra) =

      {(dat, bod) | dat : CLAN-Bo.bod **and** (dat, bod) : CLAN-Tr.tra}

A type with the empty clan is called an *empty type*. Now we may construct an *algebra of types* named TypAlg with three carriers:

   ide   : Identifier

   tra   : Transfer = CompositeE ⟼ CompositeE

   typ   : TypeE   = Type | Error

This algebra will become a fundament for the future algebra of the denotations of type expressions. The constructors of the algebra of types are the following:

1. all constructors of identifiers,

2. all constructors of the algebra of transfers,

3. type constructors which are defined below.

Many type constructors, similarly as the constructors of composites, refer to body constructors and are derived from operations on data. By Yc[ope] we denote the constructor of types associated with the operation ope. However, unlike with composites' constructors, where all operations on data were involved, now we shall use only those which we need for the creation of new types. E.g. we shall use the constructor of list creation, but not the constructor of adding an element to a list since the latter does not create a new type.

Below we see the subset of the signature of types-algebra restricted to constructors of the third group:

   Yc[create-bo].boo   :                                    ⟼ TypeE for boo : Boolean
   Yc[create-nu].num   :                                    ⟼ TypeE for num : Number
   Yc[create-wo].wor   :                                    ⟼ TypeE for wor : Word

   Yc[create-li]           : TypeE                          ⟼ TypeE

---

[52] Why here again arbitrary transfers, I was trying to explain in a foot-note of Sec. 5.2.4 concerning the function CLAN-Tr.

| | | |
|---|---|---|
| Yc[create-ar] | : TypeE | ↦ TypeE |
| Yc[create-re] | : TypeE x Identifier | ↦ TypeE |
| Yc[put-to-re] | : TypeE x Identifier x TypeE | ↦ TypeE |
| Yc[cut-from-re] | : TypeE x Identifier | ↦ TypeE |
| | | |
| replace-ty-tr | : TypeE x Transfer | ↦ TypeE |

The name of the last constructor does not have the context Yc[…] since it does not correspond to a data operation. As we are going to see, this constructor modifies a type be replacing its transfer by a new one.

Constructors of the first group serve to create simple yokeless types:

Yc[create-bo.boo].() = (Bc[create-bo.boo].(), TT) = ((boo, ('Boolean')), TT)

Yc[create-nu.num].() = (Bc[create-nu.num].(), TT) = ((num, ('number')), TT)

Yc[create-wo.wor].() = (Bc[create-wo.wor].(), TT) = ((wor, ('word')), TT)

Constructors of the second group refer to data- and yoke-constructors:

**The creation of a list type**

Yc[create-li] : TypeE ↦ TypeE

Yc[create-li].typ =

    typ : Error       ➔ typ

    **let**

       (bod, tra)  = typ

       new-bod   = Bc[create-li].bod

       new-tra    = all-on-li.tra

    new-bod : Error   ➔ new-bod

    **true**            ➔ (new-bod, new-tra)

Since the resulting type corresponds to lists with all elements being of input-list type, they all have the same body ('L', bod) and satisfy the same yoke tra. Of course, if tra is not a yoke, then the created type is empty. An array type is created analogously:

**The creation of an array type**

Yc[create-ar] : TypeE ↦ TypeE

Yc[create-ar].typ =

typ : Error             ➜ typ

**let**

   (bod, tra) = typ

   new-bod   = Bc[create-ar].bod

   new-tra = all-in-ar.tra

new-bod : Error    ➜ new-bod

**true**                    ➜ (new-bod, new-tra)

**The creation of a record type with one attribute**

Yc[create-re] : TypeE x Identifier ↦ TypeE

Yc[create-re].(typ, ide) =

   typ : Error             ➜ typ

   **let**

     (bod, tra)  = typ

     new-bod   = Bc[create-re].(ide, bod)

     new-tra    = Tc[get-from-re].ide ● tra

   new-bod : Error    ➜ new-bod

   **true**                    ➜ (new-bod, new-tra)

The body of the resulting type is [ide/bod] and its transfer new-tra is satisfied, if:

1. its input composite is a record composite and caries attribute ide,

2. composite assigned to ide satisfies transfer tra.

Notice that transfer Tc[get-from-re] selects the composite assigned to ide which is then passed (operation ●) to transfer tra. In other words, the value of the unique attribute od the record should satisfy the transfer indicated by the type which is an argument of the constructor.

**Expanding a record type by a new attribute**

Yc[put-to-re] : TypeE x Identifier x TypeE ↦ TypeE

Yc[put-to-re].(typ-r, ide, typ-n) =                          (r – record, n — new)

   typ-i : Error             ➜ typ-i          for i = r, n

   **let**

     (bod-i, tra-i)  = typ-i             for i = r, n

     new-bod      = Bc[put-to-re].(bod-r, ide, bod-n)

     new-tra       = and-T.(tra-r, Tc[get-from-re].ide ● tra-i)

   new-bod : Error    ➜ new-bod

   **true**                    ➜ (new-bod, new-tra)

The new body is created by body constructor that checks if bod-r is a record body and if ide does not appear in that body. New yoke guarantees that the new record satisfies the initial yoke tra-r on initial attributes and the new yoke on the new attribute (cf. comment to the construction Tc[create-re].ide ● tra).

**Removal of an attribute from a record**

    Yc[cut-from-re] : TypeE x Identifier ⟼ TypeE

    Yc[cut-from-re].(typ, ide) =

        typ : Error          ➜ typ

        **let**

           (bod, yok) = typ

           new-bod   = Bc[cut-from-re].(ide, bod)

        new-bod : Error   ➜ new-bod

        **true**             ➜ (new-bod, yok)

The removal of an attribute from a record type removes this attribute from the corresponding body, but it does not change the yoke. Of course, after such a change the yoke does not need to be satisfied, unless it does not refer to the removed attribute. In practice, one has to first modify the yoke using replace-ty-tr (see below) and only then remove an attribute.

**The replacement of a transfer in a type**

    replace-ty-tr : TypeE x Transfer  ⟼ TypeE

    replace-ty-tr.(typ, tra) =

        typ : Error ➜ typ

        **let**

           (bod, tra-f) = typ                            (f – former)

        **true**     ➜ (bod, tra)

This constructor replaces a transfer by a new one. This is, of course, a very general operation, hence for practical reasons one should think about a more specific constructor, e.g. adding a new yoke conjunctively to an existing one. I postpone this problem, however, to avoid going too deep into technical details of our model. It is worthwhile noticing in this place that replace-ty-tr is the only constructor that changes a transfer without modifying the corresponding body. It is also the only constructor which forces to include transfers among the carrier of the algebra of types.

# 5.3    The algebra of expression denotations

## 5.3.1    Values and memory states

As was already mentioned in Sec. 4.1, to define functions plying the role of the denotation of expressions, of declarations and of instructions, one has to define the concept of a *memory state* or simply a *state*. In a simple programming-language states might be just valuations, i.e. mappings from identifiers into data. However, in the majority of programming languages identifiers may "store" more than just data but:

- data with their types,

- types as independent beings,

- procedures.

and this requires a richer concept of a state.

In our model states will bind data-identifiers called *data variables* with *typed data* consisting of a data or *pseudo data* Ω and a type. The domain of typed data is therefore defined as follows:

    tda : TypDat = (Data | {Ω}) x Type

A typed data (dat, (bod, yok)) may also be regarded as a pair ((dat, bod), yok) consisting of a composite (dat, bod) and a yoke as well as a triple (dat, bod, yok). In the sequel, we shall refer to each these forms according to the need. Notice that typed data do not carry errors. A typed data (dat, typ) will be said to be *well-typed*, if:

    dat = Ω or

    dat : CLAN-Ty.typ

Well-typed data are called *values*. We introduce therefore a domain:

    val : Value = {(dat, typ) | dat = Ω **or** dat : CLAN-Ty.typ}

A pair of the form (Ω, typ) is called a *pseudo value* and a composite of the form (Ω, bod) — a *pseudo composite*. Pseudo values will be assigned to variables by declarations. A value that is not a pseudo value is called a *proper value*. Function sort is extended to pseudo composites:

    sort.(Ω, bod) = Ω

which means that I accordingly expand its codomain by adding Ω to it. I extend it also to values:

    sort.(com, yok) = sort.com

Our *states* will store:

- values assigned to *data variables* (identifiers),

- types assigned to *type constants* (identifiers),

- procedures (and functional procedures) assigned to *procedure names* (identifiers).

Formally the domain of states is defined by the following domain equations:

| | | | |
|---|---|---|---|
| sta : State | = Env x Store | | (state) |
| env : Env | = TypeEnv x ProEnv | | (environment) |
| sto : Store | = Value x (Error | {'OK'}) | | (store) |
| vat : Valuation | = Identifier ⟹ Value | | (valuation) |

tye  : TypeEnv  = Identifier $\Longrightarrow$ Type                    (type environment)

pre : ProEnv = Identifier $\Longrightarrow$ Procedure            (procedure environment)

The split of a state into two pairs in the place of one four-tuple is not accidental. As we shall see in the sequel it will be justified on the ground of the model of procedures (Sec. 7). The domain Procedure will be defined there too.

A *store* is that component of a state which stores values by binding them to identifiers in valuations. The principle that valuations store only well-typed data shall be assured by the assignment instruction and by the rules of passing parameters to procedures.

An error message, when generated, becomes a component of a store and since then is passed to all subsequent states. However, as long as this is not the case, the store is carrying 'OK' (no error). If the message is different from 'OK', then we say that the *state (store) carries an error*. We assume again that the set of errors contains all error messages that will appear in future definitions of the constructors of denotations. We shall also ensure that all imperative denotations, i.e. denotations that transform states, do not change states that carry an error (transparency) and that all applicative denotations, i.e. denotations that transform states into composites, generate an error whenever a state carries an error[53].

*Environments* constitute these components of states which store user-defined types, procedures and functions (functional procedures).

In order to describe the mechanism of errors at the level of states, we introduce three auxiliary functions:

error : State $\longmapsto$ Error | {'OK'}                    (error-selection operator)

error.(env, (vat, err)) = err

is-error : State $\longmapsto$ Boolean                (error-detection predicate for states)

is-error.sta =

    error.sta ≠ 'OK' ➔ tt

    **true**                ➔ ff

is-error : Store $\longmapsto$ Boolean                (error-detection predicate for stores)

is-error.(vat, err) =

    err ≠ 'OK' ➔ tt

    **true**        ➔ ff

◄ : State x Error $\longmapsto$ State                    (error-insertion operator)

    (env, (vat, err)) ◄ err-1 =

        (env, (vat, err-1))

---

[53] This principle shall not be observed when we introduce error-handling mechanisms (Sec. 12.7.6.4).

We introduce also an operator that will serve to ensure that an identifier declared in a valuation cannot be at the same type declared in an environment and vice-versa. Denotationally such a situation would be acceptable since — as we are going to see — every referring to an identifier will explicitly point to the state component where the identifier should be found. From a programmer's view, however, such solution may contribute to errors in programs.

declared : Identifier ↦ State ↦ BooleanE

declared.ide.((tye, pre), (vat, err)) =

    tye.ide = ! **or** pre.ide = ! **or** vat.ide = ! ➔ tt

    **true**                             ➔ ff

The predicate declared is satisfied for an identifier in a state which does not carry an error, if this identifier has been bound in that state with a value, type or procedure.

One methodological comment at the end. All transfer-constructors defined in this section build transfers that describe the properties of composites concentrating on the properties of data whereas for bodies they only check if bodies are of an appropriate sort. E.g. the transfer that is constructed by all-on-li only checks if the input composite carriers a list. This restriction is, of course, an engineering decision rather than a mathematical necessity. It has been adopted for the sake of the simplicity of the model and also since non-trivial transfers are used only in **Lingua-SQL** and only in a way as described in the present section.

## 5.3.2 The denotations of data expressions

Since *data expressions* are "usual" expressions, they correspond to functions that map states into composites or errors. Their denotations — also called *data expression denotations* — constitute one of the carriers of the future algebra of denotations of our language:

ded : DatExpDen = State → Composite | Error

It is to be emphasised that the results of data expression computations may be only composites but never pseudo composites. Notice also that data expression denotations are partial functions since in the future (Sec. 7.5) data expressions will include procedure calls that may generate infinite executions.

Since the program-termination problem is not decidable (see Sec. 3.4), we cannot assume that in case of an infinite execution an error signal will be generated. We have to assume therefore that in such cases the value of the executed denotation will be undefined.

Here we should explain why the denotations of data expressions map state into composites rather than into values. This solution has been chosen since in executing expressions we usually create new values and sometimes also types, but never new yokes. The yokes are associated with variable identifiers, and their role is to guarantee that new composite assigned to a variable satisfies the yoke associated with that variable.

A denotation of a data expression is said to be *transparent* wrt errors, if

ded.(env, (vat, err)) = err       whenever err ≠ 'OK'.

A constructor of data expression denotations is said to be *diligent* if it transforms transparent denotations into transparent denotations. All our constructors of data expression denotations will be defined in such a way as to be *diligent*.

Besides expression denotations that transform states into composites, errors and types — which will be called *applicative denotations* — we are going to have denotations of instructions and declarations that transform states into states and are called *imperative denotations*.

The class of constructors of data expression denotations consists of three categories:

1. one constructor of variables,

2. many constructors derived from composite constructors,

3. one constructor that corresponds to conditional expressions.

The first constructor builds *data-variable denotations*:

dat-variable : Identifier ⟼ DatExpDen

dat-variable.ide.sta =

   is-error.sta      ➜ error.sta

   **let**

      (env, (vat, 'OK')) = sta

   vat.ide = ?      ➜ 'undeclared-variable'

   **let**

      ((dat, bod), yok) = vat.ide

   dat = Ω      ➜ 'uninitialized-variable'

   **true**         ➜ (dat, bod)

The calculations of the value of a variable starts from checking if its identifier **ide** has been declared and initialized. If that is not the case, then an error signal is generated. In the opposite case the composite assigned to **ide** becomes the final result. Notice that yoke is neglected since expressions return composites rather than values. Yokes shall come to the play in assignment instructions in Sec. 6.1.4.

At the level of implementation, the appearance of an error means that this error is displayed on the monitor and program execution halts. Notice that in this way

> **we eliminate a possible pseudo value from further computations which means
> that it is never "sent" to a composite constructor as an argument**

Constructors of the second category are derived from the constructor of composites. We shall start with constructors transparent wrt errors. Let

Cc[ope] : ComIde-1 x … x ComIde-n ⟼ CompositeE

be such a constructor for n ≥ 0. The corresponding constructor of denotations of data expressions that we denote by **Cdd[Cc[ope]]** is defined by the following schema, where of course **DatExpDenIde-i** is either **DatExpDen** or **Identifier**.

Cdd[Cc[ope]] : DatExpDenIde-1 x … x DatExpDenIde-n ⟼ DatExpDen

Cdd[Cc[ope]].(arg-1,…,arg-n).sta =                                          (5.3.2-1)

   is-error.sta               ➜ error.sta

   arg-i.sta = ?            ➜ ?            for arg-i from outside of Identifier

   **let**

com-i =        for i = 1;n

    arg-i : Identifier    ➜ arg-i

    **true**              ➜ arg-i.sta

**true**                  ➜ Cc[ope].(com-1,…,com-n)

If the input state carries an error, then program execution is interrupted and the carried error becomes output value of the computation, i.e. it is displayed on a monitor.

Otherwise, an attempt is made to evaluate each of these argument-expressions which are not identifiers. It should be emphasised in this place that the single-identifier expression, i.e. a variable, is not an identifier in the sense as understood in (5.3.2-1). In our scheme, an identifier "as such" — i.e. not as the denotation of expression variable.ide — may appear only as an attribute of a record.

If any of these attempts result an infinite execution, then, of course, the execution of the whole expression does not terminate. In this place, we take an engineering decision by assuming that Cdd[Cc[ope]] is transparent not only elative to errors but also to infinite executions. As we may expect, this is not going to be the case for Boolean expressions.

If none of the argument executions is infinite then the resulting composites, identifiers or errors are "passed" to the constructor of composites. Its transparency assures the transparency of the constructor of denotations. Of course Cc[ope].(com-1,…,com-n) may be an error message.

Notice now that if the scheme (5.3.2-1) would be applied to Boolean constructors, then they would be lazy only wrt errors but not wrt infinite computations. Since, however, we want them to be lazy also in the latter case, we have to define them independently. For that reason, we write and-ded rather than Cdd[Cc[and]] and similarly for other Boolean constructors.

and-ded : DatExpDen x DatExpDen ⟼ DatExpDen

and-ded.(ded-1, ded-2).sta =                                   (5.3.2-2)

   is-error.sta           ➜ error.sta

   ded-1.sta = ?          ➜ ?

   **let**

      com-1 = ded-1.sta

   com-1 : Error         ➜ com-1

   **let**

      (dat-1, bod-1)   = ded-1.sta

   bod-1 ≠ ('Boolean')   ➜ 'Boolean-expected'

   dat-1 = ff             ➜ ff                                   (*)

   ded-2.sta = ?          ➜ ?

   **let**

      com-2 = ded-2.sta

   com-2 : Error         ➜ com-2

> **let**
>> (dat-2, bod-2) = com-2
>
> bod-2 ≠ ('Boolean')    ➔ 'Boolean-expected'
> **true**    ➔ (('Boolean'), dat-2)

Notice that the computation starts from an attempt of computing the value of the first argument and if this value is ff, then the computation terminates with this value (clause (*)). In this way, we avoid the computation of the second argument and hence a potentially infinite execution or error message. For the remaining Boolean operations, the corresponding constructors are defined analogously.

The unique constructor of the third category corresponds to conditional expressions[54]:

> when : DatExpDen x DatExpDen x DatExpDen ⟼ DatExpDen
> when.(ded-1, ded-2, ded-3).sta =
>> is-error.sta    ➔ error.sta
>> ded-1.sta = ?    ➔ ?
>> **let**
>>> com-1 = ded-1.sta
>>
>> com-1 : Error    ➔ com-1
>> **let**
>>> (dat-1, bod-1) = com-1
>>
>> bod-1 ≠ ('Boolean')    ➔ 'Boolean-expected'
>> dat-1 : Error    ➔ dat-1
>> dat-1 = tt    ➔ ded-2.sta
>> dat-1 = ff    ➔ ded-3.sta

To shorten this definition, we assume that two last clauses cover the case, where the computation of **ded-2.sta** or **ded-3.sta** does not terminate. In this case, we have to do with lazy evaluation since in evaluating **ded-2** we do not care if the evaluation of **ded-3** is infinite or results with an error message and analogously for **ded-3**[55].

## 5.3.3    The direct form of the definitions of constructors

From the viewpoint of a language implementor, the scheme (5.3.2-1) applied in the definition of **Cdd[Cc[ope]]** may be regarded as a procedure declaration where **Cc[ope]** is called as an internal procedure and which in turn calls **ope** and **round** (schema 5.2.3). This way corresponds to a bottom-up building process of an interpreter. From a user's viewpoint however, more convenient may be a definition written in a *direct form*, where only ope and round are called. Here is an example of such a definition for the constructor of adding an element to a list (-ed stands for "expression denotation").

---

[54] We call it **when** rather than **if** since the latter is reserved for conditional instructions.
[55] The acceptance of lazy evaluation in this place is a significant decision of language constructor, since it allows for the use of partial functions without the risk of error messages. Notice that if sqr(x) denotes square root of x, then the expression **if** x>0 **then** sqr(x) **else** sqr(-x) **fi** evaluated eagerly would generate an error signal for any x different from 0.

push-li-ed.(ded-l, ded-e).sta =                              (l – list, e – element)

    is-error.sta                    ➔ error.sta

    ded-i.sta = ?                    ➔ ?            for i = l,e

    ded-i.sta : Error                ➔ ded-i.sta     for i = l,e

    **let**

        (lis, bod-l)     = ded-l.sta

        (ele, bod-e)   = ded-e.sta

    sort.bod-l ≠ 'L'                ➔ 'list-expected'

    **let**

        ('L', bod-rl) = bod-l

    bod-rl ≠ bod-e                ➔ 'inconsistent-bodies'

    **let**

        new-list        = push-li.(lis, ele)

        new-com-lis  = (new-list, bod-l)

    oversized.new-com-lis  ➔ 'overload'

    **true**                        ➔ new-com-lis

## 5.3.4     The denotations of type expressions

The *denotations of type expressions* — called also *type-expression denotations* — are total functions mapping states into types and errors:

    ted : TypExpDen = State ⟼ TypeE

In the sequel they are used in type constant definitions (Sec. 6.1.3), data-variable declarations (Sec. 6.1.2) and procedure declarations (Sec. 7.3.4). In particular they allow to define types in a bottom-up style by referring to types stored earlier in type environment.

The denotations of type expressions are similar to the denotations of data expressions, but in this case, identifiers refer to type environments rather than to valuations and denotations are total functions.

The first constructor to be defined corresponds to *type constants*. In this case, we are talking about constants rather than variables since their values once established are never changed during program execution.

    typ-constant : Identifier ⟼ TypExpDen

    typ-constant.sta =

        is-error.sta   ➔ error.sta

        **let**

((tye, pre), sto) = sta

tye.ide = ?   ➜ 'type-constant-undefined'

**true**         ➜ tye.ide

Notice that unlike for data variables, in this case, we do not have a situation where a constant has been defined but not initialised. This is the consequence of the fact that type definitions (Sec. 6.1.3) always assign concrete types to constants.

The remaining constructors in this group are defined by referring to type-algebra constructors in a similar way as in the case of data expressions we were referring to composite constructors. Let Cdt (constructor of denotations of type expressions) denote a constructor that from constructors of types and transfers creates constructors of denotations.

Cdt[Yc[create-bo.boo]].().sta =

is-error.sta   ➜ error.sta

**true**         ➜ Yc[create-bo.boo]

and analogously for number- and word constants. Denotation constructor for expressions that build list-types is as follows:

Cdt[Yc[create-li]] : TypExpDen ⟼ TypExpDen

Cdt[Yc[create-li]].ted.sta

is-error.sta   ➜ error.sta

**let**

typ = ted.sta

typ : Error   ➜ typ

**true**         ➜ Yc[create-li].typ

As we see, the (internal) type which is used to construct the list type is computed from the state, and then an appropriate type constructor is applied. In an analogous way we define the remaining constructors:

| | | |
|---|---|---|
| Cdt[Yc[create-ar]] | : TypExpDen | ⟼ TypExpDen |
| Cdt[Yc[create-re]] | : TypExpDen x Identifier | ⟼ TypExpDen |
| Cdt[Yc[put-to-re]] | : TypExpDen x Identifier x TypExpDen | ⟼ TypExpDen |
| Cdt[replace-ty-tr] | : TypExpDen x Transfer | ⟼ TypExpDen |

Similarly as in the case of data-denotation constructors also in the case of type-denotation constructors we can write their definitions in a direct form.

## 5.3.5    The algebra of denotations of data-, type- and transfer expressions

The *algebra of expression denotations* — let us denote it by AlgExpDen — contains four carriers

ide   : Identifier

ded  : DatExpDen   = State → CompositeE

tra   : TraExpDen   = Transfer

ted    : TypExpDen    = State ⟼ TypeE

and all corresponding constructors defined in preceding sections.

The fact that denotations of transfer expressions are just transfers rather than functions from states to transfers is a consequence of the fact that in our model transfers cannot be "stored" in states, as it is in the case for data and types. This is, of course, an engineering decision rather than a mathematical must. It has been assumed only for the sake of simplicity.

At the end we can explain why the algebra of data was assumed to be two-sorted and as a consequence two-sorted became the algebras of bodies, composites and values. Well, the cause of that decision can be seen only on the level of the algebra of denotations. If in that algebra we would introduce separate carriers for number-, Boolean-, wordy-, list- etc. expression denotations, then we would need to introduce a separate variable constructor for each of these carriers. Consequently, at the level of syntax, we would have to somehow "label" variables with sorts. Technically this is possible, but would be rather unpractical and probably has never been applied in real languages.  As a consequence, since we decided to make the algebra of expression denotations two-sorted, there was no reason to assume that the algebras from which it was derived were more than two-sorted.

## 5.3.6    Six steps to the algebra of expression denotations

The way of passing from data algebra to the algebra of denotations as described in sections 5.2 and 5.3 will be referred to as *leveraging data algebras to the level of algebras of denotations*. Let us now sum up that way (see Fig. 5.3-1)

1. We start by defining a two-sorted algebra of data DatAlg. Even though there are different sorts of data in that algebra, they are all combined into the common carrier Data. This decision is because in the algebra data expression denotations we have only one sort of such denotations (explanations at the end of Sec. 5.3.5).

2. In the next step, we construct an algebra of bodies BodAlg which is similar to the former and has constructors adequate to data constructors, i.e. creating bodies reflecting the structures of data.

3. Over the two algebras, we construct an algebra of composites ComAlg which are pairs consisting of a data and its body. For every data operation, we define an associated to it composites constructor.

4. Over the algebra of composites, we construct a (not similar to it) two-sorted algebra of transfers TraAlg which are function mapping composites to composites. A particular case of transfers constitute boolean transfers called *yokes*.

5. Over the algebras of bodies and transfers (yokes), we construct a three-sorted algebra TypAlg of types which are pairs consisting of a body and a yoke.

6. Over the algebras of composites, transfers and types we construct a four-sorted algebra ExpDenAlg of the denotations of data-, transfer- and type expression and of identifiers. For each composites constructor and type constructor, we define in a certain standard way a constructor of corresponding denotations. To these constructors, we add data-variable and type-constant constructors and possibly some other constructors which do not correspond to composite constructors — in our case the when constructor.

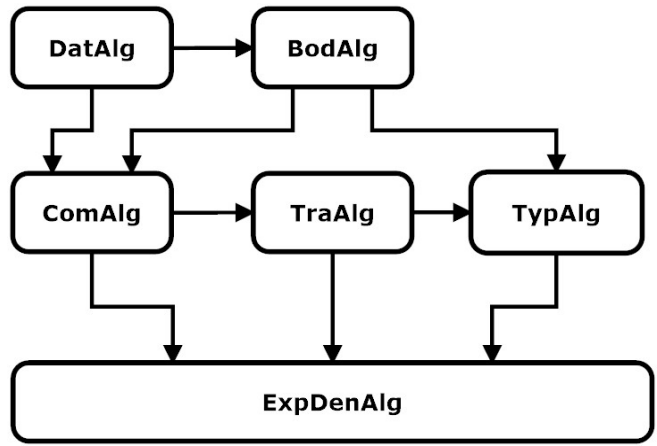

**Fig. 5.3-1 Six algebras from data to denotations**

Steps 1, 2 and 4 have a creative character. Here we are taking up basic decisions concerning the applicative part of our language. The remaining steps are rather standard what should allow for their — at least partial — algorithmization.

# 5.4 The algebras of syntax

## 5.4.1 The abstract syntax of Lingua-A

According to the five-step method of building a denotational model of an applicative part of a programming language (see Sec. 4.5) what we have to do now is to build abstract, concrete and colloquial syntax for **Lingua-A**.

As we already know from Sec. 2.12, starting from the signature of the algebra of denotations we can "algorithmically generate" an equational grammar of abstract syntax. That task will be performed for the expression denotations algebra ExpDenAlg in building the corresponding algebra of abstract syntax of expressions AlgExpA. To the four carriers of the former algebra, we shall assign now the corresponding carriers of abstract syntax (Tab. 5.4-1). Suffix A stands for "abstract".

| denotations | syntaxes | description |
|---|---|---|
| Identifier | Identifier | identifiers |
| DatExpDen | DatExpA | data expressions |
| Transfer | TraExpA | transfer expressions |
| TypExpDen | TypExpA | type expressions |

**Tab. 5.4-1 The carriers of syntactic algebras**

The equational grammar which describes our abstract-syntax algebra will be written with notational conventions introduced in Sec. 2.14. For each syntax category there is one domain equation of our grammar. First equation defines the domain of identifiers.

**Identifiers**

ide : Identifier =

{create-id.ide.() | ide : Identifier)}

If we would like to write this equation in the standard notation of abstract-syntax grammars as in Sec. 2.12, we had to explicitly list all identifiers acceptable in a given implementation. Instead, on a formal level we use the above abbreviation, and on manual's level we only show the set of characters available for identifiers, and we fix the maximal length of the latter. On the implementation level, we write a simple program checking if a given identifier is not too long and if it does not contain forbidden characters.

# Data expressions

dae : DatExpA =

**constants**

{Cdd[Cc[create-bo.boo.()]]    | boo : Boolean}          |

{Cdd[Cc[create-nu.num.()]]    | num: NumberS}         |

{Cdd[Cc[create-wo.wor.()]]    | wor : WordS}           |

**variables**

variable-dat (Identifier)                               |

**Boolean expressions**

and-ded (DatExpA , DatExpA)                          |

or-ded (DatExpA , DatExpA)                           |

no-ded (DatExpA)                                    |

Cdd[Cc[less]] (DatExpA , DatExpA)                     |

**numeric expressions**

Cdd[Cc[add]] (DatExpA , DatExpA)                      |

Cdd[Cc[divide]] (DatExpA , DatExpA)                   |

**word expressions**

Cdd[Cc[glue]](DatExpA, DatExpA)                       |

**list expressions**

Cdd[Cc[create-li]] (DatExpA)                          |

Cdd[Cc[push-li]] (DatExpA, DatExpA)                   |

Cdd[Cc[top-li]] (DatExpA)                             |

Cdd[Cc[pop-li]] (DatExpA)                             |

**array expressions**

Cdd[Cc[create-ar]] (DatExpA)                          |

Cdd[Cc[put-to-ar]] (DatExpA, DatExpA)                 |

Cdd[Cc[change-in-ar]] (DatExpA, DatExpA, DatExpA)     |

Cdd[Cc[get-from-ar]] (DatExpA, DatExpA)               |

**record expressions**

    Cdd[Cc[create-re]] (Identifier, DatExpA)          |

    Cdd[Cc[put-to-re]] (Identifier, DatExpA, DatExpA)     |

    Cdd[Cc[get-from-re]] (DatExpA, Identifier)        |

    Cdd[Cc[cut-from-re]] (Identifier, DatExpA)        |

    Cdd[Cc[change-in-re]] (DatExpA, Identifier, DatExpA)   |

**conditional expressions**

    when (DatExpA , DatExpA , DatExpA)

The abstract syntax of transfer- and type expressions is described by two following equations.

# Transfer expressions

tre : TraExpA =

**processing expressions**

    {Tc[create-nu.num.()] | num   : NumberS}  |

    {Tc[create-wo.wor.()]     | wor : WordS}    |

    Tc[add] (TraExpA, TraExpA)         |

    Tc[divide] (TraExpA, TraExpA)       |

    Tc[sum] (TraExpA)

    Tc[max] (TraExpA)            |

**yoke expressions**

    Tc[equal-nu] (TraExpA, TraExpA)      |

    Tc[less-nu] (TraExpA, TraExpA)       |

    Tc[small-nu] (TraExpA)          |

    Tc[increasing-nu] (TraExpA)      |

    {Tc[create-bo.boo.()] | boo : Boolean}  |

    and-T (TraExpA, TraExpA)         |

    or-T (TraExpA, TraExpA)          |

    not-T (TraExpA, TraExpA)         |

**structural expressions**

    create-for-li (TraExpA)          |

    create-for-ar (TraExpA)          |

    create-for-re (Identifier, TraExpA)    |

**selection expressions**

    Tc[get-tr-li]              |

Tc[get-from-ar] (TraExpA)                    |

Tc[get-from-re] (Identifier)                   |

**systemic expressions**

pass ()                                        |

# Type expressions

tex : TypExpA =

{Cdt[Yc[create-bo.boo.()]] | boo : Boolean}          |

{Cdt[Yc[create-nu.num.()]]        | num    : NumberS}  |

{Cdt[Yc[create-wo.wor.()]] | wor : WordS}             |

type-constant (Identifier)                           |

Cdt[Yc[create-li]] (TypExpA)                         |

Cdt[Yc[create-ar]] (TypExpA)                         |

Cdt[Yc[create-re]] (Identifier, TypExpA)             |

Cdt[Yc[put-to-re]] (TypExp, Identifier, TypExpA)     |

Cdt[replace-ty-tr]] (TypExpA, TraExpA)

In this place, it is worth returning again to Sec. 2.12 to recall notational convention assumed there. Namely, metaexpressions that appear in abstract-syntax-grammars, e.g.

Cdd[Ct[create-li]](TypExpA)

should be regarded as metanames of syntactic functions, hence in our case of a function which given an arbitrary word represented by metavariable tex returns the word

'Cdd[Ct[create-li]](' © tex © ')'

## 5.4.2    Concrete syntax of Lingua-A

As has been explained in Sec. 2.14 and in Sec. 4.5, *concrete syntax* was historically meant as a syntax which was provided to the user. In our approach concrete syntax constitutes only "denotational approximation" of the future programmer's syntax, i.e. such a syntax for which a denotational semantics exists. The final programmer's syntax is the result of introducing notational conventions called *colloquialisms* to concrete syntax (Sec. 4.5). Along with colloquial syntax, we define a function called *restoring transformation* that maps colloquial syntax into concrete syntax (see Fig. 4.5-2 in Sec. 4.5).

The present section contains a draft of concrete syntax of **Lingua-A** essentially devoted to illustrating the idea of concrete-syntax construction rather than to provide a well-elaborated syntax of a real language[56]. The corresponding algebra will be denoted by ExpAlg. Its carriers are defined explicitly by the equational grammar which is below, and its constructors are implicit in the equations of the grammar.

---

[56] As I have already mentioned earlier, I do not try to build a concrete **Lingua** but only to illustrate some general principles of building such a language.

The modifications of abstract syntax described below correspond to the homomorphism Co from Fig. 4.5-2, which in our case glues only expressions comprising the operator glue for words, both in the category of data expressions as well as in the category of transfer expressions. Everywhere beyond these two cases Co acts as an isomorphism, i.e. will not glue (will be one-to-one). Basic changes to be introduced are the following:

1. for concrete and colloquial syntax, we shall use the font `Courier New` whereas the symbols of not-zero-argument constructors commonly known as *keywords* shall be written with boldface characters in **`Courier New.`**

2. for Boolean constants, we take `true` and `false`,

3. numeric constants are written similarly with a colon as a separator between integer and fractional parts,

4. word constants are written in apostrophes, e.g. `'salary'`,

5. in case of data variables and type constants instead of variable-dat(abc) and type-constant(abc) we shall write `abc` in both cases; this glueing does not destroy homomorphism (in fact isomorphism) since the glued expressions belong to different carriers of the algebra,

6. numeric operators and predicates are written with infix notation and with "common" symbols +, /, <, hence we write, e.g. `(x + y)` and `(x < y)` instead of `add(x,y)` and `less(x,y)`; the "superfluous" parenthesis shall be dropped only at the level of colloquial syntax, since this transformation is not homomorphic,

7. the glueing operator for words is written as **`glue`** and also in this case we use infix notation; additionally we drop parentheses in writing, e.g. a **`glue`** b **`glue`** c instead of `(a `**`glue`**` b) `**`glue`**` c`; this homomorphism is safe in the sense of Theorem 2.13-1, which is due to the associativity of glueing constructor (more on that subject in Sec. 6.2.2 where we discuss concrete syntax of **Lingua-1**),

8. for Boolean constructors we use common names **`or, and, not`** written in an infix notation; in the context of data expressions they denote McCarthy's operators and in the context of transfer expressions — Kleene's operators; this does not lead to inconsistency since context always indicates the appropriate meaning,

9. conditional expressions are written with an infix notation:

   **`if`** DatExp **`then`** DatExp **`else`** DatExp **`fi,`**

   and similar conventions are assumed for list-, array- and record variables (see below),

10. data- and type expressions if written with infix notations are closed with the parenthesis **`ee`** which stands for *end-of-expression*.

Since the new algebra is homomorphic to the former and the corresponding homomorphism Co does not glue too much, the existence of a (unique) denotational semantics that maps new syntax into denotations follows from Theorem 2.13-1.

Our new grammar is described below. In this case the names of syntactic categories are written without a suffix, since we have to do with a grammar addressed to the user, who does not need to know about abstract and concrete syntaxes at all.

ide : Identifier =

    `ide` | …                               (for every syntactically acceptable ide)

In this clause we apply a not-too-formal notational convention instead of a set-theoretic convention applied in the definition of abstract syntax in Sec. 5.4.1. In this case `ide` denotes the syntactical representation of ide and the meta-expression:

> `ide |` ...                                                         (for every ide : Identifier)

means:

> {`ide` | ide : Identifier}

An analogous convention will be used in the definition of the syntax of data expressions:

## Data expressions

> dae : DatExp =

> **constants**

>> `true` | `false`                                                   |

>> `num`                                                       | (for every num : NumberS)

>> `wor`                                                       |    (for every wor : WordS)

> **variables**

>> Identifier                                                  | (constructor's name is omitted)

> **Boolean expressions**

>> (DatExp **and** DatExp)                                     |

>> (DatExp **or** DatExp)                                      |

>> (**not** DatExp)                                            |

>> (DatExp < DatExp)                                           |

> **numeric expressions**

>> (DatExp + DatExp)                                           |

>> (DatExp / DatExp)                                           |

> **word expressions**

>> DatExp **glue** DatExp                                      | (parentheses have been dropped!)

> **list expressions**

>> **list** DatExp **ee**                                      |

>> **push** DatExp **on** DatExp **ee**                        |

>> **top** (DatExp)                                            |

>> **pop** (DatExp)                                            |

> **array expressions**

>> **array** DatExp **ee**                                     |

>> **add-to-arr** DatExp **new** DatExp **ee**                 |

>> **change-arr** DatExp **at** DatExp **by** DatExp           |

    **arr** DatExp **at** DatExp **ee**            |

**record expressions**

    **record** Identifier **of-value** DatExp **ee**    |

    **add-atr** Identifier **of-value** DatExp **to** DatExp **ee**    |

    **rec** DatExp **at** Identifier **ee**    |

    **remove-atr** Identifier **from** DatExp **ee**    |

    **change-rec** DatExp **at** Identifier **by** DatExp **ee**    |

**conditional expression**

    **if** DatExp **then** DatExp **else** DatExp **fi**

# Transfer expressions

    tre : TraExp =

**processing expressions**

    num    | for every num : NumberS

    wor    | for every wor : WordS

    (TraExp + TraExp)    |

    (TraExp / TraExp)    |

    **sum** (TraExp)    |

    **max** (TraExp)    |

    TraExp **glue** TraExp    |    (parentheses dropped)

**transfer-yoke expressions**

    true | false    |

    (TraExp = TraExp)    |

    (TraExp < TraExp)    |

    **small-number** (TraExp)    |

    **increasing** (TraExp)    |

    (TraExp **and** TraExp)    |

    (TraExp **or** TraExp)    |

    (**not** TraExp)    |

**quantifier expressions**

    **all-list** TraExp **ee**    |

    **all-array** TraExp **ee**    |

**selection expressions**

    **top**    |

> `array`[TraExp]                           |
>
> `record.`Identifier                       |
>
> **passing expression**
>
> `value`                                    |

## Type expressions

tex :TypExp =

> `boolean`                                              |
>
> `number`                                               |
>
> `word`                                                 |
>
> Identifier                                             |
>
> `list-type`  TypExp **ee**                             |
>
> `array-type`  TypExp **ee**                            |
>
> `record-type`  Identifier **as** TypExp **ee**         |
>
> `expand-record-type`  TypExp **at** Identifier **by** TypExp  **ee**  |
>
> `replace-transfer-in` TypExp **by** TraExp **ee**

As was already pointed out, the only "gluing action" of abstract syntax into concrete syntax has a place in dropping parentheses associated with the operation `glue` which corresponds to the concatenations of words. Apparently, it might seem that glueing also appears when a data expression and a transfer expression are "glued" into the same concrete expression, e.g. when

> Cdd[Cc[create-bo.tt]] and
>
> Tc[create-bo.tt]

are both transformed into

> `true,`

Notice, however, that in this case, we do not have to do with a glueing homomorphism since these abstract expressions do not belong to different carriers.

### 5.4.3    The colloquial syntax of Lingua-A

The definition of a colloquial syntax is a very important step in the process of language construction since it makes our language more user-friendly. We free ourselves from the algebraic rigor of concrete syntax without losing anything of mathematical precision but gaining on clarity.

We shall assume that colloquial syntax includes all concrete syntax which means that the use of colloquialisms is optional. On the level of an algebra, each colloquialism is a new constructor, which makes the new syntactic algebra not similar to the former.

Below I show examples of colloquialisms associated with operations on simple data and on array- and record data. They are not necessarily the best possible solutions since the only aim here is to show the method rather than to construct a real language. All colloquialisms are defined informally based on examples, which should, however, indicate the way to both — grammatical clauses and a restoring transformation.

### 5.4.3.1   Universal rules

These rules concern all sorts of expressions:

1. we allow spaces and carriage returns which will be removed by the restoring transformation,

2. none of the keywords `true, false, if, then,`… cannot be used as an identifier; in this case, restoring transformation does not modify a program but only generates an error message; in traditional parsers, this analysis is performed at the lexical level.

### 5.4.3.2   Boolean data-expressions

For Boolean expressions we allow the omission of the "unnecessary" parentheses and assume the priority of conjunction over alternative. E.g.

- instead of writing (x **or** (y **or** z)) we write x **or** y **or** z        and

- instead of writing (x **or** (y **and** z)) we write x **or** y **and** z

In the first case the restoring transformation may add parentheses (i.e. may be defined) in an arbitrary way, which is due to the associativity of the alternative. In the second — it has to observe the assumed priority.

### 5.4.3.3   Numeric data-expressions

The case of numeric expressions is a little more complicated since in real situations, i.e. where we have four arithmetic operations (rather than two as in our simplified language), then the addition and the multiplication are not associative. This is due to the effect of overloading. E.g., if the maximal size of a number is 10, then

((-4 + 9) + 2) = 7            but

(-4 + (9 + 2)) = 'overload'

A usual practice is therefore that parentheses-free expressions are evaluated from left to right in using the priorities between operations. E.g., the expression:

```
x + y + z + x*y
```

is restored to

```
((x + y) + z) + (x*z)
```

### 5.4.3.4   Array data-expressions

In this category we are going to have four colloquialisms. The first of them concerns the constructor of an array. For instance, the colloquial expression

**array** [x, x+y,  3*y]

unfolds to the concrete expression:

**add-to-arr**                                    (add value 3*y to the array)

   **add-to-arr**                                 (add value x+y to the array)

      **array** x **ee**                          (create one-element array with value x)

```
    new x+y ee
  new 3*y ee
```

Of course, each simple numerical expression may be replaced here by an arbitrary expression. If `measurement-data` is an array variable, then the colloquial expression

```
    measurement-data.[x+1]
```

unfolds to concrete expression

```
    arr measurement-data at x+1 ee
```

and

```
    measurement-data.[x+1].[y-1]
```

unfolds to:

```
    arr arr measurement-data at x+1 ee at y-1 ee
```

The case of adding a new element to an array may be treated analogously:

```
    add-to-arr measurement-data new [x, x + y, 3*y] ee
```

and in the case of array modification (here we introduce a new symbol „<="):

```
    change-arr measurement-data by
        s   <= x,
        s+1 <= x+y,
        3*p <= z-1
    ee
```

which unfolds to:

```
    change-arr
      change-arr
        change-arr measurement-data at s by x ee
        at s+1 by x+y ee
      at 3*p by z-1 ee
```

### 5.4.3.5    Record data-expression

Examples for records may be similar to these for arrays. For instance, we may assume that a colloquial expression:

```
    record
        ch-name    <= 'John',
        fa-name    <= 'Smith',
        birth-date <= 1968,
        award-years <= award-years-Smith
    ee
```

corresponds to the concrete:

```
    add-atr award-years        of-value award-years-Smith to
```

```
add-atr birth-date      of-value 1968 to
   add-atr fa-name       of-value 'Smith' to
      set-record ch-name of-value 'John'
         ee
      ee
   ee
ee
```

and a colloquial expression

```
employee.(fa-name)
```

corresponds to the concrete:

**rec** employee **at** fa-name **ee**

Notice that despite a similarity between selection expression from an array and from a record, there is no ambiguity since array indices are closed in bracket parenthesis and record indices in ordinary parenthesis. Therefore, if `employee` is an array variable, then the corresponding selection expression would have the form

```
employee.[fa-name]
```

### 5.4.3.6    Array transfer-expressions

We introduce school rules for dropping parentheses with corresponding priorities between operations. For instance in the place of:

```
(2+value)< 10
```

we write

```
2+value < 10
```

In the place of

**get-from-array** x+1 **ee**

we write

**array.**[x+1]

It is to be recalled that in this case **array** is not an array variable — as, e.g. in the expression `measurement-data.[x+1]` — but a keyword that means that the input composite of this transfer should carry an array and our expression selects from this array an element with index `x+1`.

### 5.4.3.7    Record type-expressions

In this case, we introduce colloquialisms analogous as for data expressions. For instance:

**record-type**

```
ch-name     as string,
fa-name     as string,
birth-date  as number,
award-years as array-of number ee
```

```
    ee
```

unfolds to a concrete written employing **record-of** and **expand-record**.

### 5.4.3.8    Record transfer-expression

In this case similarly as for arrays we write:

```
    record.fa-name
```

instead of

```
    get-from-record fa-name ee
```

In the first expression, **record** is a keyword as **array** in case of arrays.

### 5.4.3.9    Type expressions

In the majority of programming languages yokes do not appear in the definitions of types, hence in such cases, concrete syntax of type definitions would be of the form:

```
    set-type TypExp with true ee
```

for example:

```
    set-type array-of number ee with true ee
```

In that case, the corresponding colloquial expression would be

```
    set-type
       array-of number ee
    ee
```

The general rule is such that if the yoke is the constant `true`, then we drop the whole phrase „**with** true".

In the case of record types, we introduce colloquialisms that allow describing yokes and bodies in one expression. For instance, we write:

```
    record-type
       ch-name      as string,
       fa-name      as string,
       birth-date   as number with small-number,
       award-years  as array-of number with small-number ee
    ee
```

which means

```
    type
    record-of
       ch-name      as string,
       fa-name      as string,
       birth-date   as number,
       award-years  as array-of number ee
       ee
```

```
with
small-number(record.birth-date) and
all-of-array record.award-years with small-number(value) ee
ee
```

The yoke of this record type is satisfied if the following three conditions are satisfied:

1. input composite carries a record with at least two attributes `birth-date` and `award-years`.

2. a small number is assigned to the attribute `birth-date`,

3. an array of small numbers is assigned to `award-years`.

The remaining information about the record are included in the type expression.

## 5.5    The tasks of a language designer

The designer of a programming language has three creative tasks to complete where he or she takes important decisions concerning the future functionality and structure of the language:

1. the construction of an algebra of denotation DenAlg that determines the unique algebra of abstract syntax AbsSynAlg,

2. the construction of a homomorphic to it algebra of concrete syntax SynAlg that is not more ambiguous than DenAlg,

3. the construction of a colloquial syntax and the corresponding restoring transformation.

If concrete syntax has been built correctly, then the semantics of the language, i.e. the homomorphism of algebras

Sem : SynAlg ⟼ DenAlg

exists and is unique.

The creation of language implementation consists in writing a procedure that each sequence of characters from colloquial syntax, let it be coll-program, will transform in three steps corresponding to the restoring transformation and two homomorphisms Co and As (see Fig. 4.5-2)

The first step performs a relatively simple transformation from colloquial program coll-program to concrete program con-program. Of course, during this transformation, an error message may be raised.

The second step is performed by a *syntax analyser,* also called a *parser,* that constructs the co-image of con-program in the abstract syntax. This coimage is nothing else but a linear description of a *parsing tree* of the concrete program. In our model, it satisfies the equation

Co.[abs-program] = con-program

If the concrete syntax is not unambiguous, i.e. Co is a glueing homomorphism, then the parser is defined in such a way that it picks up just one of many coimages of con-program. If concrete syntax glues not more that denotations, the choice of the coimage is irrelevant for the final denotation.

If the attempt of building parsing tree fails, then the user is informed that the elaborated program contains syntax errors, which means that it does not belong to the language defined by the concrete grammar. Most frequently the parsing procedure also points out the location of the error.

If the second step terminates properly, then the third one consists in program execution. In our model, it corresponds to the construction of denotation `program-den` that satisfies the equation

program-den = As.[abs-program]

At the level of implementation, it corresponds to the execution of the program by an *interpreter* or to the generation of machine code by a *compiler*. In the sequel, we shall talk about interpreters, as closer to programmer's intuition, but all remarks about interpreters will equally concern compilers.

The implementator of a programming language has therefore to create three basic software tools:

1. a syntax analyser that transforms colloquial syntax into concrete syntax,

2. a parser of concrete syntax into abstract syntax,

3. an interpreter (or compiler) of abstract syntax.

The second and third task should be algorithmizable starting from the grammar of concrete syntax and from the definitions of denotations' constructors.

A language should be constructed in such a way that as many of potential errors as possible are detectable at the level of syntax analysis since it is much faster than program execution. We try therefore to describe possibly many language features at the syntactic level, which results in creating maximally many carriers in the algebra of denotations. For instance, a well-constructed grammar should detect a syntactic error in the program

**if** y > 0 **then** y+1 **else list-type** number **ee fi**

where **else** is followed by a type expression rather than by a data expression[57]. On the other hand, on the syntactic level, we are not able to check if a given variable is e.g. of a numeric type. This analysis must be performed, therefore, at the level of execution, i.e. of semantics[58].

## 5.6   Two forms of a manual

A denotational model of a programming language is a starting point not only for the development of an implementation but also for writing a user manual. Since manuals written in that way have not appeared yet, there are no practical experiences available in that field. It seems however rather evident that such a manual should describe a language in three steps and in that order:

1. concrete syntax described by equational grammar and illustrated by examples,

2. colloquial syntax illustrated by examples of restoring transformations (e.g. as in Sec. 5.4.3),

---

[57] In some languages, e.g. in C, such a construction is acceptable.

[58] As a matter of fact type errors may be detected on the level of co called *static semantics*, where we compute only types (in our case bodies) without computing values. Such a solution was applied in the semantics of programming language Ada [12] in the framework of VDM methodology (Vienna Development Method) [10]. More abut Ada in a foot note of Sec. 4.1.

3. the semantics of concrete syntax, i.e. the association of concrete programs to their denotations without referring to abstract syntax.

In defining semantics, one has to choose between two forms of definitions (cf. Sec. 5.3.3):

A. definitions that refer to denotation-algebra constructors defined earlier; such definitions will be referred to as *algebraic*,

B. definitions that describe constructors explicitly; such definitions will be called *direct*.

Which of these definitions we choose, depends on its addressee.

For implementators, the algebraic form seems more convenient. The definitions of denotations' constructors may be written as mutually recursive procedures and the definition of semantics as a mutually recursive set of procedures that call the former procedures.

In turn, for a language user (a programmer) direct semantics seems more convenient since the meaning of each syntactic constructions is describe explicitly and totally in one definition.

## 5.7    A sketch of the semantics of Lingua-A

Let us recall that ExpAlg and ExpDenAlg denote respectively the algebras of concrete syntax and of denotations of **Lingua-A**. Since the former is not more ambiguous than the latter, there exists a unique homomorphism:

Cs : ExpAlg ⟼ ExpDenAlg

with five components:

Sid   : Identifier ⟼ Identifier

Sde  : DatExp   ⟼ DatExpDen

Stre : TraExp    ⟼ TraDenExp

Ste   : TypExp   ⟼ TypExpDen

Below some examples of the definitions of these components are shown in two versions: algebraic and direct. With `Courier,` we shall write not only concrete syntactic elements but also corresponding metavariables. I recall that Cc[ope] denotes a composite constructor that corresponds to a data operation ope (Sec. 5.2.3), whereas Cdd[Cc[ope]] denotes a corresponding constructor of data expression denotations.

### Identifiers

Sid : Identifier ⟼ Identifier

Sid.[`ide`] = create-id.ide.()       for every **ide**                                    (algebraic form)

Sid.[`ide`] = ide                         for every **ide**                                              (direct form)

### Data expressions

Sde : DatExp ⟼ DatExpDen     hence

Sde : DatExp ⟼ State → CompositeE

Sde.[`true`]      = Cdd[create-bo-co.tt].()                              (algebraic form)

Sde.[`true`].sta =                                                          (direct form)

 is-error.sta    ➔ error.sta

 **true**            ➔ (tt, ('Boolean'))

Sde.[`ide`] = variable-dat.(Sid.[`ide`])        for ide : Identifier

Sde.[`ide`].sta =

is-error.sta       ➔ error.sta

 **let**

  (env, (vat, 'OK')) = sta

 vat.ide = ?    ➔ 'undeclared-variable'

 **let**

  ((dat, bod), yok) = vat.ide

 dat = Ω    ➔ 'uninitialized-variable'

 **true**         ➔ (dat, bod)

Sde.[`(dae-1 + dae-2)`] = Cdd[add-co].(Sde.[`dae-1`], Sde.[`dae-2`])

Sde.[`(dae-1 + dae-2)`].sta =

 is-error.sta            ➔ error.sta

 Sde.[`dae-i`].sta = ?  ➔ ?                          for i = 1, 2

 **let**

  (dat-i, bod-i) = Sde.[`dae-i`].sta            for i = 1, 2

 bod-i ≠ ('number')          ➔ 'number-expected'      for i = 1, 2

 **let**

  num = round.(add.(dat-1, dat-2))[59]

  com  =  (num, ('number'))

 oversized.com      ➔ 'overload'

 **true**                ➔ (num, ('number'))

## Transfer expressions

It is to be recalled that transfer denotations are transfers themselves, hence in the following definitions we have metavariables com.

---

[59] Here we use the fact that composites are well-structured, hence if bod-I = ('number') for I = 1,2, then dat-i : Number for I = 1,2.

Stre : TraExp ⟼ Transfer

Stre : TraExp ⟼ CompositeE ⟼  CompositeE

Stre.[`true`] = create-tr-bo.tt.()

Stre.[`true`].com =

   com : Error   ➔ com

   **true**           ➔ (tt, ('Boolean'))

Stre.[**`value`**] = pass.()

Stre.[**`value`**].com = com

Stre.[**`all-list` vat-l `with` vat-b `ee`**] =

   wszystkie-na-li.(Stre.[`vat-l`], Stre.[`vat-b`])

Stre.[**`all-list` vat-l `with` vat-b `ee`**].com =

com : Error                        ➔ com

   **let**

      com-l =Stre.[`vat-l`].com

   com-l : Error                  ➔ com-l

   sort.com-l ≠ 'L'            ➔ 'list-expected'

**let**

   (dat-1,…,dat-n) = data.com-l

   ('L', bod)        = body.com-l

   com-i            = Stre.[`vat-b`].sta.(dat-i, bod)     for i = 1;n

com-i : Error                ➔ com-i

(∀ i = 1 ;n) com-i = (tt, ('Boolean'))   ➔ (tt, ("Boolean'))

**true**                            ➔ (ff, ("Boolean'))

## Type expressions

The denotations of type expressions refer to the types memorised in type environments.

   Sty : TypExp ⟼ TypExpDen

   Sty : TypExp ⟼ State ⟼ TypeE

   Sty.[`ide`] = type-constant.`ide`

Sty.[`ide`].sta

    is-error.sta    ➜ error.sta

    **let**

        ((tye, pre), sto) = sta

    tye.`ide` = ?   ➜ 'type-constant-undefined'

    **true**        ➜ tye.`ide`

The remaining definitions are left to the reader.

# 6  Lingua-1 — an imperative language without procedures

Starting from this section, we shall develop successive languages from **Lingua** series by extending each of them with new mechanisms. **Lingua-1** emerges from **Lingua-A** by adding the mechanisms type definitions, variable declarations and instructions. Procedures are postponed to Sec. 7.

## 6.1    Denotations

### 6.1.1    Denotational domains

The denotational domains of **Lingua-1** correspond to its future nine syntactic categories:

1. identifiers,
2. data expressions,
3. transfer expressions,
4. type expressions,
5. declarations of data variables,
6. definitions of type constants,
7. instructions,
8. preambles that are sequences of declarations and definitions,
9. programs that are pairs composed of a preamble and an instruction.

Consequently the carriers of the future algebra of denotations are the following:

| | | | |
|---|---|---|---|
| ide | : Identifier | | (6.1.1-1) |
| ded | : DatExpDen | = State → CompositeE | (data-expression denotations) |
| tra | : TraExpDen | = Transfer | (transfer-expression denotations) |
| ted | : TypExpDen | = State ↦ TypeE | (type-expression denotations) |
| vdd | : VarDecDen[60] | = State ↦ State | (variable-declaration denotations) |
| tdd | : TypDefDen | = State ↦ State | (type-constant denotations) |
| ind | : InsDen | = State → State | (instruction denotations) |
| pde | : PreDen | = State → State | (preamble denotations) |

---

[60] We use here the notion of variable declaration rather than just declaration, since in further versions of **Lingua** we are going to have declarations of procedures and functions.

prd   : ProDen        = State → State                                    (program denotations)

As was already mentioned earlier, denotations of data expressions are partial functions. Although in **Lingua-1** reachable denotations of such expressions are total function, in **Lingua-2** due to procedures they may be partial as well. In the case of instructions and programs partiality is already there, since a **while** instruction may generate an infinite execution.

The first four domains cover applicative denotations and have been discussed in Sec.5. The remaining concern *imperative denotations* and are discussed below.

## 6.1.2    The declarations of data variables

As we already know (Sec. 5.3.2) *data variables* or simply *variables* are identifiers with values or pseudo-values assigned to them in valuations. Variable's declaration assigns a pseudo-value to an identifiers, i.e. assigns a type leaving the data temporarily undefined. Values are assigned to variables by assignment instructions (Sec. 6.1.4).

declare-dat-var : Identifier x TypExpDen ↦ VarDecDen

declare-dat-var.(ide, ted).sta =

   is-error.sta          �ít sta

   declared.ide.sta     ➡ sta ◄ 'variable-declared'

   **let**

     (env, (vat, 'OK'))   = sta

     typ                        = ted.sta

   typ : Error          ➡ sta ◄ typ

   **true**                  ➡ (env, (vat[ide/(Ω, typ)], 'OK'))

If a state caries an error, then the declaration does not change the state. Otherwise, if in the current state the identifier has been already declared, then an error signal is "loaded" to the state. As we see, no identifier can be declared twice in one program.

If the type expression generates an error, then this error is passed to the state. Otherwise, the valuation is modified by assigning a pseudo-value $(\Omega, \text{typ})$ to **ide**. As we shall see in the sequel, variable's declarations are the only imperative constructs that introduce pseudo-values to states. An identifier with assigned value or pseudo-value (dat, typ) is said *to be of type* typ.

Variable declarations can be combined sequentially by the following constructor:

sequence-vde : VarDecDen x VarDecDen ↦ VarDecDen

sequence-vde.(vdd-1, vdd-2) = vdd-1 ● vdd-2

## 6.1.3    The definitions of type constants

Type constants are identifiers with types assigned in type environments. We call them *constants* rather than *variables* since a type once assigned to an identifier remains unchanged during the whole execution of a program.

The following constructor creates a denotation of a type constant declaration:

define-typ-con : Identifier x TypExpDen ↦ TypDefDen

define-typ-con.(ide, ted).sta =

is-error.sta        ➡ sta

declared.ide.sta    ➡ sta ◄ 'identifier-not-free'

**let**

  typ = ted.sta

  ((tye, pre), sto) = sta

typ : Error         ➡ sta ◄ typ

**true**             ➡ ((tye[ide/typ], pre), sto)

As we see, type definitions modify only type environments and possibly generate an error message.

Similarly to variable declarations also type definitions may be combined sequentially:

sequence-tde : TypDefDen x TypDefDen ↦ TypDefDen

sequence-tde.(tdd-1, tdd-2) = tdd-1 ● tdd-2

## 6.1.4    Assignment instruction

To define assignment instructions, we introduce the notion of a *coherence-relation* between bodies. We say that body bod-1 *is coherent* with bod-2, in symbols

bod-1 coherent bod-2

whenever:

1. bod-1 = bod-2 or

2. both bodies are record-bodies, and the set of attributes of one of them is a subset of the set of attributes of the other, and on the common set of attributes they coincide.

In other words, two bodies are coherent, if they are identical, or if they are record bodies and one of them results from the other by adding or by removing an attribute. (Sec. 5.2.2). As is easy to see, the relation of coherence is reflexive and symmetric, but not transitive.

An imperative denotation is said to be *conservative* if the two following conditions are satisfied:

1. dim is transparent wrt states carrying errors, i.e. if a state sta carries an error, then dim.sta is defined and carries the same error; notice that sta and dim.sta are not necessarily the same (this is going to be the case if exception handling is involved as described in Sec. 6.1.8),

2. if dim does not generate an error, then the bodies of all data variables declared in the input-state are coherent with their bodies in the output-state; in particular, this means that not-record-bodies assigned to variables are not changed.

Observe that the denotations of variable declarations and of type definitions are conservative. As we shall see, in **Lingua** all reachable imperative denotations will be conservative. Moreover, the all of them except error-handling will not change a state if it carries an error.

A constructor of imperative denotations is said to be *decent* if it transforms conservative denotations into conservative denotations. In the sequel, we shall make sure that all such constructors are decent.

Now we are prepared to define a constructor corresponding the assignment instruction:

assign : Identifier x DatExpDen ⟼ InsDen

assign.(ide, ded).sta =

    is-error.sta                ➔ sta

    **let**

        ((tye, pre), (vat, 'OK')) = sta

    vat.ide = ?            ➔ sta ◄ 'identifier-not-declared'

    ded.sta = ?            ➔ ?           (an infinite execution)

    ded.sta : Error        ➔ sta ◄ ded.sta

    **let**

        ((dat-f, bod-f), tra) = vat.ide               (f – former)

        (dat-n, bod-n)    = ded.sta              (n – new)

        com             = tra.(dat-n, bod-n)

    com : Error            ➔ sta ◄ com

    **not** bod-n coherent bod-f    ➔ sta ◄ 'no-coherence'

    **not** com : BooComposite    ➔ sta ◄ 'a-yoke-expected'

    com ≠ (tt, ('Boolean')      ➔ sta ◄ 'yoke-not-satisfied'

    **let**

        val-n = ((dat-n, bod-n), tra)

    **true**              ➔ ((tye, pre), (vat[ide/val-n], 'OK'))

For an assignment instruction to be executable, the variable which is going to have a new value, must be previously declared. Assignment may change variable's type, but only in such a way that the new body is coherent with the former and that the new composite satisfies the current yoke.

As we see, in **Lingua-1** the type of a record-variable may be changed in the course of program execution whereas the types of the remaining variables cannot be changed. As we are going to see in Sec. 12.7.6.11, the same will be true for table-variables in **Lingua-SQL**.

Some comments are needed about record-type variables. There are basically two strategies for constructing records. The first consist of declaring a record-variable with one attribute and then adding more attributes in successive assignments. Anticipating our future colloquial syntax, this strategy may be illustrated with the following example:

```
set type-employee as
   record-type
      ch-name of type string
   ee
tes ;
let employee as type-employee tel ;
employee := record ch-name <= 'John' ee ;
employee := add-atr fa-name <= 'Smith' to employee ee ;
```

```
employee := add-atr birth-year <= 1968 to employee ee
```

Notice that `'Smith'` is a concrete expression which corresponds to an abstract expression

Cdd[Cc[create-wo.'Smith'.()]]

The denotation of that expression creates the following composite (assuming that the state does not carry an error):

Cdd[Cc[create-wo.'Smith'.()]].sta =

(create-wo.'Smith'.(), Bc[create-wo.'Smith'].() ) =

('Smith', ('word'))

From there, according to the definition of assignment-denotation, the second of the three assignments expands record type by a new attribute `fa-name` and assigns to that attribute the value `'Smith'`. The third assignment acts analogously.

An alternative for such a construction consists in declaring a record-variable already with the "target" type and then assigning an appropriate value to it:

```
set type-employee as
  record-type
    ch-name       of type string,
    fa-name       of type string,
    birth-year of type number
  ee
tes ;
let employee as type-employee tel ;
employee := record
             ch-name      <= 'John',
             fa-name      <= 'Smith',
             birth-year <= 1968
           ee
```

## 6.1.5    The instruction of transfer-replacement

This instruction is analogous to the former with the difference that this time we do not change a composite but a transfer:

replace-tr : Identifier x TraExpDen ↦ InsDen

replace-tr.(ide, tra-n).sta =                                          (n - new)

is-error.sta                    ➜ sta

**let**

((tye, pre), (vat, 'OK')) = sta

vat.ide = ?                         ➜ 'identifier-not-declared'

**let**

((com, tra-f)  = vat.ide                                                       (f - former)

tra-n.com ≠ (tt, ('Boolean') ➜ 'yoke-not-satisfied'

**let**

   val-n = (com, tra-n)

**true**                          ➜ ((tye, pre), vat[ide/val-n], 'OK')

The new value has the old composite but a new transfer. This transfer must be satisfied by the current composite. This instruction has been introduced mainly for the sake of **Lingua-SQL** (Sec. 12.7.6)[61].

Transfer replacement has been rated to the group of instructions, rather than declarations, because it can appear at any position in a program, unlike declarations that can appear only in preambles (see Sec. 6.1.9).

It is worth noticing in this place that in the algebra of types we have a similar constructor replace-ty-tr, which, however, is a constructor of types rather than of instruction denotations.

## 6.1.6     Trivial instruction

*Trivial instruction* is an identity transformation of a state into itself. As we are going to see, it will be useful in defining the declarations of functional procedures (Sec. 7.5). The denotation of this instruction is created by the following constructor:

create-trivial-ins : ↦ InsDen

create-trivial-ins.().sta = sta

## 6.1.7     Structured instructions

*Structured instructions* are built by four constructors. Three basic constructors — sequential composition, conditional composition and loop — and a special error-handling constructor. Let's start with the sequential composition:

sequence-ins : InsDen x InsDen ↦ InsDen

sequence-ins.(ind-1,ind-2) = ind-1 ● ind-2

Sequentially composed instructions are executed one after another. Conditional composition is defined as follows:

if : DatExpDen x InsDen x InsDen ↦ InsDen

if.(ded, ind-1, ind-2).sta =

   is-error.sta                    ➜ sta

   ded.sta = ?                     ➜ ?

   ded.sta : Error                 ➜ sta ◄ ded.sta

   **let**

      (dat, bod) = ded.sta

---

[61] This very general form of transfer-replacement has been chosen for the sake of simplicity. In real situations one should think about more specific replacements, as e.g. by conjunctively adding a yoke to a transfer.

sort.bod ≠ ('Boolean')  ➜ sta ◄ 'Boolean-expected'

dat = tt  ➜ ind-1.sta

**true**  ➜ ind-2.sta

It is to be emphasised that in our language due to while-loops (see below) the execution of each of these instructions may be infinite, which means that the states ind-1.sta or ind-2.sta may be undefined. If dat = pp and ind-1.sta is undefined then the terminal state of the conditional instruction is undefined as well, and in the opposite case the final state is undefined if ind-2.sta is undefined.

while : DatExpDen x InsDen ⟼ InsDen

while.(ded, ind).sta =

(1) is-error.sta  ➜ sta

(2) ded.sta = ?  ➜ ?

(3) ded.sta : Error  ➜ sta ◄ ded.sta

**let**

(dat, bod) = ded.sta

(4) sort.bod ≠ ('Boolean')  ➜ sta ◄ 'Boolean-expected'

(5) dat = ff  ➜ sta

(6) **true**  ➜ (ind ● [while.(ded, ind)]).sta

In this definition we have to do with recursion which means that our constructor is defined by a fixed-point equation. Due to that fact denotations of instructions may be partial functions. In the sequel, when imperative and functional procedures are introduced, the partiality of while.(ded, ind) will may have three different sources:

1. if the Boolean expression represented by ded includes a function-procedure the call of which generates an infinite execution,

2. if the instruction represented by ind includes a local while, i.e. if ind.sta = ?,

3. if „the main" loop runs indefinitely.

---

**Comment 6.1.6-1** The definition of **while** is recursive since this operator appears in "its' own" defini-tion. Therefore we have to do with a fixed-point equation in a CPO of partial functions (Sec. 2.6), which is

**InsDen = State → State**

However the solution of our fixed-point equation is not the **while**-constructor, but the effect of its application to the pair **(ded, ind)**, hence the function:

**while.(ded, ind) : State → State**

Of course, for every such a pair **(ded, ind)** we have to do with a different equation. To be sure that solutions of such equations exist, we have to prove that the right-hand sides of such equations are continuous in the CPO **State → State**. To do that let us introduce the following notations:

**NotOK**   = {(sta, sta) | (1) satisfied}

**ExpEr**   = {(sta, sta ◄ ded.sta) | (3) not (1) and (2)}

**NotBoo**  = {(sta, sta ◄ Boolean-expected') | (4) not (1) and (2) and (3)}

**FF**      = {(sta, sta) | (5) not (1), (2), (3) and (4)}

**TT**      = {(sta, sta) | not (1), (2), (3), (4) and (5)}

Our definition may be now written as an equation of a form as in the Theorem 2.6-1:

> **X = NotOK | ExpEr | NotBoo | FF | TT•ind•X**
> Since the operators | and • are continuous, the least solution of that equation exists, and since the coefficients of that equations have mutually disjoint domains, from the quoted theorem we may conclude that its solution is a function and has the form:
> **X = (TT• Din)\* • (NotOK | ExpEr | NotBoo | FF)**

## 6.1.8   Error handling

The last structured constructor concerns *error-handling mechanism*. It allows running a chosen instruction whenever an indicated error-message is generated.

if-error : DatExpDen x InsDen → InsDen

if-error.(ded, ind).sta =

    **not** is-error.sta    ➔ sta

    **let**

        (env, (vat, err))  = sta

        sta-1           = sta ◀ 'OK'

    ded.sta-1 = ?      ➔ ?

    **let**

        com = ded.sta-1

    com : Error        ➔  sta ◀ com @ 'error-handling-not-executed'

    sort.com ≠ ('word')    ➔ 'sta ◀ 'word-expected' @

                                   'error-handling-not-executed'

    **let**

        (wor, ('word')) = com

    wor ≠ err         ➔ sta

    ind.sta-1 = ?      ➔ ?

    **let**

        sta-2 = ind.sta-1

    is-error.sta-2      ➔ sta ◀ error.sta-2 @ 'error-handling-not-executed'

    **true**              ➔ sta-2 ◀ err @ 'error-handling-executed'

If the input-state does not carry an error, then it becomes the output state.

In the opposite case, a temporary state sta-1 is created by the removal of the error from sta. In this state, we compute the value of the expression ded which indicates the handled error. If this computation does not terminate, then the execution of the whole instruction does not terminate either. Otherwise, if the result of that computation is an error or a composite, which does not carry a word, then an appropriate error message is generated together with the additional massage 'error-handling-not-executed'.

In the opposite case, if the word carried by com — the error to be handled — is different from the initial error, then the output state is again identical with the input state.

In the opposite case, the error-handling instruction ins is executed in the temporary state sta-1. If during this execution an error is generated, then it is signalled together with the information 'error-handling-not-executed'.

In the opposite case, the message 'error-handling-executed' is loaded to the output state of our handling-error instruction ind.

As we see, the expression that appears in an error-handling instruction must evaluate to a word. If that word coincides with a current error message, then the "internal" instruction is executed.

It is to be stressed that the above constructor should be regarded only as an example showing that error-handling mechanisms may be described in our model. In no way, it should be regarded as a pattern for error handling. Another examples of such mechanism are shown in sections 7.3.3 and 12.7.6.4.

## 6.1.9    Preambles and programs

Preambles are sequences of „arbitrarily mixed" type-constant definitions, data-variable declarations, and trivial preambles. Their denotations are built by four constructors:

| | | |
|---|---|---|
| create-preamble-from-type-def | : TypDefDen | ↦ PreDen |
| create-preamble-from-variable-dec | : VarDecDen | ↦ PreDen |
| create-trivial-preamble | : | ↦ PreDen |
| sequence-pre | : PreDen x PreDen | ↦ PreDen |

First two constructors are *insertions,* i.e. identity functions that allow treating elements of one carrier of an algebra as the element of another. On the ground of the algebra of denotations, this construction means that each type definition and variable declaration may be "treated" as a preamble. Notice that without these constructors the reachable part of the carrier of preambles would be empty, and consequently, the corresponding carriers of the algebra of syntax would be empty as well.

**Comment!** Notice that in that case, the whole carrier would be empty rather than only its reachable subset because the algebra of syntax is always reachable. In other words, the corresponding equational grammar would not allow generating preambles. ∎

The third constructor creates an identity state-to-state function analogously as in the case of instruction:

create-trivial-preamble.().sta = sta

Also, this constructor has only a technical character. It permits to define a program as a pair consisting of a preamble (maybe empty) followed by an instruction (maybe empty).

create-program : PreDen x InsDen ↦ ProDen

create-program.(pde, ins) = pde ● ins

Programs with trivial preambles — if executed "without a context" — will always generate an error. However, I allow such programs because in **Lingua-2** they will constitute the bodies of procedures. In turn, programs with trivial both preambles and instruction will be allowed in the declarations of functional procedures[62].

## 6.1.10    A summary about the role of types in programs

Every type constant is defined in a program only once and remains unchanged during the execution of the program. This is why we are talking about type constants rather than type-variables. These constants are used in variable declarations and in type definitions. In the latter case, they serve in the bottom-up building of complex types.

Every variable in a program is declared only once, and this results with assigning to it a pseudo-value $(\Omega, (bod, tra))$. In the course of program's execution the type of that variable that is initially $(bod, tra)$ may be changed in two cases only:

- by an assignment instruction which changes the body of the type to a coherent one; so far this is possible for record-variables only, but in **Lingua-SQL** a similar rule will apply to table-variables,

- by transfer replacement but where the new transfer has to be satisfied by the current composite.

## 6.2    Syntax

The assumption that **Lingua-1** is an extension of **Lingua-A** concerns — of course — not only denotations but also abstract, concrete and colloquial syntax. Consequently, in the subsequent sections, we shall describe only these elements of the syntax of **Lingua-1** that do not appear in **Lingua-A**.

### 6.2.1    Abstract syntax

**Variable declarations**

   vde : VarDecA =

      declare-dat-var (Identifier , TypExpA) |

      sequence-vde (VarDecA ; VarDecA)

**Type definitions**

   tde : TypDefA =

      define-typ-con (Identifier , TypExp)    |

      sequence-tde (TypDefA ; TypDefA)

**Instructions**

   ins : InstructionA =

      assign (Identifier , DatExpA)                |

      replace-tr (Identifier, TraExpA)            |

      create-trivial-ins ()                              |

---

[62] Both these solutions, although in a slightly different form, have been suggested to me by Andrzej Tarlecki.

        if (DatExpA , InstructionA , InstructionA)    |

        if-error (DatExpA , InstructionA)        |

        while (DatExpA , InstructionA)          |

        sequence-ins (InstructionA , InstructionA)

**Preambles**

  pam : PreambleA =

      create-preamble-from-type-def (TypDefA)     |

      create-preamble-from-variable-dec (VarDecA)  |

      create-trivial-pre ()                    |

      sequence-pre (PreambleA , PreambleA)

**Programs**

  prg : ProgramA =

      create-program (PreA , InsA)

## 6.2.2    Concrete syntax

Concrete syntax of the applicative part of **Lingua-1** is taken from **Lingua-A**. New concretisations are described by the following grammar:

**Variable declarations**

  vde : VarDec =

     **let** Identifier **be** TypExp **tel** |

     VarDec ; VarDec

**Type definitions**

  tde : TypDef =

     **set** Identifier **as** TypExp **tes**   |

     TypDef ; TypDef

**Instructions**

  ins : Instruction =

     Identifier **:=** DatExp                   |

     **yoke** Identifier**:=** TraExp         |

     **skip**                            |

     **if** DatExp **then** Instruction **else** Instruction **fi** |

     **if-error** DatExp **then** Instruction **fi**     |

     **while** DatExp **do** Instruction **od**       |

     Instruction ; Instruction

**Preambles**

pam : Preamble =

    TypDef    |

    VarDec    |

    **`skip`**    |

    Preamble ; Preamble

**Programs**

prg : Program =

    **`begin-program`** Preamble ; Instruction **`end-program`**

The only change between abstract and concrete syntax which is substantially homomorphic, i.e. gluing and therefore non-isomorphic, is the omission of parentheses for sequential compositions of variable declarations, type definitions, instruction and preambles.

In this place a comment is necessary which concerns an "apparent glueing" in two cases:

1. the use of the same context **`if then else fi`** in the cases of conditional expressions and conditional instructions,

2. the use of the same keyword **`skip`** in the cases of preambles and instructions.

Although in both cases we have to do with a certain type of a "gluing effect", none of these cases contradicts the existence of a denotational semantics (cf. Theorem 2.13-1 from Sec. 2.13) since this may only happen if a homomorphism maps two different abstract phrases of the same carrier into one concrete phrase. In both our cases it is, however, not the case.

The fact that the double role of **`skip`** does not contradict the existence of a denotational semantics means in practice that a parser — which maps concrete syntax into abstract syntax — will recognise whether **`skip`** is an instruction or a preamble, from a context[63].

Of course, on the abstract-syntax level, we had to introduce two different phrases since they correspond to two different functions of the algebra of denotations.

Despite these facts our grammar is, of course, ambiguous due to the omission of parentheses associated with the operator of sequential composition " ; ". We have to show, therefore, that the algebra of concrete syntax is not more ambiguous than the algebra of denotations. The proof is, of course, based on the fact that the composition of functions is associative but this simple observation cannot itself stand for a proof. Observe that addition and multiplication are also associative, but not "mutually associative" and therefore in the expression $((x + y) + z) * w$ only some parentheses may be dropped without causing ambiguity.

Formally speaking we have to prove that our two semantics, abstract and concrete, satisfy together the implication (2.13-1) from Sec. 2.13. Such proof must be carried by structural induction wrt the grammar of abstract syntax. Below is a sketch of such proof or better — as mathematicians use to say "agitation for a proof" — restricted to the case of instructions. Let us introduce the following notations:

---

[63] From a formal viewpoint instead of **`skip`** we could use a space but in my opinion **`skip`** is a safer solution since a programmer has to write this word intentionally, whereas a space may be left in a program by mistake. In the former case a parser analysing **`if`** x>0 **`then`** x := x+1 **`else fi`** will signalise an error and in the latter — will treat it as **`if`** x>0 **`then`** x := x+1 **`else skip fi`** which may contradict with programmer's intention.

| Ain | : InstructionA | ⟼ Instruction | — | our homomorphism from abstract into concrete instructions |
|-----|----------------|---------------|---|---|
| Awd | : DatExpA | ⟼ DatExp | — | our homomorphism from abstract into concrete expressions, |
| Din | : InstructionA | ⟼ InsDen | — | the unique homomorphism from abstract instructions into their denotations, i.e. the semantics of abstract instructions. |

Now we have to show that for any two abstract instructions

ins-1, ins-2 : InstructionA

the following implication is true:

if

Ain.[ins-1] = Ain.[ins-2]                                                                 (1)

then

Din.[ins-1] = Din.[ins-2]                                                                 (2)

In the first induction step observe that if ins-1 is an assignment, then (1) implies that ins-1 = ins-2, and therefore (2). Now assume the equality

ins-1 = if(dae-1, ins-11, ins-12)

Then

Ain.[ins-1] = **if** Awd.[dae-1] **then** Ain.[ins-11] **else** Ain.[ins-12] **fi**

hence from (1) there must be such dae-2, ins-21, ins-22 that

Ain.[ins-2] = **if** Awd.[dae-2] **then** Ain.[ins-21] **else** Ain.[ins-22] **fi**

and that

Awd.[dae-1]  = Awd.[dae-2]

Ain.[ins-11]   = Ain.[ins-21]

Ain.[ins-12]   = Ain.[ins-22]

Therefore, based on inductive assumption, we can conclude the equality Din.[ins-1] = Din.[ins-2].

Analogous proofs may be carried out for other constructors except sequence-ins. In that case from the equality:

Ain.[sequence-ins(ins-11, ins-12)] = Ain.[sequence-ins(ins-21, ins-22)]

we cannot conclude that

Ain.[ins-11] = Ain.[ins-21] and

Ain.[ins-12] = Ain.[ins-22],

hence we can't use the inductive assumption. Let us consider therefore the subcases of that case. The first subcase concerns the situation where ins-11 is not of the form

sequence-ins(…)

In that case (1) implies that ins-21 cannot be of this form either, hence we can apply the inductive assumption. Let then:

ins-1 = sequence-ins(sequence-ins(ins-111, ins-112), ins-12)

Hence and from (1)

Ain.[ins-1] = Ain.[ins-111] ; Ain.[ins-112] ; Ain.[ins-12]

Ain.[ins-2] = Ain.[ins-211] ; Ain.[ins-212] ; Ain.[ins-22]

and

Ain.[ins-111] = Ain.[ins-211]

Ain.[ins-112] = Ain.[ins-212]

Ain.[ins-12]   = Ain.[ins-22]

Therefore from inductive assumption

Din.[ins-111] = Din.[ins-211]

Din.[ins-112] = Din.[ins-212]

Din.[ins-12]   = Din.[ins-22]

Now we have to consider two cases. Let in the first case:

ins-2 = sequence-ins(sequence-ins(ins-211, ins-212), ins-22)

In that case we can conclude (2) based on the inductive assumption. In the second case let:

ins-2 = sequence-ins(ins-211, sequence-ins(ins-212, ins-22))

In that case

Din.[ins-1] = ( Din.[ins-111] ● Din.[ins-112] ) ● Din.[ins-12]

Din.[ins-2] =   Din.[ins-211] ● ( Din.[ins-212] ● Din.[ins-22] )

Now the expected thesis follows from inductive assumption and from the associativity of the sequential composition of functions "●". For the remaining categories of our language the proof is analogous.

From user's viewpoint the principle of writing preambles, instruction, definitions and declarations without parentheses is very simple, but from the perspective of an implementor, it leads to the necessity of building an algorithm that will unambiguously transform parentheses-free concrete programs into abstract programs with parentheses. In this case, one may take an arbitrary strategy, e.g. assuming that parentheses are inserted from left to right. In that case an instruction:

ins-1 ; ins-2 ; ins-3

is transformed into:

(ins-1 ; (ins-2 ; ins-3))

In this place, a historical comment is needed. In prehistoric times, when user manuals used grammars to describe the syntax of languages, only one grammar was defined for both the user and the implementor of the language. In that case, users were given grammars adequate for implementors, but rather complicated and discouraging to read them by programmers. As a consequence formal grammars were completely abandoned and replaced by (usually very unclear) examples. In this way, the baby was thrown out with the bathwater which has led the situation described at the beginning of the book.

## 6.2.3    Colloquial syntax

Here we add only four new colloquialisms to those already known from **Lingua-A**. First, in type definitions and variable declarations with trivial yokes `true`, we can skip the yoke. For instance instead of writing

    `set` `list-of-names` `as` **`type`** **`list-of`** `string` **`ee`** **`with`** `true` **`tes`**

we may write

    `set` `list-of-names` `as` **`list-of`** `string` **`ee`** **`tes`**

and analogously instead of

    **`let`** `names` **`be`** `list-of-names` **`ee`** **`with`** `true` **`tel`**

we write

    **`let`** `names` **`be`** `list-of-names` **`tel`**

Second, variables-declarations of the same type may be grouped into one declaration with many variables, e.g. instead of writing

    **`let`** `x` **`be`** `number` **`tel`**`;`

    **`let`** `y` **`be`** `number` **`tel`**`;`

    **`let`** `z` **`be`** `number` **`tel`**

we write

    **`let`** `x, y, z` **`be`** `number` **`tel`**

and analogously for type definitions.

Third, in programmes, we can write comments which are removed by restoring transformation. Each comment starts with „#" and ends with a carriage return.

Forth, we can freely use carriage returns and spaces in programs. They will be removed by the restoring transformation.

---

**Comment 6.2.2-1** In building concrete syntax for **Lingua-1** I have applied some notational conventions known to me from the "old times" which in my opinion improve the clarity of programs and thus contribute to a less number of errors made by programmers. They are the following:

1. For an assignment I use „:=" rather than an equality „=" as in some languages.

2. I use closing parentheses **`fi`** for **`if`** and **`while`** since my experience proves that this contributes to a better clarity of programs.

3. Hierarchical carriage returns and spaces help in exposing the structure of programs, however using them as parentheses (as, e.g. in Phyton) may be error-prone resulting from an erroneous use of the Del-key. As a mathematician, I also cannot accept the fact that a hidden formatting-sign is an element of syntax. It is, however, convenient to use carriage returns and indentations arbitrarily, i.e. without interfering into the meanings of programs. In **Lingua** the are removed by the restoring transformation.

## 6.2.4     An example of a simple program

Here is an example of a simple program that creates a database in the form of an array of employee's records. Under each part of our program, we give an explanation of its meaning.

```
set register_type as

    array-type string ee

tes
```

Identifier `register_type` is declared as an array-type constant with type (('A', ('word')), TT), where TT is the denotation of transfer expression `true`.

```
set employee_type as

    record-type

        ch-name, fa-name of type string

        birthyear of type real,

        awards of type register_type

    ee

tes;
```

Identifier `employee_type` is declared as a type constant with record type ('R', ['ch-name' / ('word'), … ), TT)

```
set hr_base_type  as

    array-type employee_type ee

tes
```

Identifier `hr_base_type`  is declared as a type constant with array-type ('A', ('R', ['ch-name' / ('word'), … ], TT)

```
let salesmen_base be hr_base_type  ee

let salesman be employee_type ee

let awards_Smith be  register_type ee

let award_1, award_2, award_3 be string ee
```

Four identifiers have been declared as data variables with pseudo-values of indicated types.

```
award_1 := 'distinguished salesman'

award_2 := 'excellent salesman'

award_3 := 'star of sales'
```

Three data variables have been given values all of the same of type ('word')'

```
awards_Smith :=

    array [award_1,award_2,award_3]
```

Data variable has been given an array-value with type (('A', ('word')), TT), i.e. with type assigned to type constant `register_type`.

```
salesman :=

    record
```

```
    ch-name      <= 'John'

    fa-name      <= 'Smith'

    birth-year   <= 1968

    awards       <= awards_Smith

  ee
```

Data variable `salesman` has been given a record-value with type assigned to type constants `employee_type`.

```
  salesmen_base := array employee [salesman]
```

Data variable `salesmen_base`  has been given an array-value with one element of types assigned to `employee_type`.

## 6.3    Semantics

The definition of **Lingua-1** semantics consists the definition of the semantics of **Lingua-A** (Sec. 5.7) extended by definitional clauses for the imperative part of the language:

| | | | |
|---|---|---|---|
| Svd | : VarDec | $\mapsto$ | VarDecDen |
| Std | : TypDef | $\mapsto$ | TypDefDen |
| Sin | : Instruction | $\mapsto$ | InsDen |
| Spre | : Preamble | $\mapsto$ | PreDen |
| Spr | : Program | $\mapsto$ | ProDen |

The definitions of these semantic functions are given in an algebraic form (cf. Sec. 5.6)

### Declarations of data variables

Svd : VarDec $\mapsto$ VarDecDen        or

Svd : VarDec $\mapsto$ State $\mapsto$ Sta

Svd.[**let** ide **be** tex]  = data-variable.(Sid.[ide], Sty.[tex])

Std.[vde-1; vde-2]   = sequence-vde.(Std.[vde-1], Std.[vde-2])

### Definitions of type constants

Std : TypDef $\mapsto$ TypDefDen

Std : TypDef $\mapsto$ State $\mapsto$ State

Std.[**set** ide **as** tex]  = define-typ-con.(Sid.[ide], Sty.[tex])

Std.[tde-1; tde-2]      = sequence-tde.(Std.[tde-1], Std.[tde-2])

## Instructions

Sin : Instruction ⟼ InsDen

Sin : Instruction  ⟼ State → State

Sin.[`ide := dae`]                    = assign.(Sid.[ide], Sw.[dae])

Sin.[**if** `dae` **then** `ins-1` **else** `ins-2` **fi**]

                                          = if.(Sw.[dae], Si.[ins-1], Si.[ins-2])

Sin.[**if-error** `dae` **then** `ins-1` **fi**] = if-error.(Sw.[dae], Si.[ins-1])

Sin.[**while** `dae` **do** `ins` **od**]         = while.(Sw.[dae], Si.[ins])

Sin.[`ins-1;ins-2`]                   = sequence-ins.(Si.[ins-1], Si.[ins-2])

## Preambles

Spre : Preamble ⟼ PreDen

Spre : Preamble ⟼ State → State

Spre.[`tde`]              = Std.[tde]

Spre.[`vde`]              = Svd.[vde]

Spre.[`pam-1;pam-2`]   = sequence-pre.(Spre.[pam-1],  Spre.[pam-2])

## Programs

Spr : Program ⟼ ProDen

Spr : Program ⟼ State → State

Spr.[`ins`]      = create-program-from-instruction.(Sin.[ins])

Spr.[`pam ; ins`] = create-program.(Spre.[pam], Sin.[ins])

# 7  Lingua-2 — procedures

## 7.1    An introduction to a model of procedures

### 7.1.1    Procedures from a historical perspective

The concept of a procedure appeared in programming languages in the decade of 1950. Initially, procedures were just lists of instructions communicating with the main program through global variables. Later, to increase the universality of procedures, they were equipped with parameter-passing mechanisms — colloquially called *parameter calls* — and with a possibility of declaring local variables.

Beside procedures understood in that way — we shall call then *imperative procedures* — another type of procedures was introduced under the name of *functional procedures* or just *functions.* These procedures correspond to expressions rather than to instructions since they return values (in our case composites) rather than states.

The most popular higher-level language of the decades 1950/1960 was Fortran. In this language, procedures could call other procedures but not themselves. The latter construction was introduced in Algol 60 and was called *recursive procedures*. The creators of Algol 60 went even one step further allowing procedures to take other procedures — and even themselves (!) — as parameters (cf. Sec. 4.1). The self-applicability of procedures as parameters was, however, abandoned rather quickly, and did not reappear in later languages. On the other hand, recursion turned out to be a very useful vehicle and today is present in many languages. In some of them, procedures may take other procedures as parameters, but not themselves (see Sec. 7.6).

It is worth mentioning in this place that at the turn of decades 1950 and 1960 Polish researchers have developed and implemented a programming language SAKO (System Automatycznego Kodowania, Eng. Automatic Coding System) that was comparable with Fortran. Its compiler was implemented on a Polish computer XYZ constructed in Zakład Aparatów Matematycznych PAN (Unite of Mathematical Apparatus) a research unite of The Institute of Mathematics of The Polish Academy of Sciences. That was the first computer which I learned to program as a student. Its first version was equipped with an operational memory of 1024 bytes, i.e. 1 KB (disks were not known yet) and was later expanded by a magnetic drum with — as far as I can remember — 5 KB.

At that time programmers were instructed that to "intellectually" control the behaviour of a program the latter should not exceed one paper-sheet. In the case of larger programs that principle was implemented by writing procedures which were calling other procedures. That style was later called *structured programming* (cf. Sec. 4.1).

Structured programming was supposed not only to help programmers in a better understanding of their programs but also in proving program-correctness by induction based on program-structure. So far, however, this idea is rather far for ma full realisation (cf. Sec. 3.1).

## 7.1.2    Procedures versus structured programming

In programming languages with procedures, the latter may call other procedures or even themselves. This mechanism allows to build programs in a structural way:

1. The whole program consists of one *main procedure* which calls *subprocedures of the first level*.

2. Subprocedures of the first level call *subprocedures of the second level*.

3. …

The number of successive levels is essentially arbitrary.

In the simplest case — which appears most frequently — procedures constitute a tree-like structure (Fig. 7.1-1). The main procedure MP calls subprocedures SP1 and SP2 the first calls in turn SP3, SP4 and SP5.

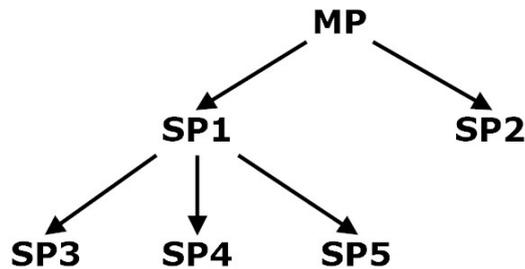

**Fig. 7.1-1 A tree of procedures without recursion**

It may also be the case that a procedure calls a higher-level procedure or itself. Such a situation is illustrated in Fig. 7.1-2.

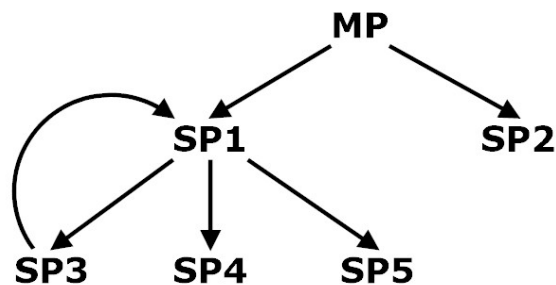

**Fig. 7.1-2 A graph of procedures with recursion**

If in the body of a procedure an interpreter encounters a call of this procedure, then basically two types of reactions are possible:

- an error message 'procedure-undeclared' is generated,

- a copy of the called procedure is activated.

The second case, which is today rather common in programming languages, is known as *recursive call* of a procedure. If a procedure calls itself directly, i.e. in its own body, then we have to do with a *simple recursion*. If, however, SP1 calls SP3 and SP3 calls SP1, then we have to do with *mutual recursion*. Of course, the cycle of procedures calling one another may have more than just two elements.

### 7.1.3   Procedures in a denotational framework

Although recursion is today a rather common standard in high-level programming languages, its technical details may differ from one language to another. For the sake of our investigations, we assume a certain more-or-less universal model chosen in such a way that it leads to relatively simple correction-proof-rules.

Imperative procedures may be regarded as named instructions with additional mechanisms which allow to use them repeatedly in many different contexts:

- they may be memorised in procedure-environments,

- they may use local variables that are *not visible* outside of a *procedure-body*,

- they may receive lists of values that are used to initialise local variables; this mechanism is known as *called-by-value actual parameters* or *actual value-parameters*,

- they may receive lists of variables known as *called-by-reference actual parameters* or just as *actual reference-parameters*; the initial values of these parameters are passed to procedures, and their terminal values are exported back to the hosting program.

Another difference between procedures and instructions (or expressions) is that procedures do not have syntactic representations. In commonly known programming languages procedures do not appear as syntactic objects and even more — they do not appear as independent concepts at all. The authors of programmer manuals talk about procedure declarations and procedure calls but not about procedures as such. This awkward situation is caused by the fact that manuals refer mainly to syntax[64].

Talking about bodies, declarations, and calls of "beings" that have not been defined not only contradicts with mathematical good-practice but may also lead to a poor understanding of language mechanisms.

In **Lingua-2** procedures constitute a carrier in the algebra of denotations. Since procedures do not appear on the syntactic level, we talk about procedures "as such", rather than on procedure denotations.

In order to include procedure-mechanisms into a denotational model, we define three sorts of objects:

| | | |
|---|---|---|
| *procedures* | — | functions that modify stores (imperative procedure) or return composites (functional procedure)[65], |
| *the denotations of procedure declarations* | — | functions that modify states by assigning a (declared) procedure to an identifier (its name) in the environment, |
| *the denotations of procedure calls* | — | functions that modify states or return composites by executing a (called) procedure. |

---

[64] To say nothing about the fact that the definitions of syntax are highly unprecise and incomplete.

[65] In this place I intentionally resign from functional procedures with so called "side-effects", i.e. from such procedures which not only compute a composite but also modify a state. Denotationally such a solution would be acceptable but it would lead to expressions with side-effects, since — as we are going to see — functional-procedure calls will expressions themselves. That solution would complicate not only our model but in the first place — the constructors of correct programs.

Since in our model, procedures may generate error messages and may "react" on them, we define them as functions on stores. As was already mentioned, we shall consider two sorts of procedures:

1. *imperative procedures* that correspond to instructions with parameters:

    ipr : ImpPro = Parameters ↦ Store ↦ Store

2. *functional procedures* that correspond to expressions with parameters

    fpr : FunPro = Parameters ↦ State ↦ CompositeE

hence

pro : Procedure = ImpPro | FunPro

Following a common wording, imperative procedures will also be called *procedures* and functional procedures — *functions*.

## 7.1.4    Denotational domains for procedures

According to our general rule about the series of **Lingua** languages (Sec. 4.3), **Lingua-2** emerges from **Lingua-1** by adding new carrier and new constructors to the algebras of denotations and syntax. Additionally, the old syntactic carriers of instruction and expressions are enriched by new elements.

As we are going to see, the declarations of imperative procedures will take two lists of formal parameters called respectively *formal value-parameters* and *formal reference-parameters*. In turn, procedure calls receive two lists of parameters called respectively *actual value-parameters* and *actual reference-parameters*. The domains of parameters are defined as follows:

fpa  : ForPar = (Identifier x TypExpDen)$^{c^*}$    (formal param. of declarations of both types)

apa : ActPar = Identifier$^{c^*}$                      (actual param. of calls of both types)

Formal parameters include type expression denotations since in the declarations of procedures we indicate the types of their future actual-parameters. Imperative procedures modify stores, and functional procedures return composites:

ipr  : ImpPro     = ActPar x ActPar ↦ Store → Store           (imperative procedures)

fpr  : FunPro     = ActPar          ↦ State → CompositeE(functional procedures)

pro : Procedure = ImpPro | FunPro                             (procedures)

Functional procedures receive at the call-time only one list of actual parameters, namely a list of value parameters. Two following domains correspond to the declarations of procedures:

idd    : IprDecDen  = State ↦ State        (denotations of imp. procedure-declarations)

fdd    : FprDecDen = State ↦ State        (denotations of fun. procedure-declarations)

Notice that in the definition of ImpPro we do not have an illegal fixed-point recursion since imperative procedures do not take as arguments states — which bind procedures in procedure-environments — but only stores[66]. This is why stores have been introduced as a separate component of a state. If we had assumed the equation

ImpPro = ActPar x ActPar ↦ State → State

---

[66] That solution was introduced in a common paper [28] of Andrzej Tarlecki and myself in 1983.

then together with the equations

State        = (TypEnv x ProEnv) x Store

ProEnv      = Identifier $\Rightarrow$ ImpPro | FunPro

we would have an illegal fixed-point equation since the operators „$\mapsto$" and „$\rightarrow$" are not continuous (Sec. 2.7). A similar situation took place in Algol 60, although not due to recursion, but because Algol 60 procedures could take themselves as parameters. Procedures which can take other procedures as parameters — but not themselves — are discussed in Sec. 7.6

For the simplicity of our model and of correct-programs construction rules, I have assumed that actual parameters of procedures must be identifiers rather than arbitrary expressions. As we shall see in the sequel, expressions will include functional-procedure calls, which in turn may contain calls of imperative procedures. If actual parameters could be arbitrary expression, then one could write a procedure that calls itself in calculating its own parameters. Denotations this should be feasible, but a corresponding program-construction rule would be pretty complicated. More on that issue in Sec. 7.5.1

After having defined domains for procedures, we have to pass to the definitions of their constructors. We assume that all constructors defined in **Lingua-1** are available in **Lingua-2.**

# 7.2    The communication between imperative procedures and programs

In the descriptions of procedure mechanisms, we shall use some concepts having to do with the fact that procedures are created when they are declared and are executed when they are called. In respect to that, we shall talk about states and their components of *declaration-time* and of *call-time* respectively[67]. Traditionally by a *procedure body,* we shall mean the program that is executed when a procedure is called.

As I have already announced, in **Lingua-2** there will be no global variables in procedures. This is not a mathematical necessity but an engineering decision[68]. The intention is that the head of a procedure-call describes explicitly and completely the communication mechanisms between a procedure and the hosting program. That solution may seem restrictive but — in my opinion — guarantees a better understanding of program functionality by programmers and also simplifies program-construction rulers.

## 7.2.1    How it works?

An execution of a procedure call may be symbolically split into four stages illustrated in Fig. 7.2-1. (technical details in Sec. 7.3).

1.  **The inspection of an *initial global state*** — that state consists of:

    a.   an *initial global environment* env-ig,

    b.   an *initial global store* sto-ig = (vat-ig, err)

---

[67] These ideas, similarly to a few others, have been borrowed from M. Gordon [44].7.3

[68] If we would like to introduced global variables, we should define the local store of a procedure call as a modification of its global store.

If err ≠ 'OK', then the initial global state is returned by procedure call and therefore becomes the terminal global state. In the opposite case, an initial local state is created.

2. **The creation of an *initial local state*** — that state consists of:

   a. *initial local environment* env-il created from the <u>declaration-time environment</u> by nesting in it the called procedure; this nesting is necessary to enable recursive calls (Sec. 7.3.3),

   b. *initial local valuation* vat-il covering only formal parameters with assigned values of corresponding actual parameters; to get the latter values, we refer to initial global valuation val-ig.

3. **The transformation of local initial state** by executing procedure body. If this execution terminates, then the local terminal state consists of:

   a. *terminal local environment* env-tl,

   b. *terminal local store* sto-tl = (val-tl, err-tl).

   If err-tl ≠ 'OK', then a global terminal state is created from the initial global-state by loading to it err-tl. Notice that in this case, terminal local environment and terminal local store are "abandoned". Otherwise the terminal global state is created.

4. **The  creation of the *terminal global state*** — that state consists of:

   a. *initial global environment* env-ig; notice that terminal local environment env-tl is "abandoned",

   b. *terminal global store* sto-tg created from initial global store sto-ig by "returning" to it the values of formal referential parameters (stored in sto-tl) and assigning them to the corresponding actual referential parameters.

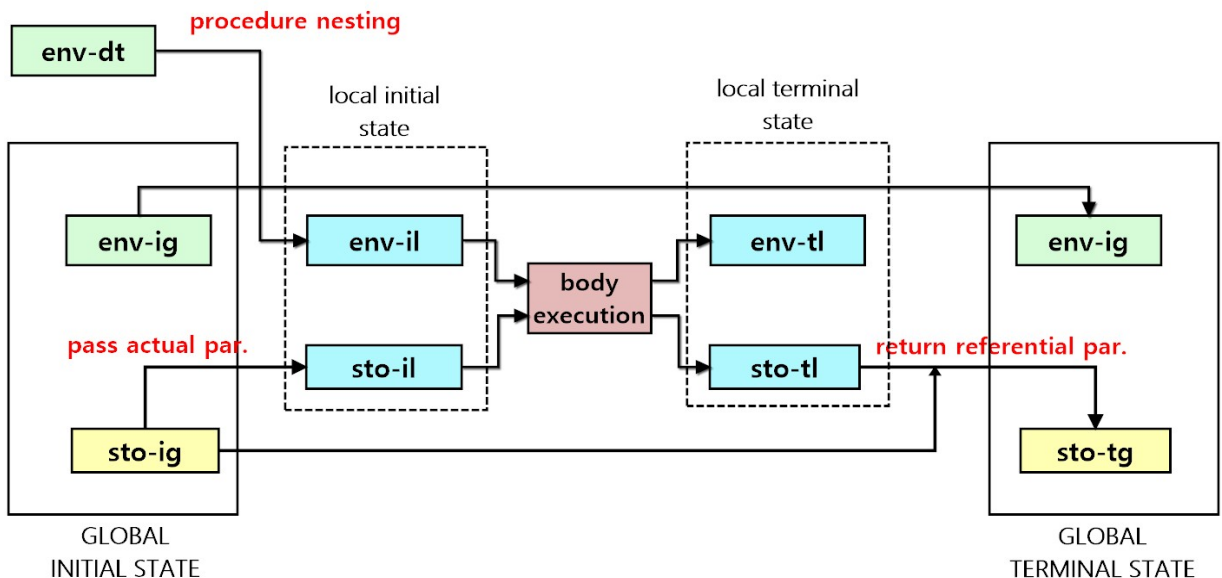

**Fig. 7.2-1 The execution of a procedure call**

Notice that initial local environment "inherits" all types and procedures from the declaration-time environment. Procedure body may keep in it its own local environment types and procedures, but after the completion of the call, they cease to exist, since the hosting program returns to the initial global environment.

It is to be underlined that procedure body may access only that part of the environment which was created <u>before procedure declaration</u>.

Of a similar character is the local valuation that is created only for procedure-execution-time, although in this case the values or reference-parameters stored in it are eventually returned to the terminal global valuation.

Summarising visibility rules concerning procedure call:

1. the only variables visible in procedure-body are formal parameters plus variables local to the body (declared in it),

2. the only types and procedures visible in procedure-body are declaration-time types and procedures plus locally declared ones,

3. variables, types and procedures declared in the procedure-body are not visible outside of procedure call.

All these choices are not mathematical necessities but pragmatic engineering decisions dictated by the intention of making our model relatively simple which should result in the simplicity of proof-rules and a better understanding of program-behaviour by language-users.

At the end one methodological remark. From an implementational view-point, the described mechanism of recursion requires that the initial global state is kept unchanged (memorised) during procedure-execution to recall it at the end. Consequently, the fact that a procedure may have many recursive calls means that each call should "memorise" its initial states. That mechanism is usually implemented by a stack of states. This is an iterative implementation of recursion. In our case, however, we do not need to use that method since — as we are going to see in Sec. 7.3.2 — the recursion in **Lingua-2** may be defined in using fixed-point recursion of **MetaSoft**.

## 7.2.2    The compatibility of parameter-lists

When an imperative procedure is called its formal parameters receive the values (typed data) of actual parameters and in this way a local valuation is created. However, in order to make such a parameters-passing possible, the list of actual parameters of procedure call must be compatible with the lists of formal parameters in procedure declaration both in their numbers and in their types. And of course, the values of actual parameters must be defined. In order to formalize these requirements we define two functions that will be used in defining constructors involved in procedural mechanisms.

statically-compatible : ForPar x ForPar x ActPar x ActPar $\longmapsto$ Error | {'OK'}

statically-compatible.(fpa-v, fpa-r, apa-v, apa-r) =

    **let**(for n, m, k, p ≥ 0)

        fpa-v    = ((ide-fv.i, ted-fv.i) | i=1;k)             (formal value-parameters)

        fpa-r    = ((ide-fr.i, ted-fr.i) | i=1;n)           (formal reference-parameters)

        apa-v    = (ide-av.i | i=1;p)                (actual value-parameters)

        apa-r    = (ide-ar.i | i=1;m)               (actual reference-parameters)

      are-repetitions.[(ide-fr.i | i=1;n) ©

(ide-fv.i | i=1;k)]    ➔ 'formal-par-repetitions'[69]

are-repetitions.apa-r    ➔ 'actual-par-repetitions

n ≠ m **or** k ≠ p    ➔ 'incompatible-numbers-of-parameters'

**true**    ➔ 'OK'

In other words, lists of formal and actual parameters of a procedure call are statically compatible if:

1. no formal parameter appears twice on a combined list of both sorts (value- and reference) parameters; a similar property of actual value-parameters is, of course, not required,

2. the mutually corresponding lists of formal and actual parameters are of the same lengths.

The defined property is called *static* since it can be checked at compilation-time, i.e. before program execution. However, "statically" does not mean "syntactically"! Moreover, the compatibility-analysis can be performed only after syntactic correctness has been checked.

Next compatibility function refers to valuations and type environments and therefore is *dynamic* since its execution is possible only at the execution of the program. Also here we compare formal parameters with actual parameters.

dynamically-compatible : ForPar x ForPar x ActPar x ActPar ↦

TypEnv x Valuation ↦  Error | {'OK'}

dynamically-compatible.(fpa-v, fpa-r, apa-v, apa-r).(tye, vat) =

   **let**

      message = statically-compatible.(fpa-v, fpa-r, apa-v, apa-r)

   message : Error    ➔ message

   **let**(for n, m, k, p ≥ 0)

      fpa-v    = ((ide-fv.i, ted-fv.i) | i=1;k)    (formal value-parameters)

      fpa-r    = ((ide-fr.i, ted-fr.i) | i=1;n)    (formal reference-parameters)

      apa-v    = (ide-av.i | i=1;p)    (actual value-parameters)

      apa-r    = (ide-ar.i | i=1;m)    (actual reference-parameters)

*checking the definedness of actual value-parameters*

   (∃ i) vat.(ide-av.i) = ?    ➔ 'value-parameter undefined'

   **let**

      ((dat-av.i, bod-av.i), tra-av.i) = vat.ide-av.i    for i = 1;k

   (∃ i) dat-av.i = Ω    ➔ 'value-parameter uninitialized'

*checking the definedness of actual reference-parameters*

   (∃ i) vat.(ide-ar.i) = ?    ➔ 'reference-parameter undeclared'

   **let**

      ((dat-ar.i, bod-ar.i), tra-ar.i) = vat.ide-ar.i    for i = 1;m

---

[69] Function **are-repetitions** (Sec. 2.1.4) has been defined for tuples, therefore its argument in this definition is a concatenation '©' of formal-reference and formal-value parameter-lists.

(∃ i) dat-ar.i = Ω        ➔ 'reference-parameter uninitialized'

*computing the types of formal value-parameters*

**let**

sta = ((tye, [ ]), (vat, 'OK'))                              (explanation below)

typ-fv.i  = ted-fv.i.sta     for i = 1 ;k            (types of formal-value-parameters)

typ-fr.i  = ted-fr.i.sta     for i = 1 ;n          (types of formal-reference-parameters)

(∃ i) typ-fv.i   : Error     ➔ 'type-error-of-formal-value-parameter'

(∃ i) typ-fr.i   : Error     ➔ ' type-error-of-formal-reference-parameter'

**let**

(bod-fv.i,   tra-fv.i) = typ-fv.i    for i = 1 ;k

(bod-fr.i,   tra-fr.i)  = typ-fr.i    for i = 1 ;n

(∃ i) bod-fv.i  ≠ bod-av.i ➔ 'incompatible-bodies-of-value-parameters'

(∃ i) bod-fr.i  ≠ bod-ar.i ➔ 'incompatible-bodies-of-reference-parameters'

(∃ i) (tra-fv.i).((dat-av.i, bod-av.i) ≠ (tt, ('Boolean')   ➔ 'yoke-not-satisfied-by-val'

(∃ i) (tra-fr.i).((dat-ar.i, bod-ar.i) ≠ (tt, ('Boolean')   ➔ 'yoke-not-satisfied-by-ref'

**true**                      ➔ 'OK'

The lists of formal and actual parameters are considered dynamically compatible, if:

1. they are statically compatible,

2. all actual parameters are declared and initialised,

3. all type expressions assigned to formal parameters of both sorts evaluate to non-errors,

4. all bodies of mutually corresponding formal and actual parameters of both sorts are identical; formal-parameter type is defined by a type expression in procedure declaration, and actual-parameter type is defined in the call-time valuation,

5. all composites carried by actual parameters satisfy the yokes of corresponding formal parameters; notice that the yokes of actual parameters are not considered at all.

To compute the types of formal parameters, a certain technical trick was applied. Since these types are defined by type expressions, to compute them the type expression denotations have to be applied to a state. Here is a problem since the function

dynamically-compatible.(fpa-v, fpa-r, apa-v, apa-r)

gets as an argument, not the whole state ((tye, pre), (vat, err)) but only two of its elements: tye and vat. To cope with this problem a "temporary" state is created

((tye, [ ]), (vat, 'OK'))

where [ ] is an empty procedure-environment. In fact, this environment might be quite arbitrary since type expression denotations do not depend on it.

Notice at the end that each of the numbers n, m, k i p may be zero, i.e. each of the corresponding parameter-lists may be empty.

### 7.2.3    Passing actual parameters to a procedure

This function is activated by a procedure call and creates local initial valuation (Fig. 7.2-1). The only identifiers bound in this valuation are formal parameters and their initial values are the current values of corresponding actual parameters.

pass-actual : ForPar x ForPar x ActPar x ActPar $\mapsto$

$\qquad\qquad\qquad\qquad\qquad\qquad$ TypEnv x Valuation $\mapsto$  Valuation | Error

pass-actual.(fpa-v, fpa-r, apa-v, apa-r).(srt, vat) =

   **let**

      message = dynamically-compatible.(fpa-v, fpa-r, apa-v, apa-r).(srt, vat)

   message ≠ 'OK'    ➜ message

   **let**(for n, k ≥ 0)

      ((ide-fv.i, ted-fv.i) | i=1;k) = fpa-v $\qquad\qquad\qquad$ (formal-value-parameters)

      ((ide-fr.i, ted-fr.i) | i=1;n) = fpa-r $\qquad\qquad\qquad$ (formal-reference-parameters)

      (ide-av.i | i=1;k)  = apa-v $\qquad\qquad\qquad\qquad$ (actual-value-parameters)

      (ide-ar.i | i=1;n) = apa-r $\qquad\qquad\qquad\qquad$ (actual-reference-parameters)

      val-w.i   = vat.(ide-av.i) for i=1;k $\qquad\qquad$ (values of actual-value-parameters)

      val-r.i   = vat.(ide-ar.i) for i=1;n $\qquad\qquad$ (values of actual-reference-parameters)

   *creating initial local valuation*

      vat-w    = [ide-fv.i/val-w.i | i=1;k] $\qquad$ (initial local valuation of value-parameters)

      vat-r    = [ide-fr.i/val-r.i | i=1;n] $\qquad$ (initial local valuation of reference-parameters)

      vat-il   = vat-v ◆ vat-r $\qquad\qquad\qquad\qquad\qquad$ (initial local valuation)[70]

   **true** $\qquad\qquad\qquad$ ➜ vat-il

The defined operator checks compatibilities of parameters and then creates local initial valuation to be later executed by procedure-body:

- formal value-parameters receive the values of actual value-parameters; the definedness of these values and compatibility of their types has been checked by the function dynamically-compatible.

- formal reference-parameters receive the values of actual reference-parameters; the definedness of these values and compatibility of their types has been checked by the function dynamically-compatible.

Similarly, as in the former definitions, the empty lists of parameters are allowed.

   Notice that the described mechanism of initial local valuations does not offer a possibility of using global variables, i.e. variable that would be visible both outside and inside procedure-body. The only communication channel of procedure call between its external world and its

---

[70] Local valuation is created as an overwriting of local reference-valuation by local value-valuation. Since their sets of identifiers are disjoint, the resulting valuation is a simple expansion of one function by another. The overwriting operation ◆ has been defined in Sec. 2.1.3..

internal world are reference-parameters that pass their values according to the following scheme:

fpa-v    := the values of apa-v

fpa-r    := the values of apa-r

　　procedure-body execution

apa-r    := value of fpa-r

## 7.2.4    Returning reference-parameters to a program

Whereas formal value-parameters play the role of local variables since they are visible only inside procedure-body, formal reference-parameters play the role of global variables the values of which are modified by procedure-body.

return-referential : ForParRef x AktParRef ↦ TypEnv x Valuation x Valuation

$$\mapsto \text{Valuation} \mid \text{Error}$$

return-referential.(fpa-r, apa-r).(tye, vat-tl, vat-ig) =

　　**let**

　　　　message = dynamically-compatible.((), fpa-r, (), apa-r).(tye, vat-ig)[71]

　　message ≠ 'OK'　　　➔ message

　　**let**

　　　　(ide-ar.i | i=1;n) = apa-r　　　　　　　　(actual reference-parameters)

　　　　((ide-fr.i, typ-fr.i) | i=1;k) = fpa-r　　　　(formal reference-parameters)

　　(∃i) vat-tl.(ide-fr.i) = ?➔ 'value-of-reference-parameter-undefined'

　　**let**

　　　　val-fr.i  = vat-tl.(ide-fr.i) for i=1;n　　　(terminal value of formal ref-parameters)

　　　　vat-tg    = vat-ig[ide-ar.i/val-fr.i | i=1;n]　　　　(terminal global valuation)

　　**true**　　　　　　　　➔ vat-tg

After procedure-body has been executed, the values of formal reference-parameters are passed to the corresponding actual reference parameters.

　　As was already mentioned this communication mechanism might be described by two symbolic assignment-instructions. Before the execution of the body:

fpa-r := the values of apa-r

and after its execution

apa-r := the value of fpa-r.

If we read these assignments literally, this means that actual-parameter values are copied to some memory-space allocated for procedure execution. If a parameter value is a small object like, e.g. a number, then, of course, such an implementation is quite acceptable, but if it is a large object, e.g. a database, such a solution would be rather absurd.

---

[71] By "()" we denote empty tuples of parameters. Doue to this trick we could use apply a four-argument function to two lists of parameters. Notice that in principle we do not need to check here the adequacy of parameters since this is checked in passing actual parameters to procedure-body (Sec. 7.2.3). However, removing this checking would make our definition incorrect.

In the majority of programming languages, this problem is solved by passing references rather than values to actual-reference parameters. These references provide access (i.e. memory-address) to the values of formal parameters. From the functional point of view such solution is equivalent to ours, but if we would like to describe it formally, we would have to introduce addresses in our model to bind identifier with addresses and addresses with data (see M. Gordon [44]). The choice between the two alternatives depend upon the addressee of our model — is he or she a language user or a language implementor. According to the philosophy assumed in this book, we address our model to users rather than to implementators, and therefore we have not introduced addresses to our model.

## 7.3    Imperative procedures with single recursion

As was already said, in the denotational model of **Lingua-2** procedures are treated as independent objects. To this end, we apply a construction described in [28] published jointly by Andrzej Blikle and Andrzej Tarlecki in 1983. Let us recall that in the case of procedures we do not talk about procedure denotations but about procedures as such, since they are denotational objects without corresponding syntax. On the syntactical side, we shall have procedure declarations and procedure calls.

### 7.3.1   The constructors of parameters

Since procedure-constructors will take parameters as arguments, we have to define parameter-constructors in the first place. Actual parameters are tuples of identifiers (possibly empty), and therefore they can be built by three following constructors:

create-empty-par : ↦ ActPar

create-empty-par.() = ()

create-empty-act-par : Identifier ↦ ActPar

create-empty-act-par.ide = (ide)

sequence-act-par : ActPar x ActPar ↦ ActPar

sequence-act-par.(apa-1, apa-2) = apa-1 © apa-2

Analogously we define constructors of formal parameters

create-empty-for-par : ↦ ForPar

create-empty-for-par.() = ()

create-for-par : Identifier x TypExpDen ↦ ForPar

create-for-par.(ide, ted) = (ide, ted)

sequence-for-par : ForPar x ForPar ⟼ ForPar

sequence-for-par.(fpa-1, fpa-2) = fpa-1 © fpa-2

In this place a comment is necessary about the constructor create-for-par, which "looks" as an identity constructor. In fact, it is not, since it transforms two arguments into a tuple. If tuples were written in square parentheses, then this fact could be better seen:

create-for-par.(ide, ted) = [ide, ted]

We remain, however, with the current notation for tuples, since this is a traditional way they have been written in mathematics.

## 7.3.2    The constructor of a procedure

We start by introducing an auxiliary concept of *imperative-procedure-components*.

ipc : IprComponents = Identifier x ForPar x ForPar x ProDen

Components of a procedure are four-tuples consisting of an identifier, two lists of formal parameters (value- and reference-parameters) and a procedure-denotation (the procedure-body). Procedure constructor takes as arguments procedure-components and an environment[72] and returns a procedure:

create-imp-proc : IprComponents x Env ⟼ ImpPro

or in an unfolded form:

create-imp-proc : ((Identifier x ForPar x ForPar x ProDen) x Env) ⟼

ActPar x ActPar ⟼ Store → Store

The environment that appears in this definition is a declaration-time environment (hence we denote it by env-dt), since a procedure is created in declaration-time[73]. The value of that constructor is a function

P = create-imp-proc.((ide, fpa-v, fpa-r, prd), env-dt)

which is a procedure, hence a function of the type

P : ActPar x ActPar ⟼ Store → Store

This function is defined as the (unique) least solution of a fixed-point equation. In order to write this equation we introduce the following notation:

(tye-dt, pre-dt)  = env-dt          — declaration-time environment

sto-ig                              — initial global store

par = (fpa-v, fpa-r, apa-v, apa-r) — the tuple of procedure-parameters

ide                                 — the name of the declared procedure

---

[72] Why the environment has not been included into procedure-components becomes clear when we discuss multiprocedures (Sec. 7.4)

[73] Observe that if procedures were supposed to be executed in call-time environments, then they would be functions of the type P : ActPar ⟼ Env → Store, i.e. they would take themselves as arguments.

prd                                    — procedure-body denotation; a program-denotation

Below the equation that defines P. It refers to the above arguments of function **create-imp-proc** and also to the arguments of P. The store **sto-ig** is, of course, a call-time store.

P.(apa-v, apa-r).sto-ig =

    is-error.sto       ➔ sto

    **let**

        (vat-ig, 'OK') = sto-ig

        vat-il = pass-actual.par.(tye-dt, vat-ig)           (initial local valuation)

    vat-il : Error       ➔ vat-il

    **let**

        sto-il = (vat-il, 'OK')                  (initial local store)

        env-il = (tye-dt, pre-dt[ide/P])        (initial local environment)

        sta-il = (env-dt, sto-il)               (initial local state)

    prd.sta-il   = ?    ➔ ?

    **let**   (procedure-body execution)

        sta-tl              = prd.sta-il          (terminal local state)

        (env-tl, (vat-tl, err))   = sta-tl

    err ≠ 'OK'       ➔ (vat-ig, err)       (*)

    **let**

        vat-tg = return-referential.(fpa-r, apa-r).(vat-tl, vat-ig)

    vat-tg : Error   ➔ (vat-ig, vat-tg)    (**)        (here vat-tg is an error)

    **true**        ➔ (vat-tg, 'OK')

In the first step we check if the initial global store carries an error and if this is the case, then it becomes the terminal global store.

In the opposite case we create the *initial local valuation* vat-il where procedure-body-execution will start (see Fig. 7.2-1). It is created form *initial global valuation* vat-ig by passing values of actual parameters to formal parameters. We recall (Sec. 7.2.3) that since the operator pass-actual checks the adequacy of parameter-lists, this part of procedure-execution may terminate with an error message.

If an error message does not appear, we create the *initial local store* sto-il with 'OK' message, and we create the *initial local environment* env-il by nesting the (just being defined) procedure P in the declaration-time environment. This nesting makes our equation a fixed-point one.

Notice that if we would not nest P in env-il, then the attempt to call P in the course of the execution of its body would result in an error message 'procedure-undeclared', since in the declaration-time environment P is not declared. It is to be underlined here that our mechanism does not cover the case where two or more procedures are calling themselves mutually. This case is discussed in Sec. 7.4.

The initial local environment with an initial local store constitute the *initial local state* sta-il.

In the next step procedure-body (represented by a program-declaration prd) is executed in sta-il and — if this execution terminates — then its terminal state sta-tl becomes the *local terminal state*.

If that state carries an error, then the terminal store consists of the <u>initial global</u> store and the <u>current</u> error.

Otherwise, we select the terminal local valuation vat-tl from that state which is then used in returning the current values of formal reference-parameters to actual reference-parameters. In this way, we create terminal global valuation vat-tg and *terminal global store* (vat-tg, 'OK').

Notice that if in the procedure-body there is no call of that procedure, then the effect of executing such a procedure in the modified environment is identical with such an effect in a non-modified environment. In other words, the definition of a recursive procedure cowers the case of a non-recursive procedure.

The procedure P is, therefore, a function which takes actual parameters and returns a function from store to store. Observe that neither the environment env-dt nor the identifier ide, nor the procedure-body (program-denotation) prd are the arguments of this function. They have been only used to define this function and are "hidden" in its definition. Formally they may be regarded as parameters of the defining equation. As we shall see later, these arguments will be "delivered" to the procedure by procedure declaration.

It is worth noticing here that the execution of our procedure involves two non-trivial error-handling cases in clauses (*) and (**). In both cases, an error message causes not only the interruption of program execution but the recovery of the initial global store. Of course, this is just one possible choice of an error-handling strategy in this place.

### 7.3.3    The instruction of a procedure call

Calling a procedure consists in getting it form an environment (where it has been declared) and "loading" to it actual parameters creating in this way an instruction denotation. Hence:

call-imp-proc : Identifier x ActPar x ActPar ⟼ InsDen

or:

call-imp-proc.(ide, apa-v, apa-r) : State → State

This instruction denotation is defined in the following way:

call-imp-proc.(ide, apa-v, apa-r).sta-ic =

    is-error.sta-ic               ➔ sta-c

    **let**

      (env-ic, sto-ic)  = sta-ic                    (initial state of the call)

      (tye-ic, pre-ic)  = env-ic             (the environment of the call)

      (vat-ic, 'OK')   = sto-ic               (initial store of the call)

    env-ic.ide = ?         ➔ sta-pw ◀ 'procedure-not-declared'

    env-ic.ide : FunPro     ➔ sta-pw ◀ 'procedure-not-imperative'

    **let**

      ipr = env-ic.ide             (the called imperative procedure)

    ipr.(apa-v, apa-r).vat-ic = ?  ➔ ?

**let**

    sto-tc = ipr.(apa-v, apa-r).sto-ic              (the terminal store of the call)

    (vat-tc, err) = sto-tc

  err ≠ 'OK'              ➔ sta ◄ err

  **true**                 ➔ (env-ic, vat-tc, 'OK')

If the call-time state does not carry an error message and the identifier ide is bound in the environment to an imperative procedure, then we apply this procedure to actual parameters getting in this way a partial function on stores:

  pro.(apa-r, apa-v) : Store → Store

This function is applied to the initial call-time store sto-ic. Notice that since our procedure carries declaration-time environment, procedure-body is executed in the state

  (env-dt, vat-ct, 'OK')

where

  env-dt   — declaration-time environment

  sto-ct   — call-time store

If the terminal store is not defined, then the result of the procedure call is not defined either. If the execution of the procedure body raises an error message, then this message is loaded into the <u>initial state of the call</u>. In the opposite case, the terminal store of the call sto-tc becomes the component of the *terminal state of the call* (env-ic, vat-tc, 'OK'). The initial environment remains unchanged.

Notice that in this definition we do not have neither parameter-adequacy check nor parameter passing since these operations are included in the procedure (Sec. 7.3.1). We check, however, the sort of the procedure since we are going to have functional procedures in the language.

## 7.3.4     Procedure declaration

Imperative-procedure declarations are constructors of the type

  declare-imp-pro : IprComponents ↦ IprDecDen

or of the type

  declare-imp-pro : IprComponents ↦ State ↦ State

defined in the following way

  declare-imp-pro.ipc.sta =

    is-error.sta       ➔ sta

    **let**

      (ide, fpa-v, fpa-r, prd)    = ipc

      (env, (vat, err))        = sta

      (tye, pre)            = env

    ide : declared.sta  ➔ sta ◄ 'variable-declared'

    **let**

>>
message = no-repetitions.(fpa-r © fpa-v)

message ≠ 'OK'    ➔ sta ◄ message

**let**

P = create-imp-proc.(ipc, env)    (proc. is created in declaration-time environment)

env-t = (tye, pre[ide/P]                                (terminal environment)

**true**                        ➔ (env-t, sto, 'OK')
>>

Procedure declaration creates a procedure and then binds it to the identifier ide in the current environment, i.e. in the declaration-time environment. This identifier and this environment become later the arguments of a procedure-constructor (Sec. 7.3.1).

# 7.4    Imperative procedures with mutual recursion

## 7.4.1    Mutual recursion

As was already explained, the recursion described so far does not cover the situation where procedure P calls procedure Q and procedure Q calls procedure P. Of course, on the syntactic lever we cannot exclude such situations, but at the denotational (implementational) level if procedural mechanisms are defined as in Sec. 7.3, mutual recursion will cause the error message 'procedure-not-declared'. Indeed, if P precedes Q, then the call of Q in the declaration-time environment of P would not find Q and analogously if Q precedes P.

To solve this problem, P and Q have to be defined jointly by one set of fixed-point equations

P = F(P,Q)

Q = G(P,Q)

and analogously for a larger number of mutually recursive procedures.

At the level of the algebra of denotations, this requires constructors creating tuples of mutually recursive procedures. Such tuples will be called *multiprocedures*. To define their constructors, we introduce three new domains.

cmp : MprComponents  = IprComponents$^{c+}$        (components of multiprocedures)

mpr : MulPro          = ImpPro$^{c+}$                      (multiprocedures)

mpd : MulProDecDen    = State ↦ State    (multiprocedure-declaration denotations)

Multi-procedure-components constitute a non-empty tuple of single procedures, and a multiprocedure is a non-empty tuple of procedures. The domain of multiprocedure-declaration denotations is identical with the corresponding domain for single procedures, but their reachable carriers are different.

## 7.4.2    Multiprocedure constructor

This constructor is a function of the type:

create-multi-pro : MprComponents x Env ↦ MulPro

and is defined analogously to the constructor of single procedures[74]. Let

cmp = ((ide.i, fpa-v.i, fpa-r.i, prd.i) | i = 1;n)            (components of multiprocedure)

Then

create-multi-pro.(cmp, env-dt) = (P.i | i=1;n)

where the tuple (P.i | i=1;n) is the solution of a fixed-point set of n equations, where the i-th equation is of the form shown below. The parameters of that equation are individual for each equation procedure-bodies prd.i and a common for all of them declaration-time environment env-dt.

P.i.(apa-v.i, apa-r.i).sto-ig =

   is-error.sto-ig   ➜ sto-ig

   **let**

      par.i          = (fpa-v.i, fpa-r.i, apa-v.i, apa-r.i)

      (vat-ig, 'OK')   = sto-ig

      vat-il = pass-actual.(par.i).(tye-dt, vat-ig)            (initial local valuation)

   vat-il : Error      ➜ vat-il

   **let**

      sto-il    = (vat-il, 'OK')            (initial local store)

      env-il   = (tye-dt, pre-dt[ide.j/P.j | j=1;n])            (initial local environment)

      sta-il    = (env-il, vat-il, 'OK')            (initial local state)

   pdr.i.sta-il = ?    ➜ ?

   **let**   (procedure-body execution)

      sta-tl             = prd.i.(sta-il)            (terminal local state)

      (env-tl, (vat-tl, err))   = sta-tl

   err ≠ 'OK'      ➜ (vat-ig, err)

   **let**

      vat-tg  = return-referential.(fpa-r.i, apa-r.i).(vat-tl, vat-ig)

   vat-tg : Error    ➜ (vat-ig, vat-tg)            (here vat-tg is an error)

   **true**       ➜ (vat-tg, 'OK')

This definition should be read in the same way as the definition of single-recursion described in Sec. 7.3.1. The only significant difference is the nesting of a vector of procedures [ide.j/P.j | j=1;n] in declaration-time procedure-environment pre-dt instead of nesting just one procedure.

At the call-time each procedure gets an initial local environment

---

[74] Now it becomes clear why environments have not been counted between procedure-components. If that were the case, then multiprocedure-constructor could get a different environment with each singe-procedure component, whereas all multiprocedures are defined in the context of a common environment.

env-il = (tye-dt, pre-dt[ide.j/P.j | j=1;n])

that is common for all procedure calls and an initial global store sto-ig which is individual for each procedure. This is why we did not count environments into procedure-components of a procedure since in the opposite case we could not pass one environment to all procedures.

Notice that since for n = 1 our definition coincides with single recursion, it can be taken as a universal constructor of a multiprocedure for an arbitrary n ≥ 1.

### 7.4.3    The instruction of an imperative-multiprocedure call

Calling each of multiprocedures is not different from calling a single procedure since the whole mechanism of multiprocedures is hidden in the multiprocedure-constructor. In this case, therefore we simply apply the definition from Sec. 7.3.3 without any modifications.

### 7.4.4    Multiprocedure declaration

The corresponding constructor is built analogously to the constructor for single procedures (Sec. 7.3.4):

declare-imp-mpr : MprComponents ↦ MulProDecDen

declare-imp-mpr : MprComponents ↦ State ↦ State

declare-imp-mpr.cmp.sta =

    is-error.sta                                    ➜ sta

    **let**

        ((ide.i, fpa-r.i, fpa-v.i, pro.i) | i = 1;n) = cmp

        (env, vat, 'OK') = sta

        (tye, pre) = env

    are-repetitions.(ide.i | i=1;n)            ➜ sta ◄ 'procedure-names-repeated'

    (∃i).vat.(ide.i) : declared.sta   ➜ sta ◄ 'procedure-declared'

    **let**

        (P.i | i=1;n) = create-multi-pro.(cmp, env)

        env-t = (tye, pre[ide.i/P.i | i=1;n])

    **true**                                        ➜ (env-t, vat, 'OK')

## 7.5    Functional procedures

The difference between imperative- and functional procedures is that the result of an imperative-procedure call is a state, whereas in the case of functional procedures — is a typed-data i.e. a composite. Imperative procedures may be regarded therefore as instructions with parameters and functional procedures — as expressions with parameters. Functional-procedure calls will belong in **Lingua-2** to the domain of expressions.

### 7.5.1   The structure of a functional-procedure declaration

Even though functional procedures correspond to expressions, the bodies of their declarations will consist of a program — maybe trivial, i.e. consisting of a trivial preamble and a trivial instruction — and an expression. The program transforms an input state and passes the resulting state to the expression which computes a composite. Below an example of a procedure (i.e. of its declaration) which computes the absolute value of a power of a number:

```
fun absolute-power(n, m real)

    let p be real

    p := 1

    while m > 0 do p := p*n ; m:=m-1 od

    return if p ≤ 0 then –p else p fi as real

end fun
```

In particular, in a functional-procedure declaration, the program that precedes **return** may be trivial, i.e. of the form **skip** and the expression that follows **return** may be reduced to a single variable[75]. The expression following **return** will be referred to as an *exporting expression*.

In some languages (e.g. in Pascal [47]) both referential parameters and global variables are admitted, which means that a functional procedure may change a state. This is frequently called a *side-effect*. In **Lingua-2** I deliberately give up this option since in my opinion, each invisible action of programs may contribute to programming errors. As a matter of fact, the authors of Pascal — although they allow side-effects — at the same time they strongly discouraged programmer to use them (see [47] page 79). One can ask, why they have not eliminated that option from their language?

In this place, I can also explain why actual parameters were assumed to be identifiers rather than arbitrary expressions (cf. Sec. 7.1.4). Notice that in the opposite case actual parameters could by functional-procedure calls and that leads to a new model of recursion and would certainly complicate construction rules for procedures (see Sec. 8).

A possible technical solution of that problem might be an assumption that actual parameters may be expressions but must not include procedure calls. Mathematically this is possible, but on the algebraic level, it leads to two sorts of expressions (with and without procedure calls) and also complicates proof rules.

Another restriction of functional procedures is the exclusion of recursion within a declaration itself, although imperative recursive-procedures may be called in their bodies. Recursion in functional procedures is denotationally possible, but I leave that issue to the reader as a useful exercise.

### 7.5.2   The domains of functional procedures

In the case of functional procedures we deal with three sorts of objects:

- *functional procedures*                    — state-to-composite functions
- *declarations of functional procedures*   — environment-to-environment functions,

---

[75] This universal form of a functional-procedure declaration was suggested to me by Andrzej Tarlecki.

- *functional-procedure calls*              — state-to-composite functions

As was already announced, functional procedures will be sometimes called simply *functions*. The extension of our language by functional procedures requires two new domains:

fdd : FprDecDen   = State ↦ State   (denotations of functional procedure-declarations)

fpr : FunPro       = ActPar ↦ State → CompositeE       (functional procedures)

At the same time, the domain of expression denotations is extended by functional-procedure calls. This is a substantial change since now expression denotations are getting access to procedure-environments to take functional procedures from them. Notice that although expressions were formally defined on states, they effectively referred to stores only.

It is also worth noticing that functional procedures operate on states rather than on stores, as it is the case for imperative procedures. This is, of course, due to the exclusion of recursion.

### 7.5.3     The constructor of a functional procedure

A functional procedure may be regarded as a sequential composition of three components:

1. a function that passes actual parameters to the call-time state of the procedure

2. an instruction,

3. an exporting expression

The domain of *functional-procedure components* is defined in the following way:

fpc : FprComponents = Identifier x ForPar x ProDen x DatExpDen x TypExpDen

As we see, the components of a functional procedure comprise the components of an imperative procedure without referential parameters plus the denotations of a data expression and a type expression:

ded  — the denotation of an exporting data-expression,

ted  — the denotation of a type expression.

In the sequel we shall need the following auxiliary operator:

export : (DatExpDen x TypExpDen) ↦ State ↦ CompositeE

export.(ded, ted).sta =

    is-error.sta                    ➔ error.sta

    **let**

        typ   = ted.sta                              (expected composite-type)

        com  = ded.sta                                 (exported composite)

    typ : Error                      ➔ typ

    com : Error                      ➔ com

    **let**

        (bod-t, yok)   = typ

        (dat, bod-c)   = com

    **not** (bod-t <u>coherent</u> bod-c)     ➔ 'bodies-not-coherent'

> yok.com ≠ (tt, ('boolean'))     ➔ 'yoke-not-satisfied'
>
> **true**                        ➔ com

This operator returns the composite computed by the denotation **ded** of the exporting expression under the condition that its body is coherent with the body of the computed type and the composite satisfies the yoke of this type.

Now we are ready to define the constructor of functional procedures that refers to the parameter-passing function and to the operator of exportation:

> create-fun-pro : FprComponents ⟼ FunPro

or in an unfolded form:

> create-fun-pro : Identifier x ForPar x ProDen x DatExpDen x TypExpDen
>
> ⟼ (AktPar ⟼ State → CompositeE)

Observe that this constructor does not „receive" an environment as a second argument (as it was the case for imperative procedures) since functional procedures act on states rather than on stores. That is, of course, the consequence of the exclusion of recursion. The value of our constructor is the function:

> F : AktPar ⟼ State → CompositeE
>
> F = create-fun-pro.(ide-n, fop-v, prd, ded, ted)

defined in the following way:

> F.apa-v.sta-gi =                                              (gi — global initial)
>
> is-error.sta-gi   ➔ error.sta-gi
>
> **let**
>
> ((tye, pre), (wrt, 'OK')) = sta-gi
>
> val-li = pass-actual.(fop-v, (), apa-v, () ).(tye, val)       (li — lokal initial)
>
> val-li : Error   ➔ val-li
>
> **let**
>
> sta-li = ((tye, pre), (val-li, 'OK'))
>
> sta-lf = prd.sta-li
>
> is-error.sta-lf   ➔ error.sta-lf
>
> **true**            ➔ export.(ded, ted).sta-lf

Employing parameter-passing operator, we create a local initial state that is passed to the program included in the procedure body. The exporting execution is evaluated in the output state of that program, and the resulting composite is the result of the procedure call. The body of this composite must be of the type indicated by the type expression, which is checked by the exporting operator. The empty tuples ( ) that appear among the arguments of this operator correspond to referential parameters respectively formal and actual.

Notice that the definitional equation of F does not have a fixed-point character. This is, of course, the consequence of the fact that we have given up recursion. Practically this means that if in a functional-procedure declaration a call of that procedure appears — which we cannot exclude syntactically — then the error message 'procedure-not-declared' will be generated.

## 7.5.4    The expressions of functional-procedure calls

A functional procedure is a function which given actual value-parameters return a data-expression denotation. A call of such a procedure is performed in four steps:

1. getting the procedure from an environment,

2. computing the values of its actual parameters,

3. applying the procedure to parameters in order to get a data-expression denotation,

4. applying this denotation to the actual state which — if the computation terminates — returns a composite or an error message.

Hence a call-constructor is of the type:

  call-fun-pro : Identifier x ActPar ⟼ DatExpDen

or:

  call-fun-pro : Identifier x ActPar ⟼ State → CompositeE

or:

  call-fun-pro.(ide, apa) : State → CompositeE

The expression denotation that is created in this way is defined as follows:

  call-fun-pro.(ide, apa).sta =

   is-error.sta          ➔ error.sta

   **let**

     ((tye, pre), sto)  = sta

   pre.ide = ?           ➔ 'procedure-not-declared'

   pre.ide : ImpPro      ➔ 'procedure-not-functional'

   **let**

     fpr = pre.ide                          (functional procedure)

   fpr.apa.sto = ?       ➔ ?

   **true**               ➔ fpr.apa.sta

If the initial state does not carry an error and in the environment a functional procedure has been declared under the name ide, then this procedure is applied to the current actual-parameters and the current state. If the application terminates, then its result is the result of the call. It may be a composite or an error.

## 7.5.5    The declaration of a functional procedure

In this case, the corresponding constructor is a function of the type:

  declare-fun-pro : FFcomponents ⟼ State ⟼ State

hence

  declare-fun-pro : Identifier x ForPar x ProDen x DatExpDen x TypExpDen ⟼

                      State ⟼ State

ff-declare-fun-pro.fpc.sta =

    is-error.sta        ➔ error.sta

    **let**

        (ide, fpa, prd, ded, ted)  = fpc

        ((tye, pre), sto)  = sta

    ide : declared.sta  ➔ sta ◄ 'variable-declared'

    **let**

      F = create-pro.(fpc, env)

    **true**              ➔ ((tye, pre[ide/F]), sto)

## 7.6     Procedures as the parameters of procedures

As we already know, the attempt to define procedures that can take other procedures as parameters lead in the general case to a non-denotational recursion of the type:

Procedure    = Parameter $\longmapsto$ Store $\rightarrow$ Store

Parameter    = Composite | Procedure

A mechanism of that sort had been implemented in Algol 60, but a mathematical description of that construction leads to non-denotational models or at least non-denotational on the ground of the classical set-theory.

However, the fact that such a solution is not denotational does not necessarily mean that in denotational models functions cannot take other functions as arguments. That is possible provided that we create a hierarchy of function, where no function can take itself as an argument neither directly or indirectly. Such a model was described in [28] and in a simplified version is as follows:

Procedure.0    = Parameter.0 $\longmapsto$ State $\rightarrow$ State

Parameter.0    = Composite

For $n > 0$:

Procedure.n  = Parameter.n $\longmapsto$ State $\rightarrow$ State

Parameter.n  = Parameter.0 | … | Parameter.(n-1)

As we see, a procedure may take as procedural arguments only procedures of a lower level than its own. To keep the description of **Lingua** of a reasonable size, this model shall not be developed further.

## 7.7     Programs

In **Lingua-1** programs are composed of a preamble and an instruction. In **Lingua-2** we keep this structure unchanged, but preambles may now include procedure declarations of all types. This is again a technical assumption which will make proof-rules simpler. The list of preamble-constructors defined in Sec. 6.1.9 is expanded by three new constructors corresponding to three types of declarations:

make-pream-from-ipd   : IprDecDen        $\longmapsto$ PreDen make-pream-from-mpd  : MulProDecDen        $\mapsto$ PreDen

make-pream-from-fpd   : FprDecDen           $\mapsto$ PreDen

All of them create preambles from corresponding declarations.

# 7.8    Syntax and semantics

## 7.8.1    The signature of the algebra of denotations

As we member from Sec. 4.2 and Sec. 6.2.2 concrete syntax of a language is derived from abstract syntax, which in turn is derived from the signature of the algebra of denotations. We shall start therefore from that signature. To a large extent it is implicit in the definition of denotations-constructors, however:

- not all of our constructors are constructors of the algebra — some of them were introduced only to define the latter,

- some constructors must be added to make the reachable parts of algebraic carriers non-empty.

The same observation concerns the domains themselves.

### 7.8.1.1    The carriers of the algebra of denotations

The list below covers all **Lingua-2** constructors, hence also the **Lingua-1** constructors.

| | | |
|---|---|---|
| ide | : Identifier | (identifiers) |
| | | |
| ded | : DatExpDen | (data-expression denotations including the calls of functions) |
| ted | : TypExpDen | (type expression denotations) |
| din | : InsDen | (instruction denotations including the calls of procedures) |
| | | |
| fpa | : ForPar | (formal parameters) |
| apa | : ActPar | (actual parameters) |
| | | |
| ipc | : IprComponents | (imperative-procedure components) |
| cmp | : MprComponents | (multiprocedure components) |
| ffc | : FprComponents | (functional procedure components) |
| | | |
| vdd | : VarDecDen | (variables-declaration denotations) |
| tdd | : TypDefDen | (type-definition denotations) |
| | | |
| idd | : IprDecDen | (imperative procedure-declarations denotations) |
| mpd | : MulProDecDen | (multiprocedure-declarations denotations) |

fdd   : FprDecDen                                  (function-declaration denotations)

dpe   : PreDen                                     (preamble denotations)

prd   : ProDen                                     (program denotations)

### 7.8.1.2    The constructors of the algebra of denotations

The list below covers only new constructors in **Lingua-2**, i.e. those that are not in **Lingua-1**. It is, however, not identical with the list of constructors defined in the present Sec.7 since:

1. it does not include auxiliary constructors defined only in order to define algebra-constructors,

2. it includes some new constructors necessary to make reachable parts of carriers non-empty.

**Function-calls as expressions**

call-fun-pro : Identifier x ActPar ↦ DatExpDen

Definition in Sec. 7.5.3

**Procedure calls as instructions**

call-imp-proc : Identifier x ActPar x ActPar ↦ InsDen

Definition in Sec. 7.3.3

**Formal and actual parameters**

create-for-par.fpa      : ↦ ForPar      for every fpa : ForPar

create-act-par.apa      : ↦ ActPar      for every apa : ActPar

In this case, we repeat the technical trick of introducing zero-argument constructors.

**Imperative-procedure components**

imp-pro-com : Identifier x ForPar x ForPar x ProDen ↦ IprComponents

imp-pro-com.(ide, fpa-r, fpa-v, prd) = (ide, fpa-r, fpa-v, prd)

This constructor looks like an identity function, but it is not. It takes four arguments from four different carriers and returns a four-tuple which is an element of another carrier. The fact that its definition is written in this way is due to a certain deficiency of our language where in principle we cannot distinguish between a function which takes four element from such which takes one four-tuple element.

**Multiprocedure components**

mul-pro-com      : IprComponents ↦ MprComponents

sequence-mpr   : MprComponents x MprComponents ↦ MprComponents

These functions create nonempty lists (tuples) of imperative-procedure components. The elements of these list (tuples) are tuples of a "lower level".

mul-pro-com.ipc                  = (ipc)                          (one-element tuple)

sequence-mpr.(mpc-1, mpc-2)   = mpc-1 © mpc-2         (the concatenation of tuples)

In the first equation we have a one-element tuple (ipc) that consists of one four-element tuple ipc.

**Functional-procedure components**

fun-pro-com : Identifier x ForPar x ProDen x DatExpDen x TypExpDen

$\mapsto$ PfcComponents

fun-pro-com.(ide-n, fpa, prd, ded, ted) = (ide-n, fpa, prd, ded, ted)

**Imperative-procedure declarations**

declare-imp-pro : Identifier x ForPar x ForPar x ProDen $\mapsto$ IprDecDen

Definition in Sec. 7.3.4

**Multiprocedure declarations**

declare-imp-mpr : MprComponents $\mapsto$ MulProDecDen

Definition in Sec. 7.4.4

**Functional-procedure declarations**

declare-fun-pro : Identifier x ForPar x ProDen x DatExpDen x TypExpDen

$\mapsto$ FprDecDen

**Preambles**

make-pream-from-ipd   : IprDecDen          $\mapsto$ PreDen

make-pream-from-mpd : MulProDecDen     $\mapsto$ PreDen

make-pream-from-fpd   : FprDecDen         $\mapsto$ PreDen

make-pream-from-tdd   : TypDefDen         $\mapsto$ PreDen

make-pream-from-vde   : VarDecDen         $\mapsto$ PreDen

sequence-pre              : PreDen x PreDen   $\mapsto$ PreDen

**Programs**

make-prog-from-ins   : InsDen              $\mapsto$ ProDen

make-prog               : PreDen x InsDen   $\mapsto$ ProDen

## 7.8.2    Concrete syntax

In the process of concrete-syntax creation, we skip the stage of abstract syntax since it has an algorithmic character and has been already described in details for **Lingua-1** (Sec. 6.2.1). In this case, we act similar to a mathematician who constructs proofs of theorems in an intuitive way rather than as formal sequences of formulas derived from each other by formalised deduction-rules. In both cases, however, we make sure that there exists a theoretical fundament that guarantees mathematical correctness of out constructions.

Contrary to Sec. 7.8.1.2 where only new constructors have been listed, here we show a full grammar of our language although without repeating these clauses which are taken from **Lingua-A** and **Lingua-1**.

ide    : Identifier = (as in **Lingua-A**)

tex    : TypExp   = (as in **Lingua-A**)

dae   : DatExp   =

   (as in **Lingua-A**) |

   Identifier (ActParameters)                              (functional-procedure call)

It is to be pointed out that ActParameters is a language which is one of the carriers of the algebra of syntax, whereas ActPar is a carrier of the algebra of denotations.

apar : ActParameters =

   **empty-ap**    |

   Identifier      |

   ActParameters , ActParameters

Here **empty-ap** is a keyword whose denotation is an empty list of actual parameters. Notice that our grammar allows for the generation of such "awkward" lists of parameters as e.g.

   **empty-ap , empty-ap**    or

   x, y, z, **empty-ap**

This is the price that we pay for the simplicity of our grammar. In order to avoid such situations we should use a grammar with two equations:

ActParameters =

   **empty-ap**             |

   NotEmptyActParameters

NotEmptyActParameters =

   Identifier   |

   NotEmptyActParameters , NotEmptyActParameters

Such a grammar, however, leads to a syntactic algebra which is not similar to our algebra of denotations. Of course, we could change the latter to make it similar, but this would mean that at the level of denotations we have to think about syntax, and this is what we actually want to avoid. We accept, therefore, our compromise grammar. Notice in this place, that our grammar allows the generation of list as we wish to have, e.g.,

   x, y, z

and on the other hand the "awkward" list have a sound denotational meaning.

   The same remark applies to the next grammatical clause:

fpar : ForParameters =

    **empty-fp**        |

    Identifier **as** TypExp |

    ForParameters , ForParameters

An example of a list of formal parameters may be:

   x **as** number, y **as** boolean, z **as** employee where employee is a user-defined type.

  vde : VarDec = (as in **Lingua-1**)

  tde  : TypDef =  (as in **Lingua-1**)

  ins : Instruction =

    (as in **Lingua-1**)    |

    **call** Identifier (**ref** ActParameters  **val** ActParameters) |

  prc : ProComponents =

    **proc** Identifier (**val** ForParameters **ref** ForParameters) Program **endproc**

  ipd : ImpProcDec =

  ProComponents

On the ground of the theorems from Sec. 2.3 on fixed-point-equations reductions we can replace the **ProComponents** in the second equation by the right-hand side of the first equation and get an explicit definition of imperative-procedures declarations

  ipd : ImpProcDec =

    **proc** Identifier (**val** ForParameters **ref** ForParameters)

      Program

    **end proc**

The components of multiprocedures are nonempty lists of imperative-procedure components, hence:

  swp : MprComponents =

ProComponents$^{c+}$

and since syntactically components are the same as declarations, we can write:

mpd : MultiProcDec =

**begin multiproc**

ImpProcDec$^{c+}$

**end multiproc**

Similarly as before, from this definition we can generate the final version of the definition of multiprocedure declaration:

mpd : MultiProcDec =

**begin multiproc**

[ **proc** Identifier (**val** ForParameters **ref** ForParameters)

Program

**endproc** ]$^{c+}$

**end multiproc**

Of course, the square-brackets belong to the metalevel. Analogously we generate the clause for functional-procedure declarations:

fpd : FunProDec =

**fun** Identifier (ForParameters) DatExp **endfun** |

**fun** Identifier (ForParameters)

Program

**return** Identifier **as** TypExp

**and fun**

pam: Preamble =

ImpProDec      |

MultiProDec     |

FunProDec      |

TypDef        |

VarDec        |

**skip**         |

Preamble ; Preamble

In this clause, we omit the names of constructors and the parentheses associated with semicolons. The first transformation has an isomorphic character because the first five categories in this clause correspond to disjoint languages which, in turn, is due to parentheses of the type **proc…end proc** and similar. The second transformation is a non-isomorphic (gluing) homomorphism but is acceptable due to the associativity of the composition of functions. More on that subject in Sec. 6.2.2.

prg : Program =

**begin-program** Instruction **end-program** |

    **begin-program** Preamble ; Instruction **end-program**

## 7.8.3    Colloquial syntax

In **Lingua-2** we allow all the colloquialisms of **Lingua-1,** and we add one concerning formal parameters in procedure declarations of both types. We allow grouping parameters into lists of variables associated with a common type as in the following example:

**proc** name(**val** w,z **as** real **ref** x,y **as** real a,b,c **as** employee)

## 7.8.4    Semantics

Since **Lingua-2** semantically coincides with **Lingua-1** wherever both languages coincide syntactically in this section, we consider these constructions only that are missing in **Lingua-2**. In the sequel, we write ide instead of Sid.[ide] since the semantics of identifiers is an identity function. We will use the algebraic style of semantics (see Sec. 5.6).

### 7.8.4.1    Actual parameters

Sapa : ActParameters $\longmapsto$ ActPar    or

Sapa : ActParameters $\longmapsto$ Identifier[c*]

Sapa.[ide] = (ide)

Sapa.[apa-1 , apa-2] = Sapa.[apa-1] © Sapa.[apa-2]

### 7.8.4.2    Formal parameters

Sfpa : ForParameters $\longmapsto$ ForPar    or

Sfpa : ForParameters $\longmapsto$ (Identifier x TypExpDen)[c*]

Sfpa.[ide **as** tex] = ((ide, Ste.[tex]))

Sfpa.[apar-1 , apar-2] = Sfpa.[apar-1] © Sfpa.[apar-2]

### 7.8.4.3    Data expressions: functional-procedure call

Sde : DatExp ⟼ DatExpDen

Sde.[`ide(apar)`] = call-fun-pro.(ide, Sapa.[`apar`])

### 7.8.4.4    Instructions: imperative-procedure call

Sin : Instructions ⟼ InsDen

Sin.[**call** `ide` (**ref** `apa-r`   **val** `apa-v`)] =

$\qquad\qquad\qquad\qquad$ call-imp-pro.(ide, Sapa.[`apa-r`], Sapa.[`apa-v`])

### 7.8.4.5    Imperative-procedure declarations

Sipd : ImpProcDec ⟼ IprDecDen

Sipd.[**proc** `ide` (**val** `fpa-v` **ref** `fpa-r`) `pro` **end proc**] =

$\qquad\qquad\qquad$ declare-imp-pro.(ide, Sfpa.[`fpa-r`], Sfpa.[`fpa-v`], Spr.[`pro`])

### 7.8.4.6    Multiprocedure declarations

Smpd : MultiProcDec ⟼ MulProDecDen

Smpd.[**begin multiproc**

$\qquad$ (**proc** `ide.i`(**val** `fpa-v.i` **ref** `fpa-r.i`)  `pro.i`  **end proc** | i=1;n)

$\quad$ **end multiproc**] =

$\qquad$ declare-imp-mpr.((ide-i, Sfpa.[`fpa-v.i`], Sfpa.[`fpa-r.i`], Spr.[`pro.i`]) | i=1;n)

### 7.8.4.7    Functional-procedure declaration

Sfpd : FprDec ⟼ FprDecDen

Sfpd.[**fun** `ide-n` (`fpa`)`pro` **return** `exp-r` **as** `tex` **and fun**] =

$\qquad$ if-declare-fun-pro.(ide-n, Sfpa.[`fpa`], Spr.[`pro`], Sde[`exp-r`], Ste.[`tex`])

### 7.8.4.8    Preambles

Spre : Preamble ⟼ PreDen

| Spre.[`vde`] | = Svd.[`vde`] | (variables declarations) |
|---|---|---|
| Spre.[`tde`] | = Std.[`tde`] | (type definitions) |

Spre.[`ipd`]         = Sipd.[`ipd`]                    (imperative-procedure declarations)

Spre.[`mpd`]         = Smpd.[`mpd`]                        (multiprocedure declarations)

Spre.[`fpd`]         = Sfpd.[`fdp`]                  (functional-procedure declaration)

Spre.[`pap ; dez`]   = Spr.[`pap`] ● Spr.[`dez`]

### 7.8.4.9    Programs

Spr : Program   $\longmapsto$ ProDen   of

Spr.[`ins`] = Sin.[`ins`]

Spr.[`pam ; ins`] = Spre.[`pam`] ● Sin.[`ins`]

# 8 Lingua-2V — validating programming

By *validating programming,* we shall mean a programming technique that guarantees the total-correctness of programs wrt program-specification created in parallel with program's code. This technique was already mentioned in Sec. 1.1 and its mathematical foundations are described in Sec. 3. The present section is devoted to general rules of equipping a language from **Lingua** family with validating tools. The rules are illustrated by examples referring to **Lingua-2**.

The general idea of validating programming was sketched (without procedures) in my papers [17], [18] and [19] published at the turn of the decades 1970 and 1980. On that ground, I came to the conclusion that to create a language with rules that guarantee program correctness, one has to start from a mathematical model of such a language. The next few years till the end of 1980. I devoted to the investigations of such models and the following 23 years (1990-2013) to run my family business (see Foreword). Therefore only in 2013, I have returned to my project, and the present book is the first step of it.

## 8.1    The structure of a validating language

Very briefly, a validating-programming language is a language of propositions that we call *metaprograms*. Each metaprogram is composed of two mutually nested layers:

3. *a programming layer* that is a program in the usual sense,

4. *a descriptive layer* which consists of pre- and post-conditions plus assertions (conditions) that are "nested" in-between instructions.

A metaprogram is said to be *correct* if its program (its programming layer) is totally-correct (Sec. 3.4) relative to its pre- and post-condition and its assertions are satisfied in the course of its execution. In the process of program creation assertions help to decide which program constructors can be applied at a given stage.

Validating programming consists in deriving correct programs from correct programs where the "initial" programs have to be proved correct in a traditional way. This situation is analogous to a formalised theory where we "derive" theorems from theorems employing deduction rules.

For every *source imperative language* **Lingua-n** we may construct a corresponding language **Lingua-nV** of validating programming which contains all of the source language plus three following (syntactic) categories of its descriptive layer:

1. *Conditions* — the denotations of which are three-valued <u>partial</u> predicates on states.

2. *Specified instructions* — the denotation of which are partial functions on states (like instruction denotations) and where the descriptive layer describes the properties of the programming layer.

3. *Propositions* — the denotations of which are <u>classical</u> Boolean values tt and ff; propositions are split into three subcategories:

3.1. *properties* that express syntactic properties of programs, e.g. that a given procedure declaration appears in a preamble,

3.2. *metaconditions* that express the semantic properties of conditions, e.g. that a given condition is never false but may be undefined,

3.3. *metaprograms* that express total-correctness properties of programs which they include.

Propositions are assumed to be closed under classical Boolean operators and classical quantifiers. This means that in constructing correct programs, we remain in the range of <u>classical logic</u>.

In dealing with properties and metaconditions, we use classical logical operators since in describing program-properties we remain in the classical logic. Non-classical predicates appear only of program-denotations, i.e. at the level of program-executions.

Contrary to our philosophy *from denotations to syntax*, in constructing a language of validating programming, we proceed from syntax to denotations. This is the consequence of the fact that this time our starting point is an "already existing" syntax of a source-language which has to be a subset of the corresponding validating language.

## 8.2   Conditions

To avoid tedious technicalities, conditions will not be defined in details. Instead, I only assume some of their properties. The description of these properties should show the way of building the category of *conditions* for each particular language from **Lingua** family.

Classes of conditions will be described and illustrated with the help to their (anticipated) concrete syntax. Hopefully, this will contribute to the readability of incoming sections without damaging (too much) the rigour of mathematical precision.

In defining the semantics of conditions we shall use the following notation for Boolean composites:

ct = (tt, ('Boolean'))

cf = (ff, ('Boolean'))

and we shall assume that McCarthy's operators are defined on such composites according to McCarthy's philosophy.

### 8.2.1   Conditions in general terms

For every **Lingua-nV** we build the following category of conditions

con : Condition =

   DatCon |                                                                    (data-conditions)

   ValCon |                                                                    (validating conditions)

   Instruction @ Condition   |                                          (algorithmic conditions)

   (Condition **and** Condition) | (Condition **or** Condition) | (**not** Condition) |

   (∀ Identifier, Condition) | (∃ identifier, Condition)

Intuitively and practically *data-conditions* may be regarded as Boolean-expressions constructed over an extended set of data-constructors which allow to express such properties of data which are not necessarily expressible in the source-language. E.g. at the level of data-conditions we may have a predicate-constructor `ordered-list` which may be not available at the level of programs. Data-conditions constitute, therefore, a certain superset of Boolean expressions. In the consequence — which follows from the equations above — also Condition constitute such a superset. However — as was explained at the end of Sec. 5.3.5 — making the set of Boolean-expressions, hence also conditions, as a separate syntactic (algebraic) cathegory, leads to solution inconvenient for programmer's perspective. We assume, therefore, that conditions constitute a superset of all data expressions, rather that only of Boolean expressions only, and hence that their semantics is a function:

Sco : Condition ↦ State → Composite | Error

Notice that condition-denotations are partial functions which is due to the fact that conditions include all data expressions.

We assume that Boolean constructors of conditions are defined according to the McCarthy's philosophy, i.e. analogously as for data expressions (Sec.5.3.2) but quantifiers are in the Kleene's style that was shortly mentioned in Sec. 2.9.

∀ : Identifier x Condition ↦ Condition

Sco.[∀`(ide, con)`].sta =

    is-error.sta                               ➔ error.sta

    **let**

        (env, (vat, 'OK')) = sta

    for every val : Value,  Sco.[`con`].(env, (vat[ide/val], 'OK')) = ct ➔ ct

    there is   val : Value,  Sco.[`con`].(env, (vat[ide/val], 'OK')) = cf ➔ cf

    **true**                                    ➔ 'never-false'

The message 'never-false' is generated in situations where the composite

Sco.[`con`].(env, (vat[ide/val], 'OK'))

is never cf, but at the same time is not always ct, i.e. if it is:

- either ct,

- or an error,

- or is undefined,

and for at least one value val it is not equal ct. The existential quantifier is defined in the following way:

∃ : Identifier x Condition ↦ Condition

Sco.[∃`(ide, con)`].sta =

    is-error.sta                                ➔ error.sta

    **let**

(env, (vat, 'OK')) = sta

there is     val : Value,  Sco.[con].(env, (vat[ide/val], 'OK')) = ct   ➔ ct

for every   val : Value,  Sco.[con].(env, (vat[ide/val], 'OK')) = cf   ➔ cf

**true**                                                              ➔ 'never-false'

Notice that the equality

[∀(ide, con)].sta = cf

holds even if for some value val, the value of con is an error and analogously in the situation where

[∃(ide, con)].sta = ct.

This choice means that quantifiers are defined according to Kleene's philosophy rather than to that of McCarthy. In the case of Boolean expressions, which are evaluated during program execution, the Kleene's philosophy was not acceptable since it would lead to non-implementable semantics. However, in the case of conditions which are not supposed to be executed, Kleene's calculus is not only acceptable but — in the case of quantifiers — even better. This claim may be justified by condition:

(∃ x)(1/x > 2)

the value of which in the calculus of McCarthy is undefined, since it is undefined for x = 0. More on the consequences of that choice in [23] and [49] [76].

## 8.2.2    Data-conditions

As was already announced, *data-conditions* describe the properties of data of the source language, and their class splits into two subclasses:

1. data expressions of the source language,

2. extended data expressions referring to composite-constructors not available in the source language.

The conditions of the second group should allow the descriptions of these data-properties which we may need in describing the properties of programs, e.g. that a given list is ordered lexicographically or that a given database satisfy given integrity constraints. We assume that the semantics of the first group of conditions coincide with the semantics of data expressions of the source language.

As we see, data-conditions do not always evaluate into a Boolean composite. In spite of that, we call them "conditions" since they belong to the domain Condition.

## 8.2.3   Validating conditions

*Validating conditions* describe properties of states that have to do with a programming context and therefore are specific for a given source language. On the other hand, they are universal as far as a source-data-algebra is concerned. In this section, we shall see a few examples of validating conditions that — in my opinion — should be available in every validating language.

---

[76] In this place one may ask a question why Kleene's calculus was assumed for quantifiers but not for Boolean operators. A spontaneous answer may be that in the case of the latter I wanted to avoid a "double semantics" that would complicate the model. However, that question maybe deserves a second thought.

Two first conditions in this group are constant-value conditions that at the syntactic level will be denoted by **TT** and **FF**. Their denotations are the following:

[**TT**].sta =

    is-error.sta   ➔ error.sta

    **true**        ➔ ct

[**FF**].sta =

    is-error.sta   ➔ error.sta

    **true**        ➔ cf

Furthermore we assume that for every data expression `dae` we have in our language a condition **defined-d**(`dae`) which is satisfied if the value of `dae` is defined:

[**defined-d**(`dae`)].sta =

    is-error.sta        ➔ error.sta

    Sde.[`dae`].sta = ?  ➔ cf

    Sde.[`dae`].sta = !  ➔ ct

    **true**             ➔ cf

Notice that since data expressions may include procedure calls, for some `dae` this condition may be not computable. This, however, does not cause any problem, since **defined-d**(`dae`) will never appear in the programming layer of the language. An analogous condition is defined for type expressions although in this case we do not need to check for definedness:

[**defined-t**(`tex`)].sta =

    is-error.sta        ➔ error.sta

    Ste.[`tex`].sta = !  ➔ ct

    **true**            ➔ cf

Among validating conditions we also have conditions describing the fact that a given identifier is a variable identifier of a given type:

[`ide` **is** `tex`].sta =

    is-error.sta           ➔ error.sta

    **let**

        (env, (vat, 'OK')) = sta

    vat.ide  = ?        ➔ cf

    Ste.[`tex`].sta : Error ➔ Ste.[`tex`].sta

    **let**

        (dat, typ)  = Sde.[`ide`].sta

        typ-e     = Ste.[`tex`].sta

```
    typ = typ-e              ➜ ct
    typ ≠ typ-e              ➜ cf
```

An example of such a condition may be:

```
  length is real
```

or

```
  employee is record-type
                c-name        as word,
                f-name        as word,
                birth-year    as number,
                award-years   as array-of number ee
            ee
```

Below we define three classes of conditions that are necessary in every validating language.

Conditions of the first of them expresses type-compatibility of the type of an identifier with the value of an expression:

```
  [ide conformant-with dae].sta =
      is-error.sta            ➜ error.sta
      let
         sta = (env, vat, 'OK')
      vat.ide = ?              ➜ cf
      Sde.[dae].sta = ?        ➜ cf
      Sde.[dae].sta : Error    ➜ Sde.[dae].sta
      let
         (bod-i, yok-i)   = vat.ide
         (dat-e, bob-e)   = Sde.[dae]
      bod-i = bod-e           ➜ ct
      true                    ➜ cf
```

As we see, an identifier is type-compatible with an expression if it is declared, the value of the expression is defined and its body is identic with the body of the identifier-value. It is assumed in this definition that the undefinability of the value of `dae` leads to **cf**, which of course makes our condition yet another non-computable case.

Notice that in contrast to the former conditions where we expect identical types, here we limit ourselves to bodies since expressions evaluate to composites (data and body) rather than to values.

The second class of condition corresponds to the function **dynamically-compatible** defined in Sec. 7.2.2 and concerning the compatibility of formal and actual parameters. Let then **fpa-v**, fpa-r, apa-v, apa-r be the list of formal parameters (value- and reference-) and the corresponding actual parameters (value- and reference-):

[ **conformant**(`fpa-v, fpa-r, apa-v, apa-r`) ].sta =

   is-error.sta                                      ➔ error.sta

      **let**

         ((tye, pre), (vat, 'OK')) = sta

      dynamically-compatible.(fpa-v, fpa-r, apa-v, apa-r).(tye, vat) = 'OK' ➔ ct

      **true**                                       ➔ cf

In order to define the third class of conditions we have to introduce some auxiliary concepts. A state shall be called *initial* if it is of the form:

   (([ ], [ ]), ([ ], 'OK'))

where [ ] is an empty mapping

A state is said to be *adequate* wrt a given preamble `pam`, if it does not carry an error and results from an initial state by the execution of an arbitrary program with `pam` as its preamble. In other words a state is adequate wrt preamble `pam`, if:

1. it does not carry an error,

2. in the environment of that state are declared all and only procedures that are in the preamble,

3. in the environment of that state are defined all and only types the definitions of which are in the preamble,

4. in the valuation of that state are declared all and only these variables the declarations of which are in the preamble,

5. the types of all declared variables have been defined in the preamble[77].

Consequently, all states of a program-execution which do not carry errors are adequate wrt the preamble of that program.

By AD.pam  we shall denote the set of all states adequate wrt to preamble `pam`.  The specific conditions of the third group are of the form **ade-for** (`pam`)  where pam is a preamble and have the following semantics:

[**ade-for**(`pam`)].sta =

    is-error.sta   ➔ error.sta

    sta : AD.pam ➔ ct

    **true**           ➔ cf.

## 8.2.4   Algorithmic conditions

*Algorithmic conditions*[78] have a syntactic form

---

[77] This condition is redundant since if follows from 1. and 4 but it has been included in the list to make it explicit.

[78] Conditions of that type are fundamental for *algorithmic logic* developed at Warsaw University in the years 1970-1980 (see [8]).

```
ins @ con
```

where `ins` is an instruction and `con` is a condition (possibly algorithmic) and where semantics is defined by the equation

Sco.[ `ins @ con` ] = Sin.[ins] ● Sco.[con]

Therefore the logical value of condition `ins @ con` in the state **sta** equals the value of `con` in the state Sin.[ins] .**sta**, i.e. in the terminal state of the execution of `ins` that starts with **sta**. As it follows form investigations of Sec. 3.4, `ins @ con` is the weakest precondition that guarantees a terminating execution of `ins` with a terminal state that satisfies `con`.

A condition which is not algorithmic is said to be in a *standard form*. Algorithmic conditions, similarly as data-conditions may assume non-Boolean composites as their values.

## 8.3    Specified instructions

Intuitively speaking *specified instructions* or just *specinstructions* are instructions with nested assertions that describe properties of states intermediate in the executions of instructions. Their grammatical clause is the following:

sin : SpecInstruction =

    Instruction                                         |

    **asr** Condition **rsa**                        |

    **if** DatExp **then** SpecInstruction **else** SpecInstruction **fi**   |

    **if-error** DatExp **then** SpecInstruction **fi**          |

    **while** DatExp **do** SpecInstruction **od**            |

    SpecInstruction ; SpecInstruction

This equation expands the grammar of **Lingua-2** by a new clause and the language by a new sort. As we see, specinstructions contain all instructions and additionally one specific construct **asr con rsa** that shall be called *assertion*.

The denotations of specinstructions belong to the same domain as the denotations of instructions, hence their semantics is a function:

Ssi : SpecInstruction  ⟼ State → State

This function is defined in the following way:

Ssi.[ins]            = Sin.[ins]

Ssi.[**off** ins **on**]    = Sin.[ins]

Ssi.[**asr** con **rsa**].sta  =

    is-error.sta          ➔ sta

Sco.[con].sta = ?        ➔ ?

Sco.[con].sta =ct       ➔ sta

**true**                        ➔ sta ◄ 'assertion-not-satisfied'

The semantics of specinstructions which are instructions coincide with the semantics of instruction. In case of assertions if their condition holds, then the state remains unchanged, and otherwise, an error message is generated. Notice that the error message 'assertion-not-satisfied' will appear in two situations:

1. when the value of the condition is (ff, ('Boolean')),
2. when the value of the condition is an error.

For the remaining four clauses our semantics is defined analogously to the semantics of instructions.

As we are going to see in subsequent sections, the described syntax and semantics of specinstructions constitute a fundament for the definitions of program-construction rules. In this context, assertions describe the properties of states that appear during program execution and — as we are going to see — are used by program-transformations that preserve program-correctness.

Quite frequently assertions are satisfied on a certain "interval" of successive *atomic instructions* (i.e. assignments and procedure calls), with the exclusion of a certain subinterval of this interval. In such a case, in order to avoid repeating the same assertion many times between successive instructions, we use two notational abbreviations of the form

**begin-asr** con; sin **end-asr**                                            (*)

**off-asr** sin **on-asr**                                                  (**)

The first of them is a colloquialism which corresponds to an instruction resulting from sin by the insertion of **asr  con  rsa** between any two atomic instructions with the exclusion of each exclusion-interval and each error-handling instruction. The specinstruction in (*) will be called the *on-range of* con, and we shall also say that in that sin the condition has been *set-on*.

The abbreviation (**) is also a colloquialism intuitively indicating that "part" of sin where all previously set-on conditions do not need to be satisfied[79].

Summing up, both (*) and (**) are not specinstructions but just notational conventions. They will be formalised as colloquialisms, i.e. by an appropriate restoring transformation. Consequently, they do not appear neither in concrete syntax nor (of course) in denotations. This choice was forced by the denotationality of our model, since in that model the denotation

Ssi.[**begin-asr** con; sin **end-asr**]

should be a composition of Sco.[con] and Ssi.[sin]. This is, however, impossible, since for example two following specinstructions:

**begin-asr** x > 0;                              **begin-asr** x > 0;

---

[79] For the sake of simplicity I assume that in the **off-on** region all previously activated conditions are not expected to be satisfied. An alternative would be, of course, a **off-on** clauses which indicates a particular condition to be off, but this more flexible form is so far left for future investigations.

```
   x := x                        x := -x ; x := -x
 end-asr                        end-asr
```

have different denotations although the denotations of their instructions are identical.

In this situation (*) and (**) have to be treated as a colloquialism described by a restoring transformation RT. This transformation is an identity function for all specinstructions without on-ranges of conditions and otherwise is defined by structured induction.

We start from the case where an on-rage is an (ordinary) instruction. Since that case requires a structured induction again we start from an assignment:

RT.[ **begin-asr** con; ide := dae **end-asr**] =
    **asr** con **rsa**; ide:= dae; **asr** con **rsa**

For yoke-assignments and procedure calls, the transformation is defined analogously. Next case is an error-handling instruction where the rule is similar to the former:

TP.[ **begin-asr** con; **if-error** wyd **then** ins **fi end-asr**] =
    **asr** con **rsa**; **if-error** wyd **then** ins **fi** ; **asr** con **rsa**

The remaining subcases with ordinary instruction are defined in the following way:

RT.[**begin-asr** con; **if** dae **then** ins-1 **else** ins-2 **fi end-asr**] =
    **asr** con **rsa**;
    **if** dae
      **then** RT.[**begin-asr** con ins-1 **end-asr**]
      **else** RT.[**begin-asr** con ins-2 **end-asr**]
    **fi**;
    **asr** con **rsa**

RT.[**begin-asr** con; **while** dae **do** ins **od end-asr**] =
    **asr** con **rsa**;
    **while** dae **do** RT.[**begin-asr** con; ins **end-asr**] **od**;
    **asr** con **rsa**

RT.[**begin-asr** con; ins-1 ; ins-2 **end-asr**] =
    **asr** con **rsa**;
    RT.[**begin-asr** con; ins-1 **end-asr**];

```
asr con rsa;
```

RT.[**begin-asr** con; ins-2 **end-asr**]

```
asr con rsa
```

RT.[**begin-asr** con; **if-error** dae **then** ins **fi end-asr**] =

   **begin-asr** con **end-asr**;

   **if-error** dae **then** RT.[**begin-asr** con; ins **end-asr**] **fi**;

   **begin-asr** con **end-asr**;

Now we have to consider the case where the on-range is a specinstruction which is not an instruction.

RT.[**begin-asr** con **off** ins **on end-asr**] = ins

As we see the assertion does not "penetrate" the instruction closed by the exclusion-brackets.

RT.[**begin-asr** con-1; **asr** con-2 **rsa end-asr**] =

   **asr** con-1 **and** con-2 **rsa**

RT.[**begin-asr** con-1; **begin-asr** con-2; **sin end-asr end-asr**] =

   RT.[**begin-asr** con-1 **and** con-2; sin **end-asr**]

The remaining cases connected to structural specinstructions are defined in a way analogous to the corresponding ordinary structured instructions.

## 8.4    Propositions

Generally speaking, conditions describe properties of states and propositions — properties of conditions, specified instructions and programs and in the case of the latter also properties of their syntax. In total we are going to deal with three classes of propositions:

- *syntactic properties*    — of programs and their components,
- *metaconditions*    — which express semantic properties of conditions,
- *metaprograms*    — which express semantic properties of programs.

Contrary to conditions, who as values assume composites, not even Boolean composites, and errors, the values of propositions may be only ct and cf. Whereas in programs we use three-valued partial predicates, in the descriptions of programs we remain in the classical logic.

The category of propositions again — i.e. as in the case of conditions and for the same reasons — we build from syntax to denotations.

## 8.4.1    Syntactic properties

*Syntactic properties* are propositions about syntactic properties of programs and their components. To define them a few auxiliary concepts and notations are necessary.

If `pam` is a preamble then by `names.pam` we denote the set of all identifiers of variables, of procedures, of functions and of types which are declared in `pam`. Since single declarations and definitions of all these types may be treated as preambles, the function `names` applies to them as well.

Two preambles are said to be *disjoint* if the corresponding sets of identifiers are disjoint.

A preamble is said to be *correct* if no identifier has been declared or defined in it twice. Notice that a preamble is correct iff its corresponding set of adequate states (Sec. 8.2.2) is not empty. We say that a declaration or a definition is *admissible* in a preamble if adding it to the preamble does not make this preamble incorrect.

In the sequel we shall use the following constructors of syntactic properties:

| | |
|---|---|
| **is-correct** `pam` | — preamble `pam` is correct |
| `dec` **is-in** `pam` | — declaration `dec` appears in `pam` |
| `dec` **allowed-in** `pam` | — declaration `dec` is admissible in `pam` |
| `ide` **is-pro-in** `pam` | — `ide` has been declared in `pam` as a procedure |
| `ide` **is-fun-in** `pam` | — `ide` has been declared in `pam` as a function |
| `ide` **is-typ-in** `pam` | — `ide` has been defined in `pam` as a type |
| `ide` **is** tex **in** `pam` | — `ide` has been declared in `pam` as a `tex`-variable |
| `ide` **not-in** `pam` | — `ide` has been not declared or defined in `pam` |
| `pam-1` **separated-from** `pam-2` | — `pam-1` and `pam-2` are disjoint |

Obvious formal definitions have been skipped.

## 8.4.2    Metaconditions

*Metaconditions* describe properties of conditions. In order to define them we shall use the following notation:

$[$`con`$]$ = Sco.$[$`con`$]$[80]

$\{$`con`$\}$ = {sta : Sco.$[$`con`$]$.sta = ct}

Metaconditions are created by means of four constructors which we shall call *metapredicates*

$\Rightarrow$ , $\sqsubseteq$ , $\Leftrightarrow$ , $\equiv$ : Condition x Condition $\longmapsto$ Proposition

The denotations of metaconditions are classical logical values tt and ff and metapredicates correspond to binary relations between conditions[81]:

| | | |
|---|---|---|
| `con-1` $\Rightarrow$ `con-2` **iff** $\{$`con-1`$\}$ $\subseteq$ $\{$`con-2`$\}$ | | *(stronger than)* |
| `con-1` $\sqsubseteq$ `con-2` **iff** $[$`con-1`$]$ $\subseteq$ $[$`con-2`$]$ | | *(less defined than)* |

---

[80] The notation for the semantics of conditions is redundant, but it will turn convenient in the investigations that follow.
[81] **iff** stands for "if and only if"

```
con-1 ⇔ con-2  iff  {con-1}={con-2}                    (weakly equivalent)

con-1 ≡ con-2  iff  [con-1]=[con-2]                    (strongly equivalent)
```

In the first case we also say that `con-2` is *weaker than* `con-1` and in the second that `con-2` is *more defined than* `con-1`. The following rather obvious relations hold between metapredicates:

```
con-1 ≡ con-2    implies         con-1 ⇔ con-2

con-1 ≡ con-2    implies         con-1 ⊑ con-2

con-1 ≡  con-2   is equivalent to  (con-1 ⊑ con-2 and con-2 ⊑ con-1)

con-1 ⇔ con-2    is equivalent to  (con-1 ⇨ con-2 and con-2 ⇨ con-1)

con-1 ⇔ con-2    implies         con-1 ⇨ con-2
```

By means of these predicates we can easily express the property of a partial- or total-correctness of an instruction `ins` wtr a precondition `pre` and a postconditions `post`.

```
pre @ ins ⇨ post          — partial correctness

pre       ⇨ ins @ post    — total correctness
```

We can also describe the properties of a *weak* and a *strong invariant* of an instruction:

```
war @ ins ⇨ war           — weak invariant

war       ⇨ ins @ war     — strong invariant
```

The weak and the strong invariants are used in correctness-proofs of respectively partial- and total-correctness of programs. Now let us examine a few examples[82]:

```
x>0 and ²√x > 2    ≡    x > 4

        ²√x > 2    ⇔    x > 4   but ≡ does not hold

        ²√x < 2    ⊑    x < 4   but neither ≡ nor ⇔ do not hold

        ²√x > 4    ⇨    x > 3   but neither ⇔ nor ⊑ do not hold
```

Notice also that[83]

con-1 ⇨ con-2 does not imply (con-1 **implies** con-2) ≡ **TT**.

Indeed, despite that the metaimplication $\sqrt[2]{x} > 4 \Rightarrow$ `x > 3` holds, the condition

$\sqrt[2]{x} > 4$ **implies** `x > 3`

is undefined for `x < 0`. As a matter of fact, the opposite-side implication is true:

if (con-1 **implies** con-2) ≡ **TT**, then con-1 ⇨ con-2.

Indeed let sta:{con-1}, which means that [con-1].sta = ct. If now [(con-1 **implies** con-2)].sta = ct and [con-1].sta = ct, then [con-2].sta = ct which means that sta:{con-2}.

---

[82] We assume that the square root of a negative number is undefined.
[83] Implication is defined in the usual way, i.e. p **implies** q means (**not** p) **or** q where the negation and the alternative belong to McCarthy's calculus.

Notice that on the ground of our non-classical calculus of conditions we have two concepts of satisfiability con ≡ **TT** — *strong satisfiability;* con is always true con ⊑ **TT** — *weak satisfiability;* con is never false.

Since our metapredicates are regarded as binary relations in the algebra of conditions, the following lemmas may be easily proved (more in [21]).

**Lemma 8.4.2-1** *Relations* ≡ *and* ⇔ *are both equivalences, i.e. they are reflexive, symmetric and transitive.* ∎

**Lemma 8.4.2-2** *The strong equivalence is a congruence, i.e. the replacement of a subcondition of a condition by a strongly equivalent one result a condition strongly equivalent to the initial one.* ∎

**Lemma 8.4.2-3** *Weak equivalence is a congruence with regard* **and** *and* **or**. ∎

Weak equivalence is not a congruence wrt negation since

con-1 ⇔ con-2 does not imply **not** con-1 ⇔ **not** con-2

For instance, although

$\sqrt[2]{x} > 2 ⇔ x > 4$

hold, the metacondition

$\sqrt[2]{x} ≤ 2 ⇔ x ≤ 4$

is not true, since for x=-1 the right-hand-side equation evaluates to ct, but on the left-hand side, we have an error.

**Lemma 8.4.2-4** *The operators* **and** *and* **or** *are strongly associative, i.e.*

(con-1 **and** con-2) **and** con-3 ≡ con-1 **and** (con-2 **and** con-3)

(con-1 **or** con-2) **or** con-3 ≡ con-1 **or** (con-2 **or** con-3) ∎

Of course, they are also weakly associative since strong equivalence implies weak equivalence.

**Lemma 8.4.2-5** *The operator* **and** *is strongly left-hand-side distributive wrt* **or** *and vice versa, i.e..*

con-1 **and** (con-2 **or** con-3) ≡

con-1**and** con-2) **or** (con-1 **and** con-3)

con-1 **or** (con-2 **and** con-3)  ≡

con-1 **or** con-2) **and** (con-1 **or** con-3)∎

However, both operators are not strongly right-hand-side distributive. Indeed:

(ct **or** ee) **and** cf = cf   but   (ct **and** cf) **or** (ee **and** cf) = ee

(cf **and** ee) **or** ct = ct   but   (cf **or** ct) **and** (ee **or** ct) = ee          (8.4.2-1)

**Lemma 8.4.2-6** The operator **and** *is weakly left-hand-side distributive wrt* **or** *i.e.*

```
(con-1 or con-2) and con-3  ⇔
```
$$(\texttt{con-1 } \textbf{and}\texttt{ con-3}) \textbf{ or } (\texttt{con-2 } \textbf{and}\texttt{ con-3}) \qquad ■$$

However, **or** is not even weakly left-hand-side distributive wrt **and** which can be seen in (8.4.2-1).

**Lemma 8.4.2-7** *The de Morgan's laws for* **and** *and* **or** *and for the negation of quantifiers are satisfied with a strong equivalence* ■

**Lemma 8.4.2-8** *Conjunction is weakly commutative.*

```
con-1 and con-2  ⇔ con-2 and con-1 ■
```

However, conjunctions are not strongly commutative, and the alternative is not even weakly commutative, since:

ct or ee = ct   but   ee or ct = ee

**Lemma 8.4.2-9** *If*

```
con-1 ⇨ con-2
```
*to*
```
con-1 and con-2  ≡ con-1  ■
```

Besides the two-argument metapredicates, we also define three-argument metapredicates which will be used in the development of correct programs:

```
con-1 ≡ con-2 whenever con iff  con and con-1 ≡ con and con-2
```

```
con-1 ⇔ con-2 whenever con iff  con and con-1 ⇔ con and con-2
```

In both cases, we say that con constitutes a *logical context* or simply a *context* for the equivalence which it follows. We shall also say that the *equivalence* con-1 ≡ con-2 *is satisfied under the condition* con and analogously for a weak equivalence. Here are two examples:

$$\texttt{n} > \texttt{x}^2 \equiv \sqrt[2]{n} > \texttt{x} \quad \textbf{whenever} \ (\texttt{n} \geq 0 \ \textbf{and} \ \texttt{x} \geq 0)$$

$$\texttt{n} > \texttt{x}^2 \Leftrightarrow \sqrt[2]{n} > \texttt{x} \quad \textbf{whenever} \ \texttt{x} \geq 0$$

This context is usually a condition in whose range we want to replace one condition by another one.

All the material presented above was published by myself in the decade 1980 in [19] and [23], and the development of these ideas towards three-valued deductive theories was investigated in a paper [49] written jointly with Beata Konikowska and Andrzej Tarlecki.

### 8.4.3    Metaprograms

*Metaprograms* are propositions with the following syntax (for the sake of simplicity we drop the program-parentheses **begin-program** and **end-program** introduced in Sec. 6.2.2):

> MetaProgram =
>> **def** Preamble
>> **pre** Condition
>>> SpecInstruction
>> **post** Condition

Metaprograms express total correctness of specinstructions (as defined in Sec. 3.6) relativized to states which are adequate for preambles. The semantics of metaprograms is then a function of the type:

> Smp : MetaProgram $\mapsto$ {tt, cf}

defined as follows:

> Smp.[**def** pam **pre** prc sin **post** poc] = tt

iff the preamble pam is correct and

> {**ade-for**(pam) **and** prc} $\subset$ Ssi.[sin] $\bullet$ {poc}          (8.4-1)

or, in other words,

> **ade-for**(pam) **and** prc  $\Rightarrow$  sin @ poc.

If the denotation of a metaprogram is tt, then we say that the metaprogram is *correct*.

A metaprogram is therefore correct if its preamble is correct and for every state adequate for that preamble, if that state satisfies the preconditions, then the execution of specinstruction terminate successfully and the terminal state satisfies the postcondition.

Notice that a successful termination of sin means that none of the assertions in sin was falsified and that the terminal state does not carry an error since otherwise post-conditions would not be satisfied.

A few useful lemmas may be formulated about metaprogram-correctness. The first follows immediately from the remark formulated above.

**Lemma 8.4.3-1** *If*

> **def** pam **pre** con-pr sin **post** con-po

*is correct and* sin-1 *has been created from* sin *by the removal of an arbitrary number of assertions or assertion-declarations, then the program*

> **def** pam **pre** con-pr sin-1 **post** con-po

*is correct as well.*   ■

**Lemma 8.4.3-2** *If*

> **def** pam **pre** con-pr sin **post** con-po

*is correct, then correct is also every program that results from the former be the replacement of conditions by weakly equivalent conditions.*   ■

The proof follows from the fact that the denotations of assertions with weakly equivalent conditions are identities on the same set of states. In particular, this lemma implies that on the level of conditions (but not of Boolean expressions of the programming layer!) we can apply all the lemmas of Sec. 8.4.2 that concern weak equivalences.

**Lemma 8.4.3-3** *If*

> **def** pam **pre** con-pr sin **post** con-po

*is correct, then correct is also each program that results from the former by replacing any Boolean data-expression* dae *that appears in* **if-then-else-fi** *or in* **while-do-od** *by an expression* dae-1 *that is stronger defined, i.e. such that* dae $\sqsubseteq$ dae-1. ■

If the source program is correct, then none of its Boolean expressions generates an error and wherever dae is defined dae-1 is also defined and has the same value.

Now let us notice that for any preamble pam the condition **ade-for**(pam) is a weak invariant of every specinstruction[84] since specinstructions do not change neither the environments nor the types of global variables. This means that the condition (8.4-1) is equivalent to the condition:

> {**ade-for**(pam) **and** prc} $\subset$ Sin.[sin] ● {**ade-for**(pam)**and** poc}

which expresses the „usual" total correctness restricted to the set of states adequate for the preamble. This is what I meant in saying earlier that the correctness of a metaprogram is relativised to its preamble[85]. In the subsequent investigations, we will assume that in writing

> **def** pam **pre** con-pr sin **post** con-po

we express the fact that this metaprogram is correct. This is as in the "everyday" mathematics where we write "x > 2" to say that "x > 2 is true".

In this place it is worth noticing that all program-constructors in **Lingua-2** are decent which implies that all program-components are conservative. This in turn means that if an execution starts from a correct state

---

[84] This conclusion is based on the fact that specinstruction-denotations are functions (rather than relations), since only in this case total-correctness implies partial-correctness.
[85] In order to have this property I have assumed that all declarations which are global in a program have to precede all instructions.

### 8.4.4   Jaco de Bakker paradox in Hoare's logic

As was noticed by Jaco de Bakker (p. 108, Sec. 4 in [5]) and later commented by Krzysztof Apt in [4], on the ground of Hoare's logic one can prove the formula:

{ **true** } a[a[2]] := 1 { a[a[2]] = 1 }

which for same arrays a is not true. To show that consider an array:

a = [2,2]

In that array

a[2] = 2

hence the execution of the assignment

a[a[2]] := 1

means the execution of

a[2] := 1

which means that the new array is a = [2,1], and therefore a[a[2]] = a[1] = 2.

Let us observe, however, that Hoare's problem does not result neither from having arrays in a language nor from the admission of expressions like a[a[2]], but from a tacit assumption that whenever such an expression appears on the left-hand-side of an assignment, then it should be treated as a variable. As a matter of fact, for many years, programmers used to talk about "subscripted variables" (Algol 60 [61]) or about "indexed variables" (Pascal [47]).

The de Bakker's problem with Hoare's logic is in the imperfect understanding of the meaning (the semantics) of array variables[86]. In our language Bakker's paradox does not appear since the instruction of the form:

```
a.(a.2) := 1
```

would be syntactically incorrect. In that place, we write

a := **change-arr** a **at** a.2 **by** 1 **ee**

or colloquially

a := **change-arr** a **by** a.2 <= 1 **ee**

On the ground of **Lingua-2** we can easily prove the correctness of the following metaprogram (which was already done a few lines above):

```
def let a be arr-type number ee
pre a.1 = 2 and a.2 = 2
    a := change-arr a by a.2 <= 1 ee
post a.1 = 2 and a.2 = 1
```

---

[86] In the denotational model described by M. Gordon in [44] array-variables or indexed-variables are admitted on the cost of a rather substantial complication of the model by distinguishing between left-values of expressions (locations) and right-values of expressions (values). In states values are assigned to locations and locations to identifiers.

This program may be also formally derived. For that sake we apply an easily provable strong equivalence:

ide := dae @ (con-1 **and** con-2)

≡

(ide := dae @ con-1) **and** (ide := dae @  con-2)

Now the Rule 8.5.2-1 in Sec. 8.5.2 guarantees the correctness of

<u>**def**</u> **let** a **be arr-type** number **ee**
<u>**pre**</u>
   a := **change-arr** a **by** a.2 <= 1 **ee** @ a.1 = 2
   **and**
   a := **change-arr** a **by** a.2 <= 1 **ee** @ a.2 = 1;
     a := **change-arr** a **by** a.2 <= 1 **ee**
<u>**post**</u> a.1 = 2 **and** a.2 = 1

In order to transform this program into the expected form we have to apply two strong equivalences that hold under the condition (guaranteed by the preamble) that a is an array of numbers:

a := **change-arr** a **by** a.2 <= 1 **ee** @ a.1 = 2

≡

a.2 ≠ 1 **and** a.1 = 2

and

a := **change-arr** a **by** a.2 <= 1 **ee** @ a.2 = 1

≡

a.2 = 2

To get the expected precondition, we apply the metaimplication

a.1 = 2 **and** a.2 = 2

⇨

a.2 ≠ 1 **and** a.1 = 2 **and** a.2 = 2 **and**

and the rule 8.5.2-5 from Sec. 8.5.2 which allows replacing a precondition with a stronger one.

# 8.5    The construction of correct metaprograms

## 8.5.1    Notational convention

As was already said, validating programming consists in building correct metaprograms from its correct components. At the level of the algebra of relations, this idea was described in Sec. 3.6. At the level of metaprograms it is based on rules which have the following general form:

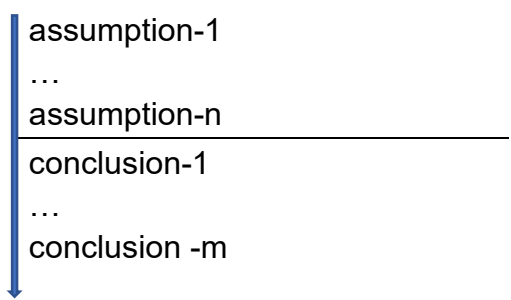

Such a rule is read as: if all assumptions are satisfied, then all conclusions are satisfied as well. If the implication is satisfied in both sides, then we use a double-direction arrow.

Now let us think what are the assumptions and the conclusions? Seemingly we might expect that they are propositions as described in Sec. 8.4. But propositions always concern concrete conditions and programs, and here we need general rules as in Sec. 3.6. In that section assumptions and conclusions are (classical) logical formulas with variables that run over domains of binary relations and sets. In case of metaprograms assumptions and conclusions are similar formulas, but now variables run over syntactic categories.

## 8.5.2    Basic rules

Before we start building rules for correct-metaprogram-construction it is worth recalling a few facts:

(1) every (reachable) instruction is conservative (Sec. 6.1.4), which means that it is transparent wrt errors and whenever it changes values assigned to variables in valuations the coherence property is observed,

(2) every instruction-constructor is decent which means that it preserves the conservativeness of its arguments,

(3) if a condition is satisfied in a state, then such a state does not carry an error,

(4) instructions do not lead out of the set of states which are adequate for preambles.

In the rules that follow we tacitly assume that all metavariables which run over preambles, conditions and instructions are bound by general quantifiers that stand before diagrams, i.e. before metaimplications that the diagrams denote.

**Rule 8.5.2-1 Assignment**

```
is-correct(pam)
def pam
pre (ide:=dae)@ con
    ide:=dae
post con
```

If `pam` is a correct preamble and the algorithmic condition `(ide:=dae)@con` is satisfied, then directly from the definition of `@` we can conclude that the assignment will terminate and the resulting state satisfies `con`.

Notice that the correctness of the preamble is a necessary and sufficient condition for the correctness of our program. This leads to an apparently paradoxical conclusion that independently of the preamble every metaprogram below the line is correct. In fact, however, the precondition implies that:

- the evaluation of `dae` terminates without an error message,

- the assignment does not generate an error which means that `ide` has been declared and its type coincides with that of `dae`.

On the base of Lemma 8.4.2-9, we do not need to include these conditions in the precondition.

Below we have an example of program-generation on the ground of this rule and of the strong equivalence:

```
x:=y+1 @ 2*x>10 ≡ 2*(y+1)>10 and y<max-num
```

which holds under the condition that `max-number` denotes the maximal representable number. From this equivalence we may conclude the correctness of two programs:

```
def let x, y be number
pre 2*(y+1) > 10
  x:= y+1
post 2*x > 10
```

Analogously we create rules for variable declarations and type definitions.

**Rule 8.5.2-2 Variable declaration**

```
is-correct(pam)
def pam
pre (let ide be tex) @ con
    let ide be tex
post con
```

**Rule 8.5.2-3 Type definition**

```
is-correct(pam)
─────────────────────────────────────
def pam
pre (set ide as tex) @ con
    set ide as tex
post con
```

**Rule 8.5.2-4 Sequential composition of instructions**

```
def pam pre prc-1 sin-1 post poc-1
def pam pre prc-2 sin-2 post poc-2
poc-1 ⇨ prc-2
───────────────────────────────────────────
def pam pre prc-1 sin-1;sin-2 post poc-2
def pam pre prc-1 sin-1;begin-asr poc-1 end-asr;sin-2
post poc-2
def pam pre prc-1 sin-1;begin-asr poc-2 end-asr ; sin-2
post poc-2
```

Given two metaprograms with a common preamble and with instructions `sin-1` and `sin-2` respectively we can construct three versions of a metaprogram with instruction `sin-1;sin-2`. In this case, our rule follows directly from the Rule 3.6.1-1. It is to be pointed out that the metaimplications go only top-to-bottom since we have skipped existential quantification of `poc-1` and `pre-2` (which in this case would not have much sense).

Observe also that in this case, we do not need to assume the correctness of `pam` above the line since it is implicit in the assumptions about the correctness of both metaprograms.

**Rule 8.5.2-5 Conditional branching if-then-else**

```
prc ⇨ dae or(not dae)
def pam pre (prc and dae) sin-1 post poc
def pam pre (prc and not dae) sin-2 post poc
──────────────────────────────────────────────
def pam pre prc if dae then sin-1 else sin-2 fi post poc
```

This rule follows directly from the Rule 3.6.1-2. In this case, the implication is two-directional since this time we do not need to construct any intermediate assertion. It is also worth noticing that the metacondition

```
prc ⇨ dae or (not dae)
```

means that whenever `prc` is satisfied, the data expression `dae` is either **ct** or **cf**, hence is defined, is not an error, and is and Boolean. Notice that in a two-valued predicate calculus this assumption would be a tautology.

**Rule 8.5.2-6 Loop while-do-od**

```
prc ⇨ inv
inv ⇨ (dae or (not dae))
inv and (not dae)) ⇨ poc
def pam pre inv and dae       sin post inv
def pam pre inv while dae do sin od post TT
```
**def** pam **pre** prc **while** dae **do** sin **od post** poc

This rule follows from Rule 3.6.2-5. The condition `inv` is the invariant of the loop-body `sin`. The last condition above the line means that the satisfaction of the invariant in the precondition guaranties program termination. Of course, in proving the termination property, we may use Lemma 3.6.2-1.

The rules that follow are immediate consequences of Rule 3.6.1-3, Rule 3.6.1-4 and Rule 3.6.1-5.

**Rule 8.5.2-7 Strengthening precondition**

```
def pam pre prc sin post poc
prc-1 ⇨ prc
```
**def** pam **pre** prc-1 sin **post** poc

**Rule 8.5.2-8 Weakening postcondition**

```
def pam pre prc  sin  post poc
poc ⇨ poc-1
```
**def** pam **pre** prc  sin  **post** poc-1

**Rule 8.5.2-9 Conjunction of conditions**

```
def pam pre prc-1  sin  post poc-1
def pam pre prc-2  sin  post poc-2
```
**def** pam **pre** (prc-1 **and** prc-2)  sin  **post** (poc-1 **and** poc-2)

**Rule 8.5.2-10 The expansion of a preamble**

```
def pam-1 pre prc sin post poc
pam-1 separated-from pam-2
```
**def** pam-1 ; pam-2 **pre** prc sin **post** poc

## 8.5.3     Imperative-procedure call

Consider a procedure declaration

```
proc procedure (val fpa-v ref fpa-r)
   pam-pro ; sin
end proc
```

and let the correctness-proposition for the call of this procedure be

| | |
|---|---|
| **def** pam-pr | (program's preamble) |
| **pre** prc-cl | (call's precondition) |
|    **call** procedure (**val** apa-v **ref** apa-r) | (8.5.3-1) |
| **post** poc-cl | (call's postcondition) |

where pam-pr is the preamble of the program in which our procedure is called. Let then

| | |
|---|---|
| **def** pam-bo | (body's preamble) |
| **pre** prc-bo | (body's precondition) |
|    sin | (8.5.3-2) |
| **post** poc-bo | (body's postcondition) |

be a correctness proposition about the procedure body.

The construction-rule for proposition (8.5.3-1) concerning the correctness of the call has to be based on five following assumptions.

<u>First</u>, the declaration of procedure — denote it by dec — must appear in the preamble pan-pr of the hosting program, i.e. the following syntactic property must hold (Sec. 8.4.1):

```
dec is-in pam-pr
```

<u>Second</u>, the proposition (8.5.3-2) for body-correctness must be satisfied.

<u>Third</u>, in every state that is adequate for preamble pam-pr (Sec. 8.2.2) and satisfies the precondition prc-cl of the call, the tuples of formal and actual parameters must be dynamically compatible (Sec. 7.2.2) which means that the following metaimplications must be true:

**ade-for**(pam-pr)**and** prc-cl ⇨

$$\textbf{conformant}(fpa\text{-}v, fpa\text{-}r, apa\text{-}v, apa\text{-}r)$$

This metacondition guaranties that parameter-passing will be executed properly.

<u>Fourth</u>, the satisfaction of the precondition prc-cl of the call must guarantee that after parameter passing the initial state of the body's execution will satisfy its precondition prc-bo. This means that the following metaimplication must be true:

```
prc-cl ⇨ prc-bo[i-fpa-v/apa-d, i-fpa-r/apa-v]
```

where `prc-bo[i-fpa-v/apa-d, i-fpa-r/apa-v]` denotes the condition `prc-bo`, where formal-parameters' identifiers were replaced with actual parameters[87].

Fifth, the satisfaction of procedure's body postcondition `poc-bo` in a state adequate for procedure's body preamble must guarantee that all formal reference-parameters will have values assigned to them — which is necessary for a parameter-passing without errors — and that the external state will satisfy procedure's postcondition `poc-cl`. This means that the following metacondition must hold:

> **ade-for**`(pam-bo)` **and** `poc-bo` ⇨
>
> > **defined**`(fpa-r)` **and** `poc-cl[apa-r/apa-v]`

The above arguments allow formulating the rule, which is shown below. This rule, however, needs a certain comment since seemingly it does not refer to the fact that the called procedure may be recursive. Formally there is no such reference indeed, but the assumption (2) will force us to cope with recursion if there are recursive calls in procedure's declaration. This issue will be investigated in Sec. 8.5.4.

**Rule 8.5.3-1 Imperative-procedure call**

```
(1) dec is-in pam
(2) def pam-bo pre prc-bo sin post poc-bo
(3) ade-for(pam-pr)and prc-cl ⇨
                    conformant(fpa-v, fpa-r, apa-v, apa-r)
(4) prc-cl ⇨ prc-bo[i-fpa-v/apa-d, i-fpa-r/apa-v]
(5) ade-for(pam-bo) and poc-bo ⇨
                    defined(fpa-r)and poc-cl[apa-r/apa-v]
```
```
def pam-pr
pre prc-cl
   call procedure (val apa-v ref apa-r)
post poc-cl
```

## 8.5.4    The case of recursive procedures

Each time we want to use Rule 8.5.3-1 we have to derive a correct program that appears in item (2) of that rule:

> **def** `pam-bo`
>
> **pre** `prc-bo`
>
> > `sin`                                                                                  (8.5.4-1)
>
> **post** `poc-bo`

---

[87] A formal definition of this transformation requires a rather laborious construction by structural induction wrt the grammar of conditions, which I omit at that stage. It is worth noticing in this place that if actual parameters could be arbitrary data expressions, then this definition would be even more complex.

In such a case we have to do with recursion whenever we have a direct or an indirect call of `procedure` in `sin`. In that case, the derivation of (8.5.4-1) depends on the context in which a procedure call will appear in `sin`. Below we consider the simplest case that corresponds to the Rule 3.6.2-4, i.e. where procedure P is the least solution of the equation:

X = H X T | E.

Of course, for P to be a function, H, T and E have to be functions as well, and additionally the domains of H and E must be disjoint. To satisfy these prerequisites let us write our equation in the form:

X = [C] H X T | [¬C] E.

that corresponds to a declaration:

```
proc procedure (val fpa-v ref fpa-r)
  def pam;
  specins
end proc
```

where `specins` is a specinstruction of the form

```
if dae
  then
    head; call procedure(apa-v, apa-r) ; tail        (8.5.4-1)
  else
    exit
fi
```

Rule 3.6.2-4 for that case has the form:

**Rule 8.5.4-1 An instruction with a recursive procedure call**

(1) (∀sin) **def** pam **pre** (prc **and** dae) sin **post** poc
        implies
    **def** pam **pre** (prc **and** dae) head ; sin ; tail **post** poc
(2) **def** pam **pre** (prc **and** (**not** dae)) exit **post** poc
────────────────────────────────────────────────────
**def** pam **pre** prc specins **post** poc

This rule should be applied in the derivation of assumption (2) in the Rule 8.5.3-1 together with the assumption that `specins` is of the form (8.5.4-1).

## 8.5.5   Functional-procedure call

To prove the correctness of a functional procedures, we can use Rule 8.5.3-1 that "serves" imperative procedures. Consider a declaration of the form:

**fun** ide-n (fpa) prg   **return** exp-r **endproc**

To prove that the value of `exp-r` has in the output state of the call a property

```
property.(exp-r)
```

one has to prove the correctness of a program of the form:

```
def pam
   pre prc
   prg
post property.(exp-r)
```

## 8.6    Transformational programming

### 8.6.1    First example

In the former section, we were dealing with rules for constructing correct programs from correct components. Expressing this in the language of the automotive industry, we were constructing tools for assembly lines. In the present section, we shall consider rules that transform programs to "enrich" their functionality. In the following examples, we show the applications of rules introduced earlier as well as new rules that are going to be introduced in the next section.

At the beginning let us consider two correct metaprograms. Let `n` and `m` denote two "concrete" positive integers, i.e. two data expressions with constant values[88]. Let `isr(n)` be the integer square-root of n, and let `iqt(n, m)` be the integer quotient of n by m.

```
def let x be number          def let x be number
pre true                     pre true
   x := 0;                      x := 0;
   while (x+1)² ≤ n             while (x+1)*m ≤ n
     do                          do
       x := x+1                    x := x+1
     od                          od
 post x = isr(n)              post x = iqt(n, m)
```

Each of these programs goes number-by-number through the set of positive integers in seeking the expected result. Going again to the automotive language we may say that both programs are driven by the same program-engine:

```
P1: def let x be number
      pre true
        x := 0;
         while x+1 ≤ k
           do
             x := x+1
             od
```

---

```
     post x = k
```

Now on that engine, we can ones "install" an "application" for `isr` and another time — for `iqt`. Formally this installation consists in applying a transformation which changes the functionality of the engine but preserves its correctness. Let us see how this can be applied to the case of a square-root.

If `isr(n)` is a constant-value expression, then the correctness of P1 implies the correctness of P2.

P2:  **let** x **be** number

    **pre** true

      x := 0;

        **begin-asr** x ≥ 0

          **while** x+1 ≤ isr(n)

            **do**

              x := x+1

            **od**

        **end-asr**

    **post** x = isr(n)

In this and in the following steps the modified parts of a program are marked with a colour.

It is to be clarified that adding the on-range for $x \geq 0$ is not a result of an application of a general rule, but a step the soundness of which has to be proved (which in this case is of course very easy).

So far our metaprogram looks a bit senseless since it refers to `isr(n)` in order to compute it. We shall, therefore, eliminate that expression from the programming layer on the strength of the strong equivalence:

    x+1 ≤ isr(n)  ≡ (x+1)² ≤ n  **whenever** x ≥ 0

and applying Lemma 8.4.3-3, which allows replacing a boolean expression by a strongly equivalent expression. In our case, this equivalence holds only in the context where $x \geq 0$, but this context is assured within its on-range.

As a result of the described transformation, we end up with a final program P3 where the (unnecessary now) assertion has been removed.

P3:  **let** x **be** number

    **pre** true

      x := 0;

      **while** (x+1)² ≤ n

        **do**

          x := x+1

        **od**

    **post** x = isr(n)

Let us now follow through a transformation to a program that again computes `isr(n)`, but this time is installed on a much faster engine. Let `po2.k` (k is a power of 2) denote a predicate which is satisfied if `k` is a power of 2, i.e.

```
po2.k ≡ (∃m≥0)  k=2^m
```

and let `mag.k` (the magnitude of k) denotes a function with values in the set of powers of 2 such that

```
mag.k ≤ k < 2*mag.k
```

In other words, if `mag.k = 2^m` then

```
2^m ≤ k < 2^{m+1}
```

Of course, for every `k` there is a unique `m` that satisfies these two inequalities.

As is well known, for any $n \geq 0$ there is a unique sequence of 0's and 1's which is a binary representation of `n`.

Now it is easy to prove the total correctness of the two following programs:

```
Q1: let z be number
       pre true
          z := 1
          begin-asr po2.z
             while z ≤ mag.k do x:=2*z od
          end-asr
       post z = 2*mag.k

Q2: let z, x be number
       pre z = 2*mag.k
          x := 0
          while z > 1
             do
                z := z/2;
                if x+z < k then x:=x+z else x:=x fi
             od
       post x = k and z = 1
```

The first program computes the successive powers of 2 until it reaches `2*mag.k`, and the second returns to `k` through successive powers $2^m$ and on that way summarises these powers of 2 that correspond to 1 in the binary representations of `k`. Now observe that the following proposition is true:

```
z ≤ mag.k  ≡  z ≤ k   whenever po2.z
```

Due to that, we can replace the Boolean expression in the **while** of the first program by the strongly equivalent `z ≤ k`. Now, if we enrich the preamble of **Q1** by the declaration of a new variable `x` and join both programs on the ground of Rule 8.5.2-3, we get our target program that computes `isr(n)` in logarithmic time. Notice that the former engine was computing in the linear time. In the same step, we can remove the unnecessary (now) assertion and move the initialisation of `x` at the beginning of the program.

```
Q3:  let z, x be number
     pre true
        z := 1
        x := 0
        while z ≤ k do z:=2*z od
        while z > 1
          do
            z := z/2;
            if x+z < k then x:=x+z else x:=x fi
          od
     post x = k and z = 1
```

If in this program we replace the expression `k` by the expression `isr(n)`, then we have a program that computes `isr(n)` but refers to that number. Now we proceed similarly as in the former example to eliminate `isr(n)`. To do that we use two strong conditional equivalences:

$$z \leq \text{isr}(n) \quad \equiv z^2 \leq n \qquad \textbf{whenever } z > 0$$

$$x+z < \text{isr}(n) \quad \equiv (x+z)^2 < n \quad \textbf{whenever} (z > 0 \textbf{ and } x \geq 0)$$

We also introduce an obvious on-range which allows for the use of both replacements and in this way we get a program that computes `isr(n)` in logarithmic time.

```
Q4:  let z, x be number
     pre true
        z := 1
        x := 0
        begin-asr z > 0 and x ≥ 0
           while z² ≤ n do z:=2*z od
           x := 0
           while z > 1
             do
               z := z/2;
               if (x+z)² < n then x:=x+z else x:=x fi
             od
        end-asr
     post x = isr(n) and z = 1
```

Now we shall optimise this program in restricting the number of variables and the number of arithmetic operations. Let us start from the observation that in each run of the first loop the program recomputes the value of $z^2$ which is not quite optimal. We introduce therefore a new variable q, and we enrich our program in such a way that the condition $q = z^2$ is satisfied.

```
Q5:  let z, x, q be number
     pre true
        z := 1;
        x := 0;
        q := 1;
```

```
      begin-asr z > 0 and x ≥ 0 and q = z²
         while q ≤ n
            do
               off z:=2*z; q:=4*q on
            od
         while z > 1
         do
            off z := z/2; q := q/4 on
            if x²+2*x*z+q ≤ n then x:=x+z else x:=x fi
         od
      end-asr
   post x = isr(n) and z = 1 and q = z²
```

Notice the double-use of **off-on** is necessary since each time when the first assignment destroys the satisfaction of $q=z^2$, the second recovers it. Now we proceed to further transformations:

1. we use the equivalence  `z>1 ≡ q>1` **whenever**  ($z>0$ **and** $q=z^2$) to modify Boolean expression in the second loop,

2. we introduce two new variables `y`  and  `p`  with the conditions $y = n-x^2$ and $p = x*z$,

3. we use the equivalence

   $x^2 + 2*x*z + q ≤ n ≡ 2*p+q ≤ y$ **whenever** ($y=n-x^2$ and $p=x*z$)

```
Q6: let z, x, q, y, p be number
    pre true
       z := 1;
       x := 0;
       q := 1;
       begin-asr z > 0 and x ≥ 0 and q = z²
          while q ≤ n
             do
                off z:=2*z; q:=4*q on
             od
          y := n;
          p := 0;
          begin-asr y = n-x² and p = x*z
             while q > 1
                do
                   off z:=z/2; q:=q/4; p:=p/2; on
                   if 2*p+q ≤ y
                      then x:=x+z; p:=p+q; y:=y-2p-q
                      else x:=x
                   fi
                od
             end-asr
          end-asr
       post x=isr(n) and z=1 and q=z² and y=n-x² and p=x*z
```

Contrary to the former introduction of a new variable which was clearly justified, now it not quite clear why p and y have been introduced. The answer to this question follows from a well-known truth that in programming, like in playing chase, we sometimes have to predict a few moves in advance. These moves will be shown a little later.

In the next transformation, we prepare our program for the removal of variable z. For that sake, we perform the following changes.

1. we apply the equivalence $q=z^2 \Leftrightarrow isr(q)=z$ **whenever** $z>0$ to change the assertion,

2. we use the condition isr(q)=z to replace z by isr(q) everywhere except the left-hand side of the assignment,

3. we make obvious changes based on the equality z=1.

```
Q7: let z, x, q, y, p be number
    pre true
      z := 1;
      x := 0;
      q := 1;
      begin-asr z > 0 and x ≥ 0 and isr(q)=z
        while q ≤ n
          do
            off z:=2*isr(q); q:=4*q on
          od

        y := n;
        p := 0;
        begin-asr y = n-x² and p = x*isr(q)
          while q > 1
            do
              off z:=(isr(q))/2; q:=q/4; p:=p/2 on
              if 2*p+q ≤ y
                then x:=x+isr(q) ; p:=p+q; y:=y-2p-q
                else x:=x
              fi
            od
        end-asr
      end-asr
    post x=isr(n) and z=1 and q=1 and y=n-x² and p=x
```

Now notice that in **Q7** the variable z does not appear neither in boolean expressions nor on the right-hand sides of assignment that do not change z. Since we do not care about the terminal value of z, we can remove that variable from our program together with the corresponding assignment (general rule will be described in Sec. 8.6.2). In this way we get:

```
Q8: let x, q, y, p be number
    pre true
```

```
    q := 1;
    x := 0;
    begin-asr x ≥ 0
      while q ≤ n
        do
          q:=4*q
        od
      y := n;
      p := 0;
      begin-asr y = n-x² and p = x*isr(q)
        while q > 1
          do
            off q:=q/4; p:=p/2 on
            if 2*p+q≤y
              then p:=p+q; y:=y-2p-q
              else x:=x fi
          od
      end-asr
    end-asr
  post x=isr(n) and q=1 and y=n-x² and p=x
```

Now we use the equivalence

```
   x=isr(n) and p=x  ≡  p=isr(n) and p=x
```

to modify the postcondition which makes variable `x` not necessary anymore. Therefore, we can remove it.

**Q9:** **let** `k, q, y, p` **be** `number`
```
      pre true
      q := 1;
      while q ≤ n
        do
          q:=4*q
        od
      y := n;
      p := 0;
      begin-asr y = n-p²/q
        while q > 1
          do
            off q:=q/4; p:=p/2 on
            if 2*p+q≤y
              then p:=p+q; y:=y-2p-q
              else x:=x fi
          od
      end-asr
    post p = isr(n) and q = 1 and y = n-p²/q
```

In the last step

1.  we remove the redundant `y = n-p²/q`, from postcondition,

2. we remove the assertion-decree for `y = n-p`$^2$`/q`,

3. due to the removal of the decree, we remove **off-on**,

4. we replace the instruction

```
p:=p/2;
if 2*p+q≤y then p:=p+q; y:=y-2p-q else x:=x fi
```

by an equivalent instruction

```
if p+q≤y then p:=p/2+q; y:=y-p-q else p:=p/2 fi
```

As a result, we get the final version of our program:

```
Q10:  let q, y, p be number
      pre true
         q := 1;
         while q ≤ n do q:=4*q od
         y := n;
         p := 0;
         while q > 1
           do
              q:=q/4; p:=p/2
              if p+q≤y
                 then p:=p/2+q; y:=y-2p-q
                 else p:=p/2
              fi
           od
      post p = isr(n)
```

This program had been written by a well-known Norwegian computer-scientist Ole-Johan Dahl in 1970. I do not know in what way he built this program, but we may suppose that he performed an optimisation similar to ours, although without formalised rules.

At the end of this section one pragmatic remark. Programmers who develop tenths or hundreds of thousands of code-lines will probably regard the discussed example with a certain scepticism. Indeed, the volume of our program is not very impressive, and the shown optimisation is rather irrelevant for the majority of applications. If however, we build microprograms that are implemented in hardware and executed hundreds millions of times by hundred millions of computers, then its correctness as well as time- and space-consumption may be quite relevant. Our example also shows a certain general — although not universal — method of building programs in three steps:

1. writing a program-engine that searches through a certain set of data,

2. installing an application on that engine which implements22 the expected functionality,

3. optimising the program.

As we are going to see in Sec. 8.6.3, program optimisation may also be used in changing the types of data elaborated by a program.

## 8.6.2   Adding a register-identifier

This section is devoted to a transformation of a metaprogram by adding to it an assertion of the form:

    `ide-r = dae-r`                                                                                      (*)

Such transformation was applied a few times in Sec. 8.6.1 e.g. in passing from Q4 to Q5.

An identifier `ide` that satisfies the condition `ide=dae` on a certain on-range that condition is called a *register-identifier* or just a *register,* the expression `dae` is called a *register-expression* and the condition `ide=dae` — a *register-condition.*

Let us start from an obvious generalisation of the operation @ (Sec. 8.2.4) from conditions to arbitrary data expressions:

Sde.[ `sin @ dae` ] = Ssi.[`sin`] ● Sde.[`dae`]

Let us consider now a correct metaprogram P of the form

P: **def** `pam`

  **pre** `prc`

    `ins-h;`                                                    (the head of the program; possibly empty)

    **asr** `con-p` **rsa ;**                                                        (initial condition)

    **begin-asr** `con-n` ;                                        (assertion condition)

      `ins`

    **end-asr**

    `ins-t`                                                    (the tail of the program; possibly empty)

  **post** `pow`

Let `dae-r` be a data expression that satisfies two metaimplications

  `con-p` ⇨ **defined**(`dae-r`)     and

  `con-n` ⇨ **defined**(`dae-r`)

Let `ide-r` be an identifier which does not appear in P and let `dez-ide-r` be such a variable declaration of `ide-r`, that the latter is typologically compatible with `dae-r`. A transformation that enriches P by introducing a register-condition `ide-r = dae-r` yields a program:

Q: **def** `pam ; dez-ide-r`

  **pre** `prc`

    `ins-h ;`

    `ide-r := dae-r ;`

    **begin-asr** `con-n` **and** `ide-r = dae-r`

```
    ins $ (ide-r=dae-r)                    (enriched instruction – see below)
  end-asr ;
  ins-t
post pow
```

where `ins $ (ide-r=dae-r)` is such an enrichment of instruction `ins` which makes Q correct, provided that P was correct. The assertion **asr** con-p **rsa** does not appear in Q (although one can leave it there), since it only served to express the fact, that in the indicated location the value of `dae-r` is defined. Now, this property is ensured by the assertion.

The syntactic operation $ is defined by structural induction wrt the structure of `ins`. Let us start with an assignment

```
ide := dae
```

where obviously ide $\neq$ ide-r, since we assumed that `ide-r` does not appear in P.

If `ide` does not appear in `dae-r`, then the execution of this assignment does not cause any change in the value of `dae-r`, and therefore we do not need to add to the instruction any actualisation.

If this is not the case, then directly after `ide:=dae` we have to add an assignment which recovers the satisfaction of the condition `ide-r=dae-r`. Therefore in such a case

```
(ide:=dae) $ (ide-r=dae-r) =
off ide := dae ; ide-r := dae-r on
```

An off-clause has been used here since `ide` appears in `con-r` and therefore the alteration of the value of `ide` may cause the alteration of the value of `con-r` and the falsification of our condition. In the case of the transformation of Q4 into Q5 with a register-assertion q=$z^2$ this leads to the enrichment of `z:=2*z` into:

```
off z:=2*z ; q:=z² on
```

This instruction may be now changed into an equivalent one:

```
off q:=((z:=2*z) @ z²) ; z:=2*z on
```

In this instruction, we eliminate @, by transforming the expression `(z:=2*z) @ z²` to a standard form:

```
off q:=4*z² ; z:=2*z on
```

Now since the assertion q=$z^2$ holds "just before" this instruction, we can replace the instruction by:

```
off z:=2*z ; q:=4*q on
```

In the general case, these transformations are as follows. First the instruction

```
off ide:=dae ; ide-r:=dae-r on
```

is replaced by an equivalent one

```
off ide-r:=((ide:=dae) @ dae-r) ; ide:=dae  on
```

Further on, the expression `((ide:=dae) @ dae-r)` is transformed to a standard form, and then we try to change it is such a way that the identifier `ide` can be eliminated due to the register-condition `ide-r = dae-r`.  This completes the transformation.

The second "atomic" case that has to be investigated is a procedure call:

**call** ide(**ref** apa-r **val** apa-v)

Let us assume that we intend to introduce the condition `ide-r=dae-r` and that procedure call appears in the program in the same context as the assignment in the former case. We again have two subcases to be considered.

If none of the actual referential parameters appears in `dae-r,` then we keep the instruction unchanged.

In the opposite case, we replace it with the instruction

**off call** ide (**ref** apa-r   **val** apa-v); ide-r:=dae-r **on.**

This completes the first step of structural instruction. The remaining steps are rather obvious:

```
[ide-1 ; ide-2] $ [ide-r=dae-r] =
   ide-1 $[ide-r=dae-r] ; ide-2 $[ide-r=dae-r]
```

```
[if dae-b then ins-1 else ins-2 fi] $[ide-r=dae-r]  =
if dae-b then ins-1 $[ide-r=dae-r]else ins-2 $[ide-r=dae-r]
   fi
```

```
[while dae-b do ins od] $[ide-r=dae-r] =
while dae-b do ins $[ide-r=dae-r] od
```

In short, after each assignment or a procedure call that changes the value of register-condition, we add a recovering assignment. The extension of $ on specinstruction is rather evident.

At the end let us observe a methodological difference between two syntactic operations @ and $. Their respective types are:

@ : Instruction x DatExp ⟼ DatExp

$ : Instruction x DatExp ⟼ Instruction

The first operation builds expressions such as e.g.

(z:=2*z) @ z$^2$

that are elements of the syntax of **Lingua-2V** and are derivable from a grammatical clause

Instruction @ DatExp.

The syntactic symbol @ is a counterpart of a denotational @ in the same sense as the semicolon ';' is a counterpart of the composition of functions '•'.

The situation with $ is different since it does not have its counterpart on the level of syntax, hence a script like

```
(z:=2*z) $ q=z²
```

does not belong to our language. It has to be transformed into an expression using the meta-operation $ as defined above.

### 8.6.3    Changing data-types

Another application of register-conditions is the replacement of one data-type used in a program, by another one. In this section we show how to transform program Q10 from Sec. 8.6.2 into a program that operates on binary representations of numbers. Let Binary be the set of binary representations of integers, i.e. a set of zero-one tuples (Sec. 2.1.4).

bin : Binary = {(1)} © {(0), (1)}$^{c*}$

On this set we define a few functions and relations:

sl : Binary ⟼ Binary                                                        (shift left)
sl.bin =
   bin = (0)    ➜ 0
   **true**       ➜ bin © (0)

sr : Binary ⟼ Binary                                                        (shift right)
sr.bin =
   bin = (0)    ➜ 0
   **true**       ➜ pop.bin

+ : Binary ⟼ Binary                                                        (addition)
− : Binary ⟼ Binary                                                        (subtraction)
< : Binary ⟼ {tt, ff}                                                       (earlier)
≤ : Binary ⟼ {tt, ff}                                                       (earlier or equal)

The addition and the subtraction of tuples are denoted by the same symbols as for numbers and we assume that they are defined in such a way that the equations (5) and (6) below are satisfied. The orderings are lexicographic and again correspond to their numeric counterparts.

b2n : Binary     ⟼ Number                      (*binary to number*; conversion function)
n2b : Number   ⟼ Binary                      (*number to binary*; conversion function)

All these function and relations are satisfied in such a way as to satisfy the following equations:

(1) b2n.(n2b.lic)     = num                    where num : Number
(2) n2b.(b2n.bin)     = bin

(3) n2b.(num*2)        = sl.(n2b.num)

(4) n2b.(num/2)        = sr.(n2b.num)        „/" denotes the integer part of division

(5) n2b.(num1 + num2) = n2b.num1 + n2b.num2

(6) n2b.(num1 − num2) = n2b.num1 − n2b.num2

(7) n2b.num1 < n2b.num2    iff        num1 < num2

(8) n2b.num1 ≤ n2b.num2    iff        num1 ≤ num2

Now we transform program **Q10** by introducing to it three new variables and three corresponding register-conditions:

```
Q = n2b(q)
Y = n2b(y)
P = n2b(p)
```

At the same time we introduce a new type `binary` into our language. The program computer in parallel on numbers and on their binary counterparts. We introduce the assertions into it and we shift all initialisations to the beginning of the program:

Q11: **def**

    **let** n, q, y, p **be** number

    **let** Q, Y, P **be** binary

  **pre** n ≥ 1

   q := 1; Q := (1);

   y := n; Y := n2b(n);

   p := 0; P := (0);

   **begin-asr** Q = n2b(q) **and** Y = n2b(y) **and** P = n2b(p)

    **while** q ≤ n **do off** q:=4*q ; Q = sl(sl(Q)) **on od**

    **while** q > 1

      **do**

       **off** q:=q/4; p:=p/2;

            Q:=sr(sr(Q)); P:=sr(P) **on**

       **if** p+q≤y

         **then off**  p:=p/2+q; y:=y-2p-q;

                  P:=sr(P)+Q; Y:=Y-sl(P)-Q **on**

         **else off** p:=p/2; P:=sr(P) **on**

       **fi**

     **od**

   **end-asr**

  **post** p = isr(n) **and** q = 1

Now we use four conditional equivalences:

```
q ≤ n        ≡ Q ≤ n2b(n) whenever Q = n2b(q)

q > 1        ≡ (1) < Q     whenever Q = n2b(q)

p+q ≤ y      ≡ P+Q ≤ Y     whenever Q = n2b(q) and Y = n2b(y)

                                and P = n2b(p)

p = isr(n)   ≡ P = n2b(isr(n)) whenever P = isr(p)
```

in order to replace Boolean numeric expressions by Boolean binary ones. Next we remove from our program all numeric variables except n with the corresponding assignments and the on-clause. Since the on-range reaches the end of the program, we can modify the postcondition in an appropriate way.

Q12: **def**

  **let** n **be** number

  **let** Q, Y, P **be** binary

**pre** n ≥ 1

  Q := (1);

  Y := n2b(n);

  P := (0);

  **while** Q ≤ N **do** Q = sl(sl(Q)) **od**

  **while** (1) < Q

    **do**

      Q:=sr(sr(Q)); P:=sr(P)

      **if** P+Q≤Y

        **then** P:=sr(P)+Q; Y:=Y-sl(P)-Q

        **else** P:=sr(P)

      **fi**

    **od**

**post** P = n2b(isr(n)) **and** Q = (1)

## 8.7    Invariants versus assertions

From a philosophical viewpoint invariants and assertions, as they have been defined in this book, are close to invariants in the sense of R. Floyd [39] and C.A.R Hoare [46]. Formally they are, however, not only quite different to each other but also belong to different linguistic categories.

Our invariants are conditions (Sec. 8.2) and the concept of an invariant concerns a relationship between a condition and an instruction. We say that a *condition* con *is an invariant of an instruction* ins if it satisfies one of two following metaconditions:

```
con @ ins ⇨ con              — weak invariant
con        ⇨ ins @ con       — strong invariant
```

Weak invariants appear in the proofs of partial correctness of programs, strong — in the proofs of total correctness.

The meaning of a *loop-invariant* for **while**, which appears in the rule 8.5.2-6:

```
prc ⇨ inv
inv ⇨ (dae or (not dae))
inv and (not dae)) ⇨ poc
def pam pre inv and dae      sin post inv
def pam pre inv while dae do sin od post TT
```
```
def pam pre prc while dae do sin od post poc
```

is slightly different. In this case for `inv` to be a loop-invariant of the loop

```
while dae do sin od,
```

`inv` must satisfy all the metaconditions above the line.

In both cases, invariants do not belong to the programming layer of a program but to its descriptive layer. As a consequence, they do not have their counterparts in the syntax and the denotations of specinstructions.

The situation with assertions is different. In the first place, they are not conditions, but specinstructions build up of conditions. A specinstruction

```
asr con rsa
```

„behaves" as a filter which does not change a state if the condition `con` is satisfied, and which generates an error (write it into a state) in the opposite case.

Whereas invariants are used in program-correctness <u>proofs</u>, assertions are used when we <u>transform</u> correct metaprograms into (optimised) correct metaprograms.

Assertions describe local properties of programs expressed by the properties of states that are intermediate in program executions. The use of assertion in program-transformations bases on the observation that if a given metaprogram is correct, then its assertions must be satisfied in every execution of that program that starts from a state which satisfies the precondition of the program. This observation allows us to decide which transformation rules may be applied to a given program[89].

Together with assertions, we have two derivative concepts that allow to decree the satisfaction of a given condition on a given range of an instruction:

```
begin-asr con; sin end-asr
```
```
off-asr sin on-asr
```

---

[89] In the examples of Sec. 8.6 assertions were applied only in transformations concerning register-identifiers. Time will show if they may have a larger scope of application.

These concepts have been defined as colloquialisms, and thus they do not belong neither to the level of concrete syntax and denotations (as assertions) nor to the meta-level (as conditions).

# 9  Lingua-3 — object-oriented programming

## 9.1    The principles of the model

The user of **Lingua-2** receives programming-tools associated with the algebras of composites, of types, and of denotations. These tools may be later enriched by user-defined types, procedures, and functions. However, this new equipment of the language is available only within the program where it has been defined.

Object-oriented programming serves the purpose of allowing one user to apply the equipment created by another one. Such a "universally available" equipment will be called an *object*. Syntactically each object is a sequence of type definitions and procedure declarations (including multiprocedures and functional procedures). In that sense, objects may be regarded as preambles without variable declarations (see Sec. 6.1.9).

Denotationally objects are state-to-state functions that modify environments by assigning types and procedures to identifiers:

obj : Object = State ⟼ State

Since objects will be stored in computer's memory we introduce the concept of *object library* where objects are assigned to identifiers:

lib  : ObjLib = Identifier ⟹ Object

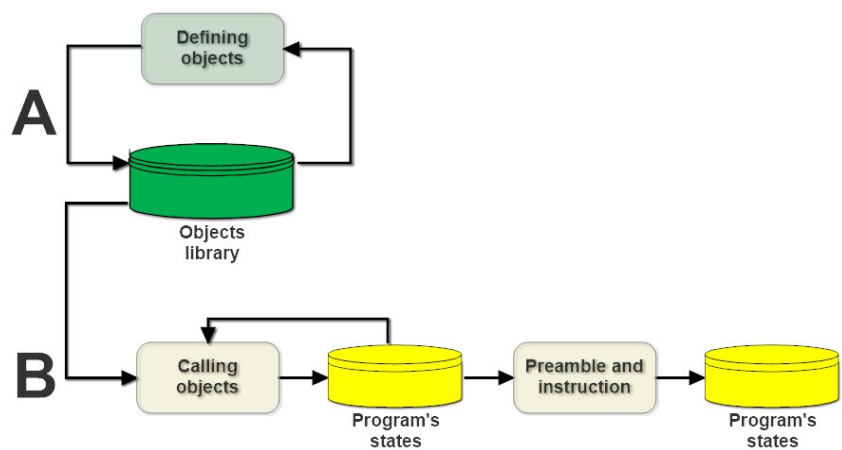

**Fig. 9.1-1 Two programming tasks in Lingua-3**

Object and libraries will be used to define programming tools permitting for the realisation of two types of programming tasks illustrated in Fig. 9.1-1:

  A.  building library, i.e. putting objects into the library and removing them

    B.  calling objects in programs.

Intuitively task A consists in defining objects and then storing them in the library under indicated names (identifiers). Objects are created by:

1. creating a type definition or a procedure declaration,

2. getting an object from the library,

3. sequential composition of objects.

Objects, once stored in the library, remain there until their possible deletion.

    Task B consists in building programs with object calls preceding preambles and instructions. Calls modify states by introducing types and procedures into their environments. Programs in **Lingua-3** consist of three successive segments: object calls, preambles, and instructions. In the preambles and programs, object calls do not appear.

    In our model, the position of objects is similar to that of data — the domain of objects is not a carrier of the algebra of denotations, and objects are values of *object expressions* which play a role analogous to that of data expressions. In turn, object libraries are similar to states hence their domain does not constitute a carrier of the algebra of denotations either.

## 9.2    Object expressions

Analogously to data expressions which generate data, *object expressions* generate objects. To enable the generation of objects also by picking them up from libraries, we assume that the denotations of object expressions are functions from libraries into objects:

oed : ObjExpDen = ObjLib ⟼ Object | Error

As we are going to see, objects will be built in three ways:

1. from the definitions of types and the declarations of procedures,

2. by getting them from the library,

3. by composing them sequentially.

The first group of constructors corresponds to expressions with constant values, i.e., with values that do not depend on libraries. The first of them creates a data-expression denotation from a type-definition denotation tdd which assumes this tdd as a value (independently of the library).

    create-obj-typ-def : TypDefDen ⟼ ObjExpDen

    create-obj-typ-def.tdd.lib = tdd

The remaining constructors of the first group are built in an analogue way from procedure declarations.

    create-oed-fpr-dec : FprDecDen ⟼ ObjExpDen       (create oed from func. proc. dec.)

    create-oed-fpr-dec.fdd.lib = fdd

    create-oed-ipr-dec : IprDecDen ⟼ ObjExpDen       (create oed from imp. proc. dec.)

create-oed-ipr-dec.idd.lib = idd

create-oed-mpr-dec : MulProDecDen ⟼ ObjExpDen      (create oed from multip. dec.)

create-oed-mpr-dec.mpd.lib = mpd

All these constructors are constant functions on libraries and are in a way similar to insertions, i.e., identity functions that allow treating elements of one carrier of an algebra as elements of another carrier (cf. Sec. 6.1.9). In the present case, the denotations of type definitions and procedure declarations "became" objects.

The next constructor creates an object-expression denotation that refers to an object in the library by indicating its name. If there is not such an object, an error message is generated.

get-obj : Identifier ⟼ ObjExpDen

get-obj.ide.lib =

> lib.ide = ?  ➔  'object-undefined'
>
> **true**        ➔  lib.ide

Notice that this construction allows to refer to earlier defined objects in object definitions which, in a sense, corresponds to the heritage mechanism.

The last constructor in this group corresponds to a sequential composition of objects:

sequence-obj : ObjExpDen x ObjExpDen ⟼ ObjExpDen

sequence-obj.(oed-1, oed-2).lib =

> oed-i.lib : Error     ➔  oed-i.lib          for i = 1,2
>
> **true**                ➔  oed-1.lib ● oed-2.lib

The composition of two oed's is either an error or the composition of objects generated by oed-1 and oed-2.

Syntactically object expressions will be sequences of type definitions, procedure declarations, and object calls. The latter allows for the construction of objects by enrichments of earlier-defined objects.

## 9.3    Object declarations

Libraries are constructed using *object declarations* that assign names (identifiers) to objects and store them in libraries. Their domain is a set of library-to-library functions that may also generate error messages:

odd : ObjDecDen = ObjLib ⟼ ObjLib | Error

The first constructor builds a denotation that assigns a name to an object and then puts the object into the library:

declare-obj : Identifier x ObjExpDen ⟼ ObjDecDen

declare-obj.(ide, oed).lib =

> lib.ide = !     ➔  'identifier-in-use'
>
> **let**

    obj = oed.lib

  obj : Error   ➔ obj

  **true**        ➔ lib[ide/obj]

The second constructor builds denotations that delete objects from libraries:

  delete-obj : Identifier ↦ ObjDecDen

  delete-obj.ide.lib =

    lib.ide = ?   ➔ 'object-undefined'

    **true**      ➔ lib[ide/?]

There is no sequential composition of object declarations, which means that each object declaration is an independent act of a programmer. This is of course an engineering decision rather than a mathematical necessity.

## 9.4    Object calls in programs

To use an object in a program, it has to be called in that program, i.e., its types and procedures have to be stored in the initial environment of the program. To introduce that mechanism we define *object calls* with denotations that are libraries-to-objects functions:

  ocd : ObjCalDen = ObjLib ↦ Object

This domain is similar to the domain of object-expression denotations, but it does not contain errors, and instead, its reachable part contains *pseudoobjects* which do not appear in ObjExpDen.

The first constructor of that domain given an identifier ide creates a denotation that calls object named ide:

  call-obj : Identifier ↦ ObjCalDen

  call-obj.ide.lib.sta =

    is-error.sta   ➔ sta

    lib.ide = ?   ➔ sta ◄ 'object-not-known'

    **true**      ➔ lib.ide.sta

If the called object is not in the library then our constructor creates a *pseudoobject* that inserts the message 'object-not-known' into each state that does not carry an error. Observe that object calls may introduce into a state not only that error message but also a message generated by lib.ide.sta and coming from a type definition or a procedure declaration.

Object calls may be composed sequentially and since atomic calls, i.e., calls generated by call-obj, are error-transparent, so are the composed calls.

  sequence-call : ObjCalDen x ObjCalDen ↦ ObjCalDen

  sequence-call.(ocd-1, ocd-2).lib = ocd-1.lib ● ocd-2.lib

## 9.5    Prefixing programs with object calls

To use, in a program, the types and procedures defined in an object, that object has to be called before the execution of the program. Analogously to the assumption that preambles precede instructors in programs (cf. Sec. 6.1.9 and Sec. 7.7) we assume now that object calls precede preambles. It is again not a mathematical necessity, but an assumption that simplifies program-construction rules. We introduce therefore the concept of a *prefixed program* with the following domain of denotations:

ppd : PreProDen = ObjLib x State → State

The basic constructor of prefixed programs creates them from object calls and non-prefixed programs:

prefix-program : ObjCalDen x ProDen ⟼ PreProDen

prefix-program.(ocd, prd).(lib, sta) =

    is-error.sta      ➔ sta

    **let**

        sta-1 = ocd.lib.sta

    is-error.sta-1   ➔ sta-1

    **true**           ➔ prd.sta-1

The execution of a prefixed program starts with the execution of its unique object call which, however, may be a sequence of several atomic calls. Then, if no error signal is raised, then the state with the objects stored in it becomes the initial state of the program. Notice that the program does not need to access the library since it will not contain object calls.

In order to include non-prefixed programs in the domain of prefixed programs we introduce the following identity insertion constructor:

no-prefix : ProDen ⟼ PreProDen

no-prefix.prd = prd

## 9.6    The extension of the algebra of syntax

The algebra of denotation of **Lingua-3** is an extension of the algebra of denotations of **Lingua-2** by the following four carriers:

oed : ObjExpDen = ObjLib ⟼ Object | Error

odd : ObjDecDen = ObjLib ⟼ ObjLib | Error

ocd : ObjCalDen = ObjLib ⟼ Object

ppd : PreProDen = ObjLib x State → State

The abstract-syntax-grammar of **Lingua-3** is therefore an extension of the abstract-syntax-grammar of **Lingua-2** by the following four clauses:

obe : ObjExp =                                        (object expressions)

    **create-oed-typ-def** ( TypDef )     |

    **create-oed-fpr-dec** ( FunProDec )  |

    **create-oed-ipr-dec** ( ImpProDec )  |

    **create-oed-mpr-dec** ( MultiProDec ) |

$\qquad$ **get-obj** ( Identifier )                            |

$\qquad$ **sequence-oed**( ObjExp , ObjExp )

ode : ObjDec =

$\qquad$ **declare-obj**  ( Identifier ,  ObjExp )|

$\qquad$ **delete-obj**  (Identifier)

pob : ObjCall =

$\qquad$ **call-obj**  ( Identifier)|

$\qquad$ **sequence-call**  (ObjCall, ObjCall)

prp : PrePro =

$\qquad$ **prefix-program**  (ObjCall; Program)|

$\qquad$ Program

From this abstract-concrete syntax, we pass to the final concrete syntax of **Lingua-3**.

obe : ObjExp =                                      (object expressions)

$\qquad$ TypDef                        |

$\qquad$ FunProDec                     |

$\qquad$ ImpProDec                     |

$\qquad$ MultiProDec                   |

$\qquad$ **get-object** ( Identifier )|

$\qquad$ ObjExp **;**  ObjExp

In this step, we omit constructor-names and parentheses associated with a semicolon. Comments about the legibility of such transformations may be found in in Sec. 6.2.2 and Sec. 7.8.2.

ode : ObjDef =

$\qquad$ **set-object** Identifier **as** ObjExp **tes-object** |

$\qquad$ **delete-obj** (Identifier)

prw: ObjectCall =

$\qquad$ **call-object**( Identifier )|

ObjectCall ; ObjectCall

prx : Program-z-Prefix =
  **begin prog**
    **call objects**
      Object-call
    **end call**
      Preamble ;
      Instruction
    **end prog**        |
  Program

## 9.7    Validating programming in Lingua-3

All constructions and investigations about validating programming in **Lingua-2** remain in force for **Lingua-3** since semantically object calls may be regarded as preambles.

# 10 External objects — a sketch of an idea

The denotational model of programming languages described in this book may also be used in situations where we intend to provide to the user of **Lingua** some resources — i.e., data structures and tools — created outside **Lingua** such as e.g. SQL databases, Excel spreadsheets, HTML scripts or control systems for physical devices. In all such situations extended **Lingua** would become a programming environment offering all standard tools of our language plus access to the tools and resources of external applications.

Formally speaking in each such situation, one has to extend the denotational model of **Lingua** by the tools of the external application, which means that one has to build a denotational model for it. It is certainly the most challenging part of the task since the existing manuals for such applications are in general unclear, incomplete and most frequently also inconsistent. However, ones such a model has been built, we can write for the user of **Lingua** a concise manual of that application**.**

The idea of accessing the tools of one application by programs in another application is, of course, not new. In all such situations, however, the "hosting" language provides access to some external software-engines, and of course, the authors of the hosting language cannot take any responsibility for the functionality of the external engine.

The situation in **Lingua** is different. Since we want to provide sound program-construction tools, we have to take responsibility not only for our programs but also for the external engines. That, in turn, requires a denotational model of the external application followed by an implementation based on that model. What then it means that in **Lingua** we provide access to an external object? How can we use that object without losing the validation features of our language? It seems that we can expect the realisation of three tasks:

A. The extension of **Lingua's** data-structures by the data structures of the external object, e.g., by SQL databases, which will be shown in Sec. 12.

B. The extension of **Lingua's** constructors by constructors "sufficiently closed" to the constructors of the external object,

C. a possible extension of **Lingua** by constructors new for the external object but applicable to the data-structures of that object.

In the case of B, by "sufficiently closed" we can mean that our constructors coincide with the external ones in "typical situations". For instance, in some implementations of SQL, the operation of arithmetic addition may accept not only numbers but also words that "resemble numbers" (cf. [38], p. 753). In such a case the implementation "guesses" that a certain word should be "treated like a number", hence, e.g. 2 + ab3 = 5. In such a situation our addition should coincide with the external one for numbers and should generate an error message in all other cases.

In Sec. 11 and Sec. 12 we show an example-application of that philosophy by expanding **Lingua** with an external SQL object in four steps:

1. We give a semi-formal description of the SQL standard without using our denotational formalism. At that stage, we identify gaps and inconsistencies in the source manuals to separate the clear and reliable from the fuzzy or differentiating from one implementation to the other.

2. We construct a denotation model for that subset of SQL that has been identified in phase one.

3. We expand **Lingua** by the new object.

4. We expand the validating rules to the new object.

In practice, the steps one and two will alternate since only in building a denotational model one may decide which informal construction can be unambiguously formalised.

# 11 Relational databases intuitively

## 11.1 Preliminary remarks

The section that follows the present one is devoted to an extension of **Lingua-3** by selected database tools of *Structured Query Language* (SQL). We shall proceed, however, in a way different than usual in typical *Application Programming Interfaces* (API) [38] or *Call Level Interfaces* (CLI)[90].

APIs have been created for programming languages C, PHP, Perl, and Phyton, and CLIs for ANSI, C, C#, VB.NET, Java, Pascal, and Fortran[91]. Each of these programming environments constitutes a programming language equipped with the mechanisms that allow running procedures of a certain existing database-engine. In the case of **Lingua-SQL,** the situation is different. Our language will base on a dedicated engine with a denotational model, and in the future, maybe, with a dedicated implementation. Such an approach is necessary if we want to provide sound program-construction rules.

This section refers to several sources since in the majority of cases one manual is not enough to determine the meaning of SQL mechanisms. The book of Lech Banachowski [7] contains a model of *Relational Databases* and a nice description of SQL standard, but some issues are missing (e.g., three-valued predicates), and some others are only sketched. On the other end of the scale of clarity is an over one thousand page long work of Paul DuBois [38]. I quote some formulations from that book just to show the scale of problems that one has to tackle in building a practical database-object for **Lingua.** Between these two extremes, but certainly closer to DuBois, are four other books [40], [45], [56], and [62].

Since all these books were published some time ago, some of the described mechanisms my look today differently. That is not much of a problem, however, since in any case all our SQL-constructions must be defined independently. Of course, I shall care to make them as close as possible to SQL standard, and — what is most important — to make them applicable to SQL databases created by existing applications.

The reader is not expected to be familiar with SQL, and therefore present section contains an informal description of selected SQL-constructions. With some of them, I associate terms that do not appear in SQL manuals, and I label them by "(my own term)".

The denotational model of **Lingua-SQL** is described in Sec. 12.

## 11.2 Simple data

Only one data-type — the type of tables (Sec. 11.3) — appears explicitly in the mentioned SQL-manuals. Several other types appear only implicitly. They include *simple data* (my own notion) that appear in the fields of database tables and *structural types* such as rows and columns of databases and the databases themselves.

---

[90] CLI refers to the standard ANSI SQL (see [62] p. 359)
[91] Access has not been mentioned on these lists since it is available only together with Microsoft Basic Access.

Simple data constitute probably one of the least standardised areas of SQL. The sorts and the types of data differ not only between different applications but also between different implementations of the same application.

In the present section, I base mainly[92] on [62], whose authors declare the compatibility with the standard ANSI SQL-2011[93]. The SQL syntax is printed in Arial Narrow.

Database-tables can carry four sorts of data which, except Booleans, split into several types:

- **Numbers** split into three subsorts: total numbers, decimal numbers, and floating-point numbers. Each of them splits again into several types differing with each other on the range of values (described by yokes), e.g. INTEGER, SMALLINT, BIGINT or DECIMAL(p, s), where p (precision) denotes the maximal number of digits and s (scale) — the maximal number of digits after decimal point.

- **Logical values** are handled as in the three-valued predicate calculus of Kleene, and in [62] they are denoted by TRUE, FALSE, and NULL whereas in [38] by 0, 1, and NULL Sometimes, e.g., in [45], instead of NULL we have UNKNOWN.

- **Strings** are in principle words in our sense, but, similarly to numbers, they are split into types according to a maximal accepted number of characters. For instance, CHARACTER(n) is the type of words of the length n. The type of a string with varying length limited to n is called in [62] CHARACTER VARYING(n), and the type of a string of an unlimited length (whatever it means) is called BLOB. There exist also binary strings, and text-strings called TEXT.

- **Times** are tuples of three types: DATE — (year, month, day), TIME — (hour, minute, second), DAYTIME — (year, month, day, hour, minute, second).

Although this is nowhere explicitly said, one may guess (cf. [62]) that all sorts of data contain NULL that essentially plays the role of an abstract error. The majority of constructors, except Boolean constructors, seem to be transparent for that error.

The constructors of simple data may be split into five following groups[94]:

1. Arithmetic operations: +, −, *, /.

2. String operations: CONCAT, UPPER, LOWER, SUBSTR, LENGTH.

3. Time operations: GETDATE, DAYNAME, DAYOFMONTH,

4. Basic predicates: =, <>, <, <=, >, >=, IS NULL, BETWEEN, LIKE.

5. Logical connectives: NOT, OR, AND.

The first group seems rather obvious. It turns out, however, that this is the case only in typical situations: 2+3=5, but if we try to add a number to a string (which is possible!), or to add two numbers whose sum exceeds the maximal allowed value, then the expected result is not clear.

---

[92] „Mainly" but not „totally" since this manual also contains gaps.
[93] ANSI is an acronym of American National Standard Institute, and SQL-2011 is a standard accepted by ANSI in December 2011.
[94] The descriptions of 1 to 4 are from [62] (pp. 129 and 180) and of 5 and 6 from [45] (pp. 191 and 201). The terminology and conceptual systematics are mine.

The source [62] does not comment on such cases at all, and in [38] p. 786 we can read the following[95]:

*If we do not provide (...) correct values to functions, we should not expect reasonable results.*

In another place of the same manual (p. 754) we read:

*(...) expressions that contain big numbers may exceed the maximal range of 64-bits computations in which case they return <u>unpredictable values</u>* (my emphasis).

It is to be emphasised as well that in the definitions of arithmetic operations NULL does not appear although it could be used as an abstract error. In this place the worth possible solution has been chosen: instead of an error message, we have an "unpredictable result" which means that the computation does not abort but simply generates a result that is contradictory to arithmetic without informing the user about that situation.

Especially many unclarities are associated with default rules for type-conversion. For instance ([38] p. 753) the following rule concerns the addition operation in the context of words as arguments:

*... + is not an operator for the concatenation of texts, as it is the case in some programming languages. Instead, before the performance of the operation textual strings are converted into numbers. Strings that <u>do not look like numbers</u> (my emphasis), are converted to 0.*

This rule has been illustrated with the following examples:

'43bc' + '21d' = 64

'abc' + 'def' = 0

It has not been explained, however, if, e.g. '43ab2c' "looks like a number", and if it does, is it converted to 43 or 432? It has not been explained either, whether these rules apply to other arithmetic operations.

Fortunately [62] treats conversion a little more seriously — although still informally — introducing four types of conversions:

1. strings to numbers,
2. numbers to strings,
3. strings to dates and times,
4. dates and times to strings.

String-operators offer fewer ambiguities but still are defined only for typical situations. For instance, I did not find information what happens if the concatenation of two strings exceeds an accepted length.

Time-operators offer another field for discrepancies between different SQL-applications concerning both, the syntax and the types of operators. However, I shall not analyse that problem further since these operators are easy to be formalised.

Predicates are typologically ambiguous since in the majority of cases they apply to all four sorts of data. E.g., the operators = and BETWEEN may be used for numbers and strings and probably also for dates. Their definitions are rather vague. E.g., in [62] p. 130 we can read:

*If in a query we use the (=) operator, the compared values must be identical, and in the opposite case, the condition is not satisfied.*

---

[95] My own translation from a Polish version of the book.

It has not been explained in "not satisfied" means "false" or "not true". E.g. is the value of the Boolean expression 12 = abc equal ff or ee?

The operator BETWEEN takes three arguments and checks if the first is between the second and the third in some default ordering.

The operator LIKE takes two string-arguments and checks, if the first coincides with the pattern described by the second. Patterns are described using letters and digits and two special symbols:

%  — an arbitrary string of characters (possibly empty)

_  — an arbitrary character

The only source where I found complete definitions of logical operators is [45], where a table-definition is given on page 191.  In our notation, this table is as in Fig. 11.2-1.

| OR | tt | ff | ee |   | AND | tt | ff | ee |   | NOT | |
|----|----|----|----|---|-----|----|----|----|---|-----|----|
| tt | tt | tt | tt |   | tt  | tt | ff | ee |   | tt  | ff |
| ff | tt | ff | ee |   | ff  | ff | ff | ff |   | ff  | tt |
| ee | tt | ee | ee |   | ee  | ee | ff | ee |   | ee  | ee |

**Fig. 11.2-1 Boolean operators in SQL**

Despite the existence of the NOT operator, special negated versions are introduced for all predicates, e.g., NOT NULL and NOT BETWEEN.

In the case of all non-Boolean operators, we have a situation typical for software-manuals. Within the area of standard ranges of arguments, everything is clear. If, however, we go beyond that, we can hardly predict what happens. With a high certainty, we can expect that in each implementation we shall encounter a different surprise.

One more remark at the end. Simple data may be assigned in SQL to table fields only but not to variables.

## 11.3   The creation of tables

An important SQL-concept is a *table*. On the ground of our denotational model, we may say that tables are tuples of records which carry simple data. In an SQL metalanguage, records that appear in tables, are called *rows*, the attributes of these records — *column-names*, and the intersections of rows and columns — *table fields*.

Tables in SQL — and precisely speaking the corresponding typed data, i.e., values as defined in Sec. 5.3.1 —  are (probably?) the only types of data that may be assigned to variables. In the sequel, variables carrying tables are called *table-variables* (my own term). To declare a data variable, we use the operator CREATE TABLE, which assigns to a variable identifier a *table type* and (we can guess) some sort of an *empty table* (my own term) whatever it means.

Table type is a record-body supplemented by some properties of attributes that may be split into two groups: yoks as defined in Sec. 5.2.4 and *default values*, which go a little beyond our model, but may be easily introduced into it (Sec. 12.2). Here is an example of two such declarations which are cited with small changes after [4] p. 14[96]:

```
CREATE TABLE Departments
    (
    Department_ID        Number(3)   PRIMARY KEY,
    Department_name      Varchar(20) NOT NULL        UNIQUE
    City                 Varchar(50)
    );
```

```
CREATE TABLE Employees
    (
    Employee_ID          Number(6)   PRIMARY KEY,
    Name                 Varchar(20) NOT NULL,
    Position             Varchar(9)  DEFAULT NULL,
    Manager              Number(6) ,
    Employment_date  Date,
    Salary               Number(8,2),
    Bonus                Number(8,2),
    Department_ID        Number(3)   REFERENCES Departments,
    CHECK (Bonus < Salary)
    )
```

The tabulation in this example shows a certain universal structure of a declaration:

- in the first column, we see column names, i.e., the attributes that are common to all the records (rows) constituting a table,

- the remaining columns carry information about data stored in table columns; in our model, they will be expressed by bodies and yoks,

- a special case is an information expressed by REFERENCES Departments which will be described in our model by a database instruction (see Sec. 12.9).,

- in the last row of the second declaration, we see a condition concerning an expected relation between the values of the fields Salary and Bonus in each row of the future table; the bonus cannot be higher than the salary; in the terminology of Sec. 5.2.4, this yok is created by the all-on-li constructor.

---

[96] In Sec. 12 I shall frequently refer to this example and also to some other examples from [4]. In both cases I keep the original notation, where Number(p) denotes a type of total numbers with p digits, and Number(p, s) denotes the type of decimal numbers of the total number of digits equal to p and the number of digits after decimal point equal to s. In turn Varchar(n) denotes the type of strings of length not exceeding n.

Declaration-columns except the first one, and also the last row, define so-called *integrity constraints*. Their meanings are the following:

1. Number(3) — The type of data in the column.

2. DEFAULT — The default value.

3. NOT NULL— All fields in the column must not be empty, i.e., none of them may be NULL. An attempt of the violation of this constraint should result in program-abortion with error-signal.

4. UNIQUE — No two identical data may appear in the column. If that happens, an error message should be generated followed by program-abortion.

5. PRIMARY KEY — This column is indicated as a *primary key*. Each primary key must be an *unambiguous key*, which means that the value of that key in a row identifies that row unambiguously. Primary keys may be defined for more than one column. The database-engine should react for each violation of the unambiguity of a primary key.

6. REFERENCES Departments — The field Department_ID in table Employees is related to the field of the same name in the table Departments. Relations between tables are used to modify tables and to set queries (see later).

7. CHECK(Bonus<Salary) — Whenever a new row is added to a table, or an existing row is modified, the engine aborts the program and generates an error message if this condition is not satisfied.

As we see from this example, when we declare a table-variable we simultaneously define its type, i.e., its body and yok[97]. This type covers five groups of properties of the future table:

1. the names of columns,

2. the types of values in all fields of a given column, e.g., Number(6),

3. restrictions concerning columns as a whole, e.g., PRIMARY KEY, NOT NULL or UNIQUE,

4. relationships between values in each row, e.g., CHECK(Bonus<Salary),

5. relationships between tables by indicating related columns in tables, e.g., REFERENCES Departments.

As was already said, the properties of columns corresponding to 2.— 5. are called integrity constraints. Another example of integrity constraints may be that e.g. some operations on a balance-sheet must not change the balance-sheet-total (an example in Sec. 11.5).

In the end, some comments about the concept of an empty table introduced at the beginning of this section. In database-literature, such a concept does not exist. I did not find either any information about the sort of data assigned to a table by its declaration.

## 11.4   The subordination relation for tables

Intuitively *relations* in this context are links between tables that allow performing operations on several linked-together tables. Using **Lingua-A** terminology, one can say that relations are yoks concerning databases.

---

[97] It seems that SQL lacks mechanisms that would allow to define a table-type independently of variable declaration.

The mechanism of establishing relations between tables appears in SQL applications in several versions. All of them are based on a common idea, although the implementations may differ from each other. In the present section, I shall try to describe this idea by introducing a few concepts that do not appear in the literature.

Consider the tables Departments and Employees from Sec. 11.3. In Employees, we have a column Department_Id which defines the association of an employee to a department. In the declaration of this column we have the constraint REFERENCES Departments expressing the fact that in the table Departments we may find information about the department where the employee is employed. Instead of storing in the table Employees the information about the department where the employee works, we only show the ID of that department that identifies the appropriate row in the table Departments. However, for this construction to have a practical sense, our two tables must satisfy three conditions:

1. the column Department_ID must appear in both tables,

2. every ID of a department which is in the table Employees must also appear in the table Departments,

3. in the table Departments the attribute Department_ID must be an unambiguous *key*.

If these conditions are satisfied, then we say that:

> *the attribute* Department_ID *links the tables* Departments *and* Employees
> *with a one-to-many relation (abbr. 1-M)*

To every department, there is associated a set (possibly empty) of employees, whereas to every employee there is associated exactly one department.

In this pair Departments is a *parent* table or a *superior table* and Employees — a *child* table or a *subordinated* table. The attribute Department_ID is a *primary key* in the table Departments and a *foreign key* in the table Employees.

If an employee's row ER and a department's row DR have the same value in the field Department_ID, then we say that the ER *points to the* DR (my own term).

By (1-M) I shall denote a ternary relation which is a subset of the Cartesian product of three domains:

(1-M) $\subset$ Table x Attribute x Table

such that (tab-1, atr, tam-2) : (1-M) iff tab-1 is a parent of tab-2 with a primary key atr.

A triple (Departments, Department_ID, Employees) is, therefore, an element of such a relation. In that case, the attribute Department_ID is called a *linking key* of our tables.

Observe now that this relation may be broken by the modification of one or both tables, e.g. whenever:

- we remove a row from Departments that is pointed by a row from Employees,

- we insert a row to Employees with department's ID that does not exist in Departments,

- in one of our tables we rename the attribute Department_ID,

- we insert to Departments a new row with an ID equal to the ID of another row, and in that way, we spoil the unambiguity of the key Department_ID.

The fact that two tables are in the relation (1-M) may be used when we generate reports or create new tables. However, checking each time, if two given tables are in the (1-M) relation, would not be very practical. It is much better to declare in advance that such a relation should hold, and then make sure that the database-engine does not allow to violate that declaration.

In our example the declaration of the (1-M) relationship between Departments and Employees is implicit in the declarations of the corresponding table-variables:

- in the declaration of Departments, the attribute Department_ID is declared as a PRIMARY KEY; notice that every primary key has to be unambiguous,

- in the declaration of Employees, the attribute Department_ID in constrained by REFER-ENCES Departments.

The establishment of a relation (1-M) between tables has consequences for operations on these tables. For instance:

- Introducing an employee employed in a non-existent department is impossible. The database-engine will force the programmer to introduce the new department in the first place.

- A department's record cannot be removed from a table until there are employees employed in that department. An alternative solution is that in such a case all employees of the deleted department are "automatically" removed.

- One can request the generation of a table with three columns that combine information from both linked tables, e.g., with columns Name, Department_name, City.

A particular case of a (1-M) relation is a (1-1) relation, where for every record in a parent table there is at most one record in the corresponding child record. Notice that "at most one" rather than "exactly one", which means that (1-1) relation does not need to be symmetric. Consequently one of these tables is a parent and another — a child.

To formalise the investigation on parent-child relations I introduce the concept of a *parent-child graph*, which is an arbitrary finite (possibly empty) set of triples of identifiers:

pcg : ParChiGra = FinSub(Identifier x Identifier x Identifier)

The elements of this set are called *parent-child edges*. Intuitively every edge (ide-c, ide, ide-p) corresponds in a database to a relation, which holds between the tables named ide-c (child), ide-p (parent) with the primary key ide.

## 11.5   The instructions of table modification

Tables that have been declared or made accessible (see Sec. 12.7.6.11) may be modified using a large class of instructions. Below a few examples:

**Entering a new column to a table**:

    ALTER TABLE Employees
        ADD COLUMN ID_number CHAR(11) DEFAULT NULL

If we add a column to a table, and we indicate a default value for that column.

**Deleting a column from a table**

    ALTER TABLE Departments
        DROP COLUMN Department_ID CASCADE (or RESTRICT)

If this instruction is executed with the option CASCADE, then the deletion of a column results in the deletion of all objects of a database (tables, perspectives,…) that refer to that column. In the case of RESTRICT, the instruction is not executed whenever such objects exist in the database.

Notice that the instructions from the group ALTER TABLE modify not only the content of a table but also its type. There are other examples of instructions altering tables ([45] p. 49):

- ALTER COLUMN — column-type is modified by SET DEFAULT or DROP DEFAULT which sets or drops a default value.

- ADD — new constraint is added to an existing column.

- DROP CONSTRAINT — the removal of a constraint from an indicated column. With this instruction, RESTRICT or CASCADE must be declared.

Another group of table-modifying instructions changes the content of a table without modifying its type. Here are some typical examples:

**The insertion of a new record** (row):

    INSERT INTO Departments

        VALUES (095, 'Marketing', 'London')

This instruction may also be written in a form where column-names are explicit (cf. [38], p. 73)

    INSERT INTO Departments (Department_ID, Dep_name, City)

        VALUES (095, 'Marketing', 'London')

**The modification of all data in a column**. E.g., the increase of salaries of all salesmen by 10%:

    UPDATE Employees

        SET Salary = Salary * 1,1

        WHERE Position = 'salesman'

**The removal of all rows that satisfy a given condition.** E.g., the removal of all employees which have no position:

    DELETE FROM Employees

        WHERE Position IS NULL

A particular situation takes place if we drop a row with a primary key which is a foreign key in a child-table, e.g.:

    DELETE FROM Departments

        WHERE Dep_name = 'production'

If in the child-table Employees the key Department_ID is — as in our case — a foreign key and there exist rows which point to the rows that are supposed to be deleted from Departments, then the operation is not executed and an error message is generated. However the operation:

    DELETE FROM Departments

        WHERE Dep_name = 'production' CASCADE

will be executed and additionally, in the table Employees, all rows that point to the row which is deleted from Departments are deleted[98].

## 11.6   Transactions

By a *transaction* we mean a sequence of instructions closed (or not) in some parentheses such as, e.g. BEGIN TRANSACTION and COMMIT TRANSACTION[99]. The mechanism of transactions that I shall call a *recovery mechanism* (my own term) stops the execution of a transaction whenever:

- the execution would violate integrity constraints, or

- the execution is impossible, e.g., we search for a non-existing element in a table.

In all such cases, the implementation returns to the initial database-state of the transaction, a state called the *roll-back value of the database*[100].

Five following instructions are used to control the recovery mechanism of transactions in SQL-programs:

| | |
|---|---|
| SAVEPOINT | — save the rollback-value of the database |
| RELEASE SAVEPOINT | — delete the rollback-value |
| ROLLBACK | — call-of the transaction |
| IF | — a conditional activation of a rollback |
| COMMIT TRANSACTION | — accept transaction. |

The instruction

SAVEPOINT *savepoint-name*

assigns the actual database value to the temporary variable *savepoint-name*. The instruction

RELEASE SAVEPOINT *savepoint-name*

deletes the variable *savepoint-name* (and its value) from the state. The instruction

ROLLBACK *savepoint-name*

brings the database to its rollback-value and deletes the variable *savepoint-name*. This instruction may also appear without a parameter, in which case the database is (probably?) rolled back to its value initial for transaction-execution[101]. That implies in turn that the execution of a transaction starts with a default SAVEPOINT which saves database value to some system-variable. It also seems that ROLLBACK aborts program execution and generates an error message.

To make the execution of ROLLBACK dependent on an error message we use the conditional IF constructor. Ben Forta ([40] p. 179) shows the following example:

IF @@ERROR <> 0 ROLLBACK *savepoint-name*

---

[98] There is a certain inconsistency in SQL that in this case there is no explicit option RESTRICT, as in the case of columns, but RESTRICT is a default option.

[99] These parentheses may differ between applications (some manuals do not mention them at all). Here we use the notation of Bena Forty ([40], p. 175) which is a standard for Microsoft SQL Server.

[100] I have to warn the reader that in all known to me manuals, transactions are described in an exceptionally unclear and incomplete way, and therefore my understanding of this construction is based more on guesses than on facts.

[101] The parameterless version of this instruction appears in the majority of manuals known to me.

It is explained there that @@ERROR is a system-variable whose value equals 0 it there is no error message, and (I guess) equals an error message in the opposite case.

This example suggests — although this has not been explicitly written — that the condition of IF might be of the form

@@ERROR = error-message

with a concrete error message. Such a solution would allow making the execution of ROLLBACK dependent on the type of an error.

The execution of COMMIT results in saving the result of the transaction and deleting all earlier declared rollback-variables.

For instance, in a database carrying data about clients of a bank, the transaction that moves 1000 $ from one account to another may have the following form:

```
 BEGIN TRANSACTION
     SAVEPOINT start
     UPDATE Accounts
         SET Balance = Balance – 1000
         WHERE ClientID = 1250 ;
         IF @@ERROR <> 0 ROLLBACK start ;
     UPDATE Accounts
         SET Balance = Balance+ 1000
         WHERE ClientID = 1260 ;
         IF @@ERROR <> 0 ROLLBACK start
     COMMIT TRANSACTION
```

The first ROLLBACK takes place if there is no client in the database with ID equal to 1250, or if its balance-value is less than 1000. The second ROLLBACK is activated if the first is not, but there is no client in the database with ID equal 1260.

Notice that after the execution of the first UPDATE, the actual sum of all deposits is not equal to the bank-balance of deposits which means that the integrity-constraints are violated. The second UPDATE "removes" this violation, but if it can't be performed because of the lack of 1260-customer, then the transaction would end with an inconsistent database. The second ROLL-BACK prevents from such a situation.

## 11.7   Queries

*Queries* are used to collect information from databases, and more precisely — from one or more database tables. The execution of a query results in the generation of a table and possibly in displaying it on a monitor. Queries are constructed by several variants of the operator SELECT. Below a few typical examples:

**The selection of indicated columns of a table:**

```
    SELECT Name, Salary, Position
        FROM Employees
```

As a result of this query, a monitor displays a three-column table with columns indicated by the parameters of SELECT.

**The selection of columns combined with the filtering of rows:**

> SELECT Name, Salary, Position
>
> > FROM Employees
> >
> > WHERE Department_ID = 10

In the WHERE clause, we may have Boolean expressions with operators on simple data described in Sec. 11.2 and Sec. 12.3.

Queries may be composed of other queries using operators called by Banachowski [7] "algebraic operators on queries". These operators may be applied to more than one table. For instance:

> SELECT Department_ID
>
> > FROM Departments
>
> EXCEPT
>
> SELECT Department_ID
>
> > FROM Employees

This query generates a one-column table of the IDs of these departments that appear in the table Departments but that do not appear in the table Employees. i.e., the IDs of departments with no employees.

Among the constructors of the same group we also have UNION, UNION ALL (union with repetitions) and  INTERSECT.

A specific group constitute queries that reach more than one table. In such a case we say that queries use the *joins of tables*. Below there is an example of such a query that selects data from two tables — Employees and Departments.

> SELECT Employee_ID, Name, Department_ID
>
> > FROM Employees, Departments
> >
> > WHERE Employees.Department_ID = Departments.Department_ID
> >
> > > AND Departments.City = 'London'

This query generates a three-column table where each row contains the ID of an employee, his/her name, and the name of the department where he/she is employed. The condition in the WHERE-clause is called *joint predicate*. In our case, it returns only such rows where employees are employed in departments located in London.

In the WHERE-clause we may contain Boolean expressions exploring basic predicates on simple data (Sec. 11.2), e.g.:

> SELECT Employee_ID, Name, Salary
>
> > FROM Employees
> >
> > WHERE Salary > 1000  AND  Salary <= 2000

or set-theoretic operators described in Sec. 12.3. For instance the query:

> SELECT Employee_ID, Name, Position, Salary

> FROM Employees
>
> WHERE Position IN ('cashier', 'salesman', 'manager').

generates a table with cashiers, salesmen, and managers. The query:

> SELECT Employee_ID, Name, Position, Salary
>
> FROM Employees
>
> WHERE Salary > ALL (
>
> SELECT Salary
>
> FROM Employees
>
> WHERE Position = 'cashier' )

generates a table that shows employees with salaries higher than the salaries of all cashiers. In this case, we have to do with a *nested query,* where the inner SELECT generates a column with the salaries of all cashiers. Let us denote:

sae  : SalEmp     — the set of values in the column Salary of the table Employees,

sac  : SalCas     — the subset of SalEmp that contains the salaries of cashiers,

shc  : SalHigCas  — the subset of SalEmp that contains salaries higher than the salaries
                    of cashiers

In that case:

> SalHigCas = { sae | sae : SalEmp and (∀ sac : SalCas) sae > sac }

therefore and according to the notation introduces in Sec. 12.3:

> SalHigCas = { sae | sae : SalEmp and <u>all</u>.(SalCas, >).sae = tt }

where > is a predicate which compares numbers and assumes the value ee whenever at least one of its arguments is not a number.

The transparency of > implies that the set SalHigCas contains numbers only, although may be empty as well. In particular, it is empty, if SalCas contains at least one not-number.

In none of the sources quoted earlier I could find information, what happens if the expression sae > sac generates an error. Will it result in program interruption and the generation of an error, or the query will generate some "unexpected" table, maybe empty?[102].

Let us consider now a query that results from the former if ALL is replaced by EXISTS, i.e., that generates the table of employees with salaries higher than the salary of at least one cashier:

> SELECT Employee_ID, Name, Position, Salary
>
> FROM Employees
>
> WHERE Salary > EXISTS (
>
> SELECT Salary
>
> FROM Employees
>
> WHERE Position = 'cashier' )

---

[102] In this case I use a notation (syntax) which is — maybe — not compatible with SQL. I have used it, however, to keep the similarity with the ALL example, whose syntax (although not the example itself) has been taken from [62] p. 139.

Denote:

  shs : SalHigSomCas — salaries higher than some salaries of cashiers.

In that case:

  SalHigSomCas = { sea | sea : SalEmp and (∃ sac : SalCas) sea > sac }

hence, according to the notation of Sec. 12.3:

  SalHigSomCas = { sea | sea : SalEmp and <u>exists</u>.(SalCas, >).sac = tt }

In that case, contrary to the former, if SalCas contains not-numbers, then the set SalHigSom-Cas does not need to be empty.

  Notice now that whenever the evaluation of sae > sac for some sac, generates an error, then

  <u>exists</u>.(SalCas, >).sac = ff

If, however, we replace EXISTS by SOME, then ee may appear in that case. This replacement does not change the table generated by our query but affects error-generation.

  Quantifiers may also appear in the context of joining tables. The query shown below generates the table of departments where at least one employee is employed.

  SELECT Department_ID

    FROM Departments

    WHERE Department_ID = EXISTS (

      SELECT Department_ID

      FROM Employees)

It was mentioned in Sec. 11.2, for every simple operator, there exists its negated version, e.g., = and <>, LIKE and NOT LIKE, etc. Similarly, we have NOT IN. In the case of set-theoretic quantifiers I have found only NOT EXISTS and only in [62] p. 147 and in [38] p. 242. Of course, none of these sources concerns the case where EXISTS generates an error.

  From the denotational perspective, queries may be regarded as expressions since they generate a value (a table) without changing a state.

## 11.8   Aggregating function

The *aggregating functions* SUM, MAX, MIN, AVG take as arguments one-column tables that are the results of queries and return a number. If the argument-table is empty then the value of an aggregating function is NULL ([45] p. 148).

  The function COUNT takes an arbitrary one-column table and returns the number of these rows where NULL does not appear. Its particular version COUNT(*) takes an arbitrary table and counts all rows including the duplicates ([62] p. 155).

## 11.9   Views

If we want to use a query more than once, we may declare it as a procedure. Such procedures are called *views*. Below we see an example of a view-declaration:

  CREATE VIEW Officials

(Employee_ID, Name, Salary)

AS SELECT Employee_ID, Name, Salary

FROM Employees

WHERE Position = 'official'

This view is named Officials and creates a three-column table by selecting columns from Employees and rows with 'official' that stands in the column Position.

Since views are procedures, they have no counterparts in syntax (cf. Sec. 7.1.3). At the syntactic level, we only have *view declarations* CREATE VIEW and *view-calls* (my own term) that refer to the names of views.

View-calls may be used in queries identically as tables and, of course, a view is executed in the call-time state rather than in the declaration-time state. In SQL-manuals views are, therefore, referred to as *virtual tables*. Views may also be called in instructions that create or modify tables. Consider the following view-declaration:

CREATE VIEW Salesmen

AS SELECT * FROM Employees

WHERE Department_ID = 20

In this declaration, the star "*" means that we chose all columns, and the number 20 is the ID of the sales department. Calling the view Salesmen we can create an instruction that modifies the table Employees by increasing the salaries of all salesmen by 10%:

UPDATE Salesmen

SET Salary = Salary * 1,1.

In the case of using vies for table-modifications, each SQL engine has its specific restrictions. E.g., MySQL requires that in SELECT-clauses only column names may appear.

A special case are *views with check option* which force the checking of a condition when views are used in instructions. Banachowski [7] shows an example of such a view:

CREATE VIEW Employees_on_not-payed_holiday

AS SELECT *

FROM Employees

WHERE Salary = 0 OR Salary IS NULL

WITH CHECK OPTION

If this view is used in the instruction:

UPDATE Employees_on_not-payed_holiday

SET Salary = 1000

WHERE Name = 'Smith'

then it is not executed if the salary of Smith is 0 or NULL.

## 11.10 Cursors

*Cursors* are used to assign selected rows of tables to data variables. This mechanism allows for processing database data using programs written in user interface programming languages such as API of CLI (see Sec. 11.1). A cursor points to a row in an indicated table and allows to get data from that row. The tables indicated by cursors are defined using queries. As a matter of fact, we should not talk about cursors as such, but about a *cursor of a table*, or maybe about a *cursor of a query*.

Cursors are created using *cursor declarations* which to a *cursor name* (an identifier) assign a cursor. Such declarations are of the form[103]:

DECLARE *cursor_name* IS

SELECT ...

After a cursor has been declared, it is not yet ready for use. To make it ready, we have to apply an *opening instruction* of the form:

OPEN *cursor_name*.

This instruction causes the execution of SELECT which appears in the declaration and (I guess) the setting of the so-called *cursor grasp* at the "position" preceding the first row of the generated table. The operation of getting data from a table is:

FETCH NEXT *cursor_name* INTO *variable*

The NEXT means getting the data of the row next to the grasp and moving the grasp one row further. It seems, therefore, that OPEN sets the cursor <u>before</u> the first row.

The FETCH NEXT instruction is usually applied in a program-loop, which means that when a grasp reaches the last row of a table, it cannot be moved further, I have found only one comment on that issue in [62] p. 353:

*In every implementation of databases, cursors are implemented in a slightly different way, but each of them enables a correct cursor-closing without an unnecessary generation of errors.*

When a cursor is temporarily not needed, we use the instruction:

CLOSE *cursor_name*

When a CLOSE instruction is executed, it leaves the cursor structure for reopening.

## 11.11 The client-server environment

So far when talking about SQL-systems, we were assuming tacitly that the user has a database to his/her excluded disposal. However, this is usually not the case. In general, there is more than one user which means that we need tools for giving them and revoking access to databases. Here is an instruction-scheme which sets a lock on a given table:

LOCK TABLE *table_name*

IN [SHARE | EXCLUSIVE]

[NOWAIT]

where the options in square-brackets mean the following:

- SHARE          — the lock covers all users,
- EXCLUSIVE     — the lock covers all users except the one who sets the lock,

---

[103] The syntax of a cursor-declaration depends upon an application. Here is the syntax of ORACLE ([62] p. 352).

- NOWAIT            —     do not wait for lock-setting, if it cannot be set at the moment.

Locks are removed by instructions COMMIT or ROLLBACK. An example of an instruction which gives permissions to a given user may be:

GRANT SELECT, UPDATE (Salary)

ON Employees

TO Smith

This instruction grants the permission of performing SELEC and UPDATE in the table Employees to the user Smith.

These mechanisms of SQL may differ between the application, but since they are relatively simple to describe, I shall not discuss them later. Therefore they are not included in my example-language **Lingua-SQL.**

# 12 Lingua-SQL

## 12.1   General assumptions about the model

As was already explained, **Lingua-SQL** does not rely on any existing SQL-engine but on its own database-operations. The denotational model of that language will be built, therefore, as an extension of the model for **Lingua-3** by adding:

1. new data domains corresponding to databases, tables rows and specific SQL simple data,

2. new operations defined on these domains and their derivatives, i.e. the corresponding operations on bodies, composites, values and expression denotations.

Of course, I do not pretend here to build a practical repertoire of SQL-tools, since my goal is just to show a denotational framework for databases, rather than to build a real API. I hope, however, that this framework will allow building such a language in the future.

## 12.2   Composites

### 12.2.1  Data, bodies and composites

So far values in **Lingua** consisted of a composite and a transfer. This principle is kept in **Lingua-SQL** for values carrying simple data, rows and tables but in the case of databases, values are records of tables supplemented by graphs of subordination relations (Sec. 12.6).

SQL data are separated from the data of **Lingua-3** in the sense that lists, records and arrays do not carry rows, tables and databases and table fields do not contain lists, records and arrays. On the other hand, the extended repertoire of simple SQL is available for the constructors of lists, records and arrays.

Simple data which are new in **Lingua-SQL** are associated with time, i.e. with calendars and clocks:

| | | |
|---|---|---|
| dat | : Date | = Year x Month x Day |
| tim | : Time | = Hour x Minute x Second |
| dti | : DateTime | = Date x Time |

where:

| | | | |
|---|---|---|---|
| yea | : Year | = {0,…,9999} | (just an example) |
| mon | : Month | = {1,…,12} | |
| day | : Day | = {1,…,31} | |
| hou | : Hour | = {0,…,23} | |
| min | : Minute | = {0,…,59} | |

sek   : Second       = {0,…,59}

Since simple data play a special role in SQL, I introduce the domain of such data:

sda : SimData = Boolean | Number | Word | Date | Time | DateTime | {Θ}

and I assume that all former constructors with simple data as arguments — e.g. that add a new attribute to a record — are extended in an obvious way to the new domain.

To include rows and tables with empty fields in our model, I introduce an *empty data* Θ[104]. This data will never appear as a value of an expression and will never be assigned to a variable.

With the extended set of simple data, we appropriately extend the set on corresponding operations, e.g. by allowing to add a number to a date. I do not define such operations explicitly assuming that their class is a parameter of our model. I assume only that they do not allow for any SQL default rules like, e.g. adding a word to a string (cf. Sec. 11.2).

The subcategories of numbers such as INTEGER, SMALLINT, BIGINT, DECIMAL(p, s), or of words CHARACTER(n), CHARACTER VARYING(n), BLOB, will correspond to yokes rather than to types.

The relation equal introduced in Sec. 5.2.1 is extended to new simple data in a natural way. As was already announced I introduce two new sorts of structural data:

row : Row      = Identifier $\Rightarrow$ SimData

tab : Table    = Row$^{c^*}$

At the level of domain equations, tables may contain rows of different length and different attributes. Of course, such tables will not be reachable in the algebra of composites. A table with an empty tuple of rows is called an *empty table*.

Data bases do not appear at the level of data. They will be defined only at the level of values (Sec.12.6)

Similarly as in **Lingua-A,** all SQL data have corresponding bodies. The bodies of new simple data are defined as one-element tuples of words, hence:

sbo : SimBody = {('Boolean'), ("number"), ('word'), ('date'), ('time'), (date-time')}

The bodies of new structural data are defined as follows:

bod : RowBody = {'Rq'} x RowRec

ror   : RowRec  = Identifier $\Rightarrow$ SimBody

bod : TabBody  = {'Tq'} x Row x RowBody

As one can guess from these definitions, the composites of rows in a table will have a common body. The row contained in a table body carries the information about default data for columns. Its list of attributes must coincide with the list of the attributes of rows' body. This property will be insured by table-body constructors.

I assume that the domain BodyE is extended by new simple bodies and the bodies of rows and tables.

The function CLAN-Bo from **Lingua-A** is extended in an obvious way on row bodies. In the case of table bodies, I assume that each row of a table must have an appropriate record structure and that in each field with a non-empty default value there is a non-empty value. Of course, it does not need to be a default value. The latter will be used when adding a new row to a table (Sec. 12.2.6) or a new column to a table (Sec. 12.2.7).

---

[104] Notice that Θ, which is assignable to fields of rows and tables, is different from Ω which is assigned to a variable at the declaration-time.

I assume that the empty table belongs to the clan of every table body. The function sort (Sec. 5.2.2 and Sec. 5.2.3) is extended in an obvious way to new bodies.

The domain CompositeE is also appropriately extended by composites associated with new simple data, row data, and table data. Additionally, I introduce an auxiliary domain of simple composites:

com : SimCom = {(dat, bod) | (dat, bod) : CompositeE **and** bod : SimBody}

I also assume that for every simple body bod

Ө : CLAN-Bo.bod

i.e. that (Ө, bod) is a composite for every simple bod.

## 12.2.2  The subordination of tables

Subordination relations describe the binary relationships that can hold between tables. Let then A and B be tables and let ide be an attribute that appears in both of them. Let A.ide and B.ide be the corresponding columns in these tables.

We say that A *is subordinate to* B *at* ide or that B *is superior to* A *at* ide — alternatively, we say that A is a *child* and B is a *parent* — that we write as

A sub[ide] B

if the following three conditions are satisfied:

1.  an ide-column appears in both tables,

2.  the column B.ide is repetition-free which means that each of its elements unambiguously identifies the row, where it appears,

3.  the column A.ide contains only the data that appear in B.ide which — together with 2. — means that each row of A unambiguously points to a row in B.

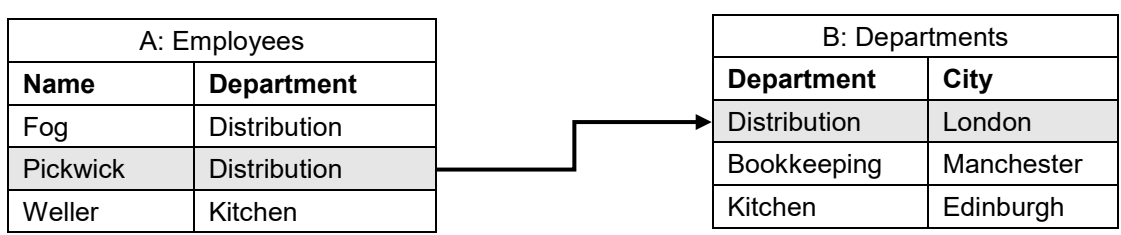

**Fig. 12.2-1 Employees is a *subordinate to* Departments *at* Department**

On Fig. 12.2-1 we see an example where the following relation holds:

Employees sub[Department] Departments

The attribute ide is called a *subordination indicator* (my own notion) for A and B. The column A.ide is said to be a *subordinated column* for B.ide. If in the column B.ide there is an element which appears in A.ide more than once (more than one employee is employed in the same department) then we say that our subordination relation is of type (1-M) (one-to-many). In the opposite case, we say that it is of (1-1) type. Notice that in both cases there may be some elements in the superior column which do not appear in subordinated column (departments with no employees). This means that a (1-1) relationship does not need to be symmetric.

Notice that the subordination relation concerns tables rather than table composites which means that to decide if that relation holds, we do not need to compare bodies.

As one may easily check, subordination relations between tables (they may be more than one) may be spoiled in four following cases:

A. if we remove a column assigned to a subordination indicator (condition 1),

B. if we add such a row to a parent table B that introduces a repetition to the indicator-column (condition 2),

C. if from a parent table B we remove a row pointed to by a row of the child table A (condition 3),

D. if to a child table A we add a row which introduces to the indicator column an element which does not appear in the indicator column of the parent table (condition 3).

The states of programs operating on databases have to carry information about the declared subordination relations. To include this mechanism in our model I use the concept of a *subordination graph* (my own notion) defined as a set of triples of identifiers:

sgr : SubGra = Sub.(Identifier x Identifier x Identifier)[105]

Each tuple (ide-c, ide, ide-p) in sgr is called an *edge of the subordination graph*, where ide-c (child) and ide-p (parent) play the role of graph nodes, and ide is a label of the edge. In the context of a given state, each edge expresses the fact that a subordination relation holds between the tables named ide-c and ide-p where ide is the subordination indicator.

About the subordination graphs, we assume only that ide-c ≠ ide-p, although such graphs may contain cycles. Notice also that there may be many edges starting in one node (one child may have many parents), and many edges may end in one node (many children may have a common parent).

## 12.2.3  The signature of new composite-constructors

SQL constructors of composites will be defined directly, i.e. with the omission of data constructors and body constructors. Their definitions will be implicit in the definitions of composite constructors. I will also make sure that they generate an error whenever — but not only in such cases (!)[106] — an error is generated by SQL-applications. The new constructors will be given names according to the rules of Sec. 5.2.3 assuming that inside the context Cc[…] we have the name of data constructor which is implicit in the definition of the composite constructor.

Set-theoretically rows are simply records but the corresponding composites are not record composites which are of the form (rec, ('R', bor)) but row composites of the form (row, ('Rq', bor)). Also, the operations on them are slightly different from record operations, although sometimes quite similar.

In creating the list of SQL constructors of composites we have to choose one of two following options:

---

[105] Notice that since the set Identifier is finite, each subordination graph is finite as well.
[106] In **Lingua-SQL** there are no situations in which other authors say that "if we do not provide correct arguments for functions, we cannot expect meaningful results". In all such situations an error message will be generated.

1. for every future language construction, we create an individual composite constructor,

2. every future language construction may be defined as a combination of some basic composite constructors.

For instance, a replacement of a data in a table one may describe as one table-to-table constructor or as a combination of the replacement of a data in a row and of a row in a table.

The first option seems to be closer to the SQL tradition, it leads, however, to long lists of constructors "one for each case" and results in a poorer understanding of language semantics. I choose, therefore, the second option which — I believe — should contribute to:

1. a simpler and shorter description of the language,

2. a shorter list of simpler program-building rules (cf. Sec. 8.4.3),

3. the restriction of interpreter's source-code for **Lingua-SQL** to basic constructors, and the definition of other constructors as procedures defined at the level of the language.

In the signature which is shown below, I omit the constructors of simple composites whose list I regard as a parameter of the model. The remaining constructors are split into groups corresponding to the sorts of data.

In the definitions of constructors that follow I refer to the concept of a transfer which in Sec. 12.4 will be extended to the new domain of composites. The extended domain of transfers I denoted again be Transfer and I assume that it contains only transparent transfers.

**The constructors of row composites**

| | | |
|---|---|---|
| Cc[create-ro] | : Identifier x CompositeE | ↦ CompositeE |
| Cc[add-to-ro] | : Identifier x CompositeE x CompositeE | ↦ CompositeE |
| Cc[cut-from-ro] | : Identifier x CompositeE | ↦ CompositeE |
| Cc[get-from-ro] | : Identifier x CompositeE | ↦ CompositeE |
| Cc[change-in-ro] | : Identifier x CompositeE x Transfer | ↦ CompositeE |

**Row constructors of table composites**

| | | |
|---|---|---|
| Cc[create-empty-table] | : CompositeE | ↦ CompositeE |
| Cc[add-ro-to-tb] | : CompositeE x CompositeE | ↦ CompositeE |
| Cc[cut-ro-from-tb] | : Transfer x CompositeE | ↦ CompositeE |
| Cc[get-ro-from-tb] | : Transfer x CompositeE | ↦ CompositeE |
| Cc[exclude-ro-from-tb] | : CompositeE x CompositeE | ↦ CompositeE |
| Cc[filter-ro-in-tb] | : Transfer x CompositeE | ↦ CompositeE |
| Cc[join-tb] | : CompositeE x CompositeE | ↦ CompositeE |
| Cc[intersect-tb] | : CompositeE x CompositeE | ↦ CompositeE |

**Column constructors of table composites**

| | | |
|---|---|---|
| Cc[add-co-to-tb] | : Identifier x CompositeE x CompositeE | ↦ CompositeE |
| Cc[cut-co-from-tb] | : Identifier x CompositeE | ↦ CompositeE |
| Cc[filter-co-from-tb] | : ActPar x CompositeE | ↦ CompositeE |
| Cc[change-co-in-tb] | : Identifier x CompositeE x Transfer | ↦ CompositeE |
| Cc[get-co-from-tb] | : Identifier x CompositeE | ↦ ColumnE |

**Table constructor creating a derivative table**

create-der-tb : CompositeE x CompositeE x Identifier x Transfer ↦ CompositeE

The last constructor is not created from a data constructor, hence its name does not contain the context Cc[…].

## 12.2.4   The constructors of simple composites

I assume that the constructors of simple composites in **Lingua-SQL** cover:

- all constructors of simple composites from **Lingua-3**,
- the zero-argument composites generating new simple composites,
- some repertoire of operations and predicates on such composites whose examples were shown in Sec. 11.2.

This set of constructors is regarded as a parameter of our model. I only assume that it contains a special constructor which to each body assigns a composite with the empty data (it corresponds to an empty field of a row or of a table).

empty : BodyE ↦ CompositeE

empty.bod =

| | |
|---|---|
| bod : Error | ➔ bod |
| **not** bod : SimpleBod | ➔ 'simple-body-expected' |
| **true** | ➔ (Ө, bod) |

Since we have assumed earlier that Ө belongs to the clan of each body, each (Ө, bod) is a correct composite.

## 12.2.5   The constructors of row composites

The SQL row constructors, although close to record constructors (Sec. 5.2.3), differ from them in two ways:

1. they allow for the construction of only such rows, whose attributes carry simple data,
2. an attribute may carry the empty data Ө.

In the second case, we have to do with an empty field which may be later filled with a data of an appropriate body.

Below the examples of the definitions of three constructors from among the five of Sec. 12.2.3

**Add an attribute to a row**

   Cc[add-to-ro] : Identifier x CompositeE x CompositeE ↦ CompositeE

   Cc[add-to-ro].(ide, com-s, com-r) =                              (s – simple, r – row)

      com-i : Error            ➔ com-i      for i = s, r

      **let**

         (dat-s, bod-s)   = com-s

         (dat-r, bod-r)   = com-r

      **not** bod-s : SimpleBod   ➔ 'simple-data-expected'

      sort.bod-r ≠ 'Rq'         ➔ 'row-expected'

      dat-r.ide = !             ➔ 'attribute-not-free'

      **let**

         ('Rq', bor) = bod-r

         new-com = (dat-r[ide/dat-s], ('Rq', bor[ide/bod-s]))

      oversized.new-com        ➔ 'overflow'

      **true**                  ➔ new-com

I recall (Sec. 5.2.2) that bor denotes the *body record* of the row body bod-r which is a record body. Adding an attribute to a row composite extends both — the row (data) and its body — which guarantees that the new composite is well-structured.

   Notice that our constructor requires a simple composite as the second argument and a row — as the third one. In this way, we restrict SQL constructors to SQL data.

**Get a data from a row**

   Cc[get-from-ro] : Identifier x CompositeE   ↦ CompositeE

   Cc[get-from-ro].(ide, com) =

      com : Error         ➔ com

      **let**

         (dat, bod)  = com

      sort.bod ≠ 'Rq'   ➔ 'row-expected'

      dat.ide = ?     ➔ 'no-such-attribute'

      dat.ide = Θ        ➔ 'empty-field'

      **let**

         ('Rq', bor) = bod

      **true**             ➔ (dat.ide, bor.ide)

**Change a data in a row conditionally**

Cc[change-in-ro] : Identifier x CompositeE x Transfer x Transfer ⟼ CompositeE

Cc[change-in-ro].(ide, com, tra, yok) =

 com : Error    ➔ com

 sort.com ≠ 'Rq'   ➔ 'row-expected'

 **let**

  (row, bod) = com

 row.ide = ?    ➔ 'no-such-attribute'

 **let**

  ('Rq', bor) = bod

 tra.com : Error   ➔ tra.com

 yok.com : Error   ➔ yok.com

 **let**

  (dat, bod)  = tra.com

  (dat-j, bod-j)  = yok.com

 bod ≠ bor.ide   ➔ 'bodies-not-compatible'

 bod-j ≠ ('Boolean') ➔ 'yoke-expected'

 **let**

  new-com =

   dat-j = tt   ➔ (row[ide/dat], bod)

   **true**     ➔ com

 oversized.new-com  ➔ 'overflow'

 **true**      ➔ new-com

The new data dat that is assigned to ide in row is created by the application of transfer tra to the row composite com. The assignment takes place under the condition that the row composite satisfies the yoke yok. Before new data is inserted into the row, it is checked if its body is compatible with the body assigned in the row to the identifier ide. SQL transfers will be described in Sec. 12.4.

## 12.2.6 Row constructors of table composites

Table constructors are used to creating table transformations, views and queries. These constructors are split into two groups: *row constructors* and *column constructors*. To define them an auxiliary concept is needed.

 We say that a *row-body* bod *is compatible with a table body* ('Tq', row, bod-r) if bod = bod-r.

 We shall also need an auxiliary function that takes two rows over the same set of attributes:

row-1 = [ide-1/dat-11,…,ide-n/dat-1n]

row-2 = [ide-1/dat-21,…,ide-n/dat-2n]

and creates a new row

   fill-in.(row-1, row-2) = [ide-1/dat-31,…,ide-n/dat-3n]

that results from the second by replacing each pair ide-i/Θ by a corresponding ide-i/dat-1i of the first row. Hence for i = 1;n:

   dat-3i =

      dat-2i ≠ Θ    ➜ dat-2i

      dat-2i = Θ    ➜ dat-1i

This function describes the rule that if we add a new row to a table and if some fields in the new row are empty then they should be filled in by default values.

In the definitions of row constructors below, we refer to the operations on tuples defined in Sec. 2.1.4.

**Create an empty table**

   Cc[create-empty-table] : CompositeE ⟼ CompositeE

   Cc[create-empty-table].com =

      com : Error        ➜ com

      sort.com ≠ 'Rq'    ➜ 'row-expected'

      **let**

         (row, bod) = com

      **true**              ➜ ( (), ('Tq', row, bod))

An empty table is created from a row composite whose row becomes the row of default values of the table and whose body indicates bodies assigned to attributes.

**Add a row to a table**

   Cc[add-ro-to-tb] : CompositeE x CompositeE ⟼ CompositeE

   Cc[add-ro-to-tb].(com-r, com-t) =

      com-i : Error          ➜ com-i      for i = w, t

      sort.com-r ≠ 'Rq'      ➜ 'row-expected'

      sort.com-t   ≠ 'Tq'    ➜ 'table-expected'

      **let**

         (row, ('Rq', bod-r)) = com-r

         (tab, bod-t)      = com-t

         ('Tq', row-d, bod-rt))  = bod-t                (rt – row-body of the table)

      bod-r ≠ bod-rt          ➜ 'bodies-not-compatible'

      **let**

    row-fi          = fill-in.(row-d, row)

    new-tab         = tab © (row-fi)

    new-com-t       = (new-tab, bod-t)

are-repetitions.new-tab        ➔ 'redundant-row'

oversized.new-com-t            ➔ 'overflow'

**true**                       ➔ new-com-t

The body of the added row must be compatible with the body of the table. Additionally, if the value of the attribute in the added row is empty then in this place we put the value of the same attribute in the row of default values (which may be empty as well). Of course, the operation of adding a row to a table does not change the body of the table.

   The table where we add a row, may be empty. In adding a row to a table, we also make sure that the new row is not redundant, i.e. equal to a row which is already in the table. The elimination of repetitions in columns — if required — will be described at the level of denotations, where yokes are available (Sec. 12.7)

**Remove a row from a table**

   Cc[cut-ro-from-tb] : Transfer x CompositeE ⟼ CompositeE

   Cc[cut-ro-from-tb].(tra, com) =

       com : Error                    ➔ com

       sort.com ≠ 'Tq'                ➔ 'table-expected'

       **let**

           (tab, ('Tq', row-d, bod)) = com

           (row-1,…,row-n) = tab

       n = 1                          ➔ 'unique-row-cannot-be-removed'

       tra.(row-i, bod) = (tt, ('Boolean'))➔ (dat[i/?], bod)      for i = 1;n

       **true**                       ➔ 'no-such-row'

This constructor removes the first row of the table which satisfies the transfer tra. If there is no such row, an error message is generated. In the definition above dat[i/?] denotes (see Sec. 2.1.3) a row tuple dat after the removal of its i-th element[107].

**Get a row from a table**

   Cc[get-ro-from-tb] : Transfer x CompositeE ⟼ CompositeE

---

[107] Users familiar with SQL are aware of the fact that the removal of a row from a table may be either blocked by integrity constraints (subordination relation) or lead to a cascade removal of rows from subordinated tables. Those mechanisms will be defined on the level of denotation-constructors where we can talk about subordination relations.

Cc[get-ro-from-tb].(tra, com) =

    com : Error                  ➔ com

    sort.com ≠ 'Tq'            ➔ 'table-expected'

    **let**

        (tab, ('Tq', row-d, bod)) = com

        (row-1,…,row-n) = tab

    tra.(row-i, bod) = (tt, ('Boolean'))➔ (row-i, bod)     for i = 1;n

    **true**                     ➔ 'no-such-row'

First, for every table row row-i, we create a row composite (row-i, bod) that consists of that row and of the row body bod of the table. Next, we select the first of such composites that satisfies the transfer tra. If there is no such composite, then an error message is generated. However, if in the course of searching for a row some of tra.(row-i, bod) turn out to be an error then the search continuous. Notice also that since ('Tq', row-d, bod) is a table body, bod must be a row body. The transfer that appears in this definition will be called a *selection transfer*.

**Exclude rows from a table**

    Cc[exclude-ro-from-tb]   : CompositeE x CompositeE ↦ CompositeE

    Cc[exclude-ro-from-tb].(com-1, com-2) =

        com-i : Error     ➔ com-i           for i = 1,2

        sort.com-i ≠ 'Tq'   ➔ 'table-expected'     for i = 1,2

        **let**

            (tab-i, bod-i) = com-i     for i = 1,2

        bod-1 ≠ bod-2     ➔ 'bodies-not-compatible'

        **let**

          new-tab    = difference.(tab-1, tab-2)            (see Sec. 2.1.4)

          new-com  = (new-tab, bod-1)

        **true**             ➔ new-com

This constructor removes all rows from the first table that belong to the second table. This may lead to an empty table.

    The next constructor generates a table consisted of all rows of a given table that satisfy a given transfer.

**Filter rows in a table**

    Cc[filter-ro-in-tb] : Transfer x CompositeE ↦ CompositeE

    Cc[filter-ro-in-tb].(tra, com) =

com : Error           ➔ com

sort.com ≠ 'Tq'     ➔ 'table-expected'

**let**

    ((row-1,…,row-n), ('Tq', bod)) = com

    in-tuple-com-row     = ((row-1, bod),…,(row-n, bod))           (in – initial)

    fi-tuple-com-row     = filter.tra.in-tuple-com-row          (fi – final)

fi-tuple-com-row = ()  ➔ 'no-row-satisfies-this-condition'

**let**

    ((row-ko-1, bod),…,(row-ko-m, bod)) = fi-tuple-com-row

**true**                 ➔ ((row-ko-1,…,row-ko-m), ('Tq', bod-r))

In the first step, we create an initial tuple of row composites that correspond to all rows of the source table. This is necessary since transfers are defined on composites, rather than on rows. In the next step, a final tuple of row composites is created by filtering the first tuple with the use of filter operation defined for tuples in Sec. 2.1.4. Tows are taken from this tuple to create the new table. Of course, this operation does not change the table's body.

### Join two tables

Cc[join-tb]  : CompositeE x CompositeE ⟼ CompositeE

Cc[join-tb].(com-1, com-2) =

    com-i : Error        ➔ com-i             for i = 1,2

    sort.com-i ≠ 'Tq'    ➔ 'table-expected'     for i = 1,2

    **let**

        (tab-i, bod-i) = com-i     for i = 1,2

    bod-1 ≠ bod-2      ➔ 'bodies-not-compatible'

    **let**

        new-tab    = join-without-repetition.(tab-1, tab-2)         (see Sec 2.1.4)

        new-com   = (new-tab, bod-1)

    oversized.new-com  ➔ 'overflow'

    **true**               ➔ new-com

Joining two tables results in adding to the first table all these rows of the second which do not lead to repetitions. The tables that are put together must have identical bodies.

### Intersect two tables

Cc[intersect-tb] : CompositeE x CompositeE ⟼ CompositeE

Cc[intersect-tb].(com-1, com-2) =

com-i : Error         ➜ com-i                    for i = 1,2

sort.com-i ≠ 'Tq'     ➜ 'table-expected'         for i = 1,2

**let**

   (tab-i, bod-i) = com-i      for i = 1,2

bod-1 ≠ bod-2        ➜ 'bodies-not-compatible'

**let**

   new-tab   = common-part.(tab-1, tab-2)                    (see Sec. 2.1.4)

   new-com  = (new-tab, bod-1)

**true**                  ➜ new-com

The resulting table contains only those rows that were common to both tables. The intersected tables must have identical bodies.

## 12.2.7    Column constructors of table composites

By a *column,* we shall mean every not-empty tuple of simple composites. We assume that columns do not contain errors, but the domain of column does. Therefore:

col : ColumnE = SimCom$^{c+}$ | Error

With columns, we associate four table constructors which are defined below. The first three are associated with columns only implicitly since none of them neither takes a column as an argument nor returns it as a value. The fourth constructor returns columns as values but is of an auxiliary character and has no syntactic counterpart. All of them are defined in three steps according to a common rule:

1. the decomposition of a table composite into a tuple of row composites,

2. a modification of every row composite by an appropriate constructor,

3. the composition of the resulting row composites into a new table composite (constructors add, cut, change) or into a column (constructor get).

**Add a column to a table**

Cc[add-co-to-tb] : Identifier x CompositeE x CompositeE ⟼ CompositeE

Cc[add-co-to-tb].(ide, com-s, com-t) =                          (s − simple, t − table)

  com-i : Error              ➜ com-i       for i = p, t

  sort.com-t ≠ 'Tq'         ➜ 'table-expected'

**let**

   (tab, ('Tq', row-d, bod-r))   = com-t

   (dat-s, bod-s)            = com-s

   (row-1,…,row-n)           = tab

   com-j                    = (row-j, bod-r)            for j = 1;n            (1)

   com-d                    = (row-d, bod-r)

new-com-j = add-to-ro.(ide, com-s, com-j)          for j = 1;n          (2)

new-com-d   = add-to-ro.(ide, com-s, com-d)

new-com-j : Error          ➔ new-com-j          for j = 1;n

new-com-d : Error          ➔ new-com-d

**let**

(new-row-j, new-bod-r)  = new-com-j  for j = 1;n

(new-row-d, new-bod-r) = new-com-d

new-tab        = (new-row-1,...,new-row-n)          (3)

new-com-t      = (new-tab, ('Tq', new-row-d, new-bod-r))          (4)

oversized.new-com-t   ➔ 'overflow'

**true**                    ➔ new-com-t

This constructor adds a new column to a table which is filled in all rows, including the row of default values, with the common (the same) data taken from the simple composite com-s. This is done in four steps:

1. After all necessary checks, we construct from the table composite com-t a family of row composites com-j with identical bodies bod-r which were taken from the table body. To that family, we add the composite com-d of the row of default values.

2. Each of these composites is extended by a new attribute in using the row constructor add-to-row (Sec. 12.2.5). This constructor also checks if com-s is a simple composite and if ide does not appear in the set of the attributes of the table. All composites constructed in this way have the common body new-bod-r.

3. The new rows new-row-j constructed in this way constitute a new table new-tab.

4. The new table composite consists of the new table and the new table body ('Tq', new-row-d, new-bor-r).

Of course, this algorithm does not need to be repeated by a future procedure implementing our constructor. It only defines the functionality of this constructor.

### Cut a column from a fable

Cc[cut-co-from-tb] : Identifier x CompositeE   ↦ CompositeE

Cc[cut-co-from-tb].(ide, com-t) =

com-t : Error              ➔ com-t

sort.com-t ≠ 'Tq'          ➔ 'table-expected'

**let**

(tab, ('Tq, row-d, bod-r) = com-t

('Rq', bor)                = bod-r

(∃ ide) dom.bor = {ide}   ➔ 'the-unique-column-cannot-be-removed'

**let**

|  |  |  |
|---|---|---|
| (row-1,…,row-n) | = tab | |
| com-j | = (row-j, bod-r) | for j = 1;n |
| com-d | = (row-d, bod-r) | |
| new-com-j | = Cc[cut-from-ro].(ide, com-j) | for j = 1;n |
| new-com-d | = Cc[cut-from-ro].(ide, com-d) | |

new-com-j : Error ➔ new-com-j         for j = 1;n

new-com-d : Error ➔ new-com-d

**let**

|  |  |  |
|---|---|---|
| (new-row-j, new-bod-r) | = new-com-j | for j = 1;n |
| (new-row-d, new-bod-r) | = new-com-d | |
| new-tab | = (new-row-1,…,new-row-n) | |
| new-com | = (new-tab, ('Tq', new-row-d, new-bod-r)) | |

**true**               ➔ new-com

This constructor is defined analogously to the former, but this time we use the constructor Cc[cut-from-ro] (Sec. 12.2.3) which also checks if **ide** is an attribute of the table. Of course, this time we do not need to check for an overflow. On the other hand, we have to check if the removed column is not the unique column of the table.

### Filter the indicated columns of a table (remove the not-indicated)

Cc[filter-co-from-tb] : ActPar x CompositeE ⟼ CompositeE

Cc[filter-co-from-tb].(apa, com) =

apa = ()        ➔ 'choose-an-attribute'

com :Error     ➔ com

sort.com ≠ 'Tq' ➔ 'table-expected'

**let**

|  |  |
|---|---|
| (ide-1,…,ide-n) | = apa |
| (tab, ('Tq', row, ('Rq', bor))) | = com |

bor.ide-i = ? ➔ 'no-attribute-ide-i'     for i = 1;n

**let**

|  |  |
|---|---|
| tab-fi | = tab trun {ide-1,…,ide-n} |
| row-fi | = row trun {ide-1,…,ide-n} |
| bor-fi | = bor trun {ide-1,…,ide-n} |

**true**          ➔ (tab-fi, ('Tq', row-fi, ('Rq', bor-fi)))

In this definition, we refer to the domain of actual parameters ActPar (Sec. 7.1.4) although now it plays a different role. Our constructor removes all columns except those whose attributes are on the list of parameters. The operator trun of the truncation of a function has been defined in Sec. 2.1.3. Notice that repetitions in the list of parameters do not affect the performance of our constructor.

**Change a column in a table conditionally**

Cc[change-co-in-tb] : Identifier x CompositeE x Transfer x Transfer ⟼ CompositeE

Cc[change-co-in-tb].(ide, com, tra) =

    com : Error       ➜ com

    sort.com ≠ 'Tq'    ➜ 'table-expected'

    **let**

        (tab, ('Tq, row, bod)  = com

        (row-1,…,row-n)     = tab

        com-j             = (row-j, bod)              for j = 1;n

        new-com-j        = Cc[change-in-ro].(ide, com-j, tra, yok)   for j = 1;n

    new-com-j : Error   ➜ new-com-j             for j = 1;n

    **let**

        (new-row-j, bod)    = new-com-j             for j = 1;n

        new-tab           = (new-row-1,…,new-row-n)

        new-com          = (new-tab, ('Tq', row, bod))

    **true**             ➜ new-com

This constructor applies Cc[change-in-ro] to each row of the table. This application does not change the table body but may generate an error message by the row constructor in the case of non-compatibility od bodies. A particular application of this constructor corresponds to the instruction:

    UPDATE Employees

        SET Salary = Salary * 1,1

        WHERE Position = 'salesman'

The last constructor of this group selects a column from a table. Although there is probably no such constructor in the SQL standard, I introduce it for later use in the definition of yokes for tables. In **Lingua-SQL** its denotational counterpart does not belong to the signature of the algebra of denotations, and therefore it is not represented at the level of syntax either.

**Get a column from a table**

Cc[get-co-from-tb] : Identifier x CompositeE  ⟼ ColumnE

Cc[get-co-from-tb].(ide, com-t) =

| | | |
|---|---|---|
| com-t : Error | ➔ com-t | |
| sort.com ≠ 'Tq' | ➔ 'table-expected' | |
| **let** | | |
| (tab, ('Tq', row-d, bod-r)) | = com-t | |
| (row-1,…,row-n) | = tab | |
| com-j | = (row-j, bod-r) | for j = 1;n |
| chosen-com-j | = Cc[get-from-ro].(ide, com-j) | for j = 1;n |
| chosen-com-j     : Error | ➔ new-com-j | for j = 1;n |
| **true** | ➔ (chosen-com-1,…, chosen-com-n) | |

This constructor creates a tuple of simple composites that correspond to the attribute ide in each of the table rows except the row of default values. Consequently, the resulting column does not contain a default composite.

## 12.2.8    A referential constructor of table composites

This constructor allows for the composition of two tables into a third. In typical applications, the source tables will be linked with a subordination relation, but in the definition of our constructor, we shall not use this fact since subordination graphs will be available only at the level of denotations[108]. We start from an auxiliary constructor that gets three arguments:

1. a row composite com-r,

2. a subordination indicator ide,

3. a table composite com-t.

and returns the row composite of the first row in table com-t that is indicated by com-r through their common value assigned the ide.

The indicated row of the table carried by com-t will be called a *superior row* for the row carried by com-r. Such a relation between rows is shown in Fig. 12.2-1. If the row com-r belongs to a table which is subordinated to com-t then the superior row always exists and is unique. In the opposite case, it may be no such row, or there may be more than one.

**A table constructor indicating a superior row**

indicate-sup-ro : CompositeE x Identifier x CompositeE ⟼ CompositeE

indicate-sup-ro.(com-r, ide, com-t) =

| | | |
|---|---|---|
| com-i : Error | ➔ com-i | for i = w, t |
| sort.com-r ≠ 'Rq' | ➔ 'row-expected' | |
| sort.com-t  ≠ 'Tq' | ➔ 'table-expected' | |

---

[108] An alternative to that solution might be a third carrier in the algebra of composites — the domain of subordination graphs. I have not chosen that solution to avoid the modification of the algebra, although, frankly speaking, I am not sur which solution is better.

**let**

   (row, ('Rq', bor-w))          = com-r

   (tab, ('Tq', row-d, ('Rq', bor-t)))  = com-t

bor-i.ide = ?        ➔ 'unknown-attribute'     for i = w, t

**let**

   (row-1,…,row-n)   = tab

row-i.ide = row.ide   ➔ (row-i, ('Rq', bor-t))

**true**               ➔ 'no-such-row'

After all necessary checks, the source-table com-t is inspected row-by-row in searching a row that under the attribute ide carries the same data as the source-row com-r. The first such row — if it exists — becomes the result of the computation. Otherwise, an error message is generated.

Now we are ready to define our target constructor which gets four arguments:

1. a table com-c that "plays the role of a child table,

2. a table com-p that "plays the role" of a parent table,

3. an indicator ide common to both tables,

4. a row-transfer tra

and creates a table of such rows in com-c that indicate the rows of com-p that satisfy tra. Of course, for such a constructor to generate a table, tra must be a yoke although this condition is, of course, not sufficient. The table created by this constructor will be called the *derivative table* of the two source tables.

**The table constructor of derivative tables**

    create-der-tb : CompositeE x CompositeE x Identifier x Transfer ⟼ CompositeE

This constructor is defined by induction on the number of rows of the child table. We start, therefore, from a table with one row only. For that case we define a separate constructor:

create-der-tb-1w.(com-c, com-p, ide, tra) =

   com-i Error               ➔ com-i    for i = p, n

   sort.com-i ≠ 'Tq'         ➔ 'table-expected'

   **let**

      ((row), ('Tq', row-d, bod-r))= com-c

      com-r               = (row, bod-r)       (composite created from a row)

      com-rs             = indicate-sup-ro.(com-r, ide, com-p)

   com-rs : Error            ➔ com-rs

   tra.com-rs = (tt, ('Boolean')    ➔ com-c

tra.com-rs : Error ➔ tra.com-rs

**true** ➔ ((), ('Tq', row-d, bod-r))

The earlier defined constructor indicate-sup-ro indicates composite com-rs that carries a row superior for the unique row com-r of the table com-c. It the superior row com-rs satisfies the transfer tra then the current table com-c becomes the result of the constructor. Otherwise:

- if the application of transfer tra leads to an error then this error becomes the result,

- otherwise, i.e., if tra generates (ff, ('Boolean')), then the result is an empty table with the body identical to the body of com-c.

The second inductive step is the following:

create-der-tb.(com-c, com-p, ide, tra) =

    com-i Error ➔ com-i   for i = p, n

    sort.com-i ≠ 'Tq' ➔ 'table-expected'

    **let**

       (tab, bod-t) = com-c

    tab = () ➔ 'empty-subordinated-table'

    **let**

       ((row-1,…,row-k), bod-t))   = com-c

       com-c-1               = ((row-1), bod-t))

       com-rs -1             = create-der-tb-1w.(com-c-1, ide, com-p)

    com-rs -1 : Error ➔ com-rs-1

    k = 1 ➔ com-rs-1

    **let**

       com-res               = ((row-2,…,row-k), bod-t)          (res – residuum)

       com-rs -res           = create-der-tb.(com-res, ide, com-p)

    com-rs -res : Error ➔ com-rs-res

    **true** ➔ Cc[join-tb].(com-rs-1, com-rs-ind)

After all necessary checks, the resulting table com-rs-1 is created for the table com-c-1 that results from com-c by reducing it to only one row.

If the table com-c has only one row, then the computation terminates. In the opposite case, we recursively apply our constructor to the residuum of the table com-c, and the resulting table is "glued" using join-tb to the table resulting from the first row.

Notice that this constructor is defined for an arbitrary pair of source tables, i.e., not necessarily linked by a subordination relation.

## 12.3   Bodies

As was announced in Sec. 12.2 the constructors of SQL composites are defined directly rather than by referring to the constructors of data and bodies. However, since some body constructors are necessary for the definitions of type constructors (Sec. 12.5), and of type constants (Sec. 12.7.4), I list therefore these constructors. It is to be stressed that they are not all constructors from composite-constructors but only those which are necessary for the definitions of type constructors. I skip their definitions since they are implicit in the definitions of composite-constructors.

**The constructors of row-bodies**

   Kr[create-ro]      : Identifier x BodyE              $\mapsto$ BodyE

   Kr[add-to-ro]      : Identifier x BodyE x BodyE  $\mapsto$ BodyE

**The constructors of table-bodies**

   Kr[create-empty-table]  : CompositeE                   $\mapsto$ BodyE

   Kr[add-ro-to-tb]           : Identifier x BodyE x BodyE  $\mapsto$ BodyE

   Kr[get-ro-from-tb]        : BodyE                          $\mapsto$ BodyE

## 12.4   Transfers

Types — as we understand them in this book — are mentioned in SQL-manuals only in the context of simple data and even in that case in a very unclear and incomplete way. The types of tables are implicit in table declarations, and the types of rows, columns and databases are totally absent. In table declarations, the descriptions of bodies are mixed with the description of yokes and even with database instructions and are called *integrity constraints* (Sec. 11.3).

Unfortunately, in none of known to me SQL manuals (their list is given in the preamble to Sec. 11), I have found a complete description of integrity constraints. Although all of them have a certain common part, besides that part, each manual offers different ideas. In this situation, I decided to construct such a model of SQL types that would cover a "sufficiently large" spectrum of types that appear in SQL applications.

Since in **Lingua-SQL** there are no database composites, there will be no database transfers either. The properties of databases will be described by:

- the yokes referring to their tables,

- subordination graphs which, however, will be seen at the level of denotations only.

We assume that in **Lingua-SQL** we have all so-far-defined transfer constructors, and in particular — Boolean constructors. New constructors will generate transfers on new simple data, that I regard as the parameters of our model, plus row- and table-transfers that are defined below.

### 12.4.1   Row transfers

In this group we have only one constructor which is analogous to the selection constructor for records:

Tc[get-from-ro] : Identifier ⟼ Transfer

Tc[get-from-ro].ide.com = Cc[get-from-ro].(ide, com)

Using this constructor, we may define a yoke constructor that creates a yoke which checks if a given attribute carries a given value:

carry-by-atr : Transfer x Identifier ⟼ Transfer

carry-by-atr.(tra, ide).com =

   Cc[equal].(tra.com, Cc[get-from-ro].(ide.com))

Notice that since com must be a row composite, tra may be either a constant-value transfer with a simple composite as a value or a transfer built using Tc[get-from-ro]. In the first case if, e.g.:

tra.com = ('board', ('word'))

then

carry-by-atr.(tra, 'department')

is the denotation of the row yoke

**row**.department = 'board'

The remaining constructors, including the Boolean ones, may be used to create row transfers in the same way as for records. An example of such a transfer, or in fact of a yoke, may be:

**row**.department = 'board' **or**

**row**.salary + **row**.bonus ≤ 7000

It expresses limits on the royalties of employees who are not board members. Such a yoke may be used in a definition of a table type as well as in a query. Notice that **row** is here a key-word rather than a variable.

## 12.4.2   Table transfers

Table transfers split into two classes. The first contains *quantified table-yokes* which describe table properties by row yokes that should be satisfied by all rows of a table. The second class contains column yokes.

In the first case, we have a situation analogous to the creation of a list yoke using all-on-li. The name of this constructor does not have the form Tc[ope] since it does not refer to any data constructor.

**Quantified table-yoke**

all-in-tb : Transfer ⟼ Transfer

all-in-tb.tra.com =

   com : Error                              ➔ com

   sort.com ≠ 'Tq'                          ➔ 'table-expected'

   **let**

      ((row-1,…,row-n), ('Tq', row-d, bod))  = com

| | | |
|---|---|---|
| com-i | = (row-i, bod) | for i = 1;n |
| tra.com-i : Error | ➔ tra.com-i | for i = 1;n |
| sort.(tra.com-i) ≠ ('Boolean') | ➔ 'yoke-expected' | |
| (∀ com-i) tra.com-i = (tt, ('Boolean')) | ➔ (tt, ('Boolean')) | |
| **true** | ➔ (ff, ('Boolean')) | |

Notice that the transit tra does not need to be satisfied by the row of default values row-d. This decision is due to the fact that some fields of row-d may be empty.

Although quantified table-yokes express properties of table rows explicitly, they express implicitly — due to quantifiers — some properties of columns. These are the properties that may be expressed by a yoke that should be satisfied by all the elements of a column. This technique does not allow, however, to express properties of columns regarded as a whole, e.g. that the column is ordered or that it does not contain repetitions. To express such properties, we need special column-dedicated constructors. Here is one example of such a constructor:

no-repetitions-tb : Identifier ⟼ Transfer

no-repetitions-tb.ide.com =

    com : Error    ➔ com

    sort.com ≠ 'Tq'    ➔ 'table-expected'

    **let**

       col = Cc[get-co-from-tb].(ide, com)        (see Sec. 12.2.7)

    col : Error    ➔ col

    **true**    ➔ no-repetitions.col

We create a tuple of composites col which represents the column of the attribute ide, and then we check if this tuple satisfies the universal predicate no-repetitions (Sec. 2.1.4). It is to be recalled that the created column does not contain the elements that corresponds to the row of default values.

Since we have Boolean constructors among the constructors of yokes (Sec. 5.2.4), we can use them to construct yokes that express properties of several columns of a table and all of its rows. Notice that contrary to the SQL standard the properties of columns and rows may be combined by arbitrary Boolean constructor rather than by conjunction only[109].

At the end, it should be emphasised that the subordination relation does not appear at the level of table yoks since the subordination of one table to another one is not a property of tables but a property of a database. Consequently, as we are going to see in Sec. 12.9, a SQL-like declaration of a table variable will correspond in our case to a colloquial declaration "unfolding" in the concrete syntax to a sequence of a table-variable declaration and a database instruction.

---

[109] To say the truth I am not sure if such a generalisation has a practical value.

# 12.5   Types

The algebra of types of **Lingua-SQL** contains four carriers:

- Identifier
- Transfer
- CompositeE
- TypeE

and results from the algebra of types of **Lingua-A** by extending it (see Sec. 2.11) by the carrier of composites and by three groups of constructors:

1. all transfer constructors described in Sec. 12.4,
2. selected constructors of row composites from those described in Sec. 12.2,
3. type constructors described below.

The presence of composites in this algebra is necessary since in order to create a table type we have to create a row composite with the default values of columns. As type constructors we are going to have (the notation analogous to that of Sec. 5.2.5):

Yc[create-ro]           : TypeE x Identifier           ↦ TypeE

Yc[add-to-ro]           : TypeE x Identifier x TypeE   ↦ TypeE

Yc[create-empty-table]  : CompositeE x Transfer        ↦ TypeE

Row types are created similarly as record types (Sec. 5.2.5) with the difference that now the added type must be simple.

**Creating of a row type with one attribute**

Yc[create-ro] : TypeE x Identifier ↦ TypeE

Yc[create-ro].(typ, ide) =

    typ : Error          ➔ typ

**let**

    (bod, tra)  = typ

    new-bod   = Bc[create-ro].(ide, bod)

    new-tra    = Tc[get-from-ro]. ide ● tra

**not** bod : SimpleBod ➔ 'simple-type-expected'

**true**                    ➔ (new-bod, new-tra)

The clans of such types contain rows where the only data has body bod and satisfies the yoke tra (see the comment to the definition of Yc[create-re] in Sec. 5.2.5).

**Adding an attribute to a row type**

Yc[add-to-ro] : TypeE x Identifier x TypeE ↦ TypeE

Yc[add-to-ro].(typ-r, ide, typ-s) =                                    (r – row, s - simple)

    typ-i : Error                 ➔ typ-i     for i = r, s

    **let**

       (bod-r, tra-r)     = typ-r

       (bod-s, tra-s)     = typ-s

    sort.bod-r ≠ 'Tq'         ➔ 'row-type-expected'

    **not** bod-s : SimpleBod ➔ 'simple-type-expected'

    bod-r.ide = !              ➔ 'attribute-not-free'

    **true**      ➔

    (add-to-ro.(bod-r, ide, bod-d), and-tr-K.(tra-r, Tc[get-from-ro].ide ● tra-s)

The rows of that type are created by extending rows of type typ-r by a new attribute whose data has a simple body bod-s and satisfies the yoke tra-s.

**Creating a table type**

Yc[create-empty-table] : CompositeE x Transfer ↦ TypeE

Yc[create-empty-table].(com, tra) =

    com : Error          ➔ com

    sort.com ≠ 'Rq'      ➔ 'row-expected'

    **true**                 ➔ (Cb[create-empty-table].com, tra)

The definition of Bc[create-empty-table] is implicit in the definition of Cc[create-empty-table] in Sec. 12.2.6  but we shall quote it explicitly for the convenience of the reader:

Cc[create-empty-table].com =

    com : Error          ➔ com

    sort.com ≠ 'Rq'  ➔ 'row-expected'

    **let**

       (row, bod) = com

    **true**                 ➔ ('Tq', row, bod)

Therefore the table type created by Yc[create-empty-table] is of the form

    (('Tq', row, bod), tra).

We do not introduce database types since database values (Sec.) are not going to be pairs consisting of a composite and a type.

## 12.6   Database values

In the former versions of **Lingua**, values have been defined as pairs consisting of a data and a type. In **Lingua-SQL** this understanding applies to simple SQL values, row values, table values but not to database values. The latter will be defined as pairs consisting of an (intuitively understood) record of table values and a subordination graph (Sec. 12.2.2). About databases we shall assume additionally the following:

- to make a database accessible in a program, its tables must be assigned to variable identifiers in valuations,

- in every state its valuation carries tables of only one database; this database is called the *active database*.

To describe this mechanism new notions are necessary.

According to our assumptions we expand the current domain of simple values and we introduce the domains of row values and table values:

RowVal = {(com, tra) | sort.com = 'Rq' **and** tra.com = (tt, ('Boolean'))}

TabVal = {(com, tra) | sort.com = 'Tq' **and** tra.com = (tt, ('Boolean'))}

By a *database record* we shall mean a mapping that maps identifiers into table values:

dbr : DatBasRec = Identifier $\Rightarrow$ TabVal

Of course, database records are not records in the sense of Sec. 5.2.1 but only in a set-theoretic sense.

We say that a database record dbr *satisfies the subordination relation* identified by a subordination graph sgr, in symbols dbr <u>satisfies</u> sgr, if for every edge  (ide-c, ide, ide-p) of the graph, the tables assigned to ide-c and ide-p are defined, i.e.

(com-c, tra-c) = dbr.ide-c

(com-p, tra-p) = dbr.ide-p

and the subordination relation holds, i.e.

com-c sub[ide] com-p

By a *database value* we mean a pair consisting of a database record and a subordination graph that describes the subordination relations satisfied by that record:

dbv : DbaVal = {(dbr, sgr) | dbr <u>satisfies</u> sgr}

We may say that in database values the role of a yoke is played by the predicate <u>satisfies</u>. Notice, however, that since a database record caries table values, the tables of the database satisfy their own yokes.

## 12.7   The algebra of denotations

As was already announced, **Lingua-SQL** will offer all the programming mechanisms of **Lingua-3** and additionally some selected tools of SQL. In the present section, I describe only new tools and even in this case many details are omitted for the sake of the clarity of the method.

## 12.7.1    States and denotational domains

Similarly as in the earlier versions of **Lingua,** states in **Lingua-SQL** bind values with variables and types with type constants. The general definitions of types and values remain as in Sec. 5.2.5 and in Sec. 5.3.1 except database values (Sec. 12.6). Consequently, the values in **Lingua-SQL**, i.e., the typed data which may be assigned to variable identifiers are all the values of **Lingua-3**, and additionally the values that carry

1. simple SQL data,

2. rows,

3. tables,

4. databases.

Of course, database values are not values in the strict sense of the word since they are not composed of a data and a type. The type of a database is implicit in the types of its tables and in the subordination graph.

In every state several data bases may be stored, i.e. assigned to identifiers, but only one base may be active at a time, i.e. the tables of only one base may be assigned to identifiers in valuations.

For states I assume the existence of four system identifiers:

sb-graph    — that binds the subordination graph of the active base in the environment,

copies      — that binds a finite sets of table names (identifiers) in the valuation,

monitor     — that binds one table in the valuations, (the table displayed on a monitor)

check       — that binds words 'yes' and 'no' in valuations.

Their role will be explained later. So far we assume only that they cannot be used as identifiers of variables, of type constants and of procedures. The identifier check is called the *security flag*. If its value is 'yes', then we say that the *flag is up*. Otherwise, we say that the flag *is down*.

The signature of the algebra of denotations of **Lingua-SQL** is an extension of the signature of **Lingua** (Sec.7.8.1.1) by new constructors. The carriers change due to new SQL-values and SQL-types.

## 12.7.2    The denotations of data expressions

In **Lingua-SQL** we allow all the denotations of data expressions of **Lingua-3** plus the denotations from outside of **Lingua-3** which are the denotations of expressions that generate simple composites, row composites and table composites. Databases will appear at the level of instructions (Sec. 12.7.6.11).

According to the assumed rules, SQL constructors of denotations will be derived from SQL constructors of composites. I recall (Sec. 5.3.2) that Cdd denotes the operator that transforms constructors of composites into constructors of expression denotations according to the scheme (5.3.2-1).

In the case of tables we assume additionally that in table expression which takes tables as arguments, tables may be represented only by identifiers, rather than by composed expressions. For instance with the constructor of table composites:

Cc[add-ro-to-tb] : CompositeE x CompositeE ↦ CompositeE

we associate the following constructor of data-expression denotations:

Cdd[Cc[add-ro-to-tb]] : DatExpDen x Identifier ⟼ DatExpDen

Cdd[Cc[add-ro-to-tb]].(ded, ide).sta =

    Cc[add-ro-to-tb].(ded.sta, dat-variable.ide.sta)

This is an engineering decision rather than a mathematical necessity. I assume it since it simplifies syntax analysis and seems to be conformant with SQL standards.

Since the definitions of all these constructors coincide with the same scheme, I do not repeat them here. I only show the signatures of these constructors which we shall need when generating the syntax of **Lingua-SQL**.

### The constructors of expression denotations that generate simple composites

These constructors are regarded as the parameters or our model. I only assume the existence of a constructor that generates empty composites. For every simple composite bod : SimpleBod I introduce the following zero-argument constructor of denotations:

Cdd[Cc[empty.bod]] : ⟼ DatExpDen

Cdd[Cc[empty.bod]].sta =

    is-error.sta   ➔ error.sta

    **true**        ➔ (∅, bod)

### The constructors of the denotations of row expressions

| | | |
|---|---|---|
| Cdd[Cc[create-ro]] | : Identifier x DatExpDen | ⟼ DatExpDen |
| Cdd[Cc[add-to-ro]] | : Identifier x DatExpDen x DatExpDen | ⟼ DatExpDen |
| Cdd[Cc[cut-from-ro]] | : Identifier x DatExpDen | ⟼ DatExpDen |
| Cdd[Cc[get-from-ro]] | : Identifier x DatExpDen | ⟼ DatExpDen |
| Cdd[Cc[change-in-ro]] | : Identifier x DatExpDen x Transfer | ⟼ DatExpDen |

### The row constructors of the denotations of table expressions

| | | |
|---|---|---|
| Cdd[Cc[create-empty-table]] | : DatExpDen x Transfer | ⟼ DatExpDen |
| Cdd[Cc[add-ro-to-tb]] | : DatExpDen  x Identifier | ⟼ DatExpDen |
| Cdd[Cc[cut-ro-from-tb]] | : Transfer x Identifier | ⟼ DatExpDen |
| Cdd[Cc[get-ro-from-tb]] | : Transfer x Identifier | ⟼ DatExpDen |
| Cdd[Cc[exclude-ro-from-tb]] | : Identifier x Identifier | ⟼ DatExpDen |
| Cdd[Cc[filter-ro-in-tb]] | : Transfer x Identifier | ⟼ DatExpDen |
| Cdd[Cc[join-tb]] | : Identifier x Identifier | ⟼ DatExpDen |
| Cdd[Cc[intersect-tb]] | : Identifier x Identifier | ⟼ DatExpDen |

**The column constructors of the denotations of table expressions**

Cdd[Cc[add-co-to-tb]]      : Identifier x DatExpDen x Identifier   $\longmapsto$ DatExpDen

Cdd[Cc[cut-co-from-tb]]    : Identifier x Identifier                    $\longmapsto$ DatExpDen

Cdd[Cc[filter-co-from-tb]]  : ActPar x Identifier                       $\longmapsto$ DatExpDen

Cdd[Cc[change-co-in-tb]]  : Identifier x Identifier x Transfer x Transfer

$\longmapsto$ CompositeE

**The constructor of the denotation of an expression that creates a derivative table**

Cdd[create-der-tb] : Identifier x Identifier x Identifier x Transfer $\longmapsto$  DatExpDen

Notice that in this clause we do not have the Cc[…] constructor since create-der-tb is a constructor of components (Sec. 12.2.8).

## 12.7.3    The denotations of type expressions and transfer expressions

In **Lingua-3** the algebra of the denotations of data-, transfer- and type expressions (Sec.5.3.5) contains four carriers:

ide   : Identifier

ded   : DatExpDen    = State $\rightarrow$ CompositeE

tra    : TraExpDen    = Transfer

ted    : TypExpDen    = State $\longmapsto$  TypeE

In **Lingua-SQL** this situation does not change.

According to the rules described in Sec.5.3.5, the denotations of transfer expressions are simply transfers. As I have already explained this is an engineering decision which means that transfers are not assigned to identifiers in a state, i.e., they are not "memorised".

To the constructors of **Lingua-3** we add, therefore:

- the constructors of SQL transfers,
- the constructors of the denotations of type expression which are derived from SQL type-constructors (Sec. 12.5).

These assumptions lead to the following list of new constructors in the algebra of denotations of **Lingua-SQL**. I recall that Cdt is a constructor which transforms type- and transfer constructors into denotation constructors.

**The constructors of denotations of transfer expressions**

Tc[get-from-ro]    : Identifier $\longmapsto$ Transfer

no-repetitions-tb  :               $\longmapsto$ Transfer

all-in-tb              : Transfer $\longmapsto$ Transfer

**The constructors of denotations of type expressions**

Cdt[Yc[create-ro]]  : TypExpDen x Identifier                  ⟼ TypExpDen

Cdt[Yc[add-to-ro]] : TypExpDen x Identifier x TypExpDen  ⟼ TypExpDen

Cdt[Yc[create-empty-table]] : DatExpDen x Transfer        ⟼ TypExpDen

The constructors of the first group have been defined in Sec. 12.4 and the next two I define according to the rules described in Sec. 5.3.4. The constructor which creates an empty table must be defined individually since the table body (hence also the table type) contains not only the bodies assigned to attributes but also their default values. Consequently, within the arguments of this constructor we must include the denotation of a data expression:

Cdt[Yc[create-empty-table]].(ded, tra).sta =

   is-error.sta   ➔ error.sta

   ded.sta = ?   ➔ ?

   **let**

      com = ded.sta

   **true**          ➔ Yc[create-empty-table].(com, tra)

## 12.7.4   The denotations of type constant definitions

New definitions of type constants in **Lingua-SQL** are the definitions that refer to new type constructors (Sec. 12.5) and to the new type of definitions related to the modifications of subordination graphs.

Since the definitions of the first group coincide with the general scheme described in Sec. 6.1.3, I shall not repeat them here.

The constructors related to the subordination graphs do not belong to that group since they do not create any type constant but only change the subordination graph assigned to the system identifier sb-graph. These constructors will appear only on the level of database instructions in Sec. 12.7.6.11.

The only constructor related to subordination graphs is therefore the following:

add-sub-type : Identifier x Identifier x Identifier ⟼ TypDefDen

add-sub-type.(ide-c, ide, ide-p).sta =

   is-error.sta                  ➔ sta

   ide-c = ide-p               ➔ sta ◀ 'reflexivity-not-allowed"

   **let**

      ((tye, pre), skł) = sta

      sgr = tye.sb-graph

   (ide-c, ide, ide-p) : sgr   ➔ 'redundant-subordination'

> **let**
>> new-sgr = sgr | {(ide-c, ide, ide-p)}
>
> **true**          ➜ ((tye[sb-graph/new-sgr], pre), skł)

This constructor extends the subordination graph and updates the type environment. At the stage of type definitions I do not introduce the constructor of removing edges from a graph since this is an operation from the level of database instructions. It will appear, therefore in Sec. 12.7.6.11.

## 12.7.5    The denotations of the declarations of data variables

Variables in **Lingua-SQL** may be bound to all values of **Lingua-3** and additionally to three groups of specific SQL-values:

1. simple SQL-values,
2. row values,
3. table values,
4. database values.

The declarations of the variables of the first two groups coincide with the general scheme of such declarations in **Lingua-3** (Sec. 6.1.2). In **Lingua-SQL** there are no declarations of database variables, and instead, we have a specific instruction of database archivation by assigning it to an indicated identifier (Sec. 12.7.6.11). The constructor of the table-variable declaration is defined in a way slightly different than in **Lingua-3**:

> declare-tab-var : Identifier x TypExpDen ⟼ VarDecDen
>
> declare-tab-var.(ide, ted).sta =
>> is-error.sta        ➜ sta
>>
>> declared.ide.sta    ➜ sta ◄ 'variable-declared'
>>
>> **let**
>>> typ = ted.sta
>>
>> typ : Error        ➜ sta ◄ typ
>>
>> **let**
>>> (bod, yok) = typ
>>
>> sort.bod ≠ 'Tq'      ➜ sta ◄ 'table-type-expected'
>>
>> **let**
>>> val               = ((), typ)
>>>
>>> (env, (vat, 'OK'))     = sta
>>
>> **true**            ➜ (env, (vat[ide/val], 'OK'))

The difference of this definition from the standard of Sec. 6.1.2 consists in the fact that in the present case a variable is bound to an empty table ((), typ), rather than to a pseudo value (Ω, typ). And, of course, we also check if typ is a table type.

## 12.7.6   Instructions

### 12.7.6.1   Categories of SQL instructions

The carrier of instruction denotations in the algebra of denotations of **Lingua-SQL** is enriched with new constructors of specific SQL instructions of three categories;

1. row assignments,

2. table assignments,

3. database instructions.

All constructors of **Lingua-3** are still available and apply to the extended carrier of instruction denotations. This rule concerns, in particular, the constructor of transfer replacement and the constructors of structural instruction, i.e., sequential composition, branching and loop. The constructors of procedure declaration and procedure call remain unchanged as well, although now they are defined on extended domains.

A particular role in SQL plays a large group of table assignments where we distinguish two categories:

1. *table-modification instruction* where on both sides of the assignment we have the name of the same table,

2. *table-creation instruction* where on the left-hand side of the instruction we may have a different table name (of the table that is being created) than on the right-hand side.

From a mathematical perspective the first category may be regarded as a particular case of the second, but denotationally they correspond to two different constructors of the algebra of denotations hence also to different constructors of the algebra of syntax. The reason for that decision will be explained later.

Independently of the described categorisation, table assignments are split into two further categories according to two ways of using subordination constraints both described in Sec. 11.5):

1. *conformist instructions* where an execution terminates with an error message whenever it would lead to a violation of subordination constraints; this category corresponds to the option RESTRICT,

2. *correcting instructions* which in the described situation introduce such changes into a database that guarantee the protection of subordination constraints; this category corresponds to the option CASCADE.

If I understood that correctly from the manuals quoted at the beginning of Sec. 11.1, the first option is (most frequently?) the default option whereas the second has to be declared explicitly and is available only for a group of chosen instructions, e.g., when a row is removed from a table.

### 12.7.6.2   Row instructions

Row instructions create and modify row values assigned to identifiers in states. We build them using the assignment constructor defined in Sec. 6.1.4 and the constructor of the denotations of data expression which return row values described in Sec. 12.7.2.

In this place a technical remark is necessary. To apply in our case the (previously defined) constructor of assignments, we have to extend the relation of coherence. This is, however, a simple task since the bodies of rows and of tables have a record structure. We shall assume therefore that since now the relation <u>coherent</u> is applicable also to rows and tables.

### 12.7.6.3   Two universal constructors of table assignment

In **Lingua-SQL** we have two assignment constructors that correspond respectively to assigning a table to a table-variable and to assigning a table to the system-identifier monitor.

In the first case, we could use the general constructor defined in Sec. 6.1.4 unless the assignment modifies an existing table in a way that violates the subordination relation. To cope with the latter case, we have to introduce a database-oriented constructor of assignments. As we are going to see, it will become a convenient tool for the definitions of many other table assignments. Since, however, there is no such constructor in the SQL standard, the issue of making it available at the level of the syntax of **Lingua-SQL** I leave open so far.

The second universal constructor of table assignments will be used in the definitions of query denotations.

To define the first constructor, we introduce two auxiliary functions called *violation-control functions*. The first of them checks if a given identifier points in the current state to a table that violates one of the declared subordination relations.

violated-sr : Identifier x State ⟼ {tt, ff} | Error

violated-sr.(ide, sta) =

   is-error.sta          ➔ error.sta

   **let**

      ((tye, pre), (vat, 'OK'))    = sta

      sgr = tye.sb-graph

   vat.ide = ?         ➔ 'no-such-table'

   sort.(vat.ide) ≠ 'Tq'   ➔ 'table-expected'

   (∃ (ide-c, ide-id, ide-p) : sgr)

      [vat.ide-i = ! **and** sort.(vat.ide-i) = 'Tq'] for i = p, n **and**

      ( (ide = ide-c **and not** vat.ide <u>sub</u>[ide-id] vat.ide-c) **or**

      (ide = ide-p **and not** vat.ide-c <u>sub</u>[ide-id] vat.ide) )

                     ➔ tt

   **true**            ➔ ff

This function returns tt if the table vat.ide does not satisfy the subordination-condition indicated by the edge (ide, ide-id, ide-p) or by (ide-c, ide-id, ide).

The second function is similar to the first one and is used to check if a given composite satisfies a given table yoke. Notice that the checking may be deactivated by setting the flag check to 'not'. In this case, we implement the mechanism of a temporary deactivation of integrity constraints described by yokes. I wish to emphasise that I have not introduced such an option for integrity constraints described by subordination relation[110].

violated-yo : Composite x Transfer x State ↦ Transfer

violated-yo.(com, tra, sta) =

    is-error.sta                ➔ error.sta

    **let**

        (env, (vat, 'OK')) = sta

    vat.check = 'not'        ➔ ff        (check is a system identifier; Sec. 12.2.1)

    tra.com = (tt, ('Boolean')   ➔ ff

    **true**                 ➔ tt

As we see, checking if a yoke has been violated is performed only if the flag is up (set into 'yes'). Notice also that this function returns tt (signalises the violation of the yoke) also if the checking result tra.com is an error. Now we are ready to define the assignment constructor for tables.

assign-tb : Identifier x DatExpDen ↦ InsDen

assign-tb.(ide, ded).sta =

    is-error.sta                  ➔ sta

    vat.ide = ?                ➔ sta ◄ 'undeclared-identifier'

    ded.sta = ?                ➔ ?

    ded.sta : Error            ➔ sta ◄ ded.sta

    **let**

        ((tye, pre), (vat, 'OK'))   = sta

        com-n           = ded.sta

        (com-d, yok)     = vat.ide

        (dat-n, bod-n)    = com-n           (n – new)

        (dat-f, bod-f)     = com-f           (f – former)

    sort.bod-n ≠ 'Tq'      ➔ sta ◄ 'table-expected'

    sort.bod-f ≠ 'Tq'      ➔ sta ◄ 'table-expected'

    **not** bod-n <u>coherent</u> bod-f  ➔ sta ◄ 'no-coherence'

    **let**

        sta-n = (env, (vat[ide/(com-n, yok)], 'OK'))

---

[110] I assume that if the violation of the subordination relation is in danger, e.g. between the deletion of one column and the insertion of another, then the programmer should introduce two instructions into the program that modify the relation accordingly.

violated-yo.(com-n, yok, sta-n)    ➔ sta ◄ 'table-yoke-violated'

violated-sr.(com-n, sta-n)    ➔ sta ◄ 'subordination-relation-violated'

**true**    ➔ stan-n

As a result of the execution of such an assignment, the identifier ide is bound to a new composite com-n under the condition that:

1. both mon-f and com-n are table composites and are mutually coherent,

2. new composite satisfies in the new state the inherited yoke yok unless the check-flag is set to 'not',

3. the new composite does not violate in the new state the current subordination relation.

Notice that the violation of a subordination relation may happen only if the assignment modifies an existing table.

The second specific assignment constructor corresponds to the situation when a table which is defined by a table expression is assigned to the system identifier monitor which physically means that it is displayed on a monitor. In that case, the new table is not restricted by any yoke which is expressed by the fact that its yoke is TT.

assign-mo : DatExpDen ⟼ InsDen

assign-mo.ded.sta =

   is-error.sta    ➔ sta

   **let**

      (env, (vat, 'OK'))    = sta

      com    = ded.sta

   com : Error    ➔ sta ◄ com

   sort.com ≠ 'Tq'    ➔ 'table-expected'

   **true**    ➔ (env, (vat[monitor/(com, TT)], 'OK')

As we see, in this case, we do not expect the identifier monitor to be declared. As a system identifier, it is always available and may be bound to an arbitrary table value. If at the time of the execution of the described assignment same value is already assigned to the monitor then it is overwritten by the new value.

### 12.7.6.4    Transactions

*Transactions*, similarly to instructions, are state transformations but contrary to the former they are total functions since they do not contain loops and procedure calls. Moreover, they do not create new tables but only modify the existing ones. Their domain is, therefore, the following:

trd : TrnDen = State ⟼ State

Transactions are regarded as a separate carrier of our algebra to avoid the use of arbitrary table instructions in the contexts of transactions.

The largest group of transactions are *table modifications* which in a traditional syntax could have the form:

```
ide := table-expression(ide)
```

where on both sides we have the same table named `ide`. The denotations of such assignments are created as combinations of a table assignment (Sec. 12.7.6.3) and some denotation of a table expression or a transfer expression. Below there is a list of such transactions that are related to data expressions described in Sec. 12.7.2. The first one corresponds to adding a row to a table:

add-ro : Identifier x DatExpDen ⟼ TrnDen

add-ro.(ide, ded-r) =

    assign-tb.( ide, Cdd[Cc[add-ro-to-tb]].(dat-variable.ide, ded-r) )

The execution of this constructor creates a transaction-denotation which to the table carried by the identifier **ide** adds a row generated by the denotation **ded-r**. Let us read that definition in details:

The table assignment constructor **assign-tb** as its first argument receives the identifier of the table that is being modified and as the second — the expression denotation generated by the constructor **Cdd[Cc[add-ro-to-tb]]** whose arguments are two expression denotations:

- the denotation of the variable **dat-variable.ide** which identifies the modified table,
- the denotation of a row expression **ded-r** which generates the row which is to be added to the table.

In an analogous way we may define constructors related to table-modifications:

cut-ro : Identifier x Transfer ⟼ TrnDen

cut-ro.(ide, tra) =

    assign-tb.(ide, Cdd[Cc[cut-ro-from-tb]].(tra, dat-variable.ide))

exclude-ro : Identifier x identyfikator ⟼ TrnDen

exclude-ro.(ide-1, ide-2) =

    assign-tb.(ide-1, Cdd[Cc[exclude-ro-from-tb]].(dat-variable.ide-1,

                                 dat-variable.ide-2))

add-co : Identifier x DatExpDen x Identifier ⟼ TrnDen

add-co.(ide-c, ded, ide-t) =                                       (c - column, t – table)

    assign-tb.(ide-t, Cdd[Cc[add-co-to-tb]].(ide-c, ded, dat-variable.ide-t))

cut-co : Identifier x Identifier ⟼ TrnDen

cut-co.(ide-c, ide-t) =

    assign-tb.(ide-t, Cdd[Cc[cut-co-from-tb]].(ide-c, dat-variable.ide-t))

filter-co : ActPar x Identifier ⟼ TrnDen

filter-co.(apa, ide) =

  assign-tb.(ide, Cdd[Cc[filter-co-from-tb].(apa, dat-variable.ide))

change-co : Identifier x Identifier x Transfer ⟼ TrnDen

change-co.(ide-c, ide-t, tra) =

  assign-tb.(ide-t, Cdd[Cc[change-co-in-tb].(ide-c, dat-variable.ide-t, tra)

The second group of transactions are *protection commands* used to protect a table against destruction.

## Create a security copy

create-security-copy: Identifier ⟼ State ⟼ State

create-security-copy.sta =

  is-error.sta    ➜ sta

  **let**

    (env, (vat, 'OK')) = sta

    base             = vat trun {ide | vat.ide : TabVal}

    sgr              = tye.sb-graph

    copy-register   = vat.copies | {ide}

  vat.ide = !    ➜ 'variable-declared'

  **true**        ➜ (env, (vat[ide/(base, sgr), copies/copy-register], 'OK'))

This function creates a database that consists of:

- all table values that appear in the current valuation,

- and of the current subordination graph,

and assigns this database to the identifier ide. This identifier is added to the register of copies assigned to the system identifier copies.

I recall that trun denotes the truncation of a function to a subset of its domain (Sec. 2.1.3).

## Remove the security copy

remove-security-copy : Identifier ⟼ TrnDen

remove-security-copy.ide.sta =

  is-error.sta         ➜ sta

  **let**

    (env, (vat, 'OK')) = sta

      copy-register    = vat.copies − {ide}

vat.ide = ?           ➔ 'unknown-identifier'

**not** vat.ide : DbaVal  ➔ 'database-expected'

**true**               ➔ (env, (vat[ide/?, copies/copy-register], 'OK'))

The copy of the base is removed from the valuation, and its name is removed from the copy register.

### Recover the security copy

    'recover-security-copy' : Identifier ⟼ TrnDen

    'recover-security-copy'.ide.sta =

        is-error.sta         ➔ sta

        **let**

            (env, (vat, 'OK')) = sta

        vat.ide = ?           ➔ 'unknown-identifier'

        **not** vat.ide : DbaVal  ➔ 'database-expected'

        **let**

            (dbr, sgr)    = vat.ide

            copy-register = vat.copies − {ide}

        **true**               ➔ (env, (vat[ide/?, copies/copy-register] ◆ dbr, 'OK'))

The database dbr carried by the identifier ide is a mapping that assigns database values to identifiers. This mapping overwrites the current valuation from which we have removed the base carried by ide. The name of the removed copy is also removed from the copy register.

### Recover the security-copy conditionally

    recover-security-copy-if : DatExpDen x Identifier ⟼ TrnDen

    recover-security-copy-if.(ded, ide) = if-error.(ded, 'recover-security-copy'.ide)

If ded generates an error, then the recovery procedure is executed. In this case, we use the constructor is-error (Sec. 6.1.6). This use is not quite formal since the second argument of is-error should be an instruction denotation whereas in our case this is a transaction denotation. However, since set-theoretically transaction denotations belong to the domain of instruction denotations, our definition makes sense.

    The following two constructors are used to set the flag assigned to the system identifier check.

**Set the security-flag down**

security-flag-down : ↦ TrnDen

security-flag-down.().sta =

    is-error.sta        ➔ sta

    **let**

        (env, (vat, 'OK')) = sta

    vat.check = 'not'   ➔ 'security-flag-is-down'

    **true**              ➔ (env, (vat[check/'not'], 'OK'))

This constructor generates an error if the flag is already down. This is, of course, not a mathematical necessity but an engineering decision which protects the programmer from committing a mistake. If he/she wants to set down a flag which is already down then maybe he/she does not quite understand the functionality of his program. The second constructor of this group sets the flag up.

**Set the security-flag up**

security-flag-up : ↦ TrnDen

security-flag-up.().sta =

    is-error.sta        ➔ sta

    **let**

        (env, (vat, 'OK')) = sta

    vat.check = 'yes'   ➔ 'security-flag-is-up'

    **true**              ➔ (env, (vat[check/'yes'], 'OK'))

Similarly as for instructions also for transactions we may apply a sequential composition:

sequence-trn : TrnDen x TrnDen ↦ TrnDen

sequence-trn.(trd-1, trd-2) = trd-1 ● trd-2

The last constructor related to transactions creates an instruction from a transaction. For its definition, we shall need a function that removes all security copies. Its definition is, of course, recursive:

remove-all-security-copies : ↦ TrnDen

    remove-all-security-copies.().sta =

        is-error.sta        ➔ sta

        **let**

      ((ten, pre), (vat, 'OK")) = sta

      sgr = ten.sb-graph

sgr = ∅               ➔ sta

{ide} = sgr          ➔ remove-security-copy.ide.sta

{ide-1,…,ide-n} = sgr ➔ remove-security-copy.ide ●

                                  remove-all-security-copies.()

The constructor which transforms a transaction into an instruction is defined as follows:

trn-into-ins : TrnDen ⟼ InsDen

trn-into-ins.trd.sta =

    is-error.sta     ➔ sta

    **true**           ➔ [remove-all-security-copies.() ● security-flag-up.()]. sta

This constructor is used to transform a block of transactions (maybe a one-element block) into an instruction. This constructor also removes all security copies and sets the security flag up.

As we see, the mechanism of transactions is used to the executions of such table modifications that allow for a temporary deactivation of integrity checks and of the mechanism of security copies.

### 12.7.6.5    Global table instructions

A *global table-instruction* is an instruction which when modifying a table modifies at the same time other tables to protect integrity constraints of a database. E.g. in SQL-standard if we remove a row from a parent table in the CASCADE mode (Sec. 11.5) then this may cause the removal of all rows from a child table which point to the removed row in the parent table. In that case "cascade" means that if a child table is a parent table for other tables, then this may result in the removals of rows from the other tables. The scheme of a definition of such a constructor is shown below:

cut-ro-cas : Identifier x Transfer ⟼ InsDen

cut-ro-cas.(ide, tra).sta =

    is-error.sta                    ➔ sta

    vat.ide = ?                 ➔ sta ◄ 'undeclared-identifier'

    **let**

        (env, (vat, 'OK')) = sta

        (com-t, yok) = vat.ide

    sort.com-t ≠ 'Tq'         ➔ 'table-expected'

    **let**

        com-n = Cc[cut-ro-from-tab].(com-t, tra)

com-n : Error                          ➔ sta ◄ com-n

**let**

   sta-n = (env, (vat[ide/(com-n, yok)], 'OK'))

violated-yo.(com-n, yok, sta-n)   ➔ sta ◄ 'table-yoke-violated'

violated-sr.(com-n, sta-n)        ➔ remove-integrity-violations.sta-n

**true**                              ➔ sta-n

This instruction removes a row from a table by using the composite constructor Cc[cut-ro-from-tab] and then, if table yoke has not been violated, but integrity constraints have, then it activates the procedure remove-integrity-violations. I do not define this procedure explicitly and regard it as a model parameter. Its definition would lead to technical considerations on searching procedures of subordination graphs which would lead out of the scope of this book.

### 12.7.6.6   Local table instructions

The instructions of this group change only the table they concern. They either create a new table or they modify an existing one using the universal table assignment (Sec. 12.7.6.3) and the denotations of table expressions (Sec. 12.7.2). In principle, we could avoid the introducing of such instructions into our model by allowing table assignments in the language. Since, however, there are no such assignments in SQL (which does not mean that we have to exclude them from **Lingua-SQL**) I give below some examples of the constructors of local table instructions.

add-ro : DatExpDen x Identifier ⟼ InsDen

add-ro.(ded, ide,) =

   assign-tb.(ide, Cdd[Cc[add-ro-to-tb]].(ded, ide))

join : Identifier x Identifier x Identifier ⟼ InsDen

join.(ide-n, ide-1, ide-2) =                                  (n – new table)

   assign-tb.(ide-n, Cdd[Cc[join-tb]].(dat-variable.ide-1, dat-variable.ide-2))

intersect : Identifier x Identifier x Identifier ⟼ InsDen

intersect.(ide-p, ide-1, ide-2) =

   assign-tb.(ide-p, Cdd[Cc[intersect-tb]].(dat-variable.ide-1, dat-variable.ide-2))

create-ref: Identifier x Identifier x Identifier x Identifier x Transfer ⟼ InsDen

create-ref.(ide-n, ide-t1, ide-t2, ide-c, tra) =                    (c – column)

   assign-tb.(ide-n, Cdd[create-der-tb].(dat-variable.ide-t1, dat-variable.ide-t2, ide-c, tra))

change-co : Identifier x CompositeE x Transfer x Transfer ⟼ CompositeE

change-co.(ide, com, tra, yok) =

      assign-tb.(ide, Cdd[Cc[change-co-in-tb]].(ide, com, tra, yok))

Notice that in the first and in the last of these constructors the identifier ide appears twice — as an argument of assignment and as an argument of the table composite. This means that each of them will be used for table modification rather than for the creation of a new table. This is, of course, an engineering decision related to the SQL standard.

### 12.7.6.7   Queries

Queries are similar to simple instructions with the difference that they always create a new table assigned to the system-identifier monitor. Consequently, we apply simplified assignments as-sign-mo that never violates any constraints since the transfer of the new value is TT.

### 12.7.6.8   Transfer-replacement instructions

The definition of the constructor of that group which has been defined in Sec. 6.1.5

    replace-tr : Identifier x TraExpDen ⟼ InsDen,

applies directly to the SQL case without any changes. Of course, we have to extend the domain of transfer-expression denotations.

### 12.7.6.9   Cursors

Cursors (Sec. 11.10) are mechanisms used to get row-by-row from tables. In our model that can be easily defined, e.g. by adding a column to a table that enumerates its rows.

### 12.7.6.10   Views

Views are essentially procedures that call table instructions. They may be introduced to our model either as predefined instruction or by providing programming mechanisms of procedures that operate on tables.

### 12.7.6.11   Database instructions

I assume that in **Lingua-SQL** an initial valuation of program execution may carry some variables assigned to database values. This is, of course, a simplification of our object-model (Sec. 10) whose full exploitation is left to the reader.

I assume additionally that in every initial state of program execution, the system identifiers are bound to the following default values:

    tye.sb-graph = ∅

    vat.copies    = ∅,

    vat.monitor   = Ω                            (interpreted as no data to be displayed)

    vat.check    = 'yes'

With these assumptions each database program in **Lingua-SQL** that is supposed to operate on tables either has to create its own tables — and a database thereof — or to import an already existing database. In **Lingua-SQL** we have therefore only two database instructions that operate on tables and besides two instruction that modify a subordination graph. Their constructions are defined below in a simplified form to avoid too many technical details.

**Database activation**

activate : Identifier ⟼ InsDen

activate.ide.sta =

 is-error.sta    ➔ sta

 **let**

  ((tye, pre), (vat, 'OK')) = sta

 tye.sb-graph = !  ➔ sta ◄ 'active-base-already-exists'

 vat.ide = ?   ➔ sta ◄ 'unknown-variable'

 **not** vat.ide : DbaVal ➔ sta ◄ 'database-expected'

 **let**

  (dbr, sgr) = vat.ide

 **true**     ➔ ((tye[sb-graph/sgr], pre), (vat ♦ dbr, 'OK'))

This instruction overwrites the current valuation by a database record which means that it stores in it table identifiers assigned to table values and to the system variable sb-graph assigns the subordination graph of the activated base. Of course, it also checks whatever has to be checked. It does not allow to create two databases at the same time (an engineering decision). I recall that DbaVal is the domain of database values defined in Sec. 12.6.

The remaining database instruction writes all current table values in the database of the given name and removes from valuation all values except database values.

archive : Identifier ⟼ InsDen

archive.ide.sta =

 is-error.sta   ➔ sta

 **let**

  ((tye, pre), (vat, 'OK')) = sta

 tye.sb-graph = ? ➔ sta ◄ 'no-base-to-be-archived'

 vat.ide = !   ➔ sta ◄ 'variable-declared'

  dbr  = tables-only.vat

  new-vat = remove-non-database.vat

  dbv  = (dbr, tye.sb-graph)

 **true**    ➔ ((tye, pre), (new-vat[ide/dbv], 'OK'))

In this definition, I use two auxiliary functions tables-only and remove-non-database whose obvious definitions are omitted. I also assume that the instruction does not allow to overwrite an existing database by a new database. This is, of course, an engineering decision.

In this definition, one might include a principle that tables which are considered as "temporary" are not subject to archivation. To do that we could assume that, e.g. their identifiers are somehow labelled.

Notice that database archivation that assigns a database to an identifier does not require that this identifier be has been declared.

Constructors that generate instructions which modify subordination graphs correspond to adding and to removing an edge of a graph.

declare-subordination : Identifier x Identifier x Identifier ⟼ InsDen

declare-subordination.(ide-c, ide, ide-p).sta =

    is-error.sta             ➔ sta

    **let**

        ((tye, pre), env, (vat, 'OK')) = sta

    vat.ide-i = ?          ➔ 'no-such-table'                for i = c, p

    **let**

        (com-i, tra-i) = vat.ide-I             for i = c, p

    sort.com-i ≠ 'Tq'      ➔ 'table-expected'

    **let**

        ((tab-i, ('Tq', row-d-i, ('Rq', ror-i)), yok-i) = vat.ide-i        for i = c, p

    ror-i.ide = ?          ➔ 'no-such-column'         for i = c, p

    **let**

        sgr = tye.sb-graph

    (ide-c, ide, ide-p) : sgr  ➔ 'redundant-declaration'

    com-c Sub[ide] com-n  ➔ sgr | {(ide-c, ide, ide-p)}

    **true**                   ➔ 'subordination-not-satisfied'

Before adding a new edge to a subordination graph this instruction checks if the subordination really holds. If the concerned tables are large, then this check may be computationally expensive. This, however, cannot be avoided if we want to protect database integrity.

The second operation does not require such a check since it only removes an edge from a subordination graph.

call-off-subordination : Identifier x Identifier x Identifier ⟼ InsDen

call-off-subordination.(ide-c, ide, ide-p).sta =

    is-error.sta         ➔ sta

    **let**

        ((tye, pre), (vat, 'OK'))   = sta

```
    sgr                        = vat.sb-graph
    (ide-c, ide, ide-p) : sgr  ➔ sgr – {(ide-c, ide, ide-p)}
    true                       ➔ 'no-such-subordination'
```

## 12.8   Concrete syntax

For a reader who reached this section designing a concrete syntax of **Lingua-SQL** should be relatively easy. Therefore I restrict further investigations to grammatical clauses related to SQL. The syntax which is described below is probably not very optimal since it contains rather long key-words. My goal is, however, not to build a „practical" language but only to show a method of building such a language. For the same reason, my concrete syntax is not very close to the SQL standard. Long key-words correspond directly to the names of constructors which should help the reader to understand their meaning.

It is worth noticing that compared to **Lingua-3** we now have a new syntactic category of transactions. The key-words **ed, et** and **ei** are read respectively as „end of declaration", „end of transaction" and „end of instruction".

**Data expressions**

dae : DatExp =

   …                                          (here stand are all clauses of **Lingua-3**)

**Expressions generating empty composites**

   **empty-bool**                                           |

   **empty-number**                                         |

   **empty-word**                                           |

   ...

**Row expressions**

   **row**  Identifier **val** DatExp **ee**                       |

   **expand-row** DatExp **at** Identifier **by** DatExp **ee**         |

   **reduce-row** DatExp **at**  Identifier **ee**                  |

   **row** DatExp **at** Identifier **ee**                          |

   **change-row** DatExp **at** Identifier **by** DatExp **ee**         |

**Row table expressions**

   **table**  DatExp **at**  Identifier **ee**                      |

   **add-row**  DatExp **to**  Identifier **ee**                    |

   **delete-row**  TraExp **from**  Identifier **ee**               |

   **remove**  Identifier **from** Identifier **ee**                |

   **clear** Identifier **with** TraExp **ee**                      |

```
        intersect Identifier with Identifier ee           |
        union Identifier with Identifier ee               |
```

**Column table expressions**

```
        add-column Identifier with DatExp to Identifier ee     |
        remove-column Identifier from Identifier ee            |
        filter-columns ActPar from Identifier ee
        remove-column Identifier from Identifier ee            |
        update-column Identifier in Identifier with  TraExp
                                        where TraExp ee   |
```

**Expression creating derivative table**

```
        table Identifier with Identifier at   Identifier
                                        where TraExp ee    |
```

**Transfer expressions**

```
    wtr : TraExp =
        …                               (here stand all clauses of Lingua-3)
        row . Identifier   |
        unique             |
        all TraExp ee
```

**Type expressions**

```
    wyt : TypExp =
        …                               (here stand all clauses of Lingua-3)
        row-type  Identifier as TypExp ee              |
        expand-row-type TypExp by Identifier as TypExp ee   |
        table-type DatExp as TraExp ee
```

**Type constant definitions**

There are no new clauses in this group. Of course, the "former clauses" refer to new type expressions.

**Data-variable declarations**

```
    vde : VarDec =
```

…                                         (here stand both clauses of **Lingua-3**)

    `create table` Identifier `as` TypExp `ed`

## Transactions

trn : Transaction =

    `add` DatExp `to` Identifier `et`                     |

    `delete` DatExp `from` Identifier `et`             |

    `exclude` Identifier `from` Identifier `et`        |

    `add column` Identifier `with` DatExp `to` Identifier `et`  |

    `drop column` Identifier `from` Identifier `et`       |

    `select columns` ActPar `from` Identifier `et`       |

    `update` Identifier `at` Identifier `with` TraExp `et`   |

    `savepoint` Identifier `et`                          |

    `release savepoint` Identifier `et`             |

    `rollback` Identifier `et`                        |

    `rollback` Identifier `if` DatExp `et`            |

    `constraints off`                            |

    `constraints on`                             |

    Transaction ; Transaction

## Instructions

ins : Instruction =

    …                                     (here stand all clauses of **Lingua-3**)

### Table instructions

    `delete cascade` TraExp `from` Identifier `ei`          |

    `add row` DatExp `to` Identifier `ei`                  |

    `union` Identifier `with` Identifier `into` Identifier `ei`     |

    `intersect` Identifier `with` Identifier `into` Identifier `ei`   |

    `create` Identifier `from` Identifier `and` Identifier `col` Identifier

                                     `where` TraExp `ei` |

    `modify column` Identifier `in` Identifier `by` TraExp

                                     `where` TraExp `ei` |

### Database instructions

    `activate` Identifier                               |

```
archive as Identifier                                    |
set reference of Identifier et Identifier to Identifier ei    |
clear reference of Identifier et Identifier to Identifier ei    |
```

**Queries**

Queries are assignments — hence instructions — which a created table assign to the system identifier monitor and do not check anything since there is no type assigned to that monitor. Consequently, their denotations are slightly different from corresponding instructions which means that their syntaxes must differ accordingly. I assume that they are created from corresponding instruction by adding a prefix **show**

```
que : Query =
    show Identifier                                           |
    show union Identifier with Identifier into Identifier ei      |
    show intersect Identifier with Identifier into Identifier ei |
    show create Identifier from Identifier and Identifier
                          col Identifier where TraExp ei
```

At the end one methodological remark. In **Lingua-SQL** we have all constructors of data expression denotations of **Lingua-3**. In particular we have all table expressions. We also have assignments where such expressions may appear. All these tools are rather far from SQL standard and may lead — with complex expressions — to hardly readable programs and difficult to formulate proof rules.

An alternative solution may consist in allowing only **Lingua-3** expressions and row expressions, in disposing of table expressions, and in using table instruction for a step-by-step construction of tables. This does not mean, however, that at the model level we cannot introduce constructors of table-expression denotations. However, when designing the syntax, we may take an engineering decision that some of these constructors are not included in the signature of the algebra of denotations but are treated as auxiliary functions used only at the level of the model. In such a case their syntactic counterparts will not appear in syntax.

## 12.9   Colloquial syntax

The majority of new syntactic constructions of **Lingua-SQL** does not seem to require the introduction of colloquialisms. They may be made more user-friendly at the level of concrete syntax. However, the introduction of colloquialisms may be worthwhile in the case of table-variable declarations to make them closer to a typical SQL-syntax. Let us consider an example of such a declaration written in an SQL style (cf. Sec.11.3):

```
create table Employees with
  Name              Varchar(20)    NOT NULL,
```

```
Position          Varchar(9),
Salary            Number(5)      DEFAULT 0,
Bonus             Number(4)      DEFAULT 0,
Department_Id     Number(3)      REFERENCES Departments,
CHECK (Bonus < Salary)
```
**ed**

The restoring transformation would change this declaration into a sequential composition of a table-variable declaration and a database instruction:

**create table** Employees **as**

  **table-type** dat_exp **with** yok_exp ee

**ed ;**

**set reference of** Employees **et** Department_Id **to** Departments **ei**

where dat_exp and tra_exp represent a type expression and a yoke expression respectively.

    Restoring the data expression by means of row-creation and row-expansion constructors and the transfer expression with transfer-expression constructors we get the following concrete version of our colloquial declaration:

| | |
|---|---|
| **create table** Employees **as** | *the beginning of the declaration* |
|   **table-type** | *the beginning of type expression* |
|     **expand-row** | *the beginning of data expression* |
|       **expand-row** | |
|         **expand-row** | |
|           **expand-row** | |
|             **row** Name **val**  **empty-word ee** | |
|           **by** Position **val empty-word ee** | |
|         **by** Salary **val** 0 **ee** | |
|       **by** Bonus **val** 0 **ee** | |
|     **by** Department_Id **by empty-number ee** | *the end of data expression* |
|   **with** | *the beginning of transfer expression (yoke expression)* |
|   **all** | |
|     varchar(20)(**row.**Name) | **and** |
|     not-null(**row.**Name) | **and** |
|     varchar(9)(**row.**Position) | **and** |
|     number(5)(**row.**Salary) | **and** |
|     number(4)(**row.**Bonus) | **and** |
|     number(3)(**row.**Department_Id) | **and** |
|     **row.**Bonus < **row.**Salary | |
| **ee** | *the end of transfer expression (yoke expression)* |
| **ee** | *the end of type expression* |
| **ed ;** | *the end of declaration* |

**set reference of** Employees **et** Department_Id **to** Departments **ei**

Of course varchar(20), varchar(9),… are the names of appropriate predicates. Notice that in this example one "syntax unite" from the colloquial lever is transformed into a sequential composition of a declaration and an instruction.

## 12.10  The rules of correct-program constructions

The enrichment of the former versions of **Lingua** to **Lingua-SQL** consists basically on the extension of data- and type-algebras whereas new instructions are table modifications that on the denotational level refer to the generalised assignment. For the author of validation rules, this means the necessity of defining new conditions and new properties (Sec. 8.2 and Sec. 8.4.1). This should be postponed, however, until some practical version of **Lingua-SQL** is created.

# 13  What remains to be done

Even though the book is already of a considerable volume, the majority of subjects has been only sketched. What remains to be done is enough for a few more books and also as a research and development area for many researchers and developers. Below a preliminary list of subjects which is certainly not complete. It covers both, the research problems as well as programming (implementational) tasks.

## 13.1  Foundations

### 13.1.1  The extension of Lingua model

All currently described languages from the **Lingua** family — maybe except **Lingua-3** (object programming) — cover mainly traditional programming tools developed in the years 1960-1980. Since they are present today in the majority of programming languages, it was rather natural to start with them, which does not mean, however, that the model of **Lingua** should not be developed further. In my opinion, the next step should be the extension of our model by newer mechanisms, e.g., by script languages of HTML type or concurrency based on Mazurkiewicz and/or Petri model.

A few minor research problems have been mentions in the main part of the book.

### 13.1.2  The completion of Lingua model

The development of a complete (a practical) model for **Lingua** covering not only denotations, syntax, and semantics but also sound program-construction rules. In the last area, a closer look to assertions (Sec. 8.3) may be worthwhile since so far this issue has been only sketched.

### 13.1.3  The principles of writing user manuals

Denotational models should provide an opportunity for the revision of current practices seen in the manuals of programming languages. New practices should on one hand base on denotational models but on the other — do not assume that todays' readers are experts in this field. A manual should therefore provide some basic knowledge and notation needed to understand the definition of a programming language written in a new style. At the same time — I strongly believe on that — it should be written for professional programmers rather than for amateurs. The role of a manual is not to teach the skills of programming. Such textbooks are, of course, necessary, but they should tell the readers what the programming is about rather than the technicalities of a concrete language.

## 13.2   Implementation

In this field I would suggest that only **Lingua-1** (appropriately completed) be implemented in some of the existing languages — my choice would by Phyton — and the remaining layers of **Lingua** as well as a programming environment, be developed in using the earlier developed layers of **Lingua**.

### 13.2.1  Tools for language developers

1.  A system generating abstract-syntax grammar from a signature (a meta-definition) of the algebra of denotations.

2.  A system supporting the development of a concrete-syntax grammar form an abstract-syntax grammar.

3.  A system supporting the generation of a restoring application from colloquial syntax into a concrete syntax.

4.  An editor supporting the writing of the definitions of denotation constructors.

5.  A generator of semantic clauses from a concrete-syntax grammar and the definitions of denotation constructors.

6.  A generator of an interpreter/compiler code from semantic clauses.

## 13.3   Tools for programmers

A system supporting program-development using correct-metaprogram development rules must be developed.

## 13.4   Manuals

To provide a practical value for the methodology which is contained in **Lingua,** there must be user manuals that follow that methodology. And, of course, they have to base on principles mentioned in Sec. 13.1.3. As a matter of fact, both these tasks should be developed in parallel. To describe rules for writing manuals, some experiments in writing manuals should take place, and experimental manuals must follow the developed general rules.

## 13.5   Programming experiments

For our idea of correct-program development to be noticed by the IT community, some convincing applications must be shown. In my opinion, an adequate field for such applications may be microprograms because:

1.  microprograms contain a relatively small number of the lines of code,

2.  their correctness is highly critical,

3.  highly critical is also the memory- and time-optimisation of such programs.

Each experimental program developed within our framework must be independently tested by usual industrial tests.

## 13.6   Building a community of Lingua supporters

Our methods of designing programming languages and constructing program may be assessed positively or negatively, but one seems to be evident — they are certainly quite far from current practices. What the book offers is a far-going change, and such changes always provoke springing up groups of opponents and supporters. The former should be convinced, and the latter must be won. And of course one has to start from the first task.

To realise that task one has to give the potential supporters some, may be very simple, but sufficiently practical, version of **Lingua** or — as an alternative — encourage them to build their own version. The first solution seems rather unrealistic since it would require finding an investor for a strange and completely unknown product. The other way that remains means that an experimental **Lingua** is built by volunteers and for volunteers as in the case of Linux, Joomla! or Drupal. However, such a product although freely available should not by an open-source product since this might lead to mathematically incorrect solutions and consequently to unsound program-construction rules.

The community of **Lingua** builders must, therefore, elaborate rules of accepting new members and of giving them rights for joining implementation teems.

# 14 ANNEXE 1 — Generalized trees

To be translated from the Polish version of the book.

# 15 ANNEXE 2 — About user manuals

To be translated from the Polish version of the book.

# 17 Indices and glossaries

## 17.1   The index of terms and authors

## 17.2   The index of notations

| | | |
|---|---|---|
| ε    : empty word | Ø    : empty set/relation | ♦    : overwriting a function |
| ⊂  : to be a subset | ⊑    : partial order | @ : algorithmic formula |
| →  : partial functions | Θ   : empty element | ■   : end of theorem/proof |
| ↦ : total functions | | |
| ⟹ : mattings | {a.i \| i=1;n} : a set | |
| •    : composition of relations | (a.i \| i=1;n) : a sequence | |
| © : concatenation | [a.i/b.i \| i=1;n] : a mapping | |
| ∃    : there exists | Rel.(A,B) : set of relations | |
| ∀   : for all | [A] : subset of identity rel. | |

## 17.3   The glossary of algebras and domains

This glossary serves mainly the authors of the book for keeping the consistency of notations.

### The algebra of data, Sec. 5.2.1

DatAlg  — the algebra of data

boo  : Boolean

num  : Number

wor  : Word

lis    : List

arr   : Array

rec   : Record

dat   : Data

dat   : SimpleData

ide   : Identifier

## The algebra of bodies, Sec. 5.2.2

BodAlg — the algebra of bodies

bod  : Body

bod  : BodyE

bor   : BodRec

err   : Error

CLAN-Bo : BodyE ↦ Sub.Data

BOD : Data → Body

sort : BodyE ↦ {('Boolean'), ('number'), ('word'), 'L', 'A', 'R'}

Bc : data-algebra-operations ↦ body-algebra-operations

## The algebra of composites, Sec. 5.2.3

ComAlg     — the algebra of composites

com : Composite

com : BooComposite

com : CompositeE

com : BooCompositeE

oversized : Composite ↦ Boolean

round : Data ↦ Data

Cc : data-algebra-operations ↦ composite-algebra-operations

## The algebra of transfers, Sec. 5.2.4

TraAlg  — transfer algebra

tra : Transfer

yok : Yoke

CLAN-Tr : Transfer ↦ Sub.Composite

TT = Tc[create-bo.tt]

FF = Tc[create-bo.ff]

Tc : data-algebra-operations ↦ transfer-algebra-operations

## The algebra of types, Sec. 5.2.5

TypAlg  — the algebra of types

typ : Type

typ : TypeE

CLAN-Ty : Type ↦ Sub.Composite

Yc[…]

## Values and memory states, Sec. 5.3.1

tda : TypDat

val  : Value

sta : State

env : Env

sto  : Store

vat  : Valuation

tye  : TypeEnv

pre : ProEnv

## The denotations of data expressions, Sec. 5.3.2

ded : DatExpDen

and-ded : DatExpDen x DatExpDen ↦ DatExpDen

Cdd[…] : composite-algebra-constructors ↦ denotation-algebra-constructors

## The denotations of type- and transfer expressions, Sec. 5.3.4

ted : TypExpDen

Cdt[…] : type-algebra-constructors ↦  denotation-algebra-constructors

# The algebra of denotations of data-, type- and transfer expressions, Sec. 5.3.5

AlgExpDen

tra : TraExpDen

# The abstract syntax of Lingua-A, Sec. 5.4.1

dae : DatExpA

tre : TraExpA

tex : TypExpA

# Concrete syntax of Lingua-A, Sec. 5.4.2

dae : DatExp

tre : TraExp

tex : TypExp

# A sketch of the semantics of Lingua-A, Sec. 5.7

Cs : ExpAlg $\longmapsto$ ExpDenAlg

with five components:

Sid   : Identifier $\longmapsto$ Identifier

Sde  : DatExp   $\longmapsto$ DatExpDen

Stre : TraExp   $\longmapsto$ TraDenExp

Ste  : TypExp   $\longmapsto$ TypExpDen

# Denotational domains, Sec. 6.1.1

ide   : Identifier

ded  : DatExpDen

tra   : TraExpDen

ted   : TypExpDen

vdd  : VarDecDen

tdd  : TypDefDen

ind   : InsDen

pde  : PreDen

prd   : ProDen

# Abstract syntax, Sec. 6.2.1

vde : VarDecA

tde : TypDefA

ins  : InstructionA

pam : PreambleA

prg  : ProgramA

## Concrete syntax of Lingua-1, Sec. 6.2.2

vde : VarDec

tde : TypDef

ins  : Instruction

pam : Preamble

prg  : Program

## Semantics, Sec. 6.3

Svd   : VarDec      $\longmapsto$ VarDecDen

Std   : TypDef      $\longmapsto$ TypDefDen

Sin   : Instruction $\longmapsto$ InsDen

Spre : Preamble    $\longmapsto$ PreDen

Spr   : Program     $\longmapsto$ ProDen

## Denotational domains for procedures, Sec. 7.1.4

fpa : ForPar   = (Indentifier x TypExpDen)$^{c*}$ (formal param. of declarations of both types)

apa : ActPar  = Identifier$^{c*}$                 (actual param. of calls of both types)

ipr  : ImpPro      = ActPar x ActPar $\longmapsto$ Store $\rightarrow$ Store      (imperative procedures)

fpr  : FunPro      = ActPar x ActPar $\longmapsto$ Store $\rightarrow$ CompositeE    (functional procedures)

pro : Procedure = ImpPro | FunPro                              (procedures)

idd   : IprDecDen  = State $\longmapsto$ State     (denotations of imp. procedure-declarations)

fdd   : FprDecDen = State $\longmapsto$ State      (denotations of fun. procedure-declarations)

## The correctness of parameter-lists, Sec. 7.2.2

statically-compatible : ForPar x ForPar x ActPar x ActPar $\longmapsto$ Error | {'OK'}

dynamically-compatible : ForPar x ForPar x ActPar x ActPar $\longmapsto$

$\qquad\qquad\qquad\qquad\qquad\qquad\qquad$ TypEnv x Valuation $\longmapsto$  Error | {'OK'}

## Passing actual parameters to a procedure, Sec. 7.2.3

pass-actual : ForPar x ForPar x ActPar x ActPar ↦

TypEnv x Valuation ↦  Valuation | Error

## Returning reference-parameters to a program, Sec. 7.2.4

return-referential : ForParRef x AktParRef ↦ TypEnv x Valuation x Valuation

↦ Valuation | Error

## The constructor of a procedure, Sec. 7.3.1

ipc : IprComponents = Identifier x ForPar x ForPar x ProDen

create-imp-proc : ((Identifier x ForPar x ForPar x ProDen) x Env) ↦

ActPar x ActPar ↦ Store → Store

## The instruction of a procedure call, Sec. 7.3.3

call-imp-proc : Identifier x ActPar x ActPar ↦ InsDen

## Procedure declaration, Sec. 7.3.4

declare-imp-pro : IprComponents ↦ IprDecDen

## Mutual recursion, Sec. 7.4.1

cmp : MprComponents  = IprComponents$^{c+}$        (components of multiprocedures)
mpr : MulPro              = ImpPro$^{c+}$                         (multiprocedures)
mpd : MulProDecDen    = State ↦ State      (multiprocedure-declaration denotations)

## Multiprocedure constructor, Sec. 7.4.2

create-multi-pro : MprComponents x Env ↦ MulPro

## Multiprocedure declaration, Sec. 7.4.4

declare-imp-mpr : MprComponents ↦ MulProDecDen

## The domains of functional procedures, Sec. 7.5.2

fdd : FprDecDen    = State ↦ State   (denotations of functional procedure-declarations)
fpr : FunPro        = ActPar ↦ Store → CompositeE          (functional procedures)

## The expressions of functional-procedures-calls, Sec. 7.5.3

call-fun-pro : Identifier x ActPar $\longmapsto$ DatExpDen

## The expressions of functional-procedure calls, Sec. 7.5.4

call-fun-pro : Identifier x ActPar $\longmapsto$ DatExpDen

or:

call-fun-pro : Identifier x ActPar $\longmapsto$ State $\rightarrow$ CompositeE

or:

call-fun-pro.(ide, apa) : State $\rightarrow$ CompositeE

## The declaration of a functional procedure, Sec. 7.5.5

ff-declare-fun-pro : FFcomponents $\longmapsto$ State $\longmapsto$ State

## Object-oriented programming, Sec. 9.1 to Sec. 9.5

obj : Object = State $\longmapsto$ State

lib  : ObjLib = Identifier $\Longrightarrow$ Object

oed : ObjExpDen = ObjLib $\longmapsto$ Object | Error

odd : ObjDecDen = ObjLib $\longmapsto$ ObjLib | Error

ocd : ObjCalDen = ObjLib $\longmapsto$ Object

ppd : PreProDen = ObjLib x State $\rightarrow$ State

## The syntax of Lingua-3, Sec. 9.6

obe : ObjExp

ode : ObjDec

pob : ObjCall

prp : PrePro

## External objects — a sketch of an idea, Sec. 10

No specific notation in this section

## Relational databases intuitively, Sec. 11

No specific notation in this section

## Lingua-SQL, Sec. 12

row : Row     = Identifier $\Rightarrow$ SimData

tab : Table    = Row$^{c*}$

sbo : SimBody = {('Boolean'), ("number"), ('word'), ('date'), ('time'), (date-time')}

bod : RowBody = {'Rq'} x RowRec

ror  : RowRec  = Identifier $\Rightarrow$ SimBody

bod : TabBody  = {'Tq'} x Row x RowBody

com : SimCom =

   {(dat, bod) | (dat, bod) : CompositeE **and** bod : SimBody}

Θ : CLAN-Bo.bod

A sub[ide] B — the subordination of tables

col : ColumnE = SimCom$^{c+}$ | Error

RowVal = {(com, tra) | sort.com = 'Rq' **and** tra.com = (tt, ('Boolean'))}

TabVal = {(com, tra) | sort.com = 'Tq' **and** tra.com = (tt, ('Boolean'))}

dbr : DatBasRec = Identifier $\Rightarrow$ TabVal

sb-graph    — that binds subordination graphs in type environments,

copies      — that binds finite sets of tables in valuations,

monitor     — that binds tables in valuations,

check       — that binds words 'yes' and 'no' in valuations.